\documentclass[aps,rmp,reprint,amsmath,amssymb,graphicx,longbibliography]{revtex4-1}

\usepackage[normalem]{ulem}
\usepackage{bm,amsmath,amssymb,braket,longtable,docmute,xcolor,graphicx,comment,grffile}
\usepackage[caption=false]{subfig}
\usepackage{bbm}
\usepackage[utf8x]{inputenc}
\newcommand{\Tr}{\ensuremath{\text{Tr}}}
\renewcommand{\Re}{\ensuremath{\text{Re}}}
\renewcommand{\Im}{\ensuremath{\text{Im}}}

\newcommand{\cN}{\mathcal{N}}

\usepackage{mathtools}
\usepackage{bbold}

\usepackage{appendix}
\usepackage{xcolor}
\usepackage{float}
\usepackage{braket}
\usepackage{slashed}
\usepackage{graphicx}
\definecolor{past}{rgb}{0,0.9,0.9}
\definecolor{darkred}{rgb}{0.4,0.0,0.0}
\definecolor{darkgreen}{rgb}{0.0,0.3,0.0}
\definecolor{darkblue}{rgb}{0.0,0.0,0.7}
\definecolor{pink}{rgb}{0.99,0.08, 0.57}
\usepackage[bookmarks,linktocpage,colorlinks,
linkcolor=darkred,
urlcolor  = darkgreen,
citecolor = darkblue]{hyperref}
\def\beq{\begin{equation}}
  \def\enq{\end{equation}}
\usepackage{lineno}
\definecolor{winered}{rgb}{0.8,0,0}
\definecolor{darkb}{rgb}{0,0,0.8}

\definecolor{green}{HTML}{4CAF50}
\definecolor{bred}{rgb}{1.0,0.0,0.0}
\definecolor{navy}{HTML}{3F51B5}

\usepackage[english]{babel}

\def\csi{\{\sigma\}}
\def\csip{\{\sigma'\}}
\def\mH{\mathcal{H}}

\def\dmu{\partial_\mu}
\def\rar{\rightarrow}

\def\cre{\hat{U}}

\def\zq{${\mathbb Z}_q\ $}
\def\tft{TLFT$\ $}
\def\ds{D_\text{cut}}

\def\lad{\mathcal{L}}
\def\tm{\mathbb T}
\def\block{{\mathfrak B}}
\def\vone{{\mathbb V}_1 }
\def\vtwo{{\mathbb V}_2 }
\begin{document}

\title{Tensor lattice field theory with applications to the
renormalization group and quantum computing}

\author{Yannick Meurice$^{1}$}
\author{Ryo Sakai$^{1}$}
\author{Judah Unmuth-Yockey$^{2,3}$}

\affiliation{$^1$ Department of Physics and Astronomy, The University of Iowa, Iowa City, IA 52242 USA }
\affiliation{$^2$ Department of Physics, Syracuse University, Syracuse, NY 13244 USA}
\affiliation{$^3$ Fermi National Accelerator Laboratory, Batavia, IL 60510 USA}
\def\lt{\lambda ^t}
\def\note{note}
\def\beq{\begin{equation}}
  \def\enq{\end{equation}}

\date{\today}

\begin{abstract}
  We discuss the successes and limitations of statistical sampling for a sequence of models studied in the context of lattice QCD and emphasize the need for new methods
  to deal with finite-density and real-time evolution.
  We show that these lattice models can be reformulated using tensorial methods where the field integrations in the path-integral formalism are replaced by discrete sums.
  These formulations involve various types of duality and
  provide exact coarse-graining formulas which can be combined with truncations to obtain  practical implementations of the Wilson renormalization group program.
  Tensor reformulations are naturally discrete and provide manageable transfer matrices. Combining truncations with the time continuum limit, we derive Hamiltonians suitable to perform quantum simulation experiments, for instance using cold atoms, or to be programmed on existing quantum computers. We review recent progress concerning
  the tensor field theory treatment of non-compact scalar models, supersymmetric models, economical four-dimensional algorithms, noise-robust enforcement of Gauss's law, symmetry preserving truncations and topological considerations. We discuss connections with other tensor network approaches.

\end{abstract}

\maketitle
\tableofcontents{}
\section{Introduction}

Quantum field theory models on space or space-time lattices play an important role in our understanding of strongly interacting particles, nuclei, superconductivity, condensed matter and phase transitions. In high-energy and nuclear physics, lattice quantum chromodynamics (QCD) provides an ab-initio theory of strong interactions. In the QCD context, the lattice is a non-perturbative ultraviolet regularization which preserves local gauge invariance. As the lattice is not a physical feature, we need to approach the continuum limit where the lattice spacing is small compared to the physical length scales involved in the problem.

{In the context of solid-state physics, lattice spacings of the order a few Angstroms are present, and atomic  physicists  can  create  optical  lattices  with  a  lattice  spacing  on  the order of the laser wavelength; however,
large correlation lengths appear near critical points and universal behavior independent of the microscopic details can be observed. It is also possible to create
actual optical lattices in laboratories by trapping cold atoms in counter-propagating laser beams, and tune the interaction in order to quantum simulate lattice models with interactions similar to  Hubbard models \cite{Bloch2008}.
This can be called an analog computing method or a quantum simulation experiment.
Again, it is possible to tune the parameters to reach universal behaviors related to quantum phase transitions with large correlation lengths.

More generally, we are getting better control in the manipulation of small quantum systems evolving in small Hilbert spaces and the idea of using physical quantum systems to study theoretical quantum models \cite{feynman82} has generated many exciting developments \cite{RevModPhys.86.153}.
As bits---that can be either on or off---can be thought of as the basic building blocks of classical computers, one can envision qubits that can each be used as a two-dimensional Hilbert space as the building blocks of a quantum computer.
If we want to use the $2^N$ dimensional Hilbert space provided
by $N$ qubits to represent the Hilbert space of a quantum field theory problem we need to apply discretizations and truncations. Discretization of space can be achieved by the lattice approximation, while the discretization of continuous field integration can be done by using character expansion, as will be discussed extensively in this review, or by other methods which include gauge magnets/quantum links \cite{orland90,horn,Brower97,wiese2013}, or
field digitization, or summation of discrete subgroups~\cite{jordan2011,klco2018b,lamm2019b,alexandru2019,hackett2018}.

General arguments \cite{lloyd96} show that for local interactions, a quantum computer will reduce the computational effort, for problems like the real-time evolution, to a polynomial in the size of the system rather than an exponential for a  classical computer.

From a purely theoretical point of view, studying models with a large number of strongly correlated degrees of freedom is very challenging. In order to deal with this situation, L. Kadanoff \cite{leo66}  suggested to consider the average field or spin in cells of variable sizes often called ``blocks". The procedure is often called ``blockspinning" and it played a crucial role in the development of the renormalization group (RG) ideas \cite{wilson73pr}.  Sometimes great theoretical intuitions can take a long time to be practically realized. Despite its visual appeal, the blockspinning procedure is not easy to implement numerically. It typically involves approximations that are difficult to improve.

Successful applications of the RG idea were made possible without requiring numerical implementations of the original blocking idea. A well-known example is
the discovery of asymptotic freedom \cite{PhysRevLett.30.1343,PhysRevLett.30.1346} which initially relied on a one-loop calculation of the
Callan-Symanzik beta function. Typically, the interplay between small and large energy scales is more easily seen in the momentum representation.
However, it is clear that $if$ it was possible to design practical methods such that 
each step of the blocking could be performed within a reasonable amount of time and with a desired  accuracy, the computational cost would scale like the number of blockings, in other words, the logarithm of the volume, since the size of a block after each blocking step doubles~\cite{leo66}.
Achieving this goal is nontrivial and  not guaranteed in general.
Below, we briefly explain the practical issues that have prevented the use of blocking for quantitative purposes and how new tensorial methods can be used to make progress in this direction.

A simple way of blocking consists in introducing 1 in the partition function in the following generic form:
\beq
\label{eq:k1}
\prod_\block\int d\Phi_\block\delta(\Phi_\block -\sum_{x\in \block} \phi_x)=1,
\enq
where the blocks $\block$ form a partition of the original lattice. For instance, on a three-dimensional cubic lattice, the blocks can be chosen as cubes with a linear size of two lattice spacings
and contain eight sites.
The $\phi_x$ are the original lattice fields and $\Phi_\block$ the block fields which inherit
new effective interactions after one performs the integration over the original fields.
Conceptually, this sounds easy, however in practice it appears to be more complicated than the original problem. As an example, one can try  to write a simple algorithm for the two-dimensional Ising model on a square lattice
by replacing four spins in a $2 \times 2$  square block by a single variable and write an effective energy function (or at least some effective measure) for the new block variables.
The procedure becomes more intricate as we proceed, and finding the effective energy function is nontrivial \cite{Liu:2013nsa}.

For this reason, approximate procedures were developed such as the Migdal-Kadanoff approximation \cite{Migdal:1975zf,kadanoff76}, the approximate recursion formula \cite{wilson73pr}  or other hierarchical approximations \cite{dyson68,baker72,hmreview}, where
no new interactions are generated by the blocking process.
However, in these examples, the lack of reference to an exact procedure to handle the original model with localized interactions makes the systematic improvement of these approximations difficult. Similar issues appear in non-perturbative functional methods based on the momentum space representation \cite{berges2000}, where the local potential approximation allows high-accuracy estimates of the critical exponents \cite{bervillier2007}, but its improvement with methods such as the derivative expansion remains difficult \cite{bervillier2013}.
For Ising models, it is possible to deal with the proliferation of
couplings generated by the blocking process by starting with the most general set of interactions \cite{kadanoff75,kadanoff75b,nvl}.
They introduce the identity in terms of probabilities $P(\csip,\csi)$ such that
\beq
\sum_{\csip}P(\csip,\csi)=1,
\label{eq:ident}
\enq
where $\csip$ are new Ising spins associated with blocks.
As we will discuss later, this special setup allows us to write formal expressions for the effective couplings as double partition functions and write RG equations. However, from a computational point of view the locality of the interactions is lost and additional assumptions are needed to proceed.

In contrast, reformulations of the partition function of classical spin models as the trace of a product of local tensors provide a new type of blocking procedure in configuration space called Tensor RG (TRG) \cite{nishinoctm,Levin:2006jai,Gu:2009dr,PhysRevLett.103.160601,Gu:2010yh,2012PhRvB..86d5139X}. 

TRG procedures can be obtained by applying truncations to exact blocking formulas. The blocking neatly separates the degrees of freedom inside the block (which are integrated over), from those kept to communicate with the neighboring blocks \cite{prb87}. However, the remaining degrees of freedom after the blocking are still microscopic and finding a truncation that captures the low-energy physics and the entanglement is a non-trivial task. 

At early stages of the TRG development, Singular Value Decomposition (SVD) methods were used extensively. This is reviewed in Ref.  \cite{efratirmp}. Some of the SVD procedures can be simplified by using character expansions \cite{Liu:2013nsa} when applied to most models studied in the context of lattice gauge theory \cite{Liu:2013nsa,pre89,prd89,Zou:2014rha,
  Shimizu:2014fsa,Takeda:2014vwa,
  Bazavov:2015kka,Shimizu:2017onf,Yoshimura:2017jpk,
  Nakamura:2018enp,Kuramashi:2018mmi,
  Kadoh:2018hqq,Unmuth-Yockey:2018xak,Kadoh:2018tis,Bazavov:2019qih,Kadoh:2019ube,
  Butt:2019uul}.
Tensorial methods are also used in the context of quantum gravity \cite{perez:2013,Dittrich_2016,Asaduzzaman:2019mtx}.

In general, tensorial methods represent a new approach to lattice field theory that we will call Tensor Lattice Field Theory (\tft).
\tft can be used for purposes more general than the blocking procedure. In particular \tft provides a very convenient, {\it discrete} framework to perform quantum computations or simulations.  There are---additionally---continuous tensor network, or tensor-like, methods such as continuous matrix product states~\cite{PhysRevB.88.085118,PhysRevB.100.195106}.

In this review article, we introduce \tft for lattice models studied in the context of lattice gauge theory and report progress made for blocking and quantum computing, two ``competing" methods that attempt to reduce the computing time logarithmically.
The models targeted are introduced in Sec.~\ref{sec:models}. We advocate a road map
starting with the Ising model and culminating with QCD, which we call the ``Kogut sequence" \cite{kogut79,kogut83}. This sequence is sometimes called a ``ladder" and has been followed successfully in situations where importance sampling methods such as the Metropolis algorithm are effective.
Lattice QCD has become a very reliable precision tool to study the static properties of hadrons. As we are writing this article, we are roughly in the middle of the sequence, \emph{i.e.} we have roughly half of the models in the sequence remaining to be studied thoroughly, with improvements and optimization on previous models still possible. However, we hope that recent progress on higher dimensional algorithms \cite{Adachi:2019paf,Kadoh:2019kqk} could be combined with the methods that we
describe to deal with gauge fields, fermions, and non-Abelian (non-commuting) symmetries in order to attempt calculations directly related to lattice QCD in the coming years.

In Sec.~\ref{sec:qc}, we discuss situations where importance sampling cannot be used and where quantum computations or simulations could provide alternate ways to perform computations. This includes unitary, real-time evolution and other situations where a sign problem is encountered. One important long-term goal with potential impact on the interpretation of high-energy collider data is doing ab-initio  real-time calculations relevant to fragmentation processes and parton distribution functions. In other words,
starting with
lattice QCD, we would like to be able to perform calculations that would ultimately replace the use of event generators such as Pythia \cite{pythia}.

The simplest starting point for the real-time evolution is 
the evolution operator $\exp(-i\hat{H}t/\hbar)$ acting on the Hilbert space of the quantum Hamiltonian $\hat{H}$.
We provide a first look at the transfer matrix which smoothly connects the ``classical" Lagrangian approach to the Hilbert space used in the Hamiltonian formalism. We discuss various types of dualities (geometrical and topological) that are often used together and taken for each other.

For the models in the Kogut sequence, the bosonic field variables and the symmetry groups are {\it compact}.
General mathematical theorems, namely the Pontryagin duality  \cite{pdual} and the Peter-Weyl theorem \cite{peter27}, guarantee that functions  over compact groups can be expanded in terms of {\it discrete} sums of representations. This is called the ``character expansion" and was exploited to calculate strong coupling expansions \cite{balian75} or introduce new variables on geometrically dual lattice elements \cite{savit80}.

The discreteness of the character expansion provides a natural starting point for building approximate reformulations of lattice models suitable for quantum computing or quantum simulation experiments.
The Ising model is an elementary example where the Hilbert space of the transfer matrix can be implemented with a set of qubits, the basic components of actual quantum computers which exist in a linear superposition of
two states, $\ket{0}$ and $\ket{1}$, rather than being just ``on'' or ``off'' like the bits of a classical computer. For models with continuous fields, character expansions allows us to perform the ``hard integrals" analytically without the need of approximate numerical discretizations which break the continuous symmetries.  Demonstrating the power of the character expansion is one of the main goals of this review. Examples of quantum computations and simulations are provided at the end of the section.
In Sec.~\ref{sec:qvc}, we clarify the use of ``classical" and ``quantum" in various contexts and make connections with other approaches \cite{uli2011,ran2020,hv2017,Cirac:2020obd}.

Sec.~\ref{sec:ising} introduces the tensor reformulation for the Ising model. SVD, truncation and the TRG method are discussed in Sec. \ref{sec:trunc}.
Spin models with an O(2) symmetry or with discrete subgroups are discussed in
Sec.~\ref{sec:abelian}. In Sec.~\ref{sec:nonabelian}, we derive expressions for local tensors in the simple case of a non-Abelian spin model with O(3) symmetry.  We also find tensor expressions for effective theories of gauge theories known as principal chiral models.

Models with local gauge symmetry are introduced in Sec.~\ref{sec:lgt}.  We first consider Abelian gauge theories and work up in complexity to tensor expressions for non-Abelian gauge theories as well.

In Sec.~\ref{sec:scalar}, tensor network expressions for the real and the complex $\phi^{4}$ theory are derived.
For models with non-compact fields such as the scalar $\phi^{4}$ theory, the Gaussian quadrature rule can be used to extract discrete degrees of freedom, just as the gauge degrees of freedom are discretized via the character expansions.
The accuracy of the tensor network approach is shown for the real-field case, and the ability to deal with a severe sign problem is shown for the complex-field case.

In Sec.~\ref{sec:fermions} we present tensor formulations for models with fermionic degrees of freedom.
In general fermions fit in well with the tensor (and discrete) approach thanks to the nilpotency of the Grassmann variables.
In the section, various models that contain fermions such as pure fermions, gauged fermions, and fermions combined with scalars are discussed.

In Sec.~\ref{sec:transfer} we re-discuss the transfer matrix using the tensor formalism and broaden the perspective. Recent \tft developments regarding symmetries, topological solutions and quantum gravity are discussed in Sec.~\ref{sec:additional}.

\section{Lattice field theory}
\label{sec:models}
\subsection{The ``Kogut sequence": from Ising to QCD}
\label{subsec:ladder}
In the early 70's, QCD appeared as a strong candidate for a theory of strong
interactions involving quarks and gluons. However the perturbative methods that provided
satisfactory ways to handle the electroweak interactions of leptons failed to explain confinement, mass gaps and chiral symmetry breaking.
A non-perturbative definition of QCD was needed.
In 1974, K. Wilson proposed
\cite{PhysRevD.10.2445} a lattice formulation of QCD where the $SU(3)$ local symmetry is
exact. As this four-dimensional model is fairly difficult to handle numerically, a certain number of research groups started considering simpler lattice models in lower dimensions and then  increased symmetry and dimensionality.
This led to a sequence of models, sometimes called the ``Kogut ladder"  which appears in J. Kogut's review articles \cite{kogut79,kogut83}, and later with small modifications in textbooks by A. Polyakov \cite{1987gauge} and
C. Itzykson with J. M. Drouffe \cite{itzykson1991statistical}.

The sequence is approximately the following:
\begin{enumerate}
\item
  $D=2$ Ising model
\item
  $D=3$ Ising model and its gauge dual
\item
  $D=2$ O(2) spin and Abelian Higgs models
\item
  $D=2$ fermions and the Schwinger model
\item
  $D=3$ and 4 $U(1)$ gauge theory
\item
  $D=3$ and 4 Non-Abelian gauge theories
\item
  $D=4$ lattice fermions
\item
  $D=4$ QCD
\end{enumerate}
This sequence should not be understood in a rigid way as if each step is necessary for the next step.  For instance steps 3. to 5. could be interchanged and the
problems involving fermions have very specific features that are not easily compared to those involving only bosonic fields.
The message that we want to convey is that there is an approximate road map that has proven to be very effective for the
``classical" approach of lattice field theory in order to deal with static problems using importance sampling (Monte Carlo) methods. We advocate to follow a similar path to develop the
quantum versions of these models and deal with real-time evolution and other problems not accessible with classical methods. The difference between quantum and classical is explained more precisely in Sec.~\ref{subsec:classicalvsquantum}. In addition, a similar path is being followed to develop numerical coarse-graining.

\subsection{Classical lattice models and path integral}
\label{subsec:lattac}

In this subsection we introduce lattice versions of classical field theory models.  At this point, we would like to point out that while we will provide definitions of the fields used, notations, and acronyms/initials, more details on basic quantum field theory and lattice field theory can be found in textbooks and review articles \emph{e.g.} \cite{peskin95,itzykson1991statistical,kogut79,montvay_munster_1994}.
We use a Euclidean time and treat space and time on the same footing. The metric is simply a Kronecker delta in $D$ dimensions. We then discretize space and time. We use a $D$-dimensional (hyper) cubic Euclidean space-time lattice. The {\it sites} are denoted $x=(x_1, x_2,\dots x_D)$, with $x_D=\tau$, the Euclidean time direction.
In lattice units, the space-time sites are labelled with integers.
In the following, the lattice units are implicit. The {\it links} between two nearest neighbor lattice sites $x$ and $x+\hat{\mu}$ are labelled by $(x,\mu )$ and the {\it plaquettes}, the smallest squares on a square or (hyper) cubic lattice,  delimited
by four sites $x$,  $x+\hat{\mu}$, $x+\hat{\mu}+\hat{\nu}$ and $x+\hat{\nu}$ are labelled by $(x,\mu \nu)$. By convention, we start with the lowest index
when introducing a conventional circulation at the boundary of the plaquette.
The total number of sites is denoted by $V$. Unless specified, periodic boundary conditions are assumed, and they preserve a discrete translational symmetry.
If we take the time continuum limit, we obtain a
quantum Hamiltonian formulation in
$D-1$ spatial dimensions.

In the continuum, the Lagrangian density for $N$ real scalar fields
with a $O(N)$ global symmetry reads
\beq
\lad^{O(N)}_{Euclidean}=
\frac{1}{2} \partial_\mu \vec{ \phi}\cdot \partial_\nu \vec{\phi}\delta^{\mu\nu}+\lambda(\vec{\phi}\cdot \vec{\phi}-v^2)^2 ,
\label{eq:euclclasson}
\enq
with $\vec{\phi}$ a $N$-dimensional vector.  Here $\lambda$ is a coupling constant, whose size determines fluctuations of the $|\vec{\phi}|$ field around some value, $v$. The potential has degenerate minima on a $N$-1 dimensional hyper-sphere $S_{N-1}$ and a local maximum at $\vec{\phi}=0$.
For $N=2$, the low energy part of the potential has a shape reminiscent of the bottom of a wine bottle. The degenerate minima form a circle at the very bottom. We can study the
small fluctuations about a given minimum on the circle. Note that the choice of a minimum breaks the $O(2)$ symmetry. There are ``soft" fluctuations along the circle that restore the symmetry and
``hard" fluctuations in the radial direction.

We can extend this analysis for arbitrary $N$.  We have one massive mode (fluctuations in the symmetry breaking direction) with
mass $2\sqrt{2\lambda} v$ and $N-1$ massless modes (``Nambu-Goldstone" (NG) modes) which is the number of broken generators~\cite{PhysRev.127.965,PhysRev.117.648,Goldstone1961,peskin95}.

We can write the Euclidean action for the NG modes on a $D$-dimensional lattice with isotropic lattice spacing $a$ as
\beq
S_{\text{NLSM}} = \frac{1}{2}\sum_{x} \sum_{\mu=1}^{D} a^{D-2}(\vec{\phi}_{x+\hat{\mu}}-\vec{\phi}_{x})\cdot(\vec{\phi}_{x+\hat{\mu}}-\vec{\phi}_{x}).
\enq
This model is called the nonlinear sigma model (NLSM)~\cite{peskin95}.  The constraint $\vec{\phi}_x\cdot\vec{\phi}_x=v^2$ (which enforces the ``nonlinear'' part of its name) can be expressed by introducing unit vectors:
$\vec{\phi}_x=v\vec{\sigma}_x$ such that
\beq
\vec{\sigma}_x\cdot \vec{\sigma}_x=1.
\enq
Redefining $ a^{D-2}v^2\equiv \beta$, we get the simple action
\beq
\label{eq:nlsm}
S_{\text{NLSM}} = \beta \sum_{x,\mu}(1-\vec{\sigma}_{x+\hat{\mu}}\cdot \vec{\sigma}_x).
\enq
These models are often called spin models as well.
The first term in the action $\beta \sum_{x,\mu} 1$ is a constant that is often dropped. However, for large
$\beta$, the configurations with almost constant $\vec{\sigma_x}$ dominate the partition function and since
under these circumstances $\vec{\sigma}_{x+\hat{\mu}}\cdot \vec{\sigma}_x \simeq 1$ it is useful to subtract the constant in order to just keep the small fluctuations.

The case $N=1$ is the well-known Ising model with $\sigma_x=\pm 1$~\cite{Ising1925}. For $N=2$, the terminology
``planar model" or ``classical XY model"~\cite{vaks1966,PhysRevLett.19.630} is common and if we use the circle parametrization
\beq
\sigma_x^{(1)}=\cos(\varphi_x), {\rm and}\  \sigma_x^{(2)}=\sin(\varphi_x),
\enq
then
\beq
\label{eq:sigma-dot}
\vec{\sigma}_{x+\hat{\mu}}\cdot \vec{\sigma}_x=\cos(\varphi_{x+\hat{\mu}}-\varphi_x),
\enq
with $\varphi_{x} \in [0,2\pi)$.

There is another class of models which breaks the $O(2)$ symmetry in Eq.~\eqref{eq:sigma-dot} into a discrete $\mathbb{Z}_{q}$ symmetry, \emph{i.e.} the possible angles are restricted to those of the $q$\textsuperscript{th} roots of unity in the complex plane.  They are called the $q$-state clock models~\cite{potts_1952}.  These models have the same action from Eq.~\eqref{eq:nlsm} with the identification of the angles as being discrete,
\begin{align}
  \varphi_{x} = \frac{2 \pi n_{x}}{q}
\end{align}
with $n_{x} = 0,1,2,\ldots,q-1$.  With this identification, the $O(2)$ model emerges as the $q \rightarrow \infty$ limit of the $q$-state clock models, and the Ising model is simply the $q=2$ model.

For $N=3$, the  symmetry becomes non-Abelian and the model is sometimes called the ``classical Heisenberg model". In the large-$N$ limit, the model becomes solvable if we take the limit in such a way that $N/\beta(N)=\lambda$ remains constant \cite{PhysRevD.10.2491}.

It is instructive to rewrite the $O(2)$ model using the complex form
\beq
\Phi_x={\rm e}^{i\varphi_x}.
\label{eq:cphase}
\enq
Dropping the constant terms, the $O(2)$ action reads
\begin{align}
  \nonumber
  S_{O(2)} &= \frac{\beta}{2} \sum_{x, \mu} (\Phi_{x+\hat{\mu}} - \Phi_{x}) \cdot (\Phi_{x+\hat{\mu}} - \Phi_{x})^{\star} \\ \nonumber
           &= -\beta \sum_{x,\mu}\cos(\varphi_{x+\hat{\mu}}-\varphi_x).
\end{align}
The $O(2)$ model has a global symmetry
\beq
\varphi_x \rar \varphi_x+\alpha.
\enq
With the complex notation, this transformation becomes
\beq
\Phi_x \rar {\rm e}^{i\alpha} \Phi_x.
\enq

We would like to promote this symmetry to a local one
\beq
\Phi_x \rar {\rm e}^{i\alpha _x} \Phi_x,
\enq
\emph{i.e.} one that is site dependent.  This can be achieved by inserting a phase $U_{x,\mu}$ between $\Phi_x^\star$ and $\Phi_{x+\hat{\mu}}$ which transforms like
\beq
\label{eq:abeliangt}
U_{x,\mu} \rar {\rm e}^{i\alpha _x} U_{x,\mu}{\rm e}^{-i\alpha _{x+\hat{\mu}} }.
\enq

The procedure can be extended for arbitrary $N$-dimensional complex vectors
${\bf \Phi}_x$ with a local transformation involving a $U(N)$ matrix $V_x$:
\beq
{\bf \Phi}_x \rar V_x {\bf \Phi}_x
\enq
In addition, we can also introduce $U(N)$ matrices ${\bf U}_{x,\hat{\mu}}$ transforming like
\beq
{\bf U}_{x,\hat{\mu}}\rar V_x {\bf U}_{x,\hat{\mu}}V_{x+\hat{\mu}}^\dagger.
\label{eq:utrans}
\enq
The action
\beq
\label{eq:unmodel}
S_{U(N)} = -\frac{\beta}{2}\sum_{x,\mu}({\bf \Phi}_x^\dagger {\bf U}_{x,\hat{\mu}}{\bf \Phi}_{x+\hat{\mu}} + \text{h.c.})
\enq
has a local $U(N)$ invariance which we call gauge invariance.
If we consider two successive links in positive directions, then the local transformation at the middle site cancels and
\beq
{\bf U}_{x,\mu}{\bf U}_{x+\hat{\mu},\nu} \rar V_x{\bf U}_{x,\mu}{\bf U}_{x+\hat{\mu},\nu} V_{x+\hat{\mu}+\hat{\nu}}^\dagger.
\enq
If the second link goes in the negative direction, we use the Hermitian conjugate and a similar property holds
\beq
{\bf U}_{x,\mu}{\bf U}_{x+\hat{\mu}-\hat{\nu},\nu}^\dagger \rar V_x{\bf U}_{x,\mu}{\bf U}_{x+\hat{\mu}-\hat{\nu},\nu}^\dagger V_{x+\hat{\mu}-\hat{\nu}}^\dagger.
\enq
We can pursue this process for an arbitrary path connecting $x$ to some $x_{\text{final}}$. The transformation on the right will be $V_{x_{\text{final}}}^\dagger$.
If we close the path and take the trace, we obtain a gauge-invariant quantity.
We call these traces of products of gauge matrices over closed loops ``Wilson loops" \cite{PhysRevD.10.2445}. In the case where the loop goes around the imaginary time direction, we often call it a ``Polyakov loop" \cite{polyakovloop}.

On a square, cubic or hypercubic lattice, the smallest path that gives a non-trivial Wilson loop is a square. We call this square a plaquette.
Claude Itzykson coined this terminology after Ken Wilson's seminar in Orsay in 1973.
The corresponding matrix is
\beq
{\bf U}_{\text{plaquette}}={\bf U}_{x,\mu \nu}={\bf U}_{x,\mu}{\bf U}_{x+\hat{\mu},\nu}{\bf U}_{x+\hat{\nu},\mu}^\dagger{\bf U}_{x,\nu}^\dagger
\enq
The simplest gauge-invariant lattice model has an action, called Wilson's action:
\beq
\label{eq:wilsonaction}
S_{\text{Wilson}} = \beta_{pl.}\sum_{\langle x,\mu \nu \rangle}\left[ 1-\frac{1}{2N}\Tr \left[ {\bf U}_{x,\mu \nu} + \text{h.c.} \right]\right].
\enq
where $\sum_{\langle x, \mu \nu \rangle}$ indicates a sum over all plaquettes. Here each $\bf{U}_{x,\mu}$ is related to the vector potential, or gauge field, in the continuum theory through \beq\mathbf{U}_{x,\mu} = e^{i \mathbf{A}_{x,\mu}}.\enq
On the lattice, both $\mathbf{U}_{x,\mu}$ and $\mathbf{A}_{x,\mu}$ are located on a link starting at $x$ and going in the $\hat{\mu}$ positive direction. 
In the Abelian case ($N=1$), the matrix reduces to a phase
\beq
U_{x,\mu}={\rm e}^{i A_{x,\mu}},
\enq
and there is no need to take the trace.

Another generalization of the $N=1$ expression of the complex phase given in Eq.~\eqref{eq:cphase}, consists of replacing $\Phi_x$ by a $SU(N)$ matrix ${\bf U}_x$. This is called the
principal chiral model~\cite{Gursey1960,GREEN1981110,SAMUEL198585,PhysRevD.52.358,peskin95},
\beq
S_{\text{PCM}} =-\frac{\beta}{2N}\sum_{x,\mu}\left[ \Tr \left[ {\bf U}^{\dagger}_{x+\hat{\mu}}{\bf U}_x \right] + \text{h.c.}\right] .
\enq
This model has a \emph{global} rotational symmetry under the $U(N)$ group, such that $\mathbf{U}_{x,\mu} \rightarrow \mathbf{U}_{x,\mu}' = V \mathbf{U}_{x,\mu} V^{\dagger}$, just as in Eq.~\eqref{eq:utrans} in the case of a uniform $V$ in all of space-time.

We can also consider the $N = 1$ case for Eq.~\eqref{eq:euclclasson} when the group is no longer compact, \emph{i.e.} $\phi_{x}$ can take on the values of any real, or complex, number.  The action on the lattice is then,
\begin{align}
  S_{\text{scalar}} = \sum_{x}\left(\frac{1}{2}\sum_{\mu = 1}^{D} |\phi_{x+\hat{\mu}} - \phi_{x} |^{2} + \lambda (|\phi_{x}|^{2} - \nu^{2})^{2} \right)
\end{align}
or equivalently,
\begin{align}
  S_{\text{scalar}}
  = \sum_{x} \left\{
  \frac{1}{2} \sum_{\mu=1}^{D} \left| \phi_{x+\hat{\mu}} - \phi_{x} \right|^{2}
  - \frac{\mu_{0}^{2}}{2} |\phi_{x}|^{2}
  + \frac{\lambda'}{4} |\phi_{x}|^{4} \right\}
\end{align}
with the substitution $\mu_{0}^{2}/2 = 2 \lambda \nu^{2}$ and $\lambda' / 4 = \lambda$, and an overall constant is ignored.

Besides scalar fields, we will also consider fermionic fields (or Grassmann fields on the lattice).  In the case of free fermions, we can write down a lattice action using a straightforward discretization due to Wilson~\cite{PhysRevD.10.2445},
\begin{align}
  S_{\text{WD}}
  = \sum_{x} \bar{\psi}_{x} \left( D\psi \right)_{x}
\end{align}
where the Wilson-Dirac operator is defined by
\begin{align}
  & D_{x x^{\prime}} = \left( a m + r D \right) \delta_{x, x^{\prime}} \nonumber \\
  & + \frac{1}{2} \sum_{\mu = 1}^{D} \left\{ \left( r - \gamma_{\mu} \right) \delta_{x', x + \hat{\mu}} + \left( r + \gamma_{\mu} \right) \delta_{x, x^{\prime}+\hat{\mu}} \right\}.
\end{align}
where $\gamma_{\mu}$ are the gamma matrices in $D$ dimensions, and $r$ is the ``Wilson parameter'' to control species doubling. This so-called ``doubling'' is the name given to the existence of extra copies of fermions---16 in the case of four dimensions---and they must be removed or gapped-out appropriately.
$\psi_{x}$ and $\bar{\psi}_{x}$ are multi-component Grassmann variables that 
\emph{anti}-commute, as opposed to classical bosonic variables 
which can be interchanged without sign change.  

There is another formulation of lattice fermions, where the different components of the fermion fields are located at different lattice sites, called staggered fermions.  This comes about from a transformation which mixes the fermion components and space-time components \cite{ks}. The action for free fermion fields is given by,
\begin{align}
  S_{F} = \frac{1}{2} \sum_{x=1}^{N} \sum_{\mu = 1}^{D} \eta_{x, \mu} &[ \bar{\psi}_{x} \psi_{x + \hat{\mu}} - \bar{\psi}_{x + \hat{\mu}} \psi_{x} ]
\end{align}
where
\begin{align}
  \eta_{x, \mu} = (-1)^{\sum_{\nu < \mu} x_{\nu}}
\end{align}
with $x_{\nu}$ the coordinate in the $\nu$th direction.  One can include gauge fields in a gauge-invariant manner by inserting $\mathbf{U}_{x,\mu}$ like,
\begin{align}
  S_{F} = \frac{1}{2} \sum_{x=1}^{N} \sum_{\mu = 1}^{D} \eta_{x, \mu} &[ \bar{\psi}_{x} \mathbf{U}_{x, \mu} \psi_{x + \hat{\mu}} - \bar{\psi}_{x + \hat{\mu}}\mathbf{U}^{\dagger}_{x, \mu} \psi_{x} ].
\end{align}

These models will appear in the rest of the review and will be very briefly reintroduced in each section.
As seen above, the matter degrees of freedom are placed on each lattice site.
When one considers a tensor network representation of a model, integer degrees of freedom\ will arise at each link on the lattice~(see Fig.~\ref{fig:convertdof}).
As for lattice gauge theories where gauge degrees of freedom are initially placed on links, the character expansion will be used for generating integer at each plaquette.
These points will be made clear in the  following sections.
The simplest example is the Ising model discussed in Sec.~\ref{sec:ising}.

\begin{figure}[htbp]
    \centering
    \includegraphics[width=0.7\hsize]{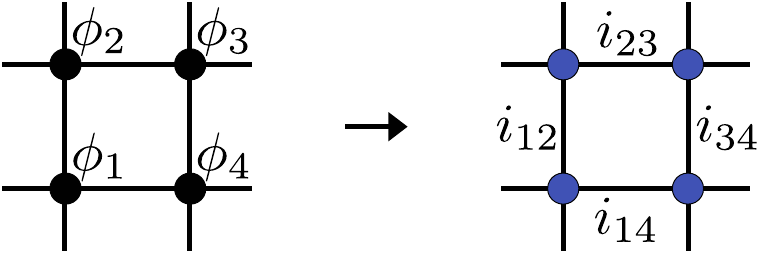}
    \caption{Reinterpretation of physical degrees of freedom ($\phi$) to tensor indices ($i$). After that, the partition function is expressed as a summation of tensor indices instead of the path integral.}
    \label{fig:convertdof}
\end{figure}

\subsection{Physical applications}
The sequence of models described in Sec.~\ref{subsec:ladder} is designed to handle lattice QCD
which is so far our best definition of the theory describing strongly interacting particles observed in a large number of experiments.
Some of the models discussed in Sec.~\ref{subsec:lattac} are also studied in condensed matter. For instance, the $O(2)$ model with a
chemical potential can be seen as an effective theory for the Bose-Hubbard model \cite{sachdev2001quantum}. $U(1)$ gauge theories with either
scalar of fermion fields are studied in the context of superconductivity \cite{herbut2007modern}. Tight-binding approximations for solids lead to interesting
lattice models, for instance a sheet of graphene can be described with fermions on a hexagonal lattice having a dispersion relation similar to the
one for a massless Dirac fermion at half-filling. \cite{RevModPhys.81.109}. 
Other interesting applications
include spin liquids and topological phases \cite{anderson73,wenbook,wenrmp,zhourmp,knolle2019}. 

We would like to point out that in the context of high-energy physics, {\it local} Lagrangian densities have played a crucial role in the development of the Standard Model. 
In the continuum limit, models compatible with relativistic
invariance, local gauge invariance and renormalizability have a very small number of free parameters provided that the interactions among the fields and their first derivatives are kept local. The standard model of electroweak and strong interactions has only
18 free parameters if we ignore the QCD vacuum angle and the additional parameters related to the masses and mixing of the  neutrinos in extensions of the standard model. It is not possible to tweak the theory each time a new experiment is completed. This makes the standard model a very predictive theory.
In the lattice formulation, the local interactions involving derivatives in the continuum are replaced by interactions involving fields located on neighboring sites, links and plaquettes and can be considered quasi-local.
For these reasons, the idea that reformulations of lattice models and their coarse-grained versions could be kept local seems to be an important consideration.

\subsection{Computational methods beyond perturbation theory}
Besides the RG methods mentioned in the Introduction, lattice models have been studied with a variety of analytical and numerical methods.
For example, expansions at small and large coupling for spin and gauge models
\cite{kogut79,kogut83,1987gauge,itzykson1991statistical,parisi1998statistical} have been investigated extensively.
As power series, they can be used to check numerical calculations in their respective limits, however, it seems hard to capture the non-perturbative effects such as the generation of a dynamical mass gap in the continuum limit.
After Wilson's original proposal \cite{PhysRevD.10.2445} a very suggestive Hamiltonian
picture was developed by Kogut and Susskind \cite{ks}, however it became clear that
the size of the Hilbert space would make numerical calculations impractical.

In many practical situations, the path integral formulation uses a real Euclidean action $S_E$. This allows importance sampling. A typical way to proceed is to start with a random
field configuration and then apply some random changes on this configuration. If the new configuration has a lower action, it is accepted. If the new configuration has an action
larger by $\delta S_E$, then it is accepted with a probability $\exp(-\delta S_E)$. The fact that the fields are only connected to a few neighboring fields makes the calculation of $\delta S_E$ easy and the exploration of the important
configurations controllable.

Mathematicians are often amazed that it is possible to obtain very reliable results with this method \cite{villani}. So far, this has been the most reliable way to capture the non-perturbative behavior
while taking the continuum limit.
M. Creutz started numerical lattice gauge theory with simulations of the gauge Ising model on a $3^4$ lattice using a HP 9830
calculator, programmed in Basic \cite{creutz2001}.
This was rapidly followed by the study of other Abelian gauge theories \cite{creutz79} and other models. This is reviewed in C. Rebbi's book \cite{claudio1983lattice}.

Several decades after its inception, numerical lattice gauge theory has become an area of research where corrections on the order of a few percent are considered important. For instance, in heavy flavor physics the determination of
the mixing angle $|V_{ub}|$ is quoted as $(3.72\pm 0.16)10^{-3}$ \cite{du2015}.
This result was obtained using ensembles of lattices of various lattice spacings and sizes, some of them as large as $64^3\times192$.
For a review of averages of numerical results with estimated errors see, for instance, the
Flavour Lattice Averaging Group (FLAG) summary \cite{flag2019}.

Given the success of importance-sampling for static problems in QCD, not much has been done to develop efficient numerical Hamiltonian methods. On the other hand, in condensed matter, 
the idea that it is possible to approximate low-energy states of the Hamiltonian with a significantly reduced Hilbert space 
has been very successful in 1+1 dimensions and has some similarities with importance sampling. It appeared in the context of the density matrix renormalization group (DMRG) method \cite{white92,schollwock2005,uli2011,hv2017} and other methods 
discussed in Sec. \ref{subsec:tn}. We expect that in the coming years great progress will be made adapting related methods in the context of lattice gauge theory.

\section{Quantum computing}
\label{sec:qc}
\subsection{Situations where importance sampling fails}
The numerical successes of lattice gauge theory can be linked to the fact that
when the action of the Euclidean-time path  integral $S_E$ is real, importance sampling works remarkably well for the selection of
configurations  with
a Boltzmann distribution $\exp(-S_E)$. However, if we return to real time or introduce a chemical potential that makes the action complex, this powerful tool becomes ineffective~\cite{deForcrand:2010j3}. This is because in these instances the Boltzmann weights become complex making their interpretation as probability weights impossible.

Real-time evolution can be set up by using a 
Hamiltonian acting on a Hilbert space which can be constructed by noticing that
the Euclidean path-integral can be recast as the trace of a transfer matrix \cite{wilson73pr,creutz77,fradkin78,kogut79,luscher88}. Outside of the transfer-matrix method, one can also extract real-time observables, such as Green's functions, from generating functional methods in  Minkowski spacetime. There are also other techniques for calculating off-equilibrium real-time observables requiring the use of the Keldysh contour~\cite{keldysh1965}.

Analysis of the exponential decays of correlations as the Euclidean time is increased is a standard tool to extract masses and form factors for momenta that are small compared to the lattice cutoff. On the other hand, real-time evolution and a
large Hilbert space involving states with large momenta are needed to describe aspects of hadron fragmentation in real space-time and deep-inelastic scattering \cite{bauer2019,DIS2019,lamm2020}. The real-time evolution operator does not provide a positive measure nor a projection onto
a small Hilbert space.
Another situation where sampling methods are challenged is the construction of interpolating operators
for nuclei as the number of Wick contractions grows rapidly with the number of light quarks involved
\cite{Detmold:2012eu}.

One could consider the possibility of abandoning the Lagrangian formulation and
directly considering
the Kogut-Susskind Hamiltonian \cite{ks} for QCD and/or the
standard many-body formalism in condensed matter and nuclear physics \cite{fetter2003quantum} where the Hilbert space is generated by creation operators on a Fock-space vacuum. For the sake of argument, let us consider the simple case of $N$  conjugate pairs of fermionic creation and annihilation operators. They generate
a Hilbert space of dimensions $2^N$. The exponential growth of this number with $N$ rapidly restricts our ability to store or manipulate the matrix elements of operators using brute force methods with classical computers.  Computations involving spatial lattices with $64^3$ sites, as made possible in the Lagrangian formulations, would be completely out of the question if we had to set up the whole quantum Hilbert space with existing classical computers. Note however, that for a broad class of many-body problems, it has been shown \cite{poulin2011} that most states of the Hilbert space can only be reached after an exponentially long time.
This suggests that  tensor networks methods should allow 
expressing Hamiltonians and ground-state without 
needing to explore the full Hilbert space.

\subsection{Qubits  and other quantum platforms}

The building blocks of an ordinary (classical) computer memory are bits taking the values 0 or 1.
For instance, in some designs of Dynamic Random Access Memory (DRAM) devices,
this is achieved by using small capacitors that are either on or off. The typical units of capacitance
are femtofarads with voltages of the order of 1V. These capacitors store a few thousand electrons and can be charged or discharged in times of order $10^{-15}$ sec. It is interesting to consider the limit of miniaturization where electrons, atoms or photons could be manipulated individually and where
the peculiarities of the quantum behavior become prominent \cite{lloyd96}.

Following the physical examples of the electron spin or the photon polarization, one could envision some
ideal and generic quantum system, where the on/off concept for bits is replaced
by the linear superposition of two states. This basic unit of quantum computing is often called a qubit \cite{qubit} and can be
represented as
\beq
\ket{\rm qubit}=\alpha_0\ket{0}+\alpha_1 \ket{1},
\enq
with two complex numbers $\alpha_0$ and $\alpha_1$ such that $|\alpha_0|^2+|\alpha_1|^2=1$.
A set of $N$ qubits spans a Hilbert space of dimension $2^N$. If we use a classical computer to apply
a dense unitary matrix representing the real time evolution on an arbitrary  state, we need to perform on the order of $2^{2N}$ arithmetic operations.
On the other hand, if the Hamiltonian is such that any qubit is only connected to a restricted number of other qubits which is fixed by the
dimensionality of space and the internal symmetries, then it is possible to design a method that performs the evolution with a number of
quantum manipulations which only scales polynomially with $N$ \cite{lloyd96}. This will be discussed in Sec.~\ref{subsec:lloyd}.
Nowadays, commercial quantum computers
provide sets of qubits and the possibility of preparing, evolving and measuring quantum states. The limitations of the current hardware are discussed in Sec. \ref{subsec:nisq}.

More generally, the idea that quantum devices could be used to perform computations for quantum problems involving many degrees of freedom is very appealing \cite{feynman82}.
Physical systems
involving cold atoms (see \cite{1998PhRvL..81.3108J} and} \cite{Bloch2008} for a review of the early developments) or trapped ions (see \cite{monroe2003} for early developments and  \cite{linke16} for a recent example) can be used to mimic the behavior of
simplified many-body models such as various types of spin chains or Hubbard models.
In addition, superconducting circuits \cite{devoret}, Rydberg atoms \cite{51qubits,rydberg} and photonic systems \cite{photonics} also provide 
interesting opportunities.

\subsection{From Euclidean  transfer matrices to Hilbert spaces}
\label{subsec:extm}
For the lattice models introduced in  Sec.~\ref{subsec:lattac}, the Euclidean time was treated on equal
footing with the $D-1$ spatial dimensions. In order to discuss real-time evolution, we first need to single-out the time direction.  Evolution then occurs along this direction according to a transfer matrix.
The key ingredient then is to connect the Lagrangian formulation to the Hamiltonian formalism and to identify the Hilbert space from the
transfer matrix introduced in Eq.~\eqref{eq:tm}.
The general idea is to organize the partition function sums or integrals into operations performed on successive time slices \cite{wilson73pr,creutz77,fradkin78,kogut79}. 
A general procedure to construct the transfer matrix of lattice models in configuration space in the general context of scattering theory is nicely presented in \cite{luscher88}. A ``dual" method based on character expansions which are at the heart of \tft are discussed in Sec.~\ref{sec:transfer}. We will illustrate these two possibilities with examples. Note that, in this subsection, Euclidean time methods are used to derive a Hamiltonian that can later be used to do real-time calculations.

Turning to the transfer matrix, with generic notations for a lattice model with $N_\tau$ sites in the Euclidean time direction,
\beq
Z=\int {\mathcal D}\Phi \, {\rm e}^{-S[\Phi]_E}=\Tr (\tm ^{N_\tau}).
\label{eq:tm}
\enq
If the lattice spacing $a_\tau$ in the Euclidean time direction is small compared to the physical time scales involved, we have
\beq
\tm \propto e^{-a_\tau \hat{H}}.
\enq
For $N_\tau$ large enough, the use of Euclidean time provides a projection in the low energy sector of the Hilbert space.
This property remains effective if we insert operators that create and destroy states with non-trivial quantum numbers.

The simplest possible example is the one-dimensional Ising model,
\begin{align}
  \label{eq:tm-ising}
  S_{\text{Ising}} &= \beta \sum_{\tau} (1 - \sigma_{\tau+1} \sigma_{\tau}) \\
                   &= \frac{\beta}{2} \sum_{\tau} (\sigma_{\tau+1} - \sigma_{\tau})^{2}
\end{align}
with partition function
\begin{align}
  Z = \sum_{\{\sigma\}} e^{-S}.
\end{align}
This is a product of exponentials which each share one spin variable with the next factor.  Then we can write the partition function as
\begin{align}
  Z = \Tr[\mathbb{T}^{N_{\tau}}]
\end{align}
with
\begin{equation}
  \mathbb{T}_{\alpha \alpha'} = \exp\left\{-\frac{\beta}{2} (\sigma^{(\alpha)}_{\tau+1} - \sigma^{(\alpha')}_{\tau})^2 \right\}
\end{equation}
and $\sigma^{(\alpha)} = 1,-1$ for $\alpha = 0,1$, respectively.  Along the diagonal of the transfer matrix we see only unity.  On the off-diagonal, a spin flip comes with weight $e^{-2\beta}$.  Then to leading order in the temporal lattice spacing, $\mathbb{T} \simeq 1 - a_{\tau} \hat{H} +\ldots$ which allows us to identify
\begin{align}
  \hat{H} = -h_{x} \hat{\sigma}^{x},
\end{align}
with $h_{x} \equiv e^{-2\beta}/a_{\tau}$ and
$\hat{\sigma}^{x}$ the $x$-Pauli matrix.  For this case, to extract a Hamiltonian from the original Lagrangian formulation, we required that the coupling, $\beta$, goes to infinity to match the temporal lattice spacing, $e^{-2\beta} \propto a_{\tau}$.

In the above we found the Hamiltonian and Hilbert space in configuration space; however, one can use a ``dual'' method as well by expanding the original Boltzmann weights.  Consider the action for the Ising model in Eq.~\eqref{eq:tm-ising} with the constant dropped,
\begin{align}
  S_{\text{Ising}} = -\beta \sum_{\tau} \sigma_{\tau+1} \sigma_{\tau}.
\end{align}
The Boltzmann weight can be expanded in the form,
\begin{align}
  \mathbb{T}_{\alpha \alpha'} &= e^{\beta \sigma^{(\alpha)}_{\tau+1} \sigma^{(\alpha')}_{\tau}} \\
                              &=  \cosh(\beta)\sum_{n=0}^{1} (\sigma^{(\alpha)}_{\tau+1} \sigma^{(\alpha')}_{\tau})^{n} \tanh^{n}(\beta).
\end{align}
which is simply the Euler identity for imaginary angles.  Ignoring the factor out front as it does not affect the Hamiltonian, in this form we can easily do the summation over the values of $\alpha$ at all lattice sites.  The transfer matrix becomes a diagonal matrix with matrix elements labeled by the integers $n$, from each Boltzmann factor,
\begin{align}
  \mathbb{T}_{n n'} =
  \begin{pmatrix}
    1 & 0 \\
    0 & \tanh(\beta)
  \end{pmatrix}.
\end{align}
In the literature, the integers $n$ are often called dual variables or characters indices. The terminology is discussed
more systematically in Sec. \ref{subsec:dualities}.
To recover a Hamiltonian, the transfer matrix must have the form $1 - a_{\tau} \hat{H} + \ldots$ for small times.  This is found by taking $\beta \rightarrow \infty$, and recalling $\tanh{\beta} = 1 - 2 e^{-2 \beta} + \ldots$
for large $\beta$.  Then the Hamiltonian in these new variables takes the form,
\begin{align}
  \hat{H} = h_{z}(1 - \hat{\sigma}^{z})
\end{align}
with $h_{z} \equiv e^{-2 \beta}/ a_{\tau}$, and $\hat{\sigma}^z$ the $z$-Pauli matrix.  These are two distinct procedures which give Hamiltonians in the time continuum limit; one, in the original configuration variables, and the second in the dual variables.

The transfer matrix of the Ising model in higher dimensions can be constructed in a similar manner \cite{kaufman49}.  The action for the Ising model in dimension $D$ is the $N=1$ case of Eq.~\eqref{eq:nlsm}.
In higher dimensions there are now two types of interactions. If we consider a particular time slice, we can first collect all the time links connected to the next time slice, each with a representation given in Eq.~\eqref{eq:tm-ising}.
In addition, we have all the spatial links in the time slice with nearest-neighbor interactions. We can represent any spin configuration in a
time slice as an element of a tensor product of eigenstates of the Pauli matrix $\sigma^z_{\bf x}$:
\beq
\label{eq:isingqbits}
\{ {\rm configurations} \} = \bigotimes _{\bf{x}}  \ket{\pm 1}_{\bf x}.
\enq
We can introduce operators $\hat{\sigma} ^{z}_{\bf x}$ and $\hat{\sigma}^{x}_{\bf x}$ acting on this
Hilbert space that can be identified with a set of qubits.
Following \cite{kaufman49}, we can collect the two types of interactions in two matrices. Using the relation
\begin{align}
  e^{\beta \sigma_{\alpha '} \sigma_{\alpha}} = (2\sinh(2\beta))^{1/2}(\exp(\tilde{\beta} \sigma^x))_{\alpha' \alpha},
\end{align}
where $\tilde{\beta}$ is the dual inverse temperature introduced by Kramers and Wannier \cite{PhysRev.60.252} which satisfies the relation
$\tanh(\tilde{\beta})=e^{-2\beta}$,
the contribution of time links can be summarized with the matrix connecting the configuration of the two time slices
\beq
\vone=(2/\sinh(2\tilde{\beta}))^{N_s^{D-1}/2} e^{\tilde{\beta}\sum_{\bf x}\hat{\sigma}^x _{\bf x}}.
\enq
On the other hand, the spatial links can be recast in the diagonal matrix
\beq
\vtwo=e^{\beta\sum_{{\bf x} ,j} \hat{\sigma} ^z_{\bf x} \hat{\sigma} ^z_{{\bf x}+\hat{j}}}
\enq
with $\hat{j}$ a unit vector pointing along one of the $D-1$ directions on the lattice.  We can now write the transfer matrix as
\beq
\tm = \vtwo ^{1/2}\vone\vtwo^{1/2},
\enq
where the matrix indices label the spatial configurations. Geometrically, the Hilbert space is located on the time slices.
Illustrations of the time-slices and the location of the Hilbert space can be found in Sec.~\ref{sec:transfer}.

Alternatively, we can work in a dual representation where the $\sigma^x _{\bf x}$ are diagonal
\beq
\tilde{\tm} = \vone^{1/2}\vtwo\vone^{1/2},
\enq
where the matrix indices label sets of group characters.
Geometrically, the Hilbert space is located between the time slices. Graphical illustrations of this situation will be provided in Sec.~\ref{sec:transfer}.
The construction generalizes easily for finite Abelian groups and in a non-trivial way for continuous and compact Abelian groups.
The advantage of using the second (dual) representation is that it remains {\it discrete} and as we will see in Sec.~\ref{subsec:symmetries}, it preserves the symmetry when truncations are applied \cite{meurice2019,meurice2020}.

\subsection{Topological and geometrical  dualities}
\label{subsec:dualities}
In Sec.~\ref{subsec:extm} we used the concept of duality in two occasions. The first was the relation between $\beta$ and $\tilde{\beta}$ which interchanges
their low and large values regimes. The second occasion was the discussion of the two ways to represent the transfer matrix.
In addition, in Secs. \ref{subsec:spindual} and \ref{subsec:gaugedual}, we will use the geometrical duality reviewed in   \cite{savit80}.
Duality is a general concept used in many branches of mathematics. According to Atiyah \cite{atiyah}
duality gives ``two different points of view of looking at the same object". In the following we clarify
the various usages of the concepts in the rest of this review.

An important notion of duality is  the so-called ``Pontryagin duality" \cite{pdual} coming in the study of topological groups. It relates an Abelian group and its characters (for instance, Fourier modes). It states that if the former is compact, the later
is discrete and vice-versa. The simplest situation is a finite group which is compact and discrete.
In the case of finite cyclic Abelian groups the characters form a finite group which is isomorphic to the group itself \cite{serre73}.
A simple example is \zq the additive group of integers
modulo $q$. If $x$ denotes an element of  $\mathbb{Z}_q$, the characters (See \ref{subsec:character-expansion}) have the form
\beq
\chi_k(x)=\exp \left( i\frac{2\pi}{q} k x \right),
\enq
and clearly satisfy the character property
\beq
\chi_k(x+x')=\chi_k(x)\chi_k(x').
\enq
The product of two characters is another character
\beq
\chi_k(x)\chi_{k'}(x)=\chi_{k+k'}(x), \enq
and one sees that they also form a $\mathbb{Z}_q$ group.
They obey the orthogonality relations
\beq
\frac{1}{q}\sum_{x=0}^{q-1}\chi_k(x)\chi^\star_{k'}(x)=\delta_{k,k'},
\enq
and
\beq
\frac{1}{q}\sum_{k=0}^{q-1}\chi_k(x)\chi^\star_{k}(x')=\delta_{x,x'}.
\enq

A simple example of a continuous (nondiscrete) and compact group is $U(1)$, the multiplicative group of complex numbers with modulus one. The group properties
can be reformulated in an additive manner by introducing the phases $z=e^{i\varphi}$ which are added modulo $2\pi$.
Topologically it is a circle which is a compact manifold. Its characters are discrete and labelled by the integers. They are the usual Fourier modes $e^{in\varphi}$. The orthogonality relations appear in an asymmetric way
\beq
\int_{-\pi}^{\pi} \frac{d\varphi}{2\pi} e^{in\varphi}(e^{in\varphi})^\star=\delta_{n,n'}
\enq

\beq
\sum_{n=-\infty}^{\infty}  e^{in\varphi}(e^{in\varphi'})^\star= 2\pi \sum_{m=-\infty}^{\infty} \delta(\varphi-\varphi' + 2\pi m)
\enq

In practice, a deep understanding of the mathematical statements made above is not necessary
and we just need to remember a few character expansions. For the Ising models we have functions over the multiplicative group $\sigma =\pm 1$,
and we can recall the expansion from  Sec.~\ref{subsec:extm} as a character expansion
\begin{align}
  \nonumber
  \bra{\sigma_{\alpha '}}\tm\ket{\sigma_{\alpha}} &= \exp(\beta \sigma_{\alpha '} \sigma_\alpha) \\
                                                  &= \frac{1}{2} \sum_{n = 0}^{1} \lambda_{n}(\beta) ( \sigma_{\alpha '} \sigma_\alpha)^{n} \\
                                                  & = \cosh(\beta)+ \sigma_{\alpha'} \sigma_\alpha \sinh(\beta).
                                                    \label{eq:kw}
\end{align}
A similar representation can be obtained for the \zq spin models.
If we replace the discrete angle variables $\frac{2\pi}{q}x$ by continuous ones $\varphi$, we obtain a ``matrix" with continuous elements which can be calculated using the standard Fourier transform
\begin{align}
  \bra{\varphi'}\tm\ket{\varphi} &= \exp(\beta \cos(\varphi'-\varphi)) \\
                                 &=\sum_{n=-\infty}^{+\infty}I_n(\beta)\exp(in(\varphi'-\varphi)),
                                   \label{eq:bessel}
\end{align}
where $I_n(\beta)$ is the modified Bessel function of order $n$. We will show in Sec.~\ref{subsec:discretesub} that finite versions of this expansion hold for
the finite \zq subgroups.

Generalizations of Pontryagin duality to compact non-Abelian groups appear in the
Peter-Weyl theorem \cite{peter27}. As an example, this translates
into expansions in spherical harmonics for problems involving the O(3) symmetry. 
Related methods were used for tensor networks with continuous symmetries \cite{celi2014,zohar2015b}.

From the point of view of quantum computing, we see that using compact fields guarantees
that we can replace the continuous integrals by discrete sums. Our strategy is to
associate the indices of these sums with quantum states. One important aspect of \tft is
that when the fields are compact, we don't need to
discretize the integrals using numerical approximations.  Instead, we can use characters expansions as Eq.~\eqref{eq:bessel}
where some integrals have been done exactly and result in Bessel functions that we can input with any desired accuracy.
In other words, the difficult part of the classical path-integral approach can be done efficiently with classical methods.
After that, the original integrals reduce to orthogonality relations and can be performed exactly.

Another notion of duality is
of a geometrical nature. It is
related to the Levi-Civita tensor $\epsilon^{\mu_1\dots \mu_D}$.
Its meaning is dimension-dependent and relates objects of dimension $d$ to objects of dimension $D-d$. For instance, in $D=4$, a dual
field-strength tensor with two indices is obtained by contracting the original field-strength tensor with the Levi-Civita tensor. This duality transformation interchanges the electric and magnetic fields and reduces to the identity
when repeated twice.
It also relates sites to four-dimensional hypercubes, and plaquettes to plaquettes.
In $D=3$ it relates the field-strength tensor to a divergenceless pseudovector $\epsilon^{ijk}\partial_j A_k$, links to plaquettes and sites to cubes.

The various notions of duality are often used simultaneously.
A common example is the ``dual formulation" of the Ising model. As we will explain in  Sec.~\ref{subsec:isingdual}, the new set of indices from Eq.~\eqref{eq:kw} which is a consequence of Pontryagin duality, leads to a representation in terms of paths.
We can then try to represent these paths as the boundaries of surfaces, which brings the geometrical duality and new
``dual variables'' together \cite{savit80}.

There are also occasions where the phrase ``dual variables'' has become associated with generic integer fields, regardless or their origin or their relation to  the concepts of duality mentioned above \cite{PhysRevD.94.114503,GATTRINGER2018344,PhysRevD.92.114508,PhysRevD.97.034508,BRUCKMANN2015495}.
These integer fields can arise from a Taylor series of the Boltzmann weight, \emph{e.g.} in the case of the Ising model,
\begin{align}
  e^{\beta \sigma \sigma'} = \sum_{q = 0}^{\infty} \frac{\beta^q}{q!} (\sigma \sigma')^{q}.
\end{align}
This expansion associates a natural number with the links of the lattice, similar to the character expansion from before.  Also just as before, this expansion allows one to perform the path-integral sums over the $\sigma$ fields, leaving one with a theory of constrained, positive, integer fields on the links of the lattice.  However, instead of two values, \emph{i.e.} $n = 0, 1$, these integers can take on an infinite number of values.  So while this expansion accomplishes similar feats, it can be seen as a less economical parametrization of the model.  Of course, the character expansion from before can be found within the Taylor expansion by summing the even and odd integers, respectively, leaving one with two terms (0 and 1).

\subsection{Real-time evolution with qubits}
\label{subsec:buildingblocks}
We are now ready to discuss the possibility of using quantum devices to represent states of the Hilbert space emerging from the
transfer matrix construction and to design methods to apply unitary transformations corresponding to the real time evolution.
For the sake of definiteness, we assume that we have at our disposal a set of qubits. For the spin Ising model, the construction of the transfer matrix
leads to a Hilbert space with an obvious qubit structure given by Eq.~\eqref{eq:isingqbits}. The method to take the time continuum limit and identify a Hamiltonian using Eq.~\eqref{eq:tm} is well-known \cite{fradkin78,kogut79}: we deform the original transfer matrix by increasing $\beta$ in the time direction
(which makes the dual value $\tilde{\beta}$ small) and decreasing $\beta$ in the spatial directions. The arguments of the exponentials in $\vone$ and $\vtwo$ become infinitesimal and provide the two non-commuting pieces of the ``quantum Hamiltonian". The role played by $\vone$ is quite special in the intuitive picture that
we are trying to convey: it only acts on
``single qubits" without connecting them together. Consequently, working in the representation where $\hat{\sigma}^{x}_{\bf x}$ is diagonal is a good starting point. Next, we can ``turn on" the terms in $\vtwo$. At lowest order in the time lattice spacing they only connect qubits which are nearest neighbors.
We will show in  Sec.~\ref{subsec:lloyd} that this type of situation permits quantum computation in a time scaling polynomially with the size of the system \cite{lloyd96}.

Character expansions and \tft provide the natural tools to perform similar constructions for the models presented in Sec.~\ref{subsec:lattac}.
A first step consists in isolating ``building blocks" which are localized in space and
have a very simple real-time evolution.
The models have interactions associated with links and plaquettes. As a first approximation we set the interactions
on spatial links and space-space plaquettes to zero.

For spin models, this results in a collection of $N_s^{D-1}$
isolated one-dimensional spin models. These isolated models are the building blocks. They are solvable and it is easy to calculate their evolution at real time. For gauge models we have, in addition, electric degrees of freedom which can be associated with the spatial links of a given time slice of the lattice. They are required to satisfy a constraint called ``Gauss's law", however, when this condition is satisfied, the real-time evolution in the isolation limit described above is straightforward as we will explain in Sec. \ref{subsec:gaugehilbert}. The Hilbert space of the isolated building blocks depends on the model considered. For the Ising spin models, a single qubit is all we need. For models with continuous symmetries, the exact treatment requires an infinite Hilbert space, however, small
size truncations provide good approximations and preserve
the symmetries of the models \cite{meurice2019,meurice2020}.
This is a very attractive feature of \tft which is discussed in Sec.~\ref{subsec:symmetries}.
Having setup a finite Hilbert space with isolated building blocks, our next step is to restore the interactions associated with the spatial links and the space-space plaquettes. This will be done for a variety of models in Sec.~\ref{sec:transfer}. Independently of the model-specific aspect of
this procedure, it is clear that each building block is only connected to a limited numbers of other building blocks.
For instance, for a spin model, there are two connections for each spatial direction.
This quasilocality is crucial to implement real-time evolution with a quantum computer.

\subsection{Lloyd-Suzuki-Trotter product formula}
\label{subsec:lloyd}

An important motivation for using a quantum computer is to calculate the real-time evolution for systems with many degrees of freedom having
quasilocal interactions in the sense discussed in  Sec.~\ref{subsec:buildingblocks}.  Ideally the time to perform computations should scale
polynomially with the size of the system rather than exponentially.
A general argument  leading to these conclusions has
been put forward by S. Lloyd in   \cite{lloyd96} where he states that: ``Feynman's 1982 conjecture, that quantum computers can be programmed to simulate any local quantum system, is shown to be correct".
The proof is based on the basic idea behind the Suzuki-Trotter product formula \cite{trotter59,suzuki76,reed1980methods}, namely that for two non-commuting operators $\hat{A}$ and $\hat{B}$ and sufficiently small $\epsilon$ \beq
e^{i\epsilon(\hat{A}+\hat{B})}\simeq e^{i\epsilon \hat{A}}e^{i\epsilon \hat{B}}+\mathcal{O}(\epsilon^2).
\enq
In the standard construction of the path integral in quantum mechanics, it is applied to the kinetic and potential energy, but it can also be applied to all the quasilocal parts of the Hamiltonian.

The argument goes as follows \cite{lloyd96}.
Consider a system
composed of $N$ variables with Hamiltonian
\beq
\hat{H}=\sum_{j=1}^\ell \hat{H}_j
\enq
where each $\hat{H}_j$ acts on
a space that involves at
most $k_{max}$ of the variables. It is assumed that $\ell$
increases linearly with $N$ but that $k_{max}$ is fixed by the dimension and the symmetries and independent of $N$. The individual $H_j$ can be represented as finite matrices in their local subspace. Under these assumptions, it is shown \cite{lloyd96} that the error associated with the approximation
\beq
e^{i \hat{H} t} \simeq (e^ {i \hat{H}_1 t/n} \dots\  e^{i \hat{H}_\ell t/n})^n+\dots
\enq
can be controlled by taking $n$ large enough. In addition, once the accuracy goal is determined, the computing time scales linearly in $N$ and $t$.

\begin{figure}
  \vskip-3.5cm
  \includegraphics[width=8.5cm]{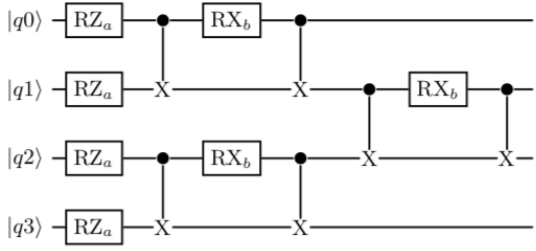}
  \vskip-3.5cm
  \caption{\label{fig:circuit}Circuit for 4 qubits with open boundary conditions used in~\cite{gustafson2019a}.  Here $RZ$ and $RX$ are rotations about the $z$ and $x$-axes, respectively.}
\end{figure}

In order to fix the ideas, for most of the available quantum computers, the $\hat{H}_j$ act on one or two qubits and can be represented
by $2\times 2$ or $4\times 4 $ matrices in this restricted space.
A simple quantum circuit for the quantum Ising model used in Refs.~\cite{gustafson2019a,Gustafson2021} is displayed in
Fig.~\ref{fig:circuit}. The basic elements are rotations generated by $\hat{\sigma} _j^x$ or $\hat{\sigma} _j^z$ and acting on the $j$-th qubit only and CNOT gates
acting on a pair of qubits and flipping the target qubit when the control qubit is in the $\ket{1}$ state. The circuit can be repeated in the space (vertical) and time (horizontal) directions and conveys the linear scalings stated above. 
More evolved circuits can be designed with the purpose of creating energy eigenstates \cite{fv2009,CerveraLierta2018exactisingmodel}. 

It is instructive to compare the
computational resources to
perform the unitary rotation $\exp(i\theta \hat{\sigma}^x_j\hat{\sigma}^x_k)$ for which the operator in the exponential flips the
$j$ and $k$-th qubit and does not act on  the rest of the system.
With a quantum computer, one expects that this operation keeps the coherence of the qubits with the rest of the system, present in the original state, but the cost is ideally independent on the size of the system. On the other hand, the same operation on an arbitrary state performed with a classical computer involves a matrix with $2^N$ non-zero and non-diagonal matrix elements and the resources necessary increase exponentially with the size of the system.

\subsection{Dealing with noise in the NISQ era}
\label{subsec:nisq}

The discussion in Sec.~\ref{subsec:lloyd} considers  algorithmic aspects in an idealized situation where
computer errors or noise can be neglected.
Quantum computing technology is still in its early development.  The current  Noisy Intermediate Scale Quantum (NISQ) computer hardware platforms can only accommodate small-depth circuits and sources of errors need to be understood in detail~\cite{Preskill2018quantumcomputingin}.  Understanding these errors and their noise, adjusting, and contriving algorithms to lessen these errors is the topic of error mitigation in quantum computing.

Various types of noise affect the single-qubit gates in a way that can be parameterized \cite{nielsen2000quantum} in terms of the density matrix $\hat{\rho}$:
\begin{align}
  \nonumber
  \label{eq:paulichannel}
  \mathcal{E}(\hat{\rho};p_x,p_y,p_z) = (1 - p) \hat{\rho} &+ p_x \hat{\sigma}^x \hat{\rho} \hat{\sigma}^x \\
                                                           &+ p_y \hat{\sigma}^y \hat{\rho} \hat{\sigma}^y + p_z \hat{\sigma}^z \hat{\rho} \hat{\sigma}^z.
\end{align}
Here $\mathcal{E}$ is a quantum operation, and the values $p_x$, $p_y$, and $p_z$ correspond to the probabilities of a $\sigma^x$, $\sigma^y$, or $\sigma^z$ error, respectively, occurring and $p = p_x + p_y + p_z$. The error channel for two qubit gates is given by $\mathcal{E}^{(2)} = \mathcal{E}\bigotimes \mathcal{E}$. Modeling the error in this way neglects spatial and temporal correlations, 2-qubit correlations, and assumes that the errors are identical and evenly distributed.  In practice, if a classical simulation of the qubit evolution is performed, each unitary evolution operation needs to be followed by applying one of the four possibilities $(\hat{1},\hat{\sigma}^x,\hat{\sigma}^y,\hat{\sigma}^z)$
with respective probabilities $1-p$, $p_x$, $p_y$, and $p_z$ on the qubits involved. The probability distribution of errors is only manifest after averaging over an ensemble of runs.

In addition, readout errors (misidentifying a $|1\rangle$ for a $|0\rangle$ or vice-versa)  should be taken into account for all current
quantum computing platforms.  Given estimates of the probabilities for these errors, it is possible to correct the actual measurements by a multiplicative factor
\cite{kandala2018}.

A common NISQ strategy for error mitigation is to increase the source of error in a controllable way and then extrapolate to the limit where the error is not present. Examples with
superconducting qubits are given in Refs. \cite{temme2017,kandala2019,klco18,gustafson2019a,Gustafson2021}.
A simple way to increase the error is to insert two successive CNOT gates. Their exact multiplication is the identity, however for a NISQ devices it increases the chances of errors. Note that mitigation methods can also be used for quantum variational methods \cite{mitigation}.

\begin{figure}[h]
  \centering
  \includegraphics[width=8cm]{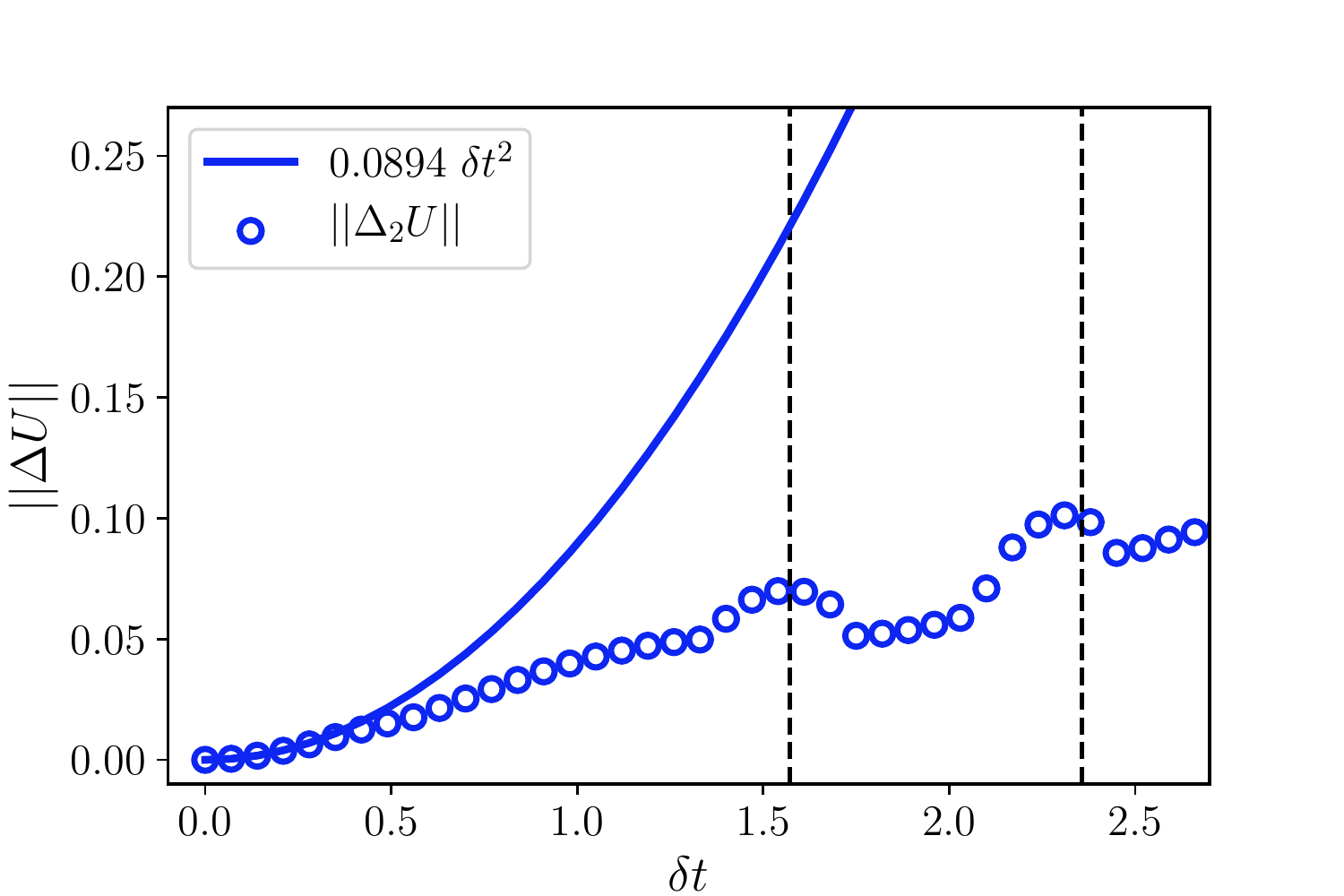}
  \caption{\label{fig:onestepu2} $||\Delta_2 U ||$ versus $\delta t$ for the quantum Ising model with $h_T=1$ and $J=0.02$.
    The vertical lines are at $\pi/2$, $3\pi/4$ .... For details see Ref . \cite{iopbook}}
\end{figure}

In this context, the choice of Trotter step $\delta t$ is crucial because the number of steps is limited by the loss of coherence and the noise. A concrete discussion can be found in  \cite{Gustafson2021} for the quantum Ising model in one spacial dimension with four sites.
If we pick a small $\delta t$ with a good control on the $(\delta t)^2$ error, we may not be able to reach a time scale relevant for what we want to learn. However, it appears that by picking a $\delta t$ significantly larger, the rigorous bound is not sharp and the empirical bound is much tighter as shown in Fig. \ref{fig:onestepu2} where we plot the operator norm of the Trotter error
\begin{eqnarray}
  \Delta_2U &\equiv&
                     e^{-i(h_T \hat{H}_T+J\hat{H}_{NN})\delta t}-e^{-ih_T \hat{H}_T\delta t}e^{-iJ\hat{H}_{NN}\delta t}\cr &\simeq &\frac{h_TJ}{2}[\hat{H}_T,\hat{H}_{NN}](\delta t)^2,
\end{eqnarray}
with 
\beq
\hat{H}_T=-\sum_{j=1}^4\hat{\sigma}^z _j, \ {\rm and} \  \hat{H}_{NN}=-\sum_{j=1}^3\hat{\sigma}^x_{j}\hat{\sigma}^x_{j+1}
\enq
The nonlinear aspects of the error are quite interesting and may involve
resonances \cite{Gustafson2021}. For other recent developments see \cite{Sewell:2021wbn,ARahman:2021ktn,Mishra:2019xbh,Gustafson:2020vqg,Gustafson:2021jtq,Gustafson:2021imb,Gustafson:2020yfe,Funcke:2021aps,Honda:2021aum,Gustafson2021}.

\subsection{Quantum computations and simulations}
The idea of using quantum devices to mimic or study theoretical quantum models has been a subject of intense activity in recent years. Here we mention theoretical and experimental studies of the Ising  model, and lattice gauge theories, in the context of quantum computation and quantum simulation.

\subsubsection{Ising model}

As stressed before, the 1+1 dimensional Ising model is the quintessential model to use to check computational tools and proposed algorithms.  This is because the model possesses non-trivial features, but has been solved exactly.  It is not surprising then that this is one of the first models to be tried and tested in different forms of quantum computation.  There have been various investigations into how to simulate the model on a quantum computer, and what interesting observables to measure (\emph{e.g.} phase shift, and thermodynamics) \cite{PhysRevE.56.3661,PhysRevLett.101.220501,gustafson2019a,PhysRevA.87.032341}.  
The following approaches have been considered: super-conducting qubit machines, trapped-ion machines, cold atoms trapped in optical lattices, and Rydberg atom simulators.

On the super-conducting qubit front, Refs.~\cite{PhysRevA.95.052339,Zhang2017} made initial simulations, and in some cases at relatively large system sizes.  Their calculations of various spin observables matched the corresponding quantities in exact diagonalization well.  In  \cite{CerveraLierta2018exactisingmodel} a simulation of the Ising model using a few spins was also carried out, and a comparison between theory and computation was made for the average magnetization. In Refs.~\cite{Gustafson2021,Smith2019}, the Ising model was simulated using a few qubits on IMB's machines.  It was shown that the Richardson extrapolation could be used to mitigate the noise 
in the regime where the nonlinear effects are not too large \cite{Gustafson2021}.
  In addition, in  \cite{Chen2019} a quantum-classical approach was taken which uses a variational algorithm to compute the ground-state wave function; the so-called Variational Quantum Computing, where the quantum computer prepares a state with a circuit ansatz depending on some number of parameters.  The expectation value of the energy is computed from this state, and circuit parameters are then tuned classically based on the quantum expectation value in order to minimize the energy.

Using trapped ions, the first work was \cite{Friedenauer2008}, followed by extensive investigations from student's work~\cite{Friedenauer2010,Korenblit2013}.  These are pioneering investigations into the trapped-ion platform.  Refs.~\cite{kim2010,PhysRevB.82.060412,Islam2011,Kim_2011} look at the phase structure of the model by simulating a few spins.  They calculate the phase diagram of the model using the  probability of the state to be in a ferromagnetic state, or moments of the magnetization.  There has also been an investigation into measuring the R\'{e}nyi entropy using digital quantum gates.  In~\cite{PhysRevA.98.052334} the authors consider a two-site antiferromagnetic Ising model, and using the SWAP gate measure the parity of two copies of the system~\cite{PhysRevB.96.195136}.  Ref.~\cite{islam2015} used bosonic many-body states and interfered the copies to extract sub-system parities.  They then calculate the R\'{e}nyi entropy from sub-system parities.

Finally, another promising approach is to use highly excited (Rydberg) states of atoms, which allows for strong atom-atom interactions across relatively large distances.  In  \cite{Kim:17} the group uses a chain of 19 Rubidium atoms whose interactions they control through tuning lattice parameters to simulate the model.  In  \cite{PhysRevX.8.021069} out-of-equilibrium dynamics are explored through a quench using an array of Lithium atoms, again placed in a Rydberg state.  Reference \cite{Simon2011} uses trapped Rubidium atoms to simulate the anti-ferromagnetic Ising model.  They are able to identify a phase transition between para- and anti-ferromagnetic phases, and observe magnetic domains using a site-resolved atomic microscope and noise correlations measurements.  In \cite{51qubits} the authors demonstrate the use of configurable and  programmable arrays of atoms, and simulate an Ising-like model on 51 qubits.  They observe a phase transition between symmetric and ordered phases and discuss out-of-equilibrium properties of spin models \cite{keesling2019}.  For reviews of this topic see  \cite{Schauss_2018,rydberg2020}.

\subsubsection{Gauge theories}
\label{subsubsec:gt}

The use of optical lattices \cite{Bloch2008} to quantum simulate lattice gauge theories has been developed extensively.
Proposals for the quantum simulation of lattice gauge theories beginning with early work on Abelian models includes~\cite{Zohar:2011cw,Zohar:2012ay,Banerjee:2012pg,celi2012,Zohar:2013zla}, and for digital quantum devices in~\cite{Zohar:2016wmo,Zohar:2016iic}.
In the case of non-Abelian models:~\cite{Banerjee:2012xg,Zohar:2012xf,Tagliacozzo:2012df}.
For reviews see  \cite{wiese2013,zohar2015,cinchy2019,banuls2020}. 
A useful early reference on quantum computing for non-Abelian gauge theories is \cite{byrnes2006}.
For recent developments combining condensed matter and gauge theory ideas see \cite{tracyli2016,kasper2017,chin2018,schweizer2019}.
Trapped ions \cite{monroe2003} have provided new opportunities to approach
lattice gauge theory models \cite{davoudi2019}. Rydberg atoms
offer a versatile platform for gauge theories \cite{celi2019,surace2020,meurice2021}.

The Schwinger model is often the first target to develop new approaches \cite{martinez2016,kharzeev2020,davoudi2019,kasper2017,klco18,surace2020,magnifico2019}.
For recent work on non-Abelian models see \cite{silvi2019,raych2018,raych2019,davoudi2020,dasgupta2020}.

\section{Quantum versus classical}
\label{sec:qvc}
In this section, we discuss the meaning of ``classical" and ``quantum" for models, phase transitions and tensor networks.

\subsection{Models}
\label{subsec:classicalvsquantum}
In textbooks on quantum mechanics, it is a common procedure to start with the Hamiltonian formulation and derive a path-integral representation which can be extended to field theory. The path-integral formalism allows for formulations that are manifestly gauge-invariant and treats space and time on equal footing. For these reasons, it can be argued that the fundamental definition of relativistic  models should be done in terms of the action and the measure of integration over all the possible configurations without reference to a Hamiltonian in the first place. Examples of such actions are given in  Sec.~\ref{subsec:lattac}.

It is very common to call models formulated with the path-integral  ``classical" while the corresponding formulation using a Hamiltonian acting on a Hilbert space is called ``quantum".
However, except for possible discretization artifacts, the two formulations describe the same quantum behavior. In the path-integral formulation for bosonic fields, the action is calculated in terms of c-numbers as in the classical formulation but the sum over all the configurations provides a quantum description which
amounts to more than
the classical equations of motion. In other words, the path-integral is an alternate method of quantization which is very convenient in Euclidean time.

Starting with a classical action, a Hamiltonian can be constructed  from the action by using the transfer matrix formalism.
This was briefly demonstrated in Sec.~\ref{subsec:extm}, and will be discussed in detail in Sec.~\ref{sec:transfer} for the models introduced in  Sec.~\ref{subsec:lattac}.
A discussion with detailed references on the connection between statistical mechanics in $D$ dimensions and quantum Hamiltonians in $D-1$ dimensions,
can be found in the classic work of Wilson and Kogut \cite{wilson73pr}.

\subsection{Phase transitions}
The actions for spin and gauge models introduced in  Sec.~\ref{subsec:lattac} contain parameters $\beta$,
or $\beta_{pl.}$ for pure gauge theories, that are often called the ``inverse
classical temperature", or ``coupling constant'', and can be associated with ``classical phase transitions". For instance the $D=2$ classical Ising model has a spontaneous magnetization when $\beta > \beta_c=(1/2)\ln(1+\sqrt{2})$.

In contrast, given a Hamiltonian $\hat{H}$, we can define a thermal quantum partition function with temperature $T_{qu.}$ in the usual way
\beq
Z(T_{qu.})=\Tr [e^{-\hat{H}/T_{qu.}}],
\enq
where $T_{qu.}$ has in general a different meaning than $1/\beta$ in the classical formulation.
With the lattice formulation at Euclidean time as a starting point, we have the identification
\beq
1/T_{qu.}=N_\tau a_\tau,
\enq
and the non-zero temperature is associated with the finite extent of the temporal dimension.
A typical situation of interest is to start in an ordered phase at $T_{qu.}=0$, corresponding to the infinite Euclidean
time limit, and induce a finite-temperature phase transition into a disordered phase by
taking a sufficiently small temporal extent. In this way, the temporal extent of the lattice is responsible for a finite-temperature phase transition, as opposed to the coupling $\beta$, which is unrelated to the temperature in the quantum partition function.  Transitions related to the couplings at $T_{qu}=0$ are quantum phase transitions. Sometimes, the transition can be understood in terms of the classical phase diagram in $D-1$ dimensions. A more detailed discussion
can be found in Cardy's monograph \cite{cardy1996scaling}

\subsection{Tensor networks}
\label{subsec:tn}

In Secs.~\ref{sec:ising} to~\ref{sec:fermions} we will introduce ``classical tensors" in
order to reformulate classical models as defined in  Sec.~\ref{subsec:classicalvsquantum}. The partition function of these models can be
visualized as an assembly obtained by ``wiring" (tracing) together objects carrying
multiple ``legs" (tensor indices) and attached to the sites, links or plaquettes of
a Euclidean space-time lattice. This type of classical constructions has been inspired \cite{Levin:2006jai,Gu:2009dr} by various quantum tensor networks
\cite{fannes92,vidal2003,vidal2004,verstraete2004,shi2005,perezgarcia2006,vidal2007,verstraete2008}, to mention a few references,
developed in various contexts often related to the density matrix renormalization group (DMRG) method \cite{white92}. There is abundant literature on the subject
reviewed for instance in \cite{schollwock2005,verstraete2008,cirac2009,schollwock2011,orus2013,silvi2017,montangero2018,ran2020}.

\begin{figure}[h]
  \centering
  \includegraphics[width=0.8\hsize]{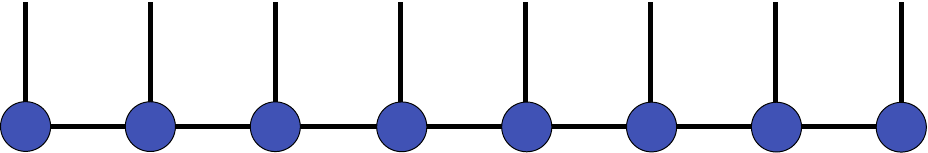}
  \caption{\label{fig:mps}Illustration of states  in the MPS approach.}
\end{figure}

\begin{figure}[h]
  \centering
  \includegraphics[width=0.8\hsize]{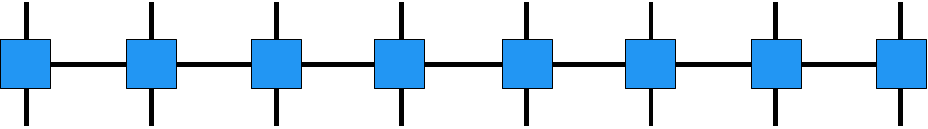}
  \caption{\label{fig:mpsop}Illustration of  operators in the MPS approach.}
\end{figure}

One important idea is the representation of quantum states by matrix product states (MPS) which appear in several references mentioned above. As an example, for a  one-dimensional quantum chain problem with $N_s$ sites, an arbitrary element of the Hilbert space can be written as
\beq
\ket{\psi}=\sum_{i_1,\dots,i_{N_s}}c_{i_1,\dots,i_{N_s}}\ket{i_1,\dots,i_{N_s}},
\enq
where
\beq
\ket{i_1,\dots,i_{N_s}}=\ket{i_1}\otimes\dots\otimes \ket{i_{N_s}},\enq
and each of the indices runs over a local Hilbert space of dimension $d_H$ attached to a site.
The dimension of the Hilbert space is $d_H^{N_s}$. It represents an exponential growth with the size of the system that rapidly becomes computationally unmanageable.
For a MPS state, one assumes the form
\beq
c_{i_1,\dots,i_{N_s}}=\Tr [A_{i_1}\dots A_{i_{N_s}}],
\label{eq:mps}
\enq
where the $A_{i_j}$ are $d_B \times d_B$ matrices for each value of $i_j$.
$d_B$ is called the bond dimension.
A graphical representation of such a state is shown in Fig.~\ref{fig:mps} for open boundary conditions. The filled circles represent the matrices,
the vertical lines represent open indices with $d_H$ values and the horizontal lines traced indices with $d_B$ values.
The size of the MPS ``subset" \footnote{The sum of two MPS is not a MPS, so we will not call it a ``subspace".} 
is only growing like $N_s \times d_B^2 \times d_H$, so linearly with the size of the system.
Similarly, one can represent operators using the trace of $N_s$  $d_B \times d_B$ matrices $A_{i_j}^{i'_{j}}$ with two indices in the one-site Hilbert space at a computational cost scaling like  $N_s \times d_B^2 \times d_H^2$. This is illustrated in Fig.~\ref{fig:mpsop}.
Objects with similar shapes will appear as ``times slices" of the classical construction. In the case where the indices take an infinite number of values and
the sums are truncated in a way that the number of indices kept in the time direction
(similar to $d_H$), is the same as the number in the space directions (similar to  $d_B$), we will use the isotropic notation $D_{\text{cut}}$ (one can of course leave the number of states in the temporal and spatial directions independent of each other). On the other
hand, we will discuss the anisotropic situation where the time direction is
singled out to define the transfer matrix and the Hamiltonian in Sec.~\ref{sec:transfer}.

MPS and the DMRG have been used in studies of 1+1 dimensional models such as the Bose-Hubbard model~\cite{bonnes}, the Schwinger model~\cite{byrnes2002,Banuls:2013jaa,buyens2014,buyens2016}, see Sec. \ref{subsec:addfermions} for more information, SU(2) gauge theory~\cite{banuls2017,kuhn2015}, and the
O(3) nonlinear sigma model~\cite{bruckmann2018}.  Fermionic tensor network studies, and the Hubbard model were discussed in \cite{PhysRevA.80.042333,corboz2010}.  Tensor network techniques for lattice gauge theories are also discussed in  ~\cite{celi2014,silvi2016,silvi2019,rico2014,pichler2016,zohar2015b} and reviewed in~\cite{banuls2018,cinchy2019}.  For a review of matrix product operators and their relations with the transfer matrix see~\cite{hv2017}.  The reader is directed to Sec.~\ref{subsec:additional} for references on tensor network studies in 2+1, and 3+1 dimensions using generalizations of MPS.

The MPS framework can also be used to perform real-time calculations based on the Suzuki-Trotter approximations. An example of method is called the time-evolving block decimation reviewed in \cite{tbde}.

The success and limitations of the MPS approach can be analyzed in terms of ``area laws" \cite{plenio2005,Verstraete:2006mgt}. Following a short pedagogical discussion \cite{schollwock2011}, if a bipartition of a system is introduced, one expects that the entanglement entropy between the two parts scales like the size of their boundary with possible logarithmic corrections. In one spatial dimension, we can separate a MPS in two parts by cutting a single bound carrying a maximal entropy $\ln_2 d_B$. Consequently, $d_B$ only needs to increase like the size of the system to capture a possibly logarithmic entanglement. On the other hand, if we fill a two-dimensional surface with a one-dimensional MPS, the entanglement grows at least like the linear size of the system which forces $d_b$ to grow exponentially with this size. Projected Entangled Pair states (PEPS)  were proposed \cite{PhysRevA.70.060302} to try to overcome this difficulty. 

It should also be mentioned that tensor methods have been applied \cite{nishinoctm} to the transfer matrix treatment of classical statistical model such as the problem of monomers and dimers on a rectangular lattice \cite{baxter68} where variational methods can be applied and compared to the DMRG 
approach. This approach is called the corner transfer matrix renormalization group method. 
Pioneering work connecting the DMRG and the transfer martix of the Ising model can be found in \cite{nishinoctm}. 
Related results and their connections with MPS are reviewed in \cite{hv2017}. 

For a very recent and comprehensive review on MPS and related topics, we recommend \cite{Cirac:2020obd}.

\section{Tensor methods explained with the Ising model}
\label{sec:ising}
\subsection{Tensor formulation}
\label{subsec:tensorising}
\begin{figure}[h]
  \includegraphics[width=4cm]{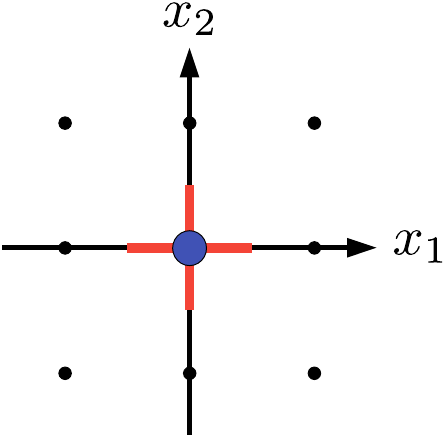}
  \caption{\label{fig:t1} Basic tensor for  $D=2$.  The indices of the tensor are shown as red ``legs'' of the purple ``body''.  Diagrams of tensors this way are ubiquitous in the literature.  The lattice is shown in black, with the centers of the plaquettes, and the half-way marks on the links, indicated by black dots. }
\end{figure}

\begin{figure}[h]
  \includegraphics[width=0.3\hsize]{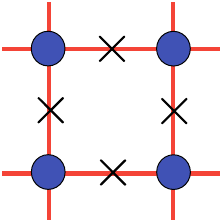}
  \caption{\label{fig:t2} Tensor assembly for $D=2$. The crosses mean contraction.}
\end{figure}

\begin{figure}[h]
  \includegraphics[width=4cm]{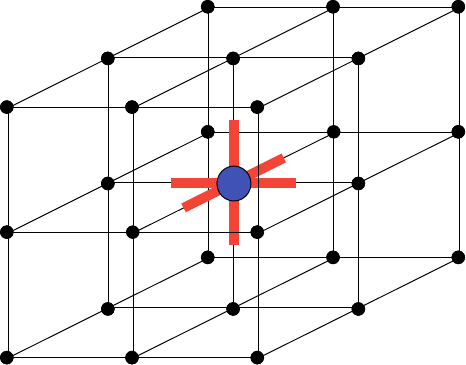}
  \caption{\label{fig:t3} Basic tensor for  $D=3$.  The six indices of the local tensor are shown as the red legs, and live along the links of the lattice.}
\end{figure}

\begin{figure}[h]
  \includegraphics[width=4cm]{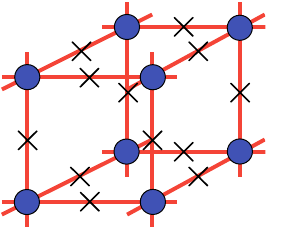}
  \caption{\label{fig:t4} Tensor assembly for $D=3$. The crosses mean contraction.}
\end{figure}

In this section we will construct a tensor formulation for the Ising model in $D$ dimensions.
The partition function for the Ising model is
\beq
Z_{\text{Ising}} = \prod_x\sum_{\sigma_x=\pm 1}e^{\beta  \sum_{x,\mu}\sigma_{x+\hat{\mu}}\sigma_x}.
\enq
For  each link $(x,\mu )$
we use the expansion
\begin{align}
  \nonumber
  &e^{\beta  \sigma_{x+\hat{\mu}}\sigma_x} = \\
  &\cosh(\beta)\sum\limits_{n_{x,\mu}=0,1} [\sigma_{x+\hat{\mu}}\sqrt{\tanh(\beta)}\sigma_x\sqrt{\tanh(\beta)}\ ]^{n_{x,\mu}} ,
    \label{eq:char}
\end{align}
This attaches an index $n_{x,\mu}$ at each link  $(x,\mu )$.
It is then possible to pull together the various factors of $(\sqrt{\tanh(\beta)}\sigma_x)^{n_{x,\mu}}$
from links coming from a single site $x$,
and sum over $\sigma_x$,
\begin{align}
  \nonumber
  &\sum_{\sigma_{x}=\pm 1} \prod_{\mu=1}^{D} (\sqrt{\tanh(\beta)} \sigma_{x})^{n_{x-\hat{\mu},\mu} + n_{x, \mu}} \\ \nonumber
  &= (\sqrt{\tanh(\beta)})^{\sum_{\mu} n_{x-\hat{\mu},\mu} + n_{x, \mu}} \\
  &\times 2 \delta({\rm mod}[\sum_{\mu} n_{x-\hat{\mu},\mu} + n_{x, \mu},2])
\end{align}
using the relation
\beq
\sum_{\sigma_{x}=\pm 1}\sigma_{x}^n =2\delta({\rm mod}[n,2]).
\enq
at every site on the lattice. The expression $\delta({\rm mod}[n,2])$ is 1 when $n$ is even (0 modulo 2) and 0 otherwise.  We can rewrite the partition function as the trace of a tensor product,
\beq
Z=2^{V} (\cosh(\beta))^{VD} \; {\Tr}\left[ \prod_x T^{(x)}_{n_{x-\hat{1},1}, n_{x,1},\dots,n_{x,D}} \right].
\label{eq:ising-trace}
\enq
$\Tr$ means contractions (sums over 0 and 1) over the link indices (the $n_{x, \mu}$s).
The local tensor $T^{(x)}$ has $2D$ indices. The explicit form is
\begin{align}
  \label{eq:isingtensor}
  \nonumber
  &T^{(x)}_{n_{x-\hat{1},1}, n_{x,1},\dots,n_{x-\hat{D},D},n_{x,D}} = \\ \nonumber
  &(\sqrt{\tanh(\beta)})^{n_{x,{\rm in}} + n_{x, {\rm out}}} \\
  &\times \delta({\rm mod}[n_{x,{\rm in}} + n_{x, {\rm out}},2])
\end{align}
with
the definitions
\begin{align}
  \label{eq:isingdefs}
  n_{x,{\rm in}} & \equiv \sum_{\mu =1}^D n_{x-\hat{\mu},\mu} \\ \nonumber
  n_{x,{\rm out}} & \equiv \sum_{\mu=1}^D n_{x,\mu}.
\end{align}
The notions of ``in" and ``out" refer to position of the link with respect to the positive directions.
The basic tensors and their assembly in two and three dimensions are illustrated in Figs. \ref{fig:t1}-\ref{fig:t4}.

Any kind of boundary condition can be accommodated by adapting the method of integration to the link configuration at the boundary. This will be discussed in subsection
\ref{subsec:bc}.

The Kronecker delta in  Eq.~(\ref{eq:isingtensor}) implies the discrete conservation law
\beq
\sum_\mu(n_{x,\mu}-n_{x-\hat{\mu},\mu})=0\ {\rm  mod}\  2,
\label{eq:isingnoether}
\enq
which we also call the ``tensor selection rule".
It implies that  only an even number of $n$s are allowed to take on the value one.
For instance, for $D=2$, there are in principle 16 tensor values, however, only eight are nonzero,
one with all four indices as zero (zeroth order in $\tanh(\beta)$), six with two zeros and two ones (linear in $\tanh(\beta)$),
and one with four ones (quadratic in $\tanh(\beta)$).
Note that if a symmetry breaking term like a magnetic field coupling to the total spin is introduced, then
all the 16 tensors elements will be nonzero generally.

There are also tensor formulations which use the singular value decomposition (See~\ref{sec:svd-review}) on each nearest-neighbor Boltzmann factor to factorize the spins. The nearest neighbor interactions can be represented as a matrix, whose indices are the spin variables themselves,
\begin{align}
  \label{eq:ebetasigma}
  e^{\beta \sigma_{x+\hat{\mu}} \sigma_{x}} =
  \begin{pmatrix}
    e^{\beta} & e^{-\beta} \\
    e^{-\beta} & e^{\beta}
  \end{pmatrix}_{(x,\mu)}.
\end{align}
One can then of course perform the singular value decomposition to get,
\begin{align}
  \label{eq:ising-nn-svd}
  e^{\beta \sigma_{x+\hat{\mu}} \sigma_{x}} = \sum_{\alpha,\beta} U_{\sigma_{x+\hat{\mu}} \alpha} \lambda_{\alpha \beta} U^{T}_{\beta \sigma_{x}},
\end{align}
and in the case of the Ising model the singular value decomposition has the same left and right unitary matrices.
This factorizes the spins and allows for the definition
of the matrix,
\begin{align}
  W_{\sigma_{x} \alpha} \equiv U_{\sigma_{x} \alpha} \sqrt{\lambda_{\alpha}}.
\end{align}
The main local tensor is then the contraction of all $W$ matrices which have a common spin,
\begin{align}
  T_{i j\cdots N} = \sum_{\sigma_{x}} W_{\sigma_{x} i} W_{\sigma_{x} j} \cdots W_{\sigma_{x} N}.
\end{align}

In fact, the singular value decomposition in this case can be completely related to the expansion in Eq.~\eqref{eq:char}.
Consider the character expansion from Sec.~\ref{subsec:tensorising} for the nearest neighbor interaction,
\begin{align}
  \label{eq:ising-nn-charac}
  e^{\beta \sigma_{x+\hat{\mu}} \sigma_{x}} = \sum_{n_{x,\mu} = 0}^{1} H_{x+\hat{\mu}} C_{n_{x,\mu}}(\beta) H_{x}
\end{align}
with matrices,
\begin{align}
  \label{eq:sigmatrix}
  H_{x} \equiv \frac{1}{\sqrt{2}}
  \begin{pmatrix}
    1 & 1 \\
    1 & -1
  \end{pmatrix}
\end{align}
whose columns are the normalized eigenvectors of the Pauli-$x$ matrix, and $C_{0} = 2 \cosh\beta$ and $C_{1} = 2\sinh\beta$.  The elements of this matrix are from the four values $n_{x,\mu}$ and $\sigma_{x}$ can take, normalized to be unitary.
Since the variables are compact, operators of them have a discrete spectrum, and the character expansion is exactly the spectrum decomposition for this matrix.  For any matrix, $M$, by definition the singular value decomposition is found from the eigenvalues and eigenvectors of $MM^{\dagger}$ and $M^{\dagger} M$.  In this case (we drop any specific spacetime lattice indices for these steps since they are completely general),
\begin{align}
  \nonumber
  \sum_{\sigma_{j}} e^{\beta \sigma_{i} \sigma_{j}}e^{\beta \sigma_{j} \sigma_{k}} &= \sum_{\sigma_{j}} \sum_{n,m} \sigma_{i}^n C_{n} \sigma_{j}^{n} \sigma_{j}^{m} C_{m} \sigma_{k}^{m} \\ \nonumber
                                                                                   &=  \sum_{n,m} \sigma_{i}^{n} C_{n} C_{m} \sigma_{k}^{m} \delta_{n,m} \\
                                                                                   &= \sum_{n} \sigma_{i}^{n} C_{n}^{2} \sigma_{k}^{n}.
\end{align}
Then the singular values are given by $\lambda_{1} = 2 |\cosh\beta|$ and $\lambda_{2} = 2 |\sinh\beta |$, as one would expect from Eq.~\eqref{eq:ising-nn-svd} and Eq.~\eqref{eq:ising-nn-charac}.

\subsection{Remarks on the forms of duality}
\label{subsec:isingdual}

The tensor representation can be used to reproduce the set of
closed paths appearing in the expansion in powers of $\tanh(\beta)$ \cite{itzykson1991statistical,parisi1998statistical} for the Ising model. The links associated with the set of indices $n_{x,\mu}=1$ form a graph (a set of sites connected by links). The selection rule means that each site is attached to an even number of nonzero links. These graphs are closed paths with specific connectivity which can in principle be enumerated order by order in their length using geometric constructions and combinatoric techniques.

One can try to construct these closed paths by assembling the most elementary contributions, namely closed loops on a single plaquette.
The way they can be assembled depends on the dimension. For instance
for $D=2$, we can decide that when two loops around two plaquettes share a link, this link is ``erased" from the path. Alternatively, one can introduce ``dual variables" originally concieved by Kramers and Wannier \cite{PhysRev.60.252} which are spins on the dual lattice
located at the centers of the plaquettes.  Each dual spin is then associated with a closed plaquette loop around it on the original lattice.

\begin{figure}[t]
  \centering
  \includegraphics[width=0.7\hsize]{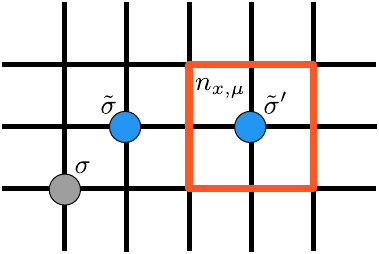}
  \caption{The distinct dual variables shown on the original lattice (black lines), and the dual lattice (gray lines).  Two dual spins are denoted with a tilde on the dual sites, an original spin, $\sigma$, on the original lattice, and the Pontryagin dual shown in red crossing  the dual link.  A closed loop of Pontryagin duals is shown all in red generally.}
  \label{fig:all-duals}
\end{figure}

Furthermore, the Pontryagin dual variables $n_{x,\mu}$ can be expressed as $(1-\tilde{\sigma} \tilde{\sigma}')/2$ with $\tilde{\sigma}$ and  $ \tilde{\sigma}'$
the two dual spin variables connected by a dual link crossing the
original link.  A picture showing the dual variables in their different locations can be seen in Fig.~\ref{fig:all-duals}. This illustrates that the notions of duality are often combined in a way that may be seen as confusing at first sight.
Note that the dual domains with a given dual spin have boundaries that can be interpreted as the closed paths of the of the original model.
The questions of completeness and multiplicity need to be addressed with specific boundary conditions.

Similarly in higher dimensions it is possible to
introduce dual spins with interactions involving $2(D-1)$ spins. For $D=3$, this leads to a gauge theory with plaquette interactions \cite{wegner71}.
It is also possible to introduce dual variables for Ising models with arbitrary all-to-all spin interactions \cite{meurice94}.
Duality questions related to Gauss's law will be discussed in  Sec.~\ref{subsec:moredual}.

\subsection{Boundary conditions}
\label{subsec:bc}

\begin{figure}[h]
  \centering
  \includegraphics[width=\hsize]{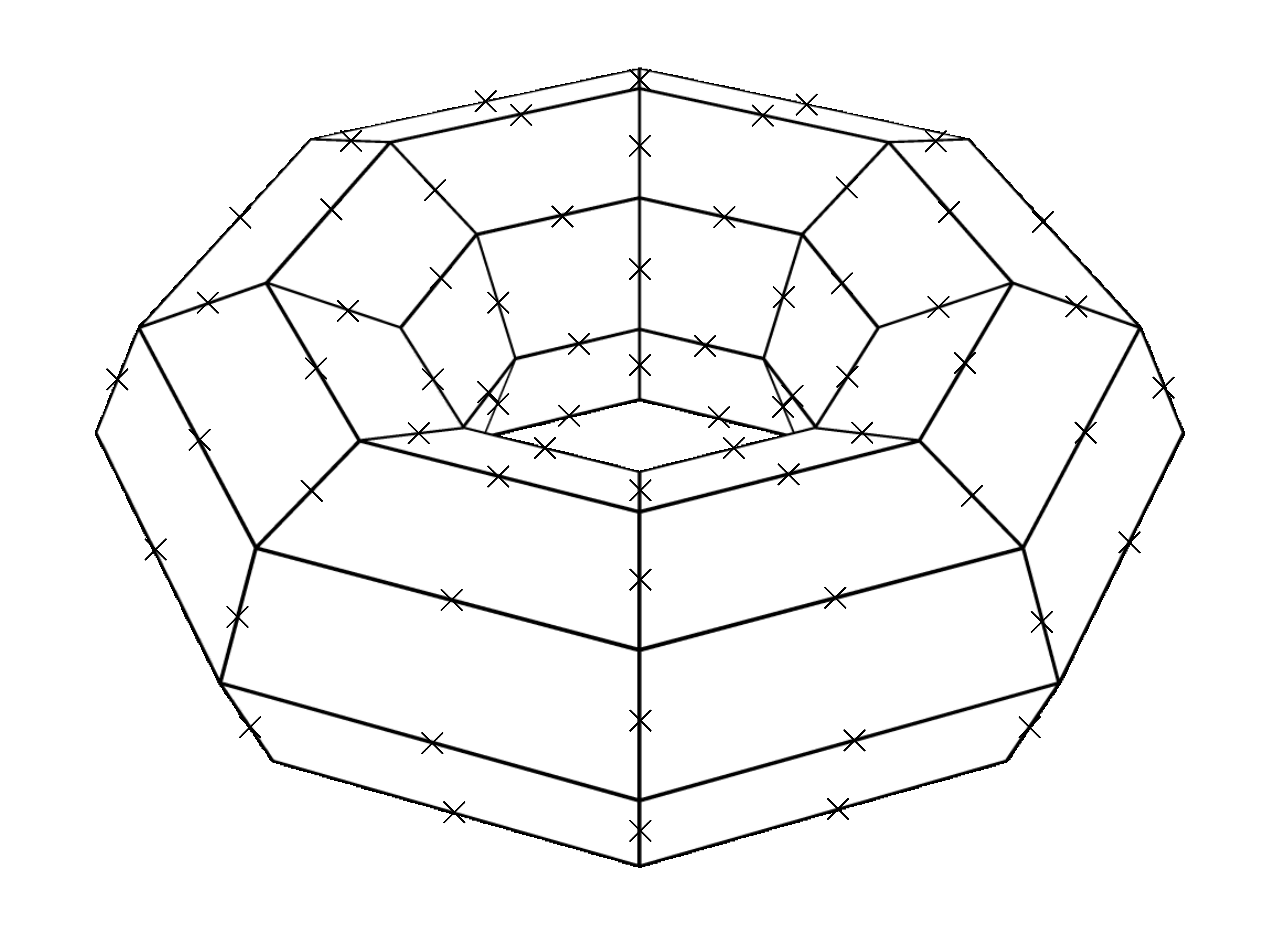}
  \includegraphics[width=0.7\hsize]{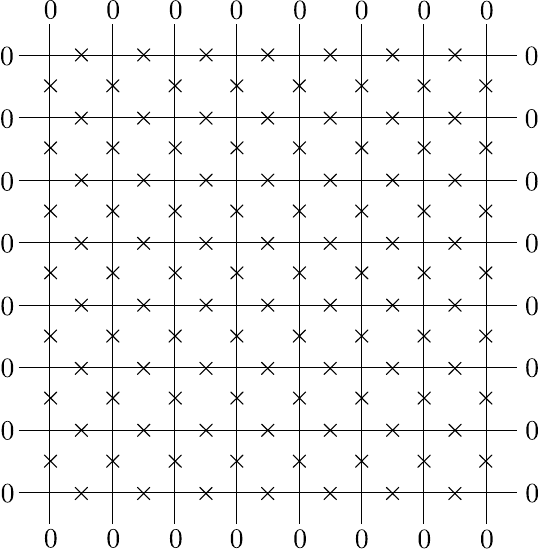}
  \caption{\label{fig:bc}Assembling the translation invariant tensor with PBC (top), or using new tensors at the boundary  for OBC (bottom).
    Tensors are assumed to be put on each lattice site.}
\end{figure}

In Eq.~\eqref{eq:ising-trace}, the trace is a sum over all the link indices. We need to specify the boundary conditions.
Periodic boundary conditions (PBC) allow us to keep a discrete translational invariance: the tensors themselves are translation invariant and
assembled in the same way at every site.
Open boundary conditions (OBC) can also be implemented by introducing new tensors that can be placed at the boundary. Their construction is similar to the tensors in the bulk. The only difference is that
the ``outside links" which would be attached at sites on the boundary have an index set to zero. With the normalization introduced in  Eq.~\eqref{eq:char} the indices carrying a zero index carry a unit weight. This construction can be understood as decoupling the
system from a larger environment by setting $\beta$ on the links connecting to this environment to zero because $\tanh(0)^0 = 1$ and $\tanh(0)=0$. This is illustrated in Fig. \ref{fig:bc}.

\subsection{Exact blocking}
\label{sec:exact-blocking}
An important feature of the tensor representation presented in Sec.~\ref{subsec:tensorising} is that it allows an {\it exact} local blocking procedure in contrast
to what can be done in configuration space \footnote{We want to make clear that this section is {\it not} describing a RG  transformation but rather an exact  reorganization of the computation of the partition function in a way that performs the integration of some degrees of freedom corresponding to increasing distance scales. As explained in Sec. \ref{sec:trunc}, truncations need to be applied to define a RG transformation.}. We divide the original lattice into cells having a linear size of two lattice spacings (``blocks") in such a way that the boundaries
are half-way between lattice sites. In other words, the boundaries are normal to links and cut them in the middle. As an example for $D=2$, the blocks are square
enclosing four sites and their four sides cross eight links. This cell partition
separates the link degrees of freedom into two disjoint categories: those completely inside the block, which can be integrated over, and those shared by neighboring blocks and kept to communicate between the blocks.
This is a generic feature of the method \cite{prb87} which motivated a systematic study of lattice models \cite{Liu:2013nsa}.  Note that when translation invariance is present, all the blocks are identical and we only need to do one calculation.
\begin{figure}[h]
  \includegraphics[width=1.9in,angle=0]{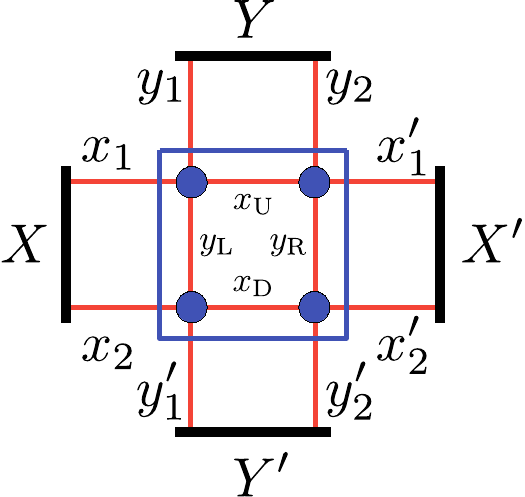}
  \caption{Graphical representation of the block (dotted square) and $T'_{XX'YY'}$
    as in   \cite{prb87}.
    \label{fig:square} }
\end{figure}

We now describe explicitly this exact  blocking for $D=2$ using generic notations inspired by those of Refs.~\cite{2012PhRvB..86d5139X,prb87}. The extension to higher dimensions on hypercubic lattices is straightforward. We contract the four tensors located at the four sites inside the block along the four indices located strictly inside the block as well.  The remaining eight indices associated with the eight links piercing the block are left as new degrees of freedom. By taking the tensor product between the two indices in each of the four directions coming out of the block, we obtain a
new rank-4 tensor $T'_{XX'YY'}$. In the case of the Ising model (See Eq.~\eqref{eq:isingtensor}),
each index now takes four values. This provides a simple isotropic formula:
\begin{align}
  \label{eq:square}
  & T'_{X({x_1} {x_2})X'(x_1' x_2')Y(y_1 y_2)Y'(y_1' y_2')} = \\ \nonumber
  & \sum_{x_U,x_D,x_R,x_L}T_{x_1 x_U y_1y_L}T_{x_Ux_1'y_2y_R}T_{x_Dx_2'y_R y_2'}T_{x_2x_Dy_Ly_1'}  ,
\end{align}
where $X(x_1 x_2)$ is a notation for the product states \emph{i.e.}, if we regroup the indices with the same parity together,
$X(0 0)=1,\  X(1 1)=2, \  X(1 0)=3,\  X(0 1)=4$.
This contraction and redefinition, relative to the block, is illustrated in Fig.~\ref{fig:square}.

After this blocking, the partition function can again be written as
\begin{equation*}
  Z=\Tr\prod_{2x}T'^{(2x)}_{XX'YY'} \ ,
  \label{eq:ZP}
\end{equation*}
where $2x$ denotes the sites of the coarser lattice with twice the lattice spacing of the original lattice.  This coarse-graining provides an exact representation of the original partition function. However, the number of states associated with each index is the square of the number of states in the original tensor.  If this exact procedure had to be carried numerically, this rapid growth would quickly run into practical limitations. Truncations are thus necessary.
It is important to appreciate that truncations are the {\it only} approximations that will be needed. We now proceed to discuss truncations.

\section{Tensor renormalization group}
\label{sec:trunc}
\subsection{Block-spinning through SVD}
\label{sec:lntrg}

Once partition functions and physical quantities are expressed as tensor networks,
one needs to contract them to obtain numerical values.
However, contracting exactly requires an extraordinary amount of computational resources.
Then, in this subsection we work through the original idea of how one can perform tensor contractions approximately, and how truncations appear in this approximation.
To do that a coarse-graining algorithm for tensor networks is introduced.
The original idea of such an approach was proposed by Levin and Nave in \cite{Levin:2006jai},
where tensor networks are simply coarse-grained by using the SVD (see Sec.~\ref{sec:svd-review}).
This method is similar in spirit to the real-space renormalization group approach,
and, in this sense, it is called the tensor renormalization group~\footnote{
There are several numerical renormalization group methods that can be regarded as ancestors of the TRG;
\textit{e.g.}~\cite{wilson74rmp,white92,White:1993zza,1995JPSJ...64.3598N,nishinoctm,1997cond.mat..5301W}. See also references the end of Sec. \ref{subsec:tn} and \cite{ueda2014}.
}

In the standard renormalization group procedure \cite{wilson73pr}, the blocking process is supplemented by
a sorting of the resulting information according to their degree of relevance. As far as universal properties characterizing the continuum limit, it is acceptable to discard the
information which only reflects the microscopic details of a specific lattice formulation.
In the context of the tensor formulation discussed above, it means that, possibly after a certain number of exact contractions, we need to restrict the number of states associated with the tensor indices to a fixed number $\ds$. We are then mapping a tensor with $\ds^{2D}$ entries into another tensor of the same shape and the question of fixed points becomes meaningful.
An important goal of the renormalization procedure is to identify fixed points.
Note that the updated tensor remains a local object which supersedes the notion of action or Hamiltonian.

Here we assume that a partition function is expressed as a two dimensional tensor network with bond dimension $\ds$:
\begin{align}
  Z = \Tr \prod_{x} T^{(x)}_{x x' y y'},
\end{align}
and that periodic boundary conditions are imposed in all directions. Hereafter in this subsection, we explain the algorithm of the original TRG proposed by Levin and Nave.\footnote{Coarse-graining approaches for tensor networks are generically referred to as TRG. Then ``TRG" does not necessarily identify a specific algorithm.}

First, the tensor $T$ can be regarded as a $\ds^{2} \times \ds^{2}$ matrix, and can be can be approximately decomposed using the SVD in two ways:
\begin{align}
  T_{x' y x y'} = M^{[13]}_{(x' y)(x y')} \approx \sum_{m=1}^{\ds} S^{[1]}_{(x' y)m} \lambda^{[13]}_{m} S^{[3]}_{m(x y')},
  \\
  T_{y' x' y x} = M^{[24]}_{(y' x')(y x)} \approx \sum_{m=1}^{\ds} S^{[2]}_{(y' x') m} \lambda^{[24]}_{m} S^{[4]}_{m (y x)},
\end{align}
where $\lambda^{[13]}$ and $\lambda^{[24]}$ are the singular values (assumed to be the descending order: $\lambda_{1} \ge \lambda_{2} \ge \cdots \ge \lambda_{\ds^2} \ge 0$), and $S^{[1]}$, $S^{[2]}$, $S^{[3]}$, $S^{[4]}$ are unitary matrices.
Here the degree of the approximation is set to be $\ds$ for simplicity.
Of course one can freely choose this parameter, and it becomes the bond dimension of the coarse-grained tensors.

\begin{figure}[htbp]
  \centering
  \includegraphics[width=\hsize]{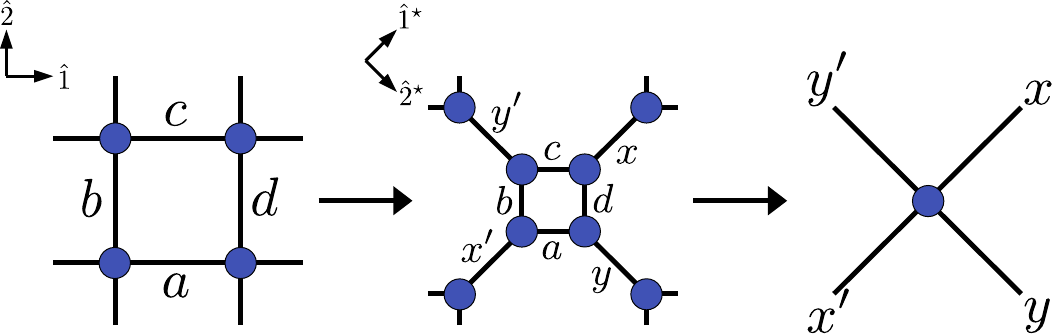}
  \caption{
    A coarse-graining step for the tensor network.
    Circles represent tensors, and closed indices should be contracted.
    The definitions of the unit vectors for the original and the coarse-grained network are also shown.
    The tensor indices are shown in the same manner as Eq.~\eqref{eq:tnew_trg}.
  }
  \label{fig:trg_trg}
\end{figure}

Using the decomposed tensors $S^{[i]}$ ($i=1,2,3,4$), a coarse-grained tensor is defined by
\begin{align}
  \label{eq:tnew_trg}
  \nonumber
  T^{\mathrm{new}}_{x x' y y'}
  &= \sqrt{\lambda^{[13]}_{x} \lambda^{[24]}_{y} \lambda^{[13]}_{x'} \lambda^{[24]}_{y'}} \\
  &\times
    \sum_{a, b, c, d = 1}^{\ds} S^{[3]}_{x (cd)} S^{[4]}_{y(da)} S^{[1]}_{(ab)x'} S^{[2]}_{(bc)y'}.
\end{align}
The number of tensors on the network is now reduced by $1/2$,
and the bond dimension of $T^{\mathrm{new}}$ is a free parameter and here set to be the same as that of $T$, $\ds$.
Repeating this procedure, one can make an effective tensor network that consists of a few tensors,
and then one can take the contraction of the tensor indices.
A graphical explanation of the TRG is given in Fig.~\ref{fig:trg_trg}.

After a coarse-graining step, the network is rotated by $45$ degrees,
and then how one defines the new unit vectors is one's choice.
One possible way is to define them by $\hat{1}^{\star} = \hat{1} + \hat{2}$ and $\hat{2}^{\star} = \hat{1} - \hat{2}$,
where $\hat{1}$ $\left( \hat{2} \right)$ and $\hat{1}^{\star}$ $\left( \hat{2}^{\star} \right)$ are the unit vector along the $\hat{1}$- ($\hat{2}$-)direction of the original lattice and that on the coarse-grained lattice, respectively (see Fig.~\ref{fig:trg_trg}).
Using this definition, the orientation of the network is recovered after every two coarse-graining steps;
\textit{i.e.} $\hat{1}^{\star\star} = \hat{1}^{\star} + \hat{2}^{\star} = 2 \cdot \hat{1}$ and $\hat{2}^{\star\star} = \hat{1}^{\star} - \hat{2}^{\star} = 2 \cdot \hat{2}$.

In this procedure the relevant degrees of freedom are decided by the SVD during the decomposition of the $T$ tensor into an intermediate sum-over-states.  In the next section we will see there are improved methods to pick out the relevant states during truncation.

\subsection{Optimized truncations}
\label{sec:truncations}
The previous section discussed the first work at a renormalization group procedure using a tensor formulation.  Here we will discuss refinements which occurred later.  These refinements incorporate an environment tensor from which relevant states are determined and kept.

\begin{figure}[h]
  \includegraphics[width=0.8\hsize]{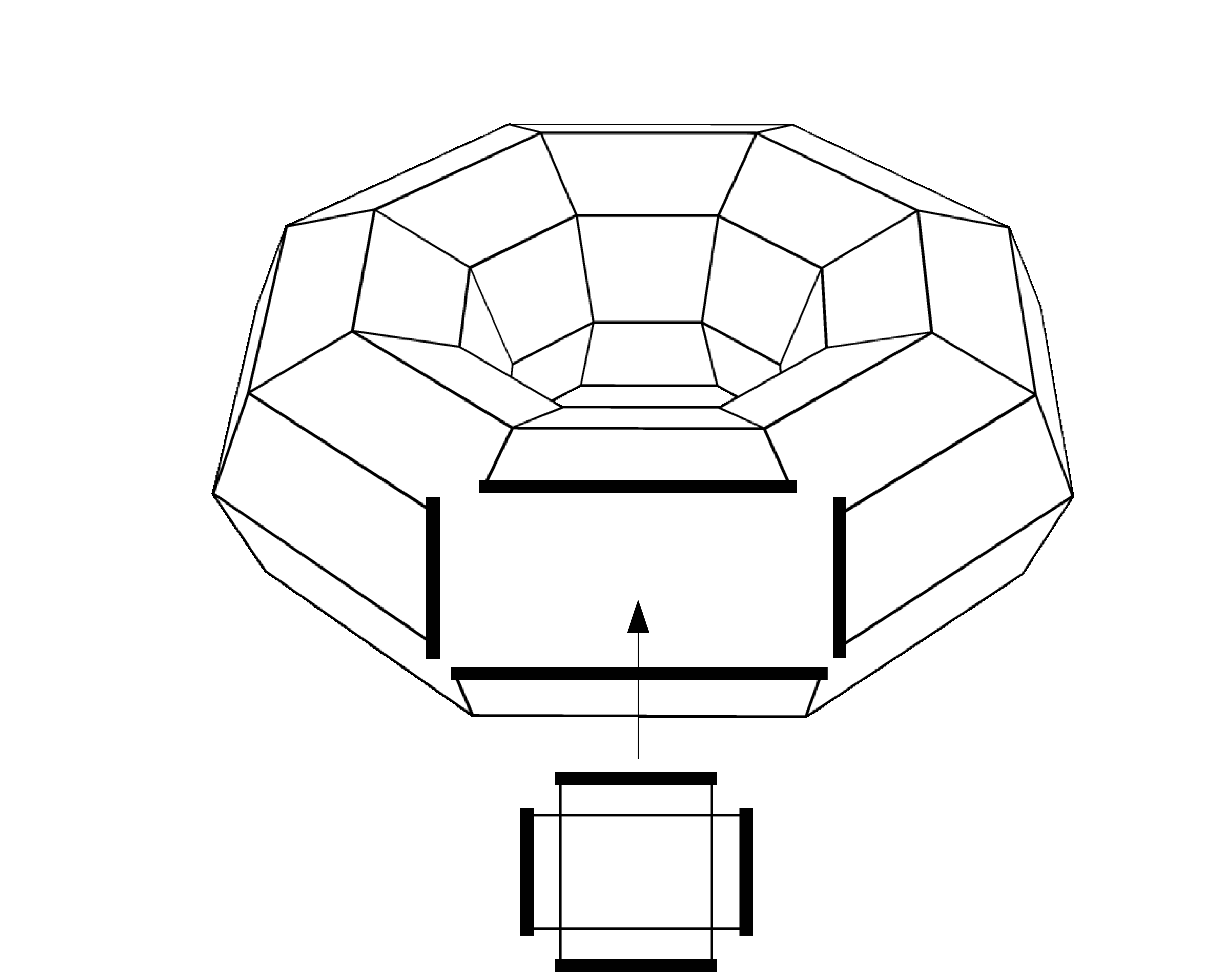}
  \caption{\label{fig:tensenv}A blocked tensor and its environment. }
\end{figure}

We begin with assuming that one has completed contraction of a single block.  Each tensor index of the blocked tensor now posses $\ds^2$ states.  The next step is to find a way to decide which of the $\ds^2$ states (or possibly a linear combination of them) corresponding to the tensor product associated with each index should be kept.  Ideally, we would like to address this question in terms of the environment of a single tensor. We can write
\beq
\Tr \prod _{x} T^{(x)}=\sum_{XX'YY'}E_{XX'YY'}T'^{(0)}_{XX'YY'},
\enq
where $E_{XX'YY'}$ is obtained by tracing all the indices except for  four pairs of indices
associated to a single block that we have located at the origin, and $T'^{(0)}$ is the first block of four tensors. This is illustrated in Fig. \ref{fig:tensenv}.
Constructing an environment from which important states are chosen is reminiscent of the finite-size density matrix renormalization group.
Obviously,  the environment tensor $E_{XX'YY'}$ is very close to what are ultimately trying to calculate and so it is not immediately available.   The only purpose of using $E_{XX'YY'}$ is to rank order the states of the tensor product and its exact form does not appear in later calculations. Consequently, the exact form may not be too important and as a first step, we can try approximations.

A very simple approximation is to ignore the details of the environment and use \cite{prb87}
\beq
\label{eq:noenv}
E^{app.}_{XX'YY}=C \delta_{XX'}\delta_{YY'},
\enq
for some positive constant $C$.
We can then optimize the truncation by maximizing the approximate partition function expressed in terms of the trace of a matrix $G$ such that
\beq
\Tr\  G=(1/C)\sum_{XX'YY'}E^{app.}_{XX'YY'}T'^{(0)}_{XX'YY'},
\enq
which can be achieved with
\beq
G_{XX'}=\sum_{Y}T'^{(0)}_{XX'YY}.
\enq
By looking at the expression in terms of the original tensors one realizes that
$G_{X X'}$ is in fact the square of another matrix \cite{prb87} and if the eigenvalues of this matrix are
real then all the eigenvalues of $G$ are positive and we can optimize the truncations by selecting the states corresponding to the largest eigenvalues of $G$.

A more refined approximation is to assume that the environment is a ``mirror image" of the
tensor itself \cite{2012PhRvB..86d5139X}:
\beq
E^{app.}_{XX'YY}=C'T'^{(0)\star}_{XX'YY'}.
\enq
The trace of $G$ can then be identified with the tensor norm:
\beq
\Tr\  G=\sum_{XX'YY'}T'^{(0)}_{XX'YY'}T'^{(0)\star}_{XX'YY'}=||T'^{(0)}||^2,
\enq
which is clearly a sum of positive terms.
This can be accomplished with the Hermitian matrix
\beq
\label{eq:gtensor}
G_{XX'}=\sum_{X''YY'}T'^{(0)}_{XX''YY'}T'^{(0)\star}_{X'X''YY'}.
\enq
The problem is then reduced to selecting the states that  provide the best approximation of
$\Tr G$ which is obvious when all the eigenvalues are positive.

The procedure that we just described is isotropic. It is however possible to coarse grain one direction at a time~\cite{2012PhRvB..86d5139X}  in order to reduce the size of the summed expressions.
For instance, the summation over the tensor product indices $Y$ and $Y'$ in Eq.~\eqref{eq:gtensor} could be first replaced by summations over single indices,
\begin{align}
  M_{X(x_1 , x_2) X'(x'_1 , x'_2)y y'} = \sum_{a} T_{x_1 x'_{1} y a} T_{x_2 x'_2 a y'}.
\end{align}
The tensor $M$ can then be used in Eq.~\eqref{eq:gtensor} in place of $T^{(0)}$ as the ``blocked'' tensor to find the most relevant states for the $X$ indices.
This provides a coarse-graining in the first direction. It is then necessary to coarse-grain in the second direction using sums over single indices in the first direction. As our discussion
is focused on $D=2$, these two steps constitute a coarse-graining that doubles the
lattice spacing in all directions. In  ~\cite{2012PhRvB..86d5139X}, these calculations were conducted using higher-order generalizations of the SVD method called the higher-order tensor renormalization group (HOTRG).

A better description of the environment can be reached by following a local truncation procedure as described above for a sufficiently large but finite number of times at which point it is assumed that there is no environment and Eq.~\eqref{eq:noenv}
can be used. In other words, by working with a finite lattice, the procedure is terminated by approximating the partition function as a trace of the last coarse-grained expression for the tensor. It is then possible to move ``backward"~\cite{PhysRevLett.103.160601,PhysRevB.81.174411} and
reconstitute the approximate environment of a single tensor coarse-grained one less time.
Explicit expressions based on the HOTRG construction can be found in~\cite{2012PhRvB..86d5139X}.
This can be pursued recursively until we reach the first coarse graining level illustrated in Fig.~\ref{fig:tensenv}.

The analogy with the backward propagation used in machine learning has been exploited to design new algorithms recently \cite{chen2020b}.
More generally, algorithmic improvement in the TRG context is a subject of active investigation \cite{bal2017,PhysRevB.98.235148,morita2020}. The idea that the bound dimension can be regarded as as a relevant direction which can be used to obtain data collapse has been discussed in \cite{Vanhecke:2019pez}. For earlier developments see \cite{luca2008,pollmann2009,pirvu2012}. We expect that these considerations will provide a more systematic understanding of the truncation errors.

\subsection{Higher dimensional algorithms}
\label{sec:highdim}

While the Levin--Nave type TRG can be applied to two dimensional systems,
higher dimensional systems are dealt with using other algorithms.
One such algorithm is the higher-order TRG (HOTRG)~\cite{2012PhRvB..86d5139X} mentioned in the previous section.
Using the HOTRG, in principal, any dimensional tensor network can be coarse-grained.
When a $D$ dimensional tensor network is build by tensors with the bond dimension $\ds$,
the computational complexity of the HOTRG is $\mathcal{O}\left( \ds^{4D-1} \right)$, and the memory complexity is $\mathcal{O}\left( \ds^{2D} \right)$.

\begin{figure}[htbp]
  \centering
  \includegraphics[width=\hsize]{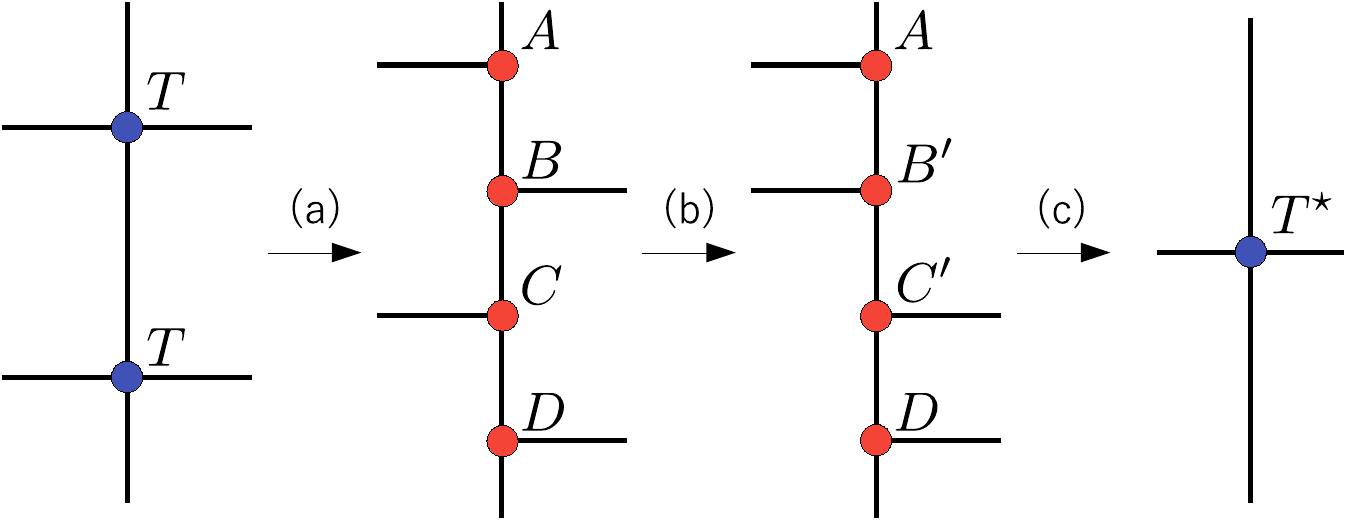}
  \caption{
    Graphical description of the ATRG.
    (a) Decompose $T$ into ($A$ and $B$) and ($C$ and $D$) with the SVD.
    (b) Swap the bonds of $B$ and $C$. The swapping can be done also by the SVD: $\sum_{i} B_{x y i} C_{i x^{\prime} y^{\prime}} = (BC)_{x y x^{\prime} y^{\prime}} = \sum_{i} B^{\prime}_{x^{\prime} y i} S_{i} C^{\prime}_{i x y_{\prime}}$.
    (c) Contract $A$, $B^{\prime}$, $C^{\prime}$, $D$ while truncating the dimensions of horizontal bonds.
  }
  \label{fig:atrg}
\end{figure}

Recently, cheaper algorithms have been invented.
The anisotropic tensor renormalization group (ATRG)~\cite{Adachi:2019paf}, whose graphical description is given in Fig.~\ref{fig:atrg}, achieved the time and the memory complexities $\mathcal{O}\left( \ds^{2D+1} \right)$ and $\mathcal{O}\left( \ds^{D+1} \right)$ that are significant reductions from the HOTRG.
The ATRG introduces an approximation of an approximation,
and indeed, when the bond dimensions are the same, the ATRG is less accurate compared to the HOTRG.
However, thanks to the cheaper complexity, the ATRG leads to better accuracy with fixed CPU time.
Using the ATRG, four dimensional systems, where the HOTRG is much more expensive, have begun investigation~\cite{Akiyama:2019chk,Akiyama:2020ntf,Akiyama:2020soe}.

Another approach is coarse-graining on a triad tensor network representation~\cite{Kadoh:2019kqk}. This paper
compares the triad tensor renormalization group approach to the original HOTRG and the ATRG algorithms.

\subsection{Observables with tensors}
\label{subsec:tensor-obs}
With the tensor formulation from Sec.~\ref{subsec:tensorising}, along with the coarse-graining algorithms from Sec.~\ref{sec:trunc} it is possible to calculate derivatives of $\ln Z$, as well as compute $n$-point correlation functions.  

Because the tensor renormalization group process requires renormalizing the tensor during iterations, in order to compute the logarithm of the partition function these normalizations must be stored.  For a coarse-graining that is isotropic with $N_{1}, N_{2}, \ldots N_{D}$ iterations in each direction and $N = \sum_{i=1}^{D} N_{i}$, let us say the normalizations at each iteration are $\cN^{(0)}$, $\cN^{(1)}$, $\cN^{(2)}$ etc. starting by normalizing the initial tensor by $\cN^{(0)}$.  Then during the coarse-graining process, $\cN^{(0)}$ appears $V = 2^{N}$ times, giving an overall factor of ${\cN^{(0)}}^V$.  Likewise each subsequent normalization appears $2^{N - n}$ where $n$ denotes the iteration number, \emph{i.e.} $1,\ldots N$.  After the final step the total normalization on the effective tensor is given by
\begin{align}
    {\cN^{(0)}}^{V} \cdot {\cN^{(1)}}^{V/2} \cdot {\cN^{(2)}}^{V/4} \cdots \cN^{(N)}.
\end{align}
The logarithm of the partition function is then given as the logarithm of this normalization, added to the logarithm of the trace of the final tensor,
\begin{align}
    \ln Z = \sum_{n=0}^{N} 2^{N-n} \ln \cN^{(n)} + \ln \Tr[ T ],
\end{align}
where $T$ is the final normalized effective tensor, and the trace is the tensor trace.

Expectation values of $N$-point correlation functions~\cite{2008PhRvB..78t5116G,Nakamoto2016} are given by ratios of partition functions which are calculated separately using the tensor renormalization group,
\begin{align}
    \langle \sigma^{(1)} \sigma^{(2)} \cdots \sigma^{(N)} \rangle 
    = \frac{Z^{(N)}}{Z}
\end{align}
with
\begin{align}
    Z^{(N)} = \sum_{\{ \sigma \}} 
    \sigma^{(1)} \sigma^{(2)} \cdots \sigma^{(N)}
    e^{-S}
\end{align}
and the positions of the $\sigma$ fields have been suppressed.  In terms of tensors this amounts to $N$ ``impure'' tensors whose namesake comes from their altered local constraint.  Since $Z^{(N)}$ contains additional spin fields located at specific sites, those spin fields alter the sum at that site over the field states, and give
\begin{align}
    &\sum_{\sigma_{x^{*}}} \sigma_{x^{*}}^{1 + \sum_{\mu=1}^{D} n_{x^{*}-\hat{\mu},\mu} - n_{x^{*},\mu}} \\ \nonumber
    &=
    \begin{cases}
    2 & \text{ if } 1 + \sum_{\mu=1}^{D} n_{x^{*}-\hat{\mu},\mu} - n_{x^{*},\mu} \text{ is even} \\
    0 & \text{ otherwise.}
    \end{cases}
\end{align}
where $x^{*}$ is the location of the additional spin.
These additional spins act as sources and sinks for the vector fields $n_{x,\mu}$ on the surrounding links as can be seen from the new constraint where the divergence of the surrounding $n$ fields must now be one modulo two.
To compute the ratio, if both $Z^{(N)}$ and $Z$ are normalized identically throughout the calculation, only the ratio of the trace of the final effective tensors is necessary for the expectation value, since the normalizations would cancel.

\begin{figure}[htbp]
    \centering
    \includegraphics[width=\hsize]{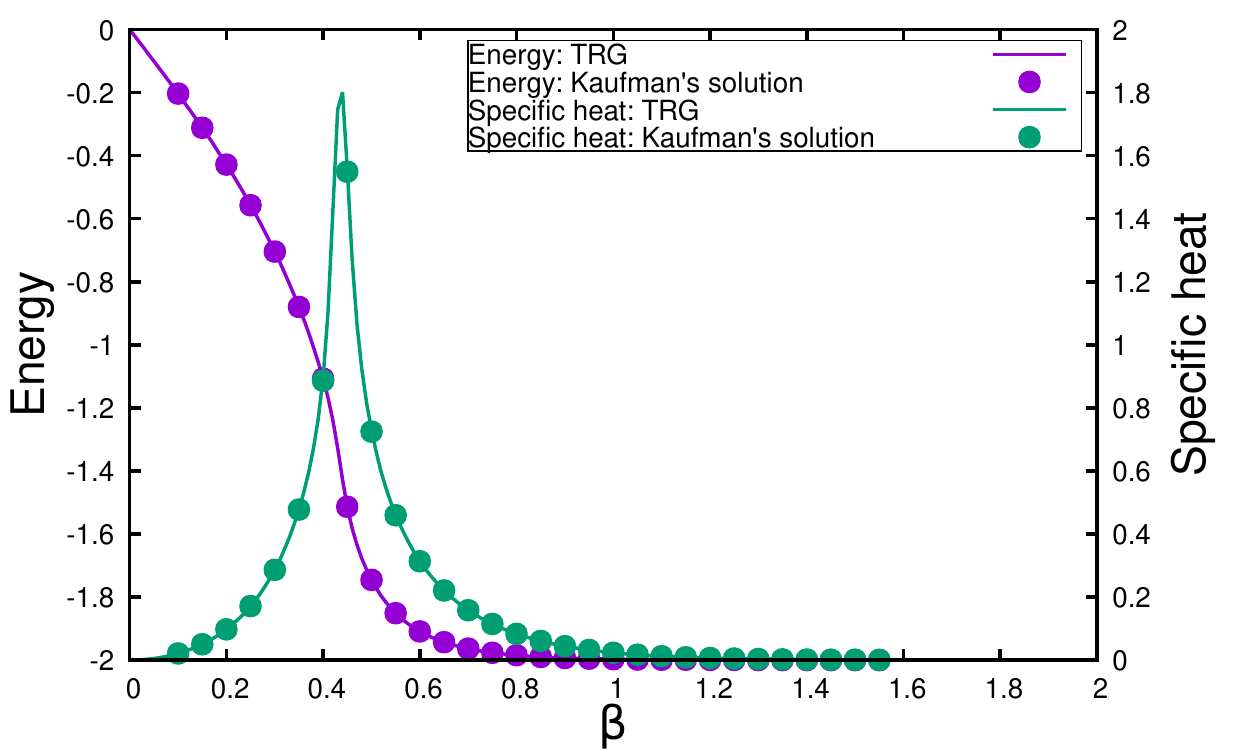}
    \caption{
    Energy and specific heat of the two dimensional Ising model on a $32\times 32$ lattice.
    $D_{\mathrm{cut}}=32$.
    }
    \label{fig:energy_specheat_2dising}
\end{figure}

\begin{figure}[htbp]
    \centering
    \includegraphics[width=\hsize]{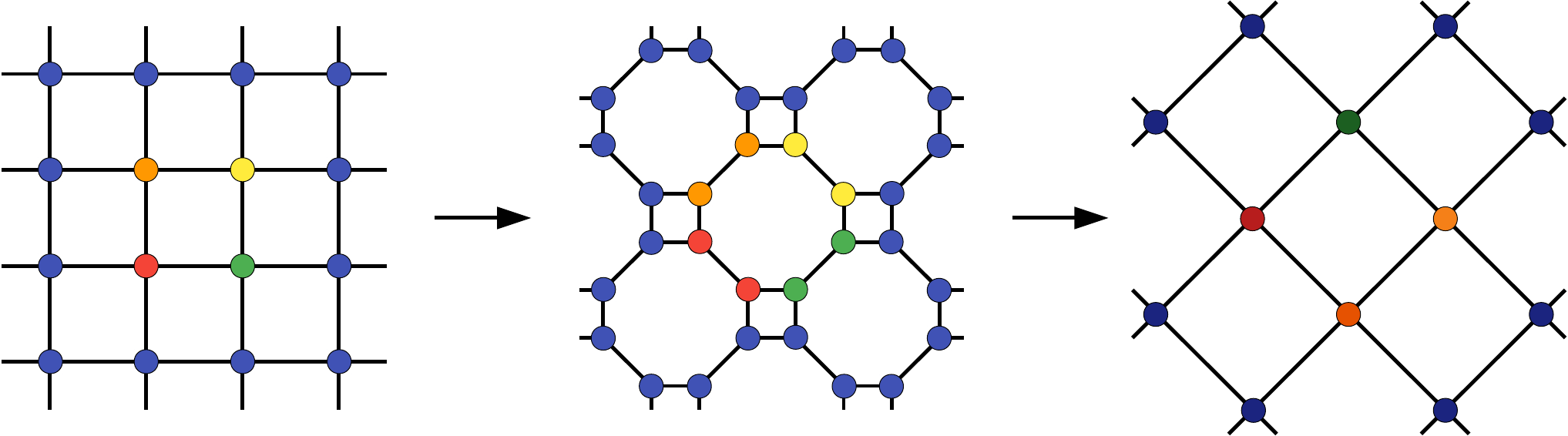}
    \caption{
    TRG process for a tensor network with four impurities on a plaquette.
    }
    \label{fig:trg_impurity}
\end{figure}

As illustrated in Fig.~\ref{fig:energy_specheat_2dising} the method can be used for the average energy and the specific heat, and are compared to the exact solutions by Kaufman's formula.
The energy is calculated using the TRG with two neighboring impurity tensors,
and the specific heat is given as the numerical $\beta$ derivative of the energy.  To show a graphical example of the contractions performed using impure tensors, four impurities on a plaquette are decomposed and contracted to form four impurities again, is illustrated in Fig.~\ref{fig:trg_impurity}.
Note that the energy can also be obtained by the numerical derivative of the logarithm of the partition function,
and then the specific heat is the second numerical derivative.
In general, the numerical derivatives cause a loss of significance,
so obtaining the energy as a primary output of the TRG using the impurity tensor method helps to improve the numerical accuracy.
The impurity tensors destroy the translational invariance, but the effect is local so that the computational complexity does not drastically increase.

\subsection{Niemeijer-Van Leeuwen equation}
In Sec.~\ref{sec:exact-blocking} we have constructed a coarse-grained tensor that can be used to give an exact expression for the partition function. Despite the fact that we integrated over microscopic degrees of freedom,  the number of tensor  indices needed for this exact representation grows exponentially with the size of the blocks. In Secs.~\ref{sec:lntrg} and~\ref{sec:truncations}, we introduced truncations where a fixed
number of indices $D_{\text{cut}}$ was kept at each step.
This procedure discards some information but allows us to compare the tensors before and after the coarse-graining.
Typically, the tensors tend to grow exponentially with the number of coarse graining steps and it is important to renormalize their absolute size
or to only consider their ratios. After such a renormalization takes place, we obtain a RG transformation.
The fixed points of this transformations are the central objects of the RG approach and it is useful to compare the TRG equations with standard RG
equations due to Niemeijer and van Leeuwen (NvL)\cite{nvl}.

NvL were aware of the difficulty of controlling the new couplings generated by the blocking procedure, so they started immediately
with the most general Ising interactions in a finite volume $V$. If no conditions are imposed, there are as many couplings as Ising configurations so this is not suitable for numerical purposes. They then introduced 1 in the partition function as in Eq.~\eqref{eq:k1},
in order to define new Ising spins $\csip$
in a volume $V'=V/b^D$ with a new lattice spacing rescaled by the linear size of the blocks $b$.  They were able to give a formal expression for
the new couplings in terms of the original ones as ${\bf K}'({\bf K})$. Strictly speaking there are less couplings after the coarse graining because they considered the most general case involving the products of spins in arbitrary domains, but they assumed that only a certain number of quasi-local couplings were important and had the same form before and after the coarse graining.
In addition they assumed that the dependence of the free energy
density $f({\bf K})$ on these couplings is the same after the coarse graining. This led to the NvL equation
\beq
f({\bf K})=g({\bf K})+b^{-D} f({\bf K}').
\label{eq:nvl}
\enq
The function $g=G/V$, comes from
\beq
G=\sum_{\csip} \ln\Big(\sum_{\csi}P(\csip,\csi)\exp\big(\mH (\csi)\big)\Big),
\enq
and is defined by the condition
\beq
\sum_{\csip}\mH'(\csip)=0,
\enq
where $\mH$ and $\mH '$ are the Hamiltonians before and after coarse graining.

Even though computing the new couplings and the functions may be very difficult in practice (the new couplings are double partition functions), NvL succeeded to obtain a formal relation which can be iterated and linearized near a fixed point. This allows us to identify the relevant directions and it is often taken as the
starting point for the introduction of the RG method in textbooks \cite{cardy1996scaling}.

For the TRG, we can factor out the increasing size of the tensors, for instance,  by imposing the
normalization condition
\beq
\label{eq:t0000}T_{0000}=1,\enq
at each step. In other words we divide all the tensors by one unnormalized tensor element.  We now have the two steps (coarse-graining and renormalization) that define a RG transformation.
We can write the exact identity
\begin{eqnarray}
  &&\ln(\Tr \prod_{sites}T^{(sites)}_{xx'yy'})/V=\\ \nonumber &&(1/4)\ln(T'_{0000})+(1/4)\ln(\Tr \prod_{sites'}T^{(sites')}_{XX'YY'})/V'. \end{eqnarray}
$T'_{0000}$ is the unnormalized tensor element that we constructed in Sec.~\ref{sec:exact-blocking}. $T^{(sites')}_{XX'YY'}$ is the renormalized tensor meaning the unnormalized tensor divided by $T'_{0000}$. Bearing in mind that
$b^{-D}=1/4$ we see the analogy with the NvL equation (\ref{eq:nvl}). $(1/4)\ln(T'_{0000})$ plays the role of $g({\bf K})$. Note that $\Tr \prod_{....}$ has a different meaning in both sides of the equation. However, if we assume that the coupling dependence of the densities are the same before and after as in NvL and we obtain a RG equation. In both cases, neglecting couplings can be justified by the fact the the RG transformation has only a small number of important directions in the space of couplings. This will be illustrated with a simple example in the next subsection. More details can be found in Ref. \cite{iopbook}.

\subsection{A simple example of TRG fixed point}
In the following we discuss the two-state truncation for the Ising model. In other words, we keep the same number of states for each index as for the initial tensor.
With the indices taking two values, the rank four tensor has in principle 16 independent entries, however
because of the Ising selection rule  the sum of the indices must be even and so eight of the tensor values are zero. In addition, if we preserve the symmetry
under  the rotation of the square lattice by $\pi/2$, this
imposes that
\beq
T_{1010}=T_{0110}=T_{1001}=T_{0101}\equiv t_1,\enq
and
\beq
T_{1100}=T_{0011}\equiv t_2.
\enq
In addition, we define
\beq
T_{1111}\equiv t_3.
\enq
For the initial tensor, we have
\beq
t_1=t_2=\tanh(\beta)\  {\rm and} \  t_3=t_1^2 .
\label{eq:ic}
\enq
The property $t_1=t_2$ is not preserved by the blocking procedure which can be expressed as a mapping of the three dimensional parameter space $(t_1,t_2,t_3)$ into itself that we denote $t'_i(t_1,t_2,t_3)$.  Under this mapping, the elements of the tensor flow towards their fixed-point values based on the bare input value of $\beta$.  The point of bifurcation in the fixed-point tensor elements can be used to determine the critical value of $\beta$.

\begin{figure}[h]

  \includegraphics[width=8cm,angle=0]{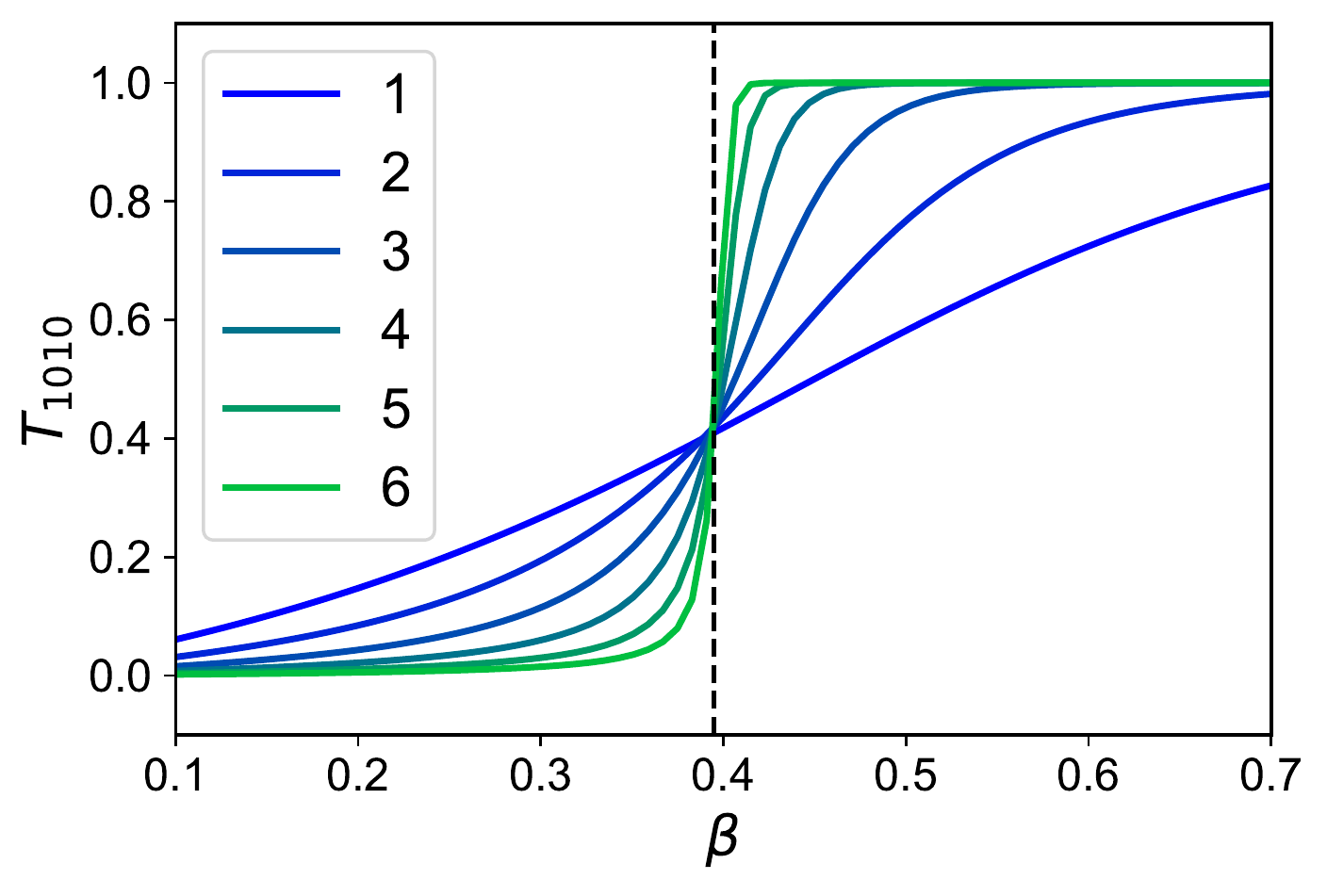}
  \includegraphics[width=8cm,angle=0]{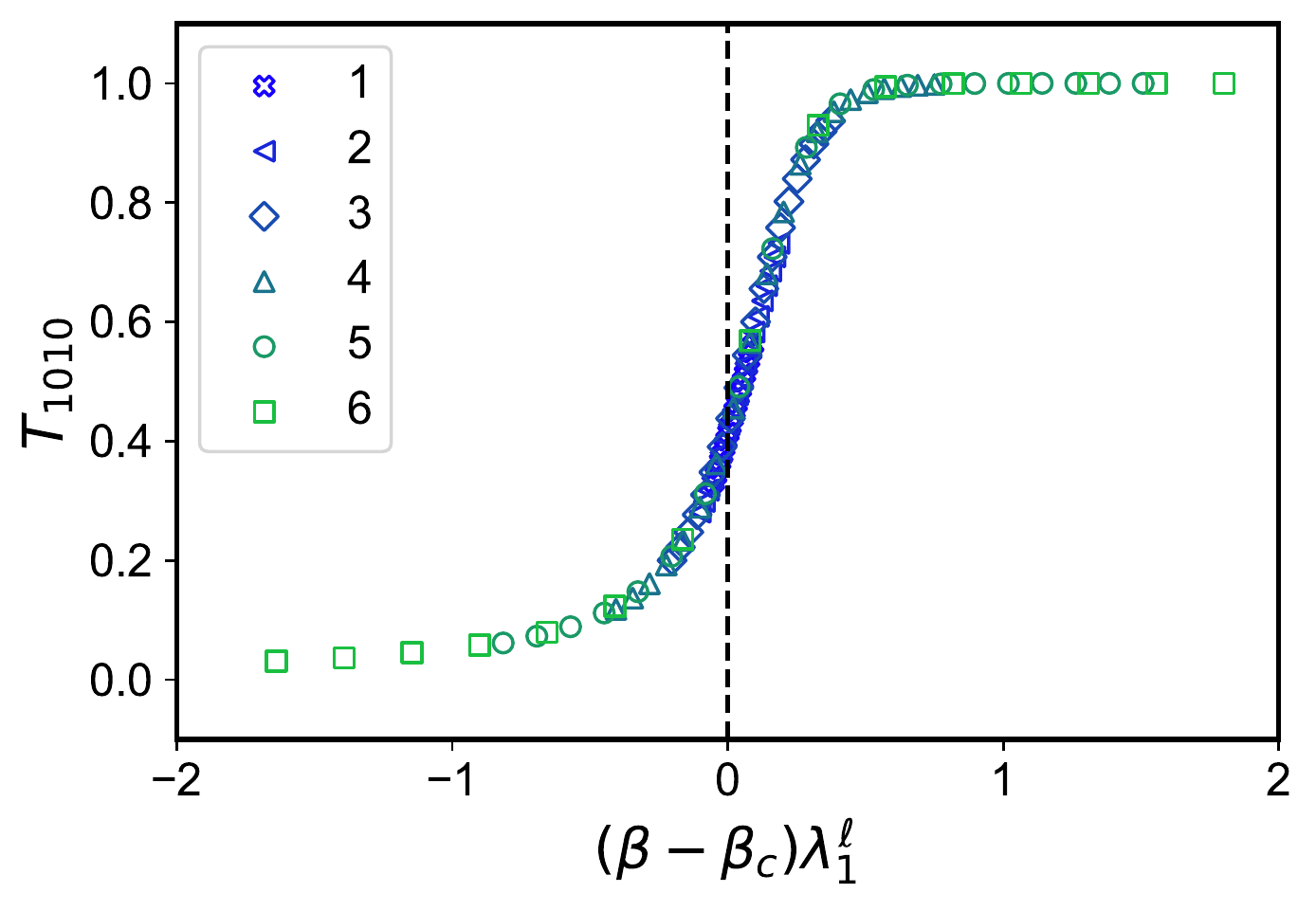}

  \caption{$t_1=T_{1010}$,  versus $\beta$ (top) and versus $(\beta-\beta_c)\lambda_1^\ell$ (bottom) for $\ell=1,\ \dots,\ 6$ iterations of the two-state approximation. The dotted line is at the critical value. As the iteration number increases, the color smoothly changes from blue to green.
    \label{fig:ti} }
\end{figure}

As a numerical example, we use the method of Eq.~\eqref{eq:noenv} and the normalization from Eq.~\eqref{eq:t0000} which is discussed after Eq.~(10) in  \cite{prb87}. The results for $t_1$
as a function of the initial $\beta$ for six iterations are shown in Fig.~\ref{fig:ti}. We see that for values of $\beta$ low enough,
$T_{1010}$ goes to zero at a faster rate as the the number of iterations increases. On the other hand, for values of $\beta$ large enough,
$T_{1010}$ goes to 1. As the number of iterations increases, the transition becomes sharper and sharper and singles out a critical value
$\beta_c=0.394867858...$ where the curves for successive iterations intersect. This also singles out a fixed point value for $t_1$ near 0.4. The graphs for $t_2$ and $t_3$ have similar features. Near $\beta_c$, the departure from the fixed point value is
approximately linear in $\beta -\beta_c$ with a slope of the form $\lambda_1^\ell$, where $\ell$ is the number of iterations.
It is possible to rescale $\beta -\beta_c$ by a factor $\lambda _1$ at each iteration to obtain a ``data collapse" shown at the bottom of  Fig.~\ref{fig:ti}. Numerically \cite{prb87}, $t_1^\star=0.42229$, $t_2^\star=0.28637$ and $t_3^\star=0.27466$ and $\lambda_1=2.00931069$ which provides a critical exponent $\nu =\log b/\log \lambda_1\simeq 0.993$ which is surprisingly close to the exact value 1 given that the truncation is quite drastic. Very similar results are obtained in a dual version of the map in  \cite{aoki2009}.

One would think that by adding a few more states, we could get even better results, however this is not the case \cite{efratirmp}.
One of the reasons is explained in the next section.

\subsection{Corner double line structure on tensor network}
\label{sec:cdl}

\begin{figure}[htbp]
  \centering
  \includegraphics[width=0.35\hsize]{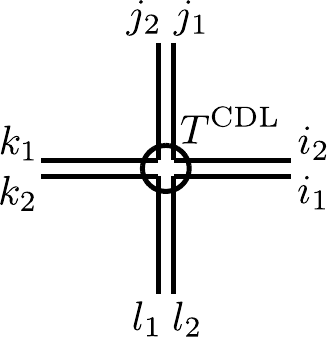}
  \caption{CDL tensor.}
  \label{fig:cdltensor}
\end{figure}

\begin{figure}[htbp]
  \centering
  \includegraphics[width=0.4\hsize]{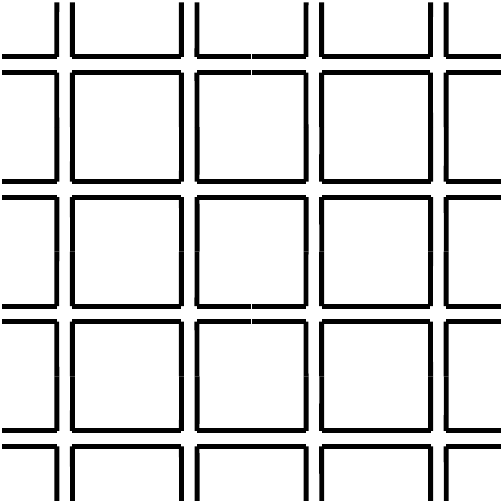}
  \caption{CDL tensor network.}
  \label{fig:cdltn}
\end{figure}

In this subsection we discuss a fixed point of the TRG~\cite{Levin:2006jai} that is called the corner double line (CDL) tensor~\cite{Gu:2009dr}.
Here let us consider a toy model of short-range correlations that is expressed as a tensor network spanned by tensors with the following form:
\begin{align}
  T^{\mathrm{CDL}}_{kijl} = \delta_{i_{1}, l_{2}} \delta_{j_{1}, i_{2}} \delta_{k_{1}, j_{2}} \delta_{l_{1}, k_{2}},
\end{align}
where each tensor index has two components: {\cal e.g.} $i = \left( i_{1}, i_{2} \right)$ (see Fig.~\ref{fig:cdltensor}).
We call $T^{\mathrm{CDL}}$ the CDL tensor, and it describes interactions on plaquettes as seen in Fig.~\ref{fig:cdltn}.
Now, let us consider a TRG step for this tensor network.

\begin{figure}[htbp]
  \centering
  \includegraphics[width=\hsize]{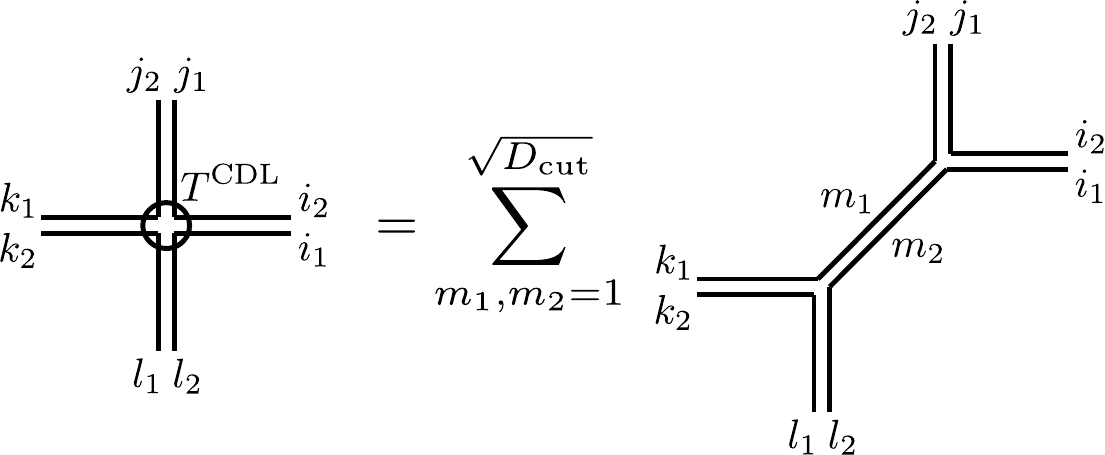}
  \caption{SVD of CDL tensor.}
  \label{fig:svd_cdltensor}
\end{figure}

\begin{figure}[htbp]
  \centering
  \includegraphics[width=\hsize]{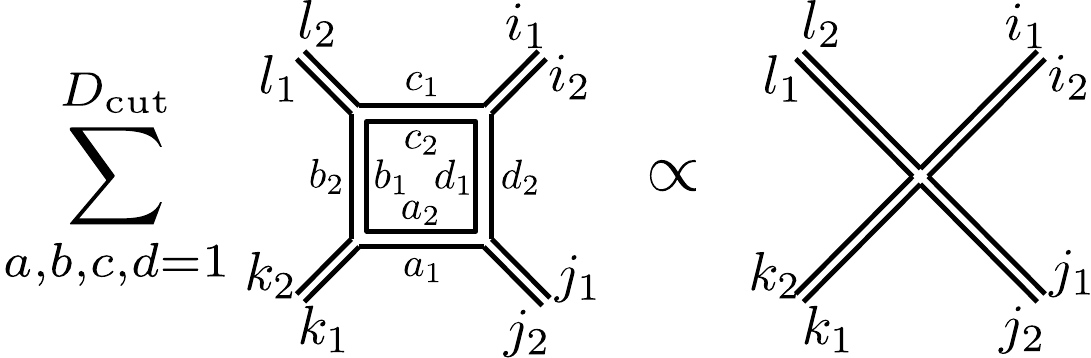}
  \caption{Contraction step for CDL tensor.}
  \label{fig:contraction_cdltensor}
\end{figure}

The SVD of the CDL tensor is uniquely given by
\begin{align}
  T^{\mathrm{CDL}}_{kijl} = \sum_{m_{1}, m_{2}=1}^{\sqrt{\ds}} \delta_{i_{1}, m_{2}} \delta_{j_{1}, i_{2}} \delta_{m_{1}, j_{2}} \delta_{k_{1}, m_{1}} \delta_{l_{1}, k_{2}} \delta_{m_{2}, l_{2}},
\end{align}
where we simply assume that all elementary components of the tensor indices run from $1$ to $\sqrt{\ds}$: {\cal e.g.} $1 \le i_{1}, i_{2} \le \sqrt{\ds}$.
$\ds$ is assumed to be a square number.
Then, by contracting the decomposed components, the coarse-grained tensor is given by
\begin{align}
  & \left( T^{\mathrm{CDL}} \right)^{\prime}_{kijl} \nonumber \\
  = & \begin{aligned}[t]
    \sum_{a, b, c, d=1}^{\ds} & \delta_{i_{1}, c_{1}} \delta_{d_{2}, i_{2}} \delta_{j_{1}, d_{2}} \delta_{a_{1}, j_{2}} \delta_{k_{1}, a_{1}} \delta_{b_{2}, k_{2}} \\
    & \cdot \delta_{l_{1}, b_{2}} \delta_{c_{1}, l_{2}} \delta_{a_{2}, b_{1}} \delta_{b_{1}, c_{2}} \delta_{c_{2}, d_{1}} \delta_{d_{1}, a_{2}}
  \end{aligned} \nonumber \\
  & \propto T^{\mathrm{CDL}}_{kijl}.
\end{align}
Each step is graphically displayed with the assignments of indices in Figs.~\ref{fig:svd_cdltensor}--~\ref{fig:contraction_cdltensor}.
Surprisingly the CDL tensor has turned out to be a fixed point of the TRG up to a constant factor.
This is not a physical but an artificial fixed point,
and unfortunately this fact means that the TRG leaves short-range correlations on coarse-grained tensors.
This is because the SVD is the best approximation of a tensor but is not the best for a network.

To avoid such an unexpected fixed point, one needs to consider more global blocking procedures.
One ideal way is to insert unknown tensors (called ``(dis)entangler'') on a network and variationally tune them to remove CDL structures.
This method is called the tensor network renormalization (TNR)~\cite{2015PhRvL.115r0405E}
and leads to more precise computations.
In principal this can be extended to three or higher dimensions although the computational complexities would be extremely demanding.
In two dimensions, more low-cost methods are invented so far~\cite{yang2017loop,Hauru:2017tne,2018PhRvB..98h5155E} although we do not mention in detail here.
The common concept in such approaches is to consider global cost functions and to remove the CDL structures on a network.

\section{Tensors for spin models with an Abelian symmetry}
\label{sec:abelian}
In this section, we discuss tensor formulations of generalizations of the Ising model. We first consider the case  of the $O(2)$ nonlinear sigma model introduced in Sec.~\ref{subsec:lattac} where the spin variables form a two-dimensional unit vector varying continuously over a circle. We will show later that the  results extend easily to the clock models with a discrete \zq symmetry.

\subsection{$O(2)$ nonlinear sigma model}
\label{subsec:o2}

The partition function for the $O(2)$ model reads
\beq
Z_{O(2)}=\prod_x\int_{-\pi}^{\pi}\frac{d\varphi _x}{2\pi}e^{-S_{O(2)}},
\enq
with
\beq
S_{O(2)}=-\beta \sum\limits_{x,\mu} \cos(\varphi_{x+\hat{\mu}}-\varphi_x).
\enq
We use the Fourier expansion to expand the Boltzmann weights,
\begin{align}
  \label{eq:fou}
  e^{\beta \cos(\varphi_{x+\hat{\mu}} - \varphi_{x})} = \sum_{n_{x,\mu}=-\infty}^{\infty} e^{i n_{x,\mu} \varphi_{x+\hat{\mu}}} I_{n_{x,\mu}}(\beta) e^{-i n_{x,\mu} \varphi_{x}}.
\end{align}
This expansion factorizes the $\varphi$ fields.
We then integrate over all the $\varphi$ fields using the
orthogonality relations of the Fourier modes, \emph{e.g.},
\begin{align}
  \nonumber
  &\int_{-\pi}^{\pi}\frac{d\varphi_{x}}{2\pi} \prod_{\mu = 1}^{D} e^{i (n_{x-\hat{\mu},\mu} - n_{x,\mu}) \varphi_{x}} \\ \nonumber
  &= \delta_{\sum_{\mu = 1}^{D} (n_{x-\hat{\mu},\mu} - n_{x,\mu}),0} \\
  &= \delta_{n_{x,\text{in}} - n_{x,\text{out}}, 0},
\end{align}
with $n_{x,\text{in}}$ and $n_{x,\text{out}}$ defined in the same way as in Sec.~\ref{sec:ising}.  We can rewrite the partition function as the trace of a tensor product:
\beq
Z = {\rm Tr} \prod_x T^{(x)}_{n_{x-\hat{1},1}, n_{x,1},\dots,n_{x,D}}.
\label{eq:trace}
\enq
The local tensor $T^{(x)}$ has $2D$ indices. The explicit form is
\begin{eqnarray}
  \label{eq:tensor}
  T^{(x)}_{n_{x-\hat{1},1}, n_{x,1},\dots,n_{x-\hat{D},D},n_{x,D}}&=&\\ \nonumber
  \sqrt{I_{n_{x-\hat{1},1}} I_{n_{x,1}},\dots, I_{n_{x-\hat{D},D}} I_{n_{x,D}}  }&\times&\delta_{n_{x, \text{out}},n_{x, \text{in}}},
\end{eqnarray}

The graphical representations of the tensors are similar to the Ising model.
The only difference is that the indices attached to the legs are integers instead of integers modulo 2.
The Kronecker delta in Eq.~\eqref{eq:tensor} enforces
\beq
\sum_{\mu=1}^D(n_{x,\mu}-n_{x-\hat{\mu},\mu})=0,
\label{eq:noether}
\enq
which is a discrete version of Noether current conservation
if we interpret the $n_{x,j}$ with $j=1,\dots D-1$ as spatial current densities and $n_{x,D}$ as a charge density.
At sufficiently small
$\beta$, the relative size of the higher order Bessel functions compared to the zeroth decay rapidly with their order, and it is justified to introduce a truncation\footnote{At large $\beta$, Bessel functions of all orders approach each other.}.
If any of the indices of a tensor element are larger in magnitude than
a certain value $n_{\text{max}}$, we approximate the tensor element by zero. The compatibility of this type of truncation with the
symmetries of the model is discussed in  Sec.~\ref{subsec:symmetries}.

The tensor formulation here can be seen from another perspective, \emph{i.e.}, by using the SVD on each nearest neighbor factor, and then doing the $\varphi$ integrals.  The Fourier expansion of the nearest neighbor interaction given by Eq.~\eqref{eq:fou} can be understood as the spectral decomposition of the Boltzmann weight.
Here the ``matrix'' $e^{i n \varphi}$ is unitary, and is parameterized by the ``indices'' $\varphi$ and $n$.  To find the singular values, we multiply by the Hermitian conjugate and diagonalize (just as in the Ising case),
\begin{align}
  \nonumber
  &\int_{0}^{2\pi} \frac{d \varphi_{j}}{2\pi} e^{\beta \cos(\varphi_{i} - \varphi_{j})} e^{\beta \cos(\varphi_{j} - \varphi_{k})} \\ \nonumber
  &= \int_{0}^{2\pi} \frac{d \varphi_{j}}{2\pi} \sum_{n,m} e^{i n \varphi_{i}} I_{n} e^{-i n \varphi_{j}} e^{i m \varphi_{j}} I_{m} e^{-i m \varphi_{k}} \\ \nonumber
  &= \sum_{n,m} e^{i n \varphi_i} I_{n} I_{m} e^{-i m \varphi_k} \delta_{m,n} \\
  &= \sum_{n} e^{i n \varphi_{i}} I_{n}^{2} e^{-i n \varphi_{k}}
\end{align}
which gives the expected result, that the singular values are the absolute value of the modified Bessel functions, $\lambda_{n}(\beta) = |I_{n}(\beta)|$.  So we see from these examples that performing the character (or Fourier) expansion of the nearest neighbor interaction during the tensor formulation is similar to the SVD of that same factor, and in fact when the coupling is positive, they are equivalent.

For convenience when dealing with the group $U(1)$ (or $O(2)$ for that matter) we factorize all the $I_0(\beta)$ factors which dominate the
small $\beta$ regime and define the ratios
\beq
\label{eq:tn}
t_n (\beta) \equiv \frac{I_n(\beta)}{I_0(\beta)} \simeq\begin{cases} 1-\frac{n^2}{2\beta}+ \mathcal{O}(1/\beta^2),\  {\rm for}\  \beta \rar \infty \\
  \frac{\beta^n}{2^n n!}+ \mathcal{O}(\beta^{n+2}), \  {\rm for}\ \beta \rar 0 \end{cases}.\enq
This factorization helps elucidate the role of the boundary conditions in the tensor formulation.  As discussed in Sec.~\ref{subsec:bc}, with open boundary conditions, one sets the boundary tensor indices to 0, which in the above definition reduces the weights on those links to value 1.

For a recent numerical investigation into the three-dimensional O(2) nonlinear sigma model comparing tensor methods, see~\cite{bloch2021tensor}.

\subsection{$q$-state clock models}
\label{subsec:discretesub}
The results of this section hold for the \zq  restrictions. The infinite sums are replaced by finite sums with $q$ values. The modified Bessel functions are
replaced by their discrete counterparts:
\beq
I_n(\beta) \rar I_n^{(q)}(\beta)\equiv (1/q)\sum_{\ell=0}^{q-1}e^{\beta \cos(\frac{2\pi}{q} \ell)} e^{-i n\frac{2\pi}{q} \ell},\enq
which in the large $q$ limit turns into the usual integral formula. In the Ising case ($q=2$), we have
\beq
I_0(\beta) \rar \cosh(\beta), \ {\rm and} \ I_1(\beta) \rar \sinh(\beta).
\enq
The selection rules in
Eq.~\eqref{eq:noether} remain valid modulo $q$.

For recent numerical TRG-inspired work on clock models and discussion of the second transition see \cite{chen2017,Chenfivestat2018,liping2020}.  For an investigation into the critical behavior of the \zq models when $q$ is fractional, see~\cite{PhysRevD.104.054505}.

\subsection{Dual reformulations with unconstrained variables}
\label{subsec:spindual}
In Secs. \ref{subsec:dualities} and \ref{subsec:isingdual}, we mentioned the possibility of
expressing the closed paths of the expansion in powers of $\tanh(\beta)$ of the
Ising model using dual variables. These ideas can be generalized for a large class of models with Abelian symmetries
\cite{savit77,banks77,einhorn77b,einhorn77a,kogut79,savit80}. In this subsection, we discuss the case of spin models,
with interactions on links. Models with interactions on plaquettes and higher dimensional simplices will be discussed in  Sec.~\ref{subsec:gaugedual}.

Consistent with the rest of the article, in this subsection we use a Euclidean metric with all lower indices as well as implicit summations of repeated indices in order to make a stronger connection with the covariant formulation of Maxwell's equations, unless specified otherwise.  The discrete form of Noether current conservation, given in Eq.~\eqref{eq:noether}, which also holds modulo-$q$ for $q$-state clock models can be written in a compact way as
\beq
\label{eq:divfree}
\nabla_\mu n_\mu=0,
\enq
where $\nabla_\mu$ is a discrete derivative
\beq
\nabla_\mu f_x=f_{x-\hat{\mu}}-f_x.
\enq
Since more indices will appear we keep the reference to the site $x$ implicit.
Following the example of Maxwell's equations written in a relativistically covariant way, we can express
a conserved current as the gradient of an antisymmetric tensor of order 2:
\beq
\label{eq:cmunu}
n_\nu=\nabla_\mu C_{\mu \nu}
\enq
This holds in arbitrary dimensions $D$.

Because of the divergenceless condition Eq.~\eqref{eq:divfree}, $n_\mu$ has $D-1$
degrees of freedom per link. On the other hand,
$C_{\mu \nu}$ has $D(D-1)/2$ degrees of freedom which is $(D-1)(D-2)/2$ more than $D-1$. The redundancy which appears for $D\geq 3$ can be made
more obvious by introducing a a dual tensor with
$D-2$ indices \cite{savit77,savit80} and also $D(D-1)/2$ components:
\beq
C_{\mu \nu}=\frac{1}{(D-2)!}
\epsilon_{\mu\nu \mu_1 \dots \mu_{D-2}}   \tilde{C}_{\mu_1 \dots \mu_{D-2}}.
\enq
The $\tilde{C}$ field is precisely the dual field which lives on the dual lattice.
If we plug this dual form into Eq.~\eqref{eq:cmunu}, we see that $\tilde{C}_{\mu_1 \dots \mu_{D-2}}$ can be shifted by antisymmetrized derivatives of lower rank tensors. For $D=3$, $\tilde{C}_\mu$ has 3 components and is defined up to a gradient so we end up with the desired 2 independent degrees of freedom. For $D\geq 4$, the redundancy becomes nested: we need to count the redundancy of the redundancy etc. For instance for $D=4$, $\tilde{C}_{\mu\nu}$ has 6 components. The shift by the gradient of a 4-vector naively subtracts 4 degrees of freedom, but this 4-vector can itself be shifted by a gradient without affecting the initial shift and we end up with $6-4+1=3$ degrees of freedom.

\subsection{Chemical potential, complex temperature, and importance sampling}
Since the tensor renormalization group---and tensor network methods generally---do not rely on sampling from probability measures, situations where sampling methods would falter or fail due to the loss of real, positive definite weights never arise.  Instead, only linear algebra is needed in the form of tensor contractions.  This allows the method to address the ``sign problem'', which can occur during the inclusion of a chemical potential, and a complex coupling.  In~\cite{prd89} this was addressed in the case of a complex temperature.  The authors studied the zeros in the complex temperature plane, \emph{i.e.} Fisher zeros, and found the tensor renormalization group is able to out perform the re-weighting method using in classical Monte Carlo studies that involve imaginary parts of the action.

In Refs.~\cite{Zou:2014rha,pre93} a purely real chemical potential, $\mu$, is added to the action of the two-dimensional $O(2)$ nonlinear sigma model in the form,
\begin{align}
  S_{\mu} = -\beta \sum_{x,\nu} \cos(\theta_{x+\hat{\nu}} - \theta_{x} - i\mu).
\end{align}
In this form, the action has a complex sign problem. The authors studied the phase diagram of the model in the $\beta$-$\mu$ plane both in the discrete time, and continuous time limits. Reference~\cite{PhysRevD.81.125007} studied this action as well using a sampling method known as the ``worm algorithm'' using a Fourier expanded form of the Boltzmann weight which eliminates the sign problem completely.

In fact, the worm algorithm \cite{worm} is intimately related to the tensor formulation. 
The beginning and the end of the worm correspond to the insertions of an impurity tensor and when the tensor elements are positive definite, it is possible to design a reformulation of the worm algorithm where tensor elements are weighted against each other in an ``accept-reject'' Metropolis style algorithm.
In this way, a lattice configuration is populated by tensor indices at their respective locations (sites, links, etc.).  These indices correspond to tensor elements, and hence weights. Moreover, the tensor interpretation---along with the exact blocking procedure \cite{Liu:2013nsa}---allows one to use the worm algorithm on exactly coarse-grained lattice models (when the coarse-grained tensor elements are again positive-definite) and therefore use the renormalization group exactly with Monte Carlo studies.  This procedure of ``tensor sampling'' is quite general for positive weights and is an interesting direction deserving more attention.

\section{Tensors for spin models with non-Abelian symmetries}
\label{sec:nonabelian}

\subsection{O(3) nonlinear sigma model}
Consider the action for the $O(3)$ nonlinear sigma model in $D$ dimensions,
\begin{align}
  \nonumber
  S &= -\beta \sum_{x=1}^{N} \sum_{\mu=1}^{D} \sum_{a=1}^{3} \sigma^{(a)}_{x+\hat{\mu}} \sigma^{(a)}_{x} \\
  &= -\beta \sum_{x=1}^{N} \sum_{\mu=1}^{D} \cos\theta_{x+\hat{\mu}} \cos\theta_{x} \nonumber \\
  &\quad \quad + \sin\theta_{x+\hat{\mu}} \sin\theta_{x} \cos(\phi_{x+\hat{\mu}} - \phi_{x}) \nonumber \\
    &= -\beta \sum_{x=1}^{N} \sum_{\mu=1}^{D} \cos\gamma_{x+\hat{\mu}, x},
\end{align}
where $\theta$ is the polar angle, $\phi$ is the azimuthal angle, and where $\gamma$ is the angle in the plane created by the two vectors. $\sigma^{(a)}_{x}$ is a unit vector in three dimensions parameterized as $\sigma^{(1)}_{x} = \sin\theta_{x}\cos\phi_{x}$, $\sigma^{(2)}_{x} = \sin\theta_{x}\sin\phi_{x}$, and $\sigma_{x}^{(3)} = \cos\theta_{x}$.

 We will discuss two different possible ways to construct a tensor network here.  
The first tensor construction is based on the global symmetry group of the model.  This construction has been explored and used successfully in Refs.~\cite{Liu:2013nsa,JUDAH.05.21.2015,bruckmann2018} for classical tensor network calculations and MPS calculations in the Hamiltonian formulation.  Here we give the classical tensor formulation.  Since each term in the action is a dot product between vectors of length-one, we can expand on basis functions for the sphere, i.e. the spherical harmonics.  First consider the partition function,
\begin{align}
  \nonumber
  Z &= \int \mathcal{D}\Omega \, e^{-S} \\
    &= \prod_{x} \int d\Omega_{x} \prod_{x,\mu} e^{\beta \cos \gamma_{x+\hat{\mu}, x}},
\end{align}
where $d\Omega$ is the normalized measure on $S_{2}$, i.e. $d\Omega = -d(\cos\theta)d\phi / 4\pi$
Since each Boltzmann factor is a function of the cosine of the angle between the vectors we can expand using Legendre polynomials straight-forwardly,
\begin{align}
  e^{\beta \cos \gamma_{x+\hat{\mu}, x}} = \sum_{l=0}^{\infty} \frac{2l+1}{4 \pi} A_{l}(\beta) P_{l}(\cos \gamma_{x+\hat{\mu}, x}).
\end{align}
This step is advantageous since it gives the $A$ coefficients only $l$ dependence.  The $A$s can be solved for by inverting the above,
\begin{align}
  A_{l}(\beta) = 4 \pi i^l j_{l}(-i \beta),
\end{align}
here $j_{n}(z)$ are the spherical Bessel function.  The Legendre Polynomials can then be rewritten in terms of spherical harmonics using the addition theorem for spherical harmonics,
\begin{align}
  \label{eq:legend-expand}
  P_{l}(\cos \gamma_{x+\hat{\mu}, x}) = \frac{4 \pi}{2l+1} \sum_{m = -l}^{l} Y^{*}_{l m}(\theta_{x+\hat{\mu}}, \phi_{x+\hat{\mu}})
  Y_{l m}(\theta_{x}, \phi_{x}).
\end{align}
This step separates the dependencies on the coupled $x+\hat{\mu}$ and $x$ sites and allows the factors to be treated individually.
With the $\theta$ and $\phi$ dependence decoupled between neighboring sites we can perform the angular integration for the field at each site.  In $D$ dimensions there are $2D$ nearest-neighbors for each site, giving an integral of the form,
\begin{align}
  \label{eq:ylm-const}
  \int d\Omega_{x} \prod_{\mu = 1}^{D} Y_{(l m)_{x, \mu}}(\theta_{x}, \phi_{x}) Y^{*}_{(l m)_{x-\hat{\mu}, \mu}}(\theta_{x}, \phi_{x}).
\end{align}
This integral can be evaluated with the use of the Clebsch-Gordan series,
\begin{align}
  \nonumber
  &Y_{l_1 m_1}(\theta, \phi) Y_{l_2 m_2}(\theta, \phi) = \\ \nonumber
  & \sum_{L=|l_2 - l_1|}^{l_1 + l_2} \sum_{M = -L}^{L} C_{l_1 m_1 l_2 m_2}^{L M} C_{l_1 0 l_2 0}^{L 0} \times \\
  &\sqrt{\frac{(2l_1 + 1)(2l_2+1)}{4 \pi (2L+1)}} Y_{L M}(\theta, \phi),
\end{align}
along with the orthogonality of the spherical harmonics.

We now restrict to $D = 2$ and continue explicitly.  The integral in Eq.~\eqref{eq:ylm-const} takes the form,
\begin{align}
  \label{eq:ylm-2d-int}
  \int d\Omega_{x}  Y_{(l m)_{x, 1}} Y_{(l m)_{x, 2}} Y^{*}_{(l m)_{x-\hat{1}, 1}} Y^{*}_{(l m)_{x-\hat{2}, 2}}(\theta_{x}, \phi_{x}).
\end{align}
If we make the change of notation for $(l m)_{x,1}$ with $l_1 m_{1}$, $(l m)_{x,2}$ with $l_2 m_{2}$ etc\ldots we find for Eq.~\eqref{eq:ylm-2d-int},
\begin{align}
  \label{eq:ylm-sol}
  \mathcal{C}_{x} &\equiv \int d\Omega_{x} \, Y_{l_1 m_1} Y_{l_2 m_2} Y^{*}_{l_3 m_3} Y^{*}_{l_4 m_4}(\theta_{x}, \phi_{x}) \\ \nonumber
  &= \sqrt{(2l_1 + 1)(2l_2+1)(2l_3+1)(2l_4+1)} \times \\ \nonumber
  &\sum_{L, M}  \frac{1}{(4\pi)^2 (2L+1)} C_{l_1 m_1 l_2 m_2}^{L M} C_{l_1 0 l_2 0}^{L 0} C_{l_3 m_3 l_4 m_4}^{L M} C_{l_3 0 l_4 0}^{L 0}.
\end{align}
The Clebsch-Gordan coefficients constrain the surrounding $l$s around a site to satisfy the triangle inequalities, and enforce a conservation law between the $m$s.  This constraint must be imposed at every site. Let us define
a composite index defined formally as $L \equiv \{l, m\}$ which has dimension $(l_{\text{max}}+1)^2$.  This index contains all the states from $l=0$ up to some $l_{\text{max}}$ given by $\sum_{l=0}^{l_{\text{max}}} (2l+1) = (l_{\text{max}}+1)^2$.  We can write down the local tensor as,
\begin{align}
  T_{L_{x-\hat{1}, 1} L_{x, 1} L_{x, 2} L_{x-\hat{2}, 2}}^{(x)} = \sqrt{A_{l_{x-\hat{1}, 1}} A_{l_{x, 1}} A_{l_{x, 2}} A_{l_{x-\hat{2}, 2}}} \mathcal{C}_{x}.
\end{align}
The constraint $\mathcal{C}$ in Eq.~\eqref{eq:ylm-sol} is somewhat complicated, but the physical content is that it simply demands the four $l$s around a site to satisfy the triangle-inequalities according to the typical addition of angular momenta, while enforcing a conservation law on the $O(2)$ subgroup $m$s.  A pleasing feature of this formulation is that the weights, $A$s, only depend on $l$.

The second tensor formulation for the $O(3)$ nonlinear sigma model uses the Taylor expansion of the Boltzmann weight to recast the model in terms of discrete fields.  This formulation follows directly from \cite{PhysRevD.94.114503,WOLFF2010254,BRUCKMANN2015495}.

Starting with the partition function,
\begin{align}
  \nonumber
  Z &= \int \mathcal{D}\Omega \, e^{-S} \\
    &= \prod_{x=1}^{N} \frac{1}{4\pi} \int \sin\theta_{x} d\theta_{x} d\phi_{x} \prod_{x=1}^{N} \prod_{\mu=1}^{2} \prod_{a=1}^{3} e^{\beta \sigma_{x}^{(a)} \sigma_{x+\hat{\mu}}^{(a)}}.
\end{align}
We now expand the Boltzmann weight in a Taylor series,
\begin{align}
  \label{eq:o3-taylor}
  e^{\beta \sigma_{x}^{(a)} \sigma_{x+\hat{\mu}}^{(a)}} = \sum_{n_{x,\mu}^{(a)} = 0}^{\infty} \frac{\beta^{n_{x,\mu}^{(a)}}}{n_{x,\mu}^{(a)} !} \left( \sigma_{x}^{(a)} \sigma_{x+\hat{\mu}}^{(a)} \right)^{n_{x,\mu}^{(a)}}
\end{align}
associating three natural numbers, $n^{(a)}$, with each link.  Reordering and collecting the same spin-field at a site we can write the partition function as,
\begin{align}
  \nonumber
  Z &= \sum_{\{ n \}} \left( \prod_{x} \prod_{\mu} \prod_{a} \frac{\beta^{n_{x, \mu}^{(a)}}}{n_{x, \mu}^{(a)} !} \right) \times \\
    &\left( \prod_{x} \prod_{\mu} \prod_{a} \frac{1}{4\pi} \int (\sigma_{x}^{(a)})^{n_{x, \mu}^{(a)} + n_{x-\hat{\mu}, \mu}^{(a)}} \sin\theta_{x} d\theta_{x} d\phi_{x} \right).
\end{align}
The first factors in parenthesis are the new weights associated with a configuration of $n$s.  The integrals inside the second factor in parenthesis must be evaluated for each site.  They are all identical, so we focus on a single site and perform the integration.  For one site the integral we must evaluate looks like
\begin{align}
  \nonumber
  \frac{1}{4\pi} \int \prod_{\mu} &(\sigma_{x}^{(1)})^{n_{x, \mu}^{(1)} + n_{x-\hat{\mu}, \mu}^{(1)}} (\sigma_{x}^{(2)})^{n_{x, \mu}^{(2)} + n_{x-\hat{\mu}, \mu}^{(2)}} \times \\
                                  &(\sigma_{x}^{(3)})^{n_{x, \mu}^{(3)} + n_{x-\hat{\mu}, \mu}^{(3)}}
                                    \sin\theta_{x} d\theta_{x} d\phi_{x}.
\end{align}
Using the explicit expressions for $\sigma^{(1)}$, $\sigma^{(2)}$, and $\sigma^{(3)}$ in terms of $\phi$ and $\theta$, we can perform the $\phi$ and $\theta$ integrals separately.  We find for $\theta$,
\begin{align}
  \nonumber
  \label{eq:thetaint}
  \Theta_{x} \equiv \frac{1}{2} \int_{0}^{\pi} &(\sin\theta_{x})^{\sum_{\mu} \sum_{b = 1}^{2} (n_{x, \mu}^{(b)} + n_{x-\hat{\mu}, \mu}^{(b)}) + 1} \times \\
                                               &(\cos\theta_{x})^{\sum_{\mu} (n_{x, \mu}^{(3)} + n_{x-\hat{\mu}, \mu}^{(3)})} d\theta_{x}
\end{align}
and for $\phi$,
\begin{align}
  \nonumber
  \label{eq:phiint}
  \Phi_{x} \equiv \frac{1}{2\pi} \int_{0}^{2\pi} &(\sin\phi_{x})^{\sum_{\mu} ( n_{x, \mu}^{(2)} + n_{x-\hat{\mu}, \mu}^{(2)} )} \times \\ &(\cos\phi_{x})^{\sum_{\mu} ( n_{x, \mu}^{(1)} + n_{x-\hat{\mu}, \mu}^{(1)} )} d\phi_{x}.
\end{align}
Eq.~\eqref{eq:thetaint} can be computed exactly and gives,
\begin{align}
  \nonumber
  \label{eq:theta-beta}
  \Theta_{x} &= \frac{1}{2} \int_{0}^{\pi} (\sin\theta_{x})^{\sum_{\mu} \sum_{b = 1}^{2} (n_{x, \mu}^{(b)} + n_{x-\hat{\mu}, \mu}^{(b)}) + 1} \times \\ \nonumber
             &\hspace{1.5cm}(\cos\theta_{x})^{\sum_{\mu} (n_{x, \mu}^{(3)} + n_{x-\hat{\mu}, \mu}^{(3)})} d\theta_{x} \\ \nonumber
             &= \frac{1}{2} \delta^{\text{mod}2}_{\sum_{\mu} (n_{x, \mu}^{(3)} + n_{x-\hat{\mu}, \mu}^{(3)}), 0} \times \\ \nonumber
             &\hspace{1cm} B \left( \frac{1}{2} \left( 1 + \sum_{\mu} (n_{x, \mu}^{(3)} + n_{x-\hat{\mu}, \mu}^{(3)}) \right) \right., \\
             & \left. \hspace{1cm}\quad\quad 1 + \frac{1}{2}\sum_{\mu}\sum_{b=1}^{2}(n_{x, \mu}^{(b)} + n_{x - \hat{\mu}, \mu}^{(b)}) \right)
\end{align}
where $B(p, q)$ is the beta function\footnote{
$$B(x,y) = \int_{0}^{1} t^{x-1}(1-t)^{y-1} dt = \frac{\Gamma(x) \Gamma(y)}{\Gamma(x+y)}$$},
and $\delta^{\text{mod}2}$ is the Kronecker delta but the indices need only be equal modulo 2.  Similarly Eq.~\eqref{eq:phiint} can be computed as well giving,
\begin{align}
  \nonumber
  \label{eq:phi-beta}
  \nonumber
  \Phi_{x} &= \frac{1}{2\pi} \int_{0}^{2\pi} (\sin\phi_{x})^{\sum_{\mu} ( n_{x, \mu}^{(2)} + n_{x-\hat{\mu}, \mu}^{(2)} )} \times \\ \nonumber &\hspace{1.5cm}(\cos\phi_{x})^{\sum_{\mu} ( n_{x, \mu}^{(1)} + n_{x-\hat{\mu}, \mu}^{(1)} )} d\phi_{x} \\ \nonumber
           &= \frac{1}{\pi}  \delta^{\text{mod}2}_{\sum_{\mu} (n_{x,\mu}^{(1)} + n_{x-\hat{\mu},\mu}^{(1)}), 0} \delta^{\text{mod}2}_{\sum_{\mu} \sum_{b=1}^{2} (n_{x,\mu}^{(b)} + n_{x-\hat{\mu},\mu}^{(b)}), 0} \times \\ \nonumber
           & \hspace{1cm} B\left( \frac{1}{2} \left( 1 + \sum_{\mu}(n_{x, \mu}^{(1)} + n_{x-\hat{\mu}, \mu}^{(1)}) \right), \right. \\
           & \left. \hspace{1cm}\quad\quad \frac{1}{2} \left( 1 + \sum_{\mu}(n_{x, \mu}^{(2)} + n_{x-\hat{\mu}, \mu}^{(2)}) \right)  \right).
\end{align}
These are two constraints which need to be imposed at each site of the lattice.  With these constraints and weights from the Taylor series expansion we can now define a tensor at every lattice site.

First, define a collective index given by $N_{x, \mu} \equiv n_{x, \mu}^{(1)} \otimes n_{x, \mu}^{(2)} \otimes n_{x, \mu}^{(3)}$, as well as a weight associated with a link,
\begin{align}
  w_{x, \mu} \equiv \frac{\beta^{\frac{1}{2}\sum_{a} n_{x, \mu}^{(a)}}}{\sqrt{n_{x, \mu}^{(1)}! n_{x, \mu}^{(2)}! n_{x, \mu}^{(3)}!}}.
\end{align}
Now the tensor at site $x$ in $D$ dimensions is given by,
\begin{align}
  T_{N_{x-\hat{1}, 1} N_{x, 1}\ldots N_{x-\hat{D},D} N_{x, D}}  = \left( \prod_{\mu = 1}^{D} w_{x-\hat{\mu}, \mu} w_{x, \mu} \right) \Theta_{x} \Phi_{x}.
\end{align}
This tensor has the nice property that---assuming $\beta > 0$---the tensor elements are positive.  This follows from the positivity of $\beta$ and that the $n$s are non-negative.  The constraints coming from the $\theta$ and $\phi$ integrals are positive as well, since the beta functions are positive for positive arguments.  This allows this formulation to be used in sampling methods, which it has been (see \cite{PhysRevD.94.114503,WOLFF2010254}).  However, there are more indices necessary in this description, which increases the cost numerically in a tensor renormalization group algorithm.

\subsection{$SU(2)$ principal chiral model}

The $SU(2)$ principal chiral model consists of a $SU(2)$ matrix associated with each site on the lattice that interacts with its nearest neighbors.  The action on a $D$-dimensional square lattices with periodic boundary conditions is given by,
\begin{align}
  \label{eq:pcm-action}
  S = -\frac{\beta}{2} \sum_{x = 1}^{N} \sum_{\mu = 1}^{D} \Tr[U_{x} U_{x+\hat{\mu}}^{\dagger}].
\end{align}
The partition function for the model is given as the Haar integration over each of the fields on the lattice,
\begin{align}
  \nonumber
  Z &= \int \mathcal{D}U \, e^{-S} \\
    &= \prod_{x} \int dU_{x} e^{\frac{\beta}{2} \sum_{x, \mu} \Tr[U_{x} U_{x+\hat{\mu}}^{\dagger}]}.
\end{align}
The partition function only depends on the trace of group elements which means the characters of the group can be expanded on.  We expand the Boltzmann weight,
\begin{align}
  \label{eq:pcm-expand}
  e^{\frac{\beta}{2} \Tr[U_{x} U^{\dagger}_{x+\hat{\mu}}]} = \sum_{r_{x,\mu} = 0}^{\infty} F_{r_{x,\mu}}(\beta) \chi^{r_{x,\mu}}(U_{x} U^{\dagger}_{x+\hat{\mu}})
\end{align}
where the sum runs over all half-integer irreducible representations of the group, and the $F$s are given in Appendix~\ref{subsec:character-expansion}.  The expansion coefficients can be solved for by inverting Eq.~\eqref{eq:pcm-expand} using the orthogonality of the characters.  The characters are traces over matrix representations of the group (See Appendix~\ref{subsec:ortho}).  This allows the group elements to be split and factorized, $\chi^{r}(U_{x} U^{\dagger}_{x+\hat{\mu}}) = \sum_{a,b} D^{r}_{ab}(U_{x}) {D^{r}}^{\dagger}_{ba}(U_{x+\hat{\mu}})$, and subsequently integrated over.  Collecting all the $D$ matrices associated with the same site we find an integral of the form,
\begin{align}
  \label{eq:pcm-int}
  \int dU_{x} \, \prod_{\mu = 1}^{D} D^{r_{x, \mu}} {D^{r_{x-\hat{\mu}, \mu}}}^{\dagger}(U_{x}).
\end{align}
where the matrix indices have been suppressed. We can perform this integral with the help of the Clebsch-Gordan series,
\begin{align}
  \nonumber
  &D^{r_1}_{m_1 n_1}(U) D^{r_2}_{m_2 n_2}(U) = \\
  &\sum_{R = |r_1 - r_2|}^{r_1 + r_2} \sum_{M = -R}^{R} \sum_{N = -R}^{R} C^{R M}_{r_1 m_1 r_2 m_2} C^{R N}_{r_1 n_1 r_2 n_2} D^{R}_{M N}(U),
\end{align}
along with the orthogonality of the $D$ matrices (See Appendix~\ref{subsec:ortho}).

We now restrict to the case of $D = 2$. Eq.~\eqref{eq:pcm-int} takes the form,
\begin{align}
  \int dU_{x} \, D^{r_{x, 1}}_{m_1 n_1} {D^{r_{x-\hat{1}, 1}}_{m_2 n_2}}^{\dagger} D^{r_{x, 2}}_{m_3 n_3} {D^{r_{x-\hat{2}, 2}}_{m_4 n_4}}^{\dagger}  (U_{x})
\end{align}
and using the steps mentioned above we find,
\begin{align}
  \label{eq:pcm-const}
  \nonumber
  \mathcal{C}_{x} &\equiv \int dU_{x} \, D^{r_{x,1}}_{m_1 n_1} D^{r_{x,2}}_{m_2 n_2} {D^{r_{x-\hat{1},1}}_{m_3 n_3}}^{\dagger} {D^{r_{x-\hat{2},2}}_{m_4 n_4}}^{\dagger}(U_{x}) = \\ \nonumber
  & \sum_{R,M,N} d_{R}^{-1} C^{R M}_{r_{x,1} m_1 r_{x,2} m_2} C^{R N}_{r_{x,1} n_1 r_{x,2} n_2} \times \\
  & \quad C^{R N}_{r_{x-\hat{1},1} m_3 r_{x-\hat{2},2} m_4}
    C^{R M}_{r_{x-\hat{1},1} n_3 r_{x-\hat{2},2} n_4}
\end{align}
where $d_r = 2r+1$ is the dimension of the representation.  This is the constraint associated with a site.  Similar to the $O(3)$ nonlinear sigma model, this constraint constrains the surrounding representation numbers on the links around a site through the triangle inequalities.  If we define 
a composite index formally as $X_{x,1} \equiv \{ r_{x,1}, m, n \}$, where $m$ and $n$ are the matrix indices naturally associated with an $r$ on a link.  Then we can define a local tensor at each site by,
\begin{align}
  \nonumber
  & T_{X_{x-\hat{1}, 1} X_{x,1} Y_{x,2} Y_{x-\hat{2},2}} = \\
  & \sqrt{F_{r_{x,1}} F_{r_{x,2}} F_{r_{x-\hat{1},1}} F_{r_{x-\hat{2},2}}(\beta)} \; \mathcal{C}_{x}.
\end{align}
By contracting this tensor with itself recursively one rebuilds the original partition function.

A possible alternative way to formulate a local tensor is to use the same discrete variables used in  \cite{GATTRINGER2018435}.  This formulation follows along the same lines as the second tensor formulation for the $O(3)$ nonlinear sigma model by expanding the Boltzmann weight in a Taylor series.  However, in this reference this formulation was used in sampling methods.  We do not attempt to give the tensor formulation using these variables, but the required steps seem straightforward, and mimic the steps in the second formulation of the $O(3)$ tensor.

\subsection{Truncations and asymptotic freedom}
An important question is how do the previous tensor formulations, and specifically the expansions before-hand which lead up to the tensor definitions, affect universality.  Looking at Eq.~\eqref{eq:legend-expand} for the $O(3)$ nonlinear sigma model it is clear that this expansion does not affect the global $O(3)$ invariance of the model, since one expands on the dot product between nearest neighbor vectors.  This interaction is $O(3)$ invariant, so long as each spin is rotated by the same rotation matrix, and so a polynomial in this interaction remains $O(3)$ invariant.  Similarly a truncation in the $l$ variable to a finite $l_{\text{max}}$ leaves the expansion $O(3)$ invariant for the same reason.  Because of this, after truncation but before integration, the model consists of a local nearest neighbor interaction which is $O(3)$ invariant and in the same number of dimensions we started with indicating, naively, that this truncated model lies in the same universality class as the original $O(3)$ nonlinear sigma model.  Indeed further evidence for this conclusion is found in~\cite{PhysRevLett.126.172001}.  Likewise, the expansion in Eq.~\eqref{eq:o3-taylor} is also globally $O(3)$ invariant.  So any truncation on the $n$ variables leaves the expansion only dependent on spins which interact with their neighbors in an $O(3)$ invariant fashion.

The $O(3)$ nonlinear sigma model in two dimensions is known to be asymptotically free \cite{HASENFRATZ1990522}.  On a two-dimensional lattice, the continuum limit is approached by taking the nearest-neighbor coupling, $\beta$, infinitely large.  In this limit one expects the mass gap to obey the continuum perturbative result which predicts \cite{HASENFRATZ1990522},
\begin{align}
  \nonumber
  a m &= \frac{8}{e}a \Lambda_{\overline{\text{MS}}} \\
      & = 128 \pi \beta \exp{(-2 \pi \beta)}.
\end{align}

The mass gap can be calculated by studying the exponential decay of the spin-spin correlation function~\cite{WOLFF1990581}.  An initial study of the asymptotic scaling of the mass gap in this model was done in  \cite{JUDAH.05.21.2015} using the tensor renormalization group.  There they compared tensor renormalization group calculations of the mass gap with results from Monte Carlo simulations.  They found a slow convergence to the expected result as a function of $l_{\text{max}}$.  A more thorough study in the Hamiltonian limit was done in  \cite{bruckmann2018} using 
MPS at different truncations and different volumes.  While they found relatively good convergence to the asymptotic result for $l_{\text{max}} > 2$ and in the large volume limit, which support this idea of universality, the lower $l_{\text{max}}$ values did not converge as well.

For the action in Eq.~\eqref{eq:pcm-action} one has the freedom to rotate all group elements by the same global matrix like $U_{x} \rightarrow U'_{x} = V U_{x} V^{\dagger}$.  This leaves the action invariant, as well as the measure.  The expansion in Eq.~\eqref{eq:pcm-expand} retains this freedom since the thing that is expanded on is the trace of group elements which is the type of interaction which allows for this freedom.  Furthermore, a truncation on the sum of representations in Eq.~\eqref{eq:pcm-expand} does nothing to this symmetry.  Again, a truncation then naively leaves the model with the same nearest neighbor interaction with the same symmetries in the same number of dimensions and one expects this truncated model to lie in the same universality class as the original principal chiral model.

\section{Tensors for lattice gauge theories}
\label{sec:lgt}
In this section we will discuss gauge theories with Abelian symmetries. The gauge Ising model is the simplest model that we can consider. However, as for the spin models, we
will start with the continuous case and then obtain the models with discrete symmetries
such as the  Ising model and the gauge clock models using the substitutions described in
Sec.~\ref{subsec:discretesub}.
\def\bpl{\beta_{pl.}}
\subsection{Pure gauge $U(1)$}
\label{subsec:pure}
\begin{figure}[h]
  \centering
  \includegraphics[width=0.8\hsize]{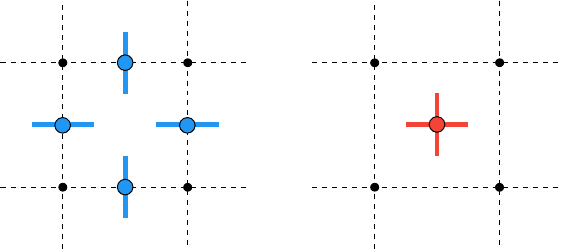}
  \vskip10pt
  \includegraphics[width=6cm]{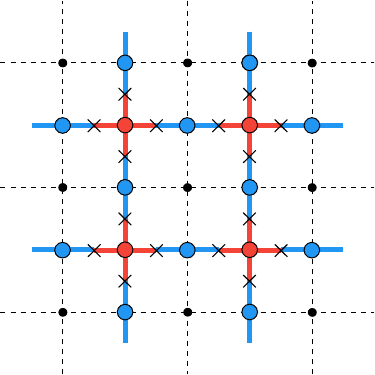}
  \caption{\label{fig:AandB}Assembly of the $A$ and $B$ tensors for $D=2$.}
\end{figure}

\begin{figure}[h]
  \centering
  \includegraphics[width=0.3\hsize]{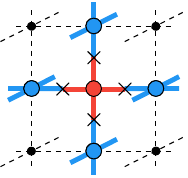}
  \hspace{0.1\hsize}
  \includegraphics[width=0.55\hsize]{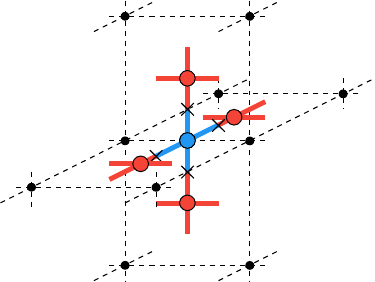}
  \caption{\label{fig:AandBd3}Assemblies of the $A$ and $B$ tensors for $D=3$.}
\end{figure}

The partition function for the pure gauge $U(1)$ model introduced in Sec.~\ref{subsec:lattac} reads
\beq
Z_{PG} =\prod_{x,\mu}\int_{-\pi}^{\pi}\frac{dA_{x,\mu}}{2\pi} e^{-S_{\text{Wilson}}},
\label{eq:u1-gaugemeasure}
\enq
with the action
\beq
S_{\text{Wilson}}=-\bpl\sum_{x,\mu<\nu} \cos(A_{x,\mu}+A_{x+\hat{\mu},\nu}-A_{x+\hat{\nu}, \mu}-A_{x,\nu}).
\label{eq:gauge}
\enq
It possesses a local symmetry
\beq
A_{x,\mu}'=A_{x,\mu}-(\alpha_{x+\hat{\mu}}-\alpha_x).
\enq
Using the Fourier expansion
\begin{eqnarray}
  &\ & {\rm e}^{\bpl  \cos(A_{x,\mu}+A_{x+\hat{\mu},\nu}-A_{x+\hat{\nu}, \mu}-A_{x,\nu})} = \\ \cr
  &\ &\sum\limits_{m_{x,\mu \nu}=-\infty}^{+\infty} {\rm e}^{i m_{x,\mu \nu}(A_{x,\mu}+A_{x+\hat{\mu},\nu}-A_{x+\hat{\nu}, \mu}-A_{x,\nu})} I_{m_{x,\mu \nu}}(\bpl)\  ,\nonumber
       \label{eq:foug}
\end{eqnarray}
to factorize the gauge fields, and integrating over $A_{x,\mu}$ using the orthogonality of the $U(1)$ elements, we obtain the selection rule
\begin{eqnarray}
  \label{eq:discrmax}
  &\ & \sum_{\nu>\mu}[m_{x,\mu \nu}-m_{x-\hat{\nu},\mu \nu}]\cr
  &+&\sum_{\nu<\mu}[ -m_{x,\nu \mu}+m_{x-\hat{\nu},\nu \mu}]\cr
  &=&0.
\end{eqnarray}
In this expression, the index $m_{x,\mu \nu}$ is associated with a plaquette starting a $x$, going in the direction of lower index $\mu$ and then the direction $\nu$.
In $D$ dimensions, there are $2(D-1)$ plaquettes attached to each link.  This selection rule constrains the $m$ values associated with those plaquettes.  It is convenient to introduce a tensor that is associated with the plaquettes of the lattice.  It has four indices which can be naturally associated with the four links bounding a plaquette.  Since each plaquette has a single $m$ value associated to it, the four tensor legs attached to a given plaquette $(x,\mu \nu)$ must carry the same index, $m$. Following the terminology of Ref. \cite{Liu:2013nsa}, we introduce a ``$B$-tensor" for each plaquette
\beq
\label{eq:ttensorB}
B^{(x,\mu \nu)}_{m_1m_2m_3m_4}
=\begin{cases}
  t_{m_1}(\bpl ),  &\mbox{if all } m_i\mbox{ are the same} \\
  0, & \mbox{otherwise}.
\end{cases}
\enq
where $t_m$ is defined in Sec.~\ref{subsec:o2}.  The $B$-tensors are assembled (traced) together with ``$A$-tensors" attached to links with $2(D-1)$ legs orthogonal to the
link $(x,\mu)$
\beq
\label{eq:u1-Atensor}
A^{(x,\mu)} _{m_1\dots m_{2(D-1)}}=
\delta_{m_{in},m_{out}},
\enq
where $\delta_{m_{in},m_{out}}$ is a short notation for Eq.~\eqref{eq:discrmax} and $m_{in} \equiv \sum_{\nu > \mu} m_{x-\hat{\nu},\mu\nu} - \sum_{\nu < \mu} m_{x-\hat{\nu}, \nu\mu}$, and $m_{out} \equiv \sum_{\nu >\mu} m_{x,\mu\nu} - \sum_{\nu < \mu} m_{x,\nu\mu}$.
Notice that in contrast to Ref. \cite{Liu:2013nsa}, the weight of the plaquettes is carried by the $B$-tensor.
The partition function with PBC can now be written as
\begin{eqnarray}
  \label{eq:traceall}
  Z&=&(I_0(\bpl))^{VD(D-1)/2} \cr
  &&\cr
  &\times &\Tr \prod_{l.}A^{(l.)}_{m_1, \dots m_{2(D-1)}}\prod_{pl.}B^{(pl.)}_{m_1m_2m_3m_4},
\end{eqnarray}
where the trace means index contraction following the geometric procedure described above.
The tensor assembly is illustrated in Fig. \ref{fig:AandB} for $D=2$ and in Fig. \ref{fig:AandBd3} for $D=3$.

\subsubsection{Discrete Maxwell equations}
\label{subsec:discrmax}
The selection rule above can be recast as a constraint on the $m$ values which surround a link.  Here we will show that Eq.~\eqref{eq:discrmax} represents a discrete version of
Maxwell's equations: $\partial _\mu F^{\mu \nu}=0$. For this purpose,
we define the ``electric integers"
\beq e_{x,j}\equiv m_{x,jD},\enq
with $j=1,\dots,D-1$, which are associated with temporal plaquettes and which can be interpreted
as electric fields.  Eq.~\eqref{eq:discrmax} for $\mu =D$ reads
\beq
\label{eq:gausslaw}
\sum_{j=1}^{D-1}(e_{x,j}- e_{x-\hat{j},j})=0.
\enq
This is a discrete form of Gauss's law, $\boldsymbol\nabla \cdot {\bf E}=0$, in the source-free model.

For $D\geq 3$, we can introduce magnetic fields in a dimension dependent way.
For $D=3$, we define
\beq
b_x\equiv m_{x,12}.
\enq
Eq.~\ref{eq:discrmax} for $\mu =1, 2$ are
\begin{eqnarray}
  e_{x,1}- e_{x-\hat{\tau},1}&=&-( b_{x}- b_{x-\hat{2}}),\cr
                                 e_{x,2}- e_{x-\hat{\tau},2}&=&( b_{x}- b_{x-\hat{1}}).
\end{eqnarray}
These are a discrete version of the $D=3$ Euclidean Maxwell's equations
\begin{eqnarray}
  \partial _1 B&=&\partial_\tau E_2\cr
                   \partial _2 B&=&-\partial_\tau E_1,
\end{eqnarray}
with $B=F^{12}$.
However, there is {\it no} discrete equation corresponding to the Maxwell equation for the dual field strength tensor
\beq
\label{eq:3dual}
\dmu\epsilon^{\mu \nu\sigma}F_{\nu \sigma}=0.
\enq
An example of a legal configuration violating the discrete version of Eq.~\eqref{eq:3dual}, also written $\partial_\tau B=-\boldsymbol\nabla \times {\bf E}$, can be constructed.

For $D=4$, we define
\beq
b_{x,j}\equiv \epsilon _{jkl}m_{x,kl},
\enq
and obtain a discrete version of
\beq
\label{eq:max2}
\partial_\tau{\bf E}=-\boldsymbol\nabla \times  {\bf B},
\enq
with the Euclidean magnetic field
\beq
F^{jk}=+\epsilon^{jkl}B^l.
\enq
Note that
the sign in Eq.~\eqref{eq:max2} is different in Euclidean and Minkowskian spaces.
Again there is no discrete version of the homogeneous equations for the dual field strength
$\partial_\tau{\bf  B}=-\boldsymbol\nabla \times {\bf E}$  and $\boldsymbol\nabla \cdot  {\bf B}=0$.

\subsubsection{Abelian gauge duality}
\label{subsec:gaugedual}
The dual construction of
Sec.~\ref{subsec:isingdual} for spin models
can be extended for models with plaquettes and higher-dimensional simplex interactions~\cite{savit77}, \emph{i.e.} interactions over higher-dimensional geometric shapes, such as cubes or tetrahedra etc.. First, if we define
$m_{\mu \nu}=-m_{\nu \mu}$ when $\mu>\nu$,
the discrete Maxwell's Eqs.~\eqref{eq:discrmax} take the obvious form
\beq
\label{eq:discovmax}
\nabla_\nu m_{\mu \nu}=0.
\enq
As explained in Sec.~\ref{subsec:spindual} this expression is divergenceless and represent $D-1$ conditions.
We can introduce a dual tensor with
$D-3$ indices \cite{savit77,savit80}
\beq
m_{\mu \nu}=\frac{1}{(D-3)!}
\epsilon_{\mu\nu\rho \mu_1 \dots \mu_{D-3}} \nabla_\rho  \tilde{C}_{\mu_1 \dots \mu_{D-3}},
\enq
which provides an automatic solution of Eq.~\eqref{eq:discovmax}. After using the $D-1$ conditions of Eq.~\eqref{eq:discovmax} we are left
with $(D-1)(D-2)/2$ independent components for $m_{\mu \nu}$. For $D=3$, there is no redundancy and we have one degree of freedom. For $D=4$, $\tilde{C}_\mu$ is defined up to a gradient and we recover the 3 degrees of freedom.

\subsection{The compact Abelian Higgs model}
\label{subsec:cahm}
The compact Abelian Higgs model (CAHM)  is a gauged version of the O(2) model where the global symmetry under a $\varphi$ shift becomes local
\beq
\varphi_x'=\varphi_x+\alpha_x.
\label{eq:varphix}
\enq
Its partition function is
\beq
Z_{CAHM}=\prod_x\int_{-\pi}^{\pi}\frac{d\varphi _x}{2\pi}\prod_{x,\mu}\int_{-\pi}^{\pi}\frac{dA_{x,\mu}}{2\pi}
e^{-S_{\text{Wilson}}-S_{U(1)}},
\label{eq:cahm-gaugemeasure}
\enq
with
\beq
\label{eq:smatter}
S_{U(1)}=-\beta_{l.} \sum\limits_{x,\mu} \cos(\varphi_{x+\hat{\mu}}-\varphi_x+A_{x,\mu})
\enq
and $S_{\text{Wilson}}$ as in Eq.~\eqref{eq:gauge}.  Using the same $U(1)$ Fourier expansions as before, the $A$-field integration can be carried out.
The integration over $A_{x,\mu}$ yields the selection rule
\begin{eqnarray}
  \label{eq:discrmax2}
  &\ & \sum_{\nu>\mu}[m_{x,\mu \nu}-m_{x-\hat{\nu},\mu \nu}]\cr
  &+&\sum_{\nu<\mu}[ -m_{x,\nu \mu}+m_{x-\hat{\nu},\nu \mu}]\cr
  &+&n_{x,\mu}\cr
  &=&0,
\end{eqnarray}
which simply inserts the $n_{x,\mu}$ in Eq.~\eqref{eq:discrmax} and corresponds to the
Maxwell equations with charges and currents
\beq
\label{eq:maxwith}\partial _\mu F^{\mu \nu}=J^\nu.
\enq
Equation~\eqref{eq:discrmax} means that the link indices $n_{x,\mu}$ can be seen as determined by unrestricted plaquette indices $m_{x,\mu \nu}$.
We write this dependence as $n_{x,\mu}(\{m\})$ as a shorthand for Eq.~\eqref{eq:discrmax2}.

Note that for $n_{x,\mu}(\{m\})$, the discrete current conservation
Eq.~\eqref{eq:noether} is automatically satisfied \cite{meurice2019}, and as long as the gauge fields are present, there is no need to enforce Eq.~(\ref{eq:noether}) independently. This is a discrete version of the fact that
Maxwell's equations with charges and currents (\ref{eq:discrmax2}) imply $\dmu J^\mu=0$.

With the introduction of the matter fields, we need to update the definition of the $A$-tensors\def\bl{\beta_{l.}}. We now have quantum numbers on the links, $n_{x,\mu}$, which are completely fixed by the plaquette quantum numbers, and they bring a weight $t_{n_{x,\mu}}(\beta _{l.})$. This translates into
\beq
\label{eq:cahm-Atensor}
A^{(x,\mu)} _{m_1\dots m_{2(D-1)}}=t_{n_{x,\mu}(\{m\})}(\bl).
\enq
since Eq.~\eqref{eq:discrmax2} gives $n_{x,\mu}$ in terms of the surrounding $m$s.
The partition function with PBC can now be written as
\begin{align}
  \label{eq:cahm-traceall}
  Z_{CAHM} &= ( I_0(\bpl))^{VD(D-1)/2}(I_0(\bl))^{VD} \\ \nonumber
           &\times \Tr \prod_{x,\mu}A^{(x,\mu)}_{m_1, \dots m_{2(D-1)}}\prod_{x,\mu \nu}B^{(x,\mu \nu)}_{m_1m_2m_3m_4}.
\end{align}

\subsection{$SU(2)$ gauge theory}
\label{sec:su2lgt}

$SU(2)$ gauge theory in $D$ dimensions is governed by an action of the form,
\begin{align}
  S_{\text{Wilson}} = -\frac{\beta_{pl.}}{2} \sum_{x=1}^{N} \sum_{\mu < \nu =1}^{D} \Re\Tr[U_{x, \mu} U_{x+\hat{\mu}, \nu} U^{\dagger}_{x+\hat{\nu}, \mu} U^{\dagger}_{x, \nu}].
\end{align}
In order to construct a local tensor we will proceed as before and use the character expansion, since the action only depends on the trace of group elements.
The partition function for this model can be written as the Haar integration over the group elements on the links of the lattice,
\begin{align}
  \nonumber
  Z &= \int \mathcal{D}U \, e^{-S_{\text{Wilson}}} \\ \nonumber
    &= \int \mathcal{D} U_{x, \mu} \prod_{x} \prod_{\mu < \nu} e^{\frac{\beta_{pl.}}{2} \Re\Tr[U_{x, \mu} U_{x+\hat{\mu}, \nu} U^{\dagger}_{x+\hat{\nu}, \mu} U^{\dagger}_{x, \nu}]} \\
    &= \int \mathcal{D} U_{x, \mu} \prod_{x} \prod_{\mu < \nu} e^{\frac{\beta_{pl.}}{2} \Re\Tr[U_{x, \mu \nu}]}
\end{align}
where $U_{x, \mu \nu}$ is the product of gauge fields around a plaquette. While this model is trivial in $D = 2$, there are no results using tensor methods in $D > 2$ for this model at the time of writing this; however, there have been tensor studies of other gauge models \cite{Kuramashi:2018mmi,Unmuth-Yockey:2018xak,Zohar:2015eda,Zohar:2016wcf}.

To proceed we expand the Boltzmann weight {(See Appendix~\ref{subsec:character-expansion} for the $F$s)},
\begin{align}
  \label{eq:ym-expand}
  e^{\frac{\beta_{pl.}}{2} \Re\Tr[U_{x, \mu\nu}]} = \sum_{r_{x,\mu\nu} = 0}^{\infty} F_{r_{x, \mu \nu}}(\beta_{pl.}) \chi^{r_{x, \mu\nu}}(U_{x, \mu\nu}).
\end{align}
This expansion associates an $r$ with each plaquette on the lattice.  The characters can be written as the trace of the product of matrix representations of the group,
\begin{align}
  \label{eq:charac-d}
  \nonumber
  \chi^{r_{x, \mu\nu}}(U_{x, \mu\nu}) = & \sum_{a,b,c,d} D^{r_{x, \mu\nu}}_{ab}(U_{x, \mu}) D^{r_{x, \mu\nu}}_{bc}(U_{x+\hat{\mu}, \nu}) \times \\
                                        &{D^{r_{x, \mu\nu}}_{cd}}^{\dagger}(U_{x+\hat{\nu}, \mu}) {D^{r_{x, \mu\nu}}_{da}}^{\dagger}(U_{x, \nu}).
\end{align}
By factorizing the group elements in this way, we can perform the link integration link-by-link, reformulating the model in terms of the discrete representations, and the matrix indices.  In $D$ dimensions, there are $2(D-1)$ plaquettes associated with each link.  The integral over the group element associated with link $(x,\mu)$ then has the form
\begin{align}
  \label{eq:su2-int}
  \int dU_{x,\mu}
  \prod_{\nu > \mu} D^{r_{x,\mu\nu}} {D^{r_{x-\hat{\nu},\mu\nu}}}^{\dagger}
  \prod_{\nu < \mu} {D^{r_{x,\nu\mu}}}^{\dagger} {D^{r_{x-\hat{\nu},\nu\mu}}}.
\end{align}
where the matrix indices have been suppressed, and the $D$-matrices are all the same $U_{x, \mu}$ rotation matrix or its Hermitian conjugate.  This integral is in general quite complicated but is simplified by using the Clebsch-Gordan series to systematically reduce Eq.~\eqref{eq:su2-int} to an integral over only two $D$-matrices.  The Clebsch-Gordan series is given by,
\begin{align}
  \label{eq:D-cgseries}
  \nonumber
  & D_{m_1 n_1}^{r_1} D_{m_2 n_2}^{r_2} = \\
  &\sum_{R = |r_1 - r_2|}^{r_1 + r_2} \sum_{M = -R}^{R} \sum_{N = -R}^{R} C_{r_1 m_1 r_2 m_2}^{R M} C_{r_1 n_1 r_2 n_2}^{R N} D^{R}_{M N}.
\end{align}
Using this we can collect the daggered, and non-daggered $D$-matrices in Eq.~\eqref{eq:su2-int}, and simplify them in pairs.  The $D$-matrices are orthogonal (See \ref{subsec:ortho}).

\begin{figure}[htbp]
  \centering
  \includegraphics[width=0.8\hsize]{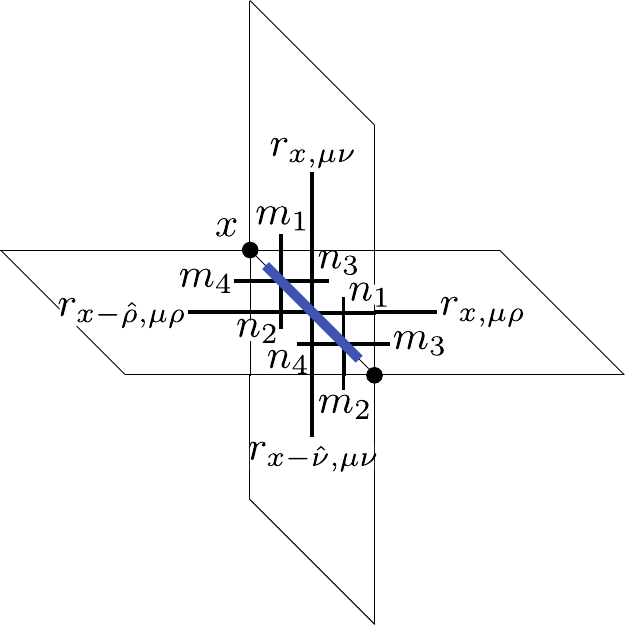}
  \caption{An illustration of the three-dimensional tensor $A$ tensor for $SU(2)$ gauge theory in three dimensions.}
  \label{fig:su2atensor}
\end{figure}

The final expression is tedious to write down, but there is nothing subtle about it.  Here we will write the final expression for $D=3$ for a link in the $\mu = 2$ direction, and subsequently write the local tensors for $D=3$ as well,
\begin{align}
  \nonumber
  \mathcal{C}^{(x,2)} &\equiv \int dU_{x,2}
    D^{r_{x,23}}_{m_1 n_1} {D^{r_{x-\hat{3},23}}_{m_2 n_2}}^{\dagger}
    {D^{r_{x,12}}_{m_3 n_3}}^{\dagger} D^{r_{x-\hat{1},12}}_{m_4 n_4} \\ \nonumber
  &= \sum_{R,M,N} d_{R}^{-1} C^{R M}_{r_{x,23}, m_1 r_{x-\hat{1},12} m_4}
    C^{R N}_{r_{x,23}, n_1 r_{x-\hat{1},12} n_4} \times \\
  & \quad \quad \quad C^{R N}_{r_{x,12}, m_{3} r_{x-\hat{3},23} m_{2}} C^{R M}_{r_{x,12}, n_3 r_{x-\hat{3},23} n_2}.
\end{align}
Then for each link in the lattice there is a constraint, $\mathcal{C}^{(x,\mu)}$, of this form.
If we define a composite index formally as $R_{x, \mu\nu} = \{ r_{x, \mu\nu}, m_1, n_1 \}$ we can define a tensor associated with the links of the lattice whose indices are associated with the shared plaquettes as,
\begin{align}
  \label{eq:su2atensor}
  A^{(x, \mu)}_{R_{x,\mu\nu} R_{x-\hat{\nu}, \mu\nu} R_{x, \mu \rho} R_{x-\hat{\rho}, \mu\rho}} = \mathcal{C}^{(x, \mu)}_{R_{x,\mu\nu} R_{x-\hat{\nu}, \mu\nu} R_{x, \mu \rho} R_{x-\hat{\rho}, \mu\rho}}
\end{align}
for $\nu \neq \rho \neq \mu$.  The $A$ tensor is defined as the constraint on a link.  An illustration of the three-dimensional tensor can be found in Fig.~\ref{fig:su2atensor}.

\begin{figure}[htbp]
  \centering
  \includegraphics[width=0.6\hsize]{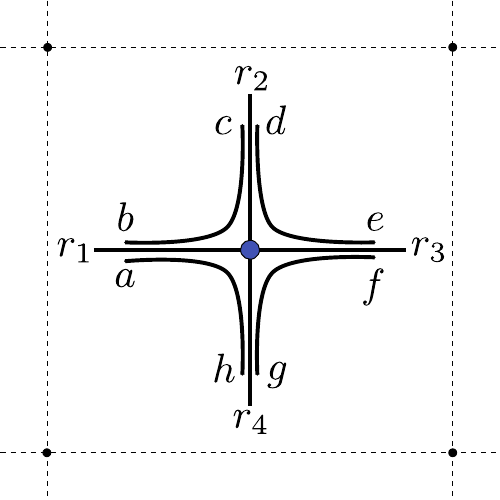}
  \caption{An illustration of $B$ tensor.}
  \label{fig:su2btensor}
\end{figure}

This is not the whole story though, since this tensor is not enough to reproduce the partition function of the original model.  The weight factors, $F_{r}(\beta_{pl.})$, still need to be accounted for.  To include the weight factors we define an additional tensor associated with the plaquettes of the lattice; however, there is a slightly subtle aspect with this tensor.  That is the circulation of the $D$-matrix indices in Eq.~\eqref{eq:charac-d} around the plaquette.  These indices---which are now a part of the $A$ tensor---are still required to be contracted in the pattern found in Eq.~\eqref{eq:charac-d}.  To enforce this circulation we assign Kronecker deltas to the new tensor in such a way that the contraction pattern of the matrix indices in Eq.~\eqref{eq:charac-d} is reproduced.  To be clear, consider Eq.~\eqref{eq:charac-d} again, rewritten,
\begin{align}
  \nonumber
  \chi^{r_{x, \mu\nu}}(U_{x, \mu\nu}) = & D^{r_{x, \mu\nu}}_{ab}(U_{x, \mu}) \delta_{bc} D^{r_{x, \mu\nu}}_{cd}(U_{x+\hat{\mu}, \nu}) \delta_{de} \times \\
                                        &{D^{r_{x, \mu\nu}}_{ef}}^{\dagger}(U_{x+\hat{\nu}, \mu}) \delta_{fg} {D^{r_{x, \mu\nu}}_{gh}}^{\dagger}(U_{x, \nu}) \delta_{ha}
\end{align}
with an implied sum over repeated indices here.  These Kronecker deltas will be moved onto the new plaquette tensor,
\begin{align}
  \label{eq:su2btensor}
  \nonumber
  &B_{ \{r_1 a b \} \{r_2 c d \} \{r_3 e f \} \{r_4 g h \}} = \delta_{bc} \delta_{de} \delta_{fg} \delta_{ha} \times \\
  &\begin{cases}
    F_{r_1}(\beta_{pl.}) & \text{if all $r$s are the same} \\
    0 & \text{otherwise.}
  \end{cases}
\end{align}
Each index of the $B$ tensor is associated with one of the four links which border the plaquette.  This makes this tensor identical for all dimensions.  An illustration of this tensor can be seen in Fig.~\ref{fig:su2btensor}.  By contracting this $B$ tensor on the plaquettes with the $A$ tensors on the links the full partition function is constructed exactly.  The contraction pattern between these indices can be seen in Fig.~\ref{fig:AandBd3}.

\subsection{The non-Abelian Higgs model}
The lattice $SU(2)$ gauge-Higgs model in $D$ dimensions consists of three main parts.  There are the pure Yang-Mills lattice action, a gauge-matter interaction term, and a matter potential term.  For the pure Yang-Mills term we use the standard Wilson action,
\begin{equation}
  \label{eq:lat-ym}
  S_{\text{Wilson}} = -\frac{\beta_{pl.}}{2} \sum_{x} \sum_{\mu < \nu} \Re\Tr[U_{x, \mu} U_{x+\hat{\mu}, \nu} U^{\dagger}_{x+\hat{\nu}, \mu} U^{\dagger}_{x, \nu}]
\end{equation}
where one takes a product of the gauge fields associated with the links around an elementary square (plaquette) for each square of the lattice.  For the gauge-matter coupling term we have,
\begin{equation}
  S_{U(2)} = -\frac{\kappa}{2} \sum_{x=1}^{N} \sum_{\mu = 1}^{D} \Phi_{x+\hat{\mu}}^{\dagger} U_{x, \mu} \Phi_{x}.
\end{equation}
The $\Phi$ field can be re-expressed in terms of a $2 \times 2$ matrix \cite{montvay_munster_1994} and the gauge-matter term becomes,
\begin{equation}
  S_{U(2)} = -\frac{\kappa}{2} \sum_{x} \sum_{\mu} \Re\Tr \left[\phi_{x+\hat{\mu}}^{\dagger} U_{x, \mu} \phi_{x}\right],
\end{equation}
where $\phi$ is now a $2 \times 2$ matrix.  Since $\phi_{x}^{\dagger} \phi_{x} = \rho_{x}^2 \mathbbm{1}$, $\phi_{x}$ can be written as $\phi_{x} = \rho_{x} \alpha_{x}$ with $\rho_{x} \in \mathbb{R}$, $\rho_{x} \geq 0$, and $\alpha_{x} \in SU(2)$.  This expresses $\phi_{x}$ in terms of the Higgs ($\rho_{x}$) and Goldstone ($\alpha_{x}$) modes, respectively. This allows the gauge-matter term to be again re-written as,
\begin{equation}
  \label{eq:gauge-matter}
  S_{U(2)} = -\frac{\kappa}{2} \sum_{x} \sum_{\mu} \rho_{x+\hat{\mu}} \rho_{x} \Re\Tr \left[ \alpha_{x+\hat{\mu}}^{\dagger} U_{x, \mu} \alpha_{x}\right].
\end{equation}
Finally, the Higgs potential,
\begin{align}
  V = \sum_{x}(|\Phi_{x}|^{2} + \lambda (|\Phi_{x}|^{2} - 1)^{2})
\end{align}
only couples same-site fields and therefore---in terms of the matrix $\phi_{x}$---only involves the Higgs mode,
\begin{equation}
  \label{eq:potential}
  V = \sum_{x} \rho_{x}^{2} + \lambda(\rho_{x}^{2} - 1)^2.
\end{equation}
The partition function for this model is then,
\begin{equation}
  Z = \int \mathcal{D}U \, \mathcal{D}\rho \, \mathcal{D}\alpha \, e^{-S_{\text{Wilson}}-S_{U(2)}-V}
\end{equation}
where the integration over $U$ and $\alpha$ is the $SU(2)$ Haar measure, and the integration measure over $\rho$ is given by $\rho_{x}^{3} d\rho_{x}$ over [0,$\infty$).

We only consider the limit in which $\lambda \rightarrow \infty$, $\rho_{x} \rightarrow 1$, which is when the Higgs mode becomes infinitely massive. In addition we perform a change of variables on the gauge fields such that $U_{x,\mu} \rightarrow U_{x, \mu}^{\prime} = \alpha_{x+\hat{\mu}}^{\dagger} U_{x, \mu} \alpha_{x}$.  Up to an overall constant this reduces the partition function to the form,
\begin{align}
  \nonumber
  Z & = \int \mathcal{D}U \, \mathcal{D}\alpha \, e^{-S_{\text{Wilson}}-S_{U(2)}} \\ \nonumber
    & = \int \mathcal{D}U \, \exp \frac{\beta_{pl.}}{2} \sum_{x} \sum_{\mu < \nu} \Re\Tr[U_{x, \mu} U_{x+\hat{\mu}, \nu} U^{\dagger}_{x+\hat{\nu}, \mu} U^{\dagger}_{x, \nu}] \\
    & +
      \frac{\kappa}{2} \sum_{x} \sum_{\mu = 1}^{D} \Re\Tr \left[ U_{x, \mu} \right].
\end{align}

The tensor formulation for this model follows very similar steps to the previous Sec.~\ref{sec:su2lgt}. 
In fact the expansion for the Yang-Mills term is identical to Eqs.~\eqref{eq:ym-expand} and~\eqref{eq:charac-d}.  The expansion for the gauge matter term is similar,
\begin{align}
  e^{\frac{\kappa}{2}\Re\Tr[U_{x, \mu}]} = \sum_{r_{x,\mu} = 0}^{\infty} F_{r_{x \mu}}(\kappa) \chi^{r_{x, \mu}}(U_{x, \mu}),
\end{align}
using the same character expansion from Sec.~\ref{sec:su2lgt}.  Similarly we know
\begin{align}
  \label{eq:chi-D}
  \chi^{r_{x, \mu}}(U_{x, \mu}) = \sum_{a} D^{r_{x, \mu}}_{a a}(U_{x, \mu}).
\end{align}
With these expansions for the gauge and gauge-matter Boltzmann weights, we find an integral for each link similar to Eq.~\eqref{eq:su2-int}; however, there is now an additional $\chi^r = \Tr[D^r]$ coming from the gauge-matter factor, giving,
\begin{align}
  \label{eq:gh-int}
  \nonumber
  \int dU_{x,\mu} \, \chi^{r_{x, \mu}}
  &\prod_{\nu > \mu} D^{r_{x,\mu\nu}} {D^{r_{x-\hat{\nu},\mu\nu}}}^{\dagger}\times \\
  & \prod_{\nu < \mu} {D^{r_{x,\mu\nu}}}^{\dagger} {D^{r_{x-\hat{\nu},\mu\nu}}}.
\end{align}
Eq.~\eqref{eq:gh-int} can again be reduced to a manageable integral over only two $D$-matrices using Eq.~\eqref{eq:D-cgseries} and the form of $\chi^r$ given in Eq.~\eqref{eq:chi-D}.

Here we will proceed setting $D = 2$ and perform the computations explicitly for the local tensors.  This was done in detail in  \cite{Bazavov:2019qih}.  Eq.~\eqref{eq:gh-int} for the $\mu = 1$ direction takes the form,
\begin{align}
  \nonumber
  & \sum_{k} \int dU_{x, 1} \, D^{r_{x, 1}}_{k k} D^{r_{x,12}}_{m_1 n_1} {D^{r_{x-\hat{2}, 12}}_{m_2 n_2}}^{\dagger} = \\
  & \sum_{k} d_{r_{x-\hat{2},12}}^{-1} C^{r_{x-\hat{2},12} n_2}_{r_{x, 1} k \, r_{x, 12} m_1} C^{r_{x-\hat{2},12} m_2}_{r_{x, 1} k \, r_{x, 12} n_1},
\end{align}
and in the $\mu = 2$ direction,
\begin{align}
  \label{eq:mu2-const}
  \nonumber
  & \sum_{k} \int dU_{x, 2} \, D^{r_{x, 2}}_{k k} {D^{r_{x,12}}_{m_1 n_1}}^{\dagger} D^{r_{x-\hat{1}, 12}}_{m_2 n_2} = \\
  & \sum_{k} d_{r_{x, 12}}^{-1} C^{r_{x, 12} n_1}_{r_{x, 2} k \, r_{x-\hat{1}, 12} m_2} C^{r_{x, 12} m_1}_{r_{x, 2} k \, r_{x-\hat{1}, 12} n_2}.
\end{align}
With these constraints on the links we can define analogous $A$ tensors on the links as well.  We again formally define a composite index $R_{x, \mu \nu} = \{ r_{x, \mu\nu}, m, n \}$ and define a tensor on a link from site $x$ in the $\mu = 1$ direction as,
\begin{align}
  \nonumber
  A^{(x, 1)}_{R_{x, 12} R_{x-\hat{2}, 12}} = &\sum_{r_{x, 1}} F_{r_{x, 1}}(\kappa) \times \\
                                             & \sum_{k} d_{r_{x-\hat{2},12}}^{-1} C^{r_{x-\hat{2},12} n_2}_{r_{x, 1} k \, r_{x, 12} m_1} C^{r_{x-\hat{2},12} m_2}_{r_{x, 1} k \, r_{x, 12} n_1}
\end{align}
and in the $\mu = 2$ direction as,
\begin{align}
  \nonumber
  A^{(x, 2)}_{R_{x, 12} R_{x-\hat{1}, 12}} = &\sum_{r_{x,2}} F_{r_{x, 2}}(\kappa) \times \\
                                             & \sum_{k} d_{r_{x, 12}}^{-1} C^{r_{x, 12} n_1}_{r_{x, 2} k \, r_{x-\hat{1}, 12} m_2} C^{r_{x, 12} m_1}_{r_{x, 2} k \, r_{x-\hat{1}, 12} n_2}.
\end{align}
As mentioned before, the tensor associated with the plaquettes from the pure Yang-Mills term is the same as Eq.~\eqref{eq:su2btensor}, and is the same regardless of dimension for the $SU(2)$ gauge-Higgs model as well.  With the $A$ and $B$ tensors mentioned here, one can contract them in the appropriate pattern to construct the partition function exactly.  This contraction pattern is shown in Fig.~\ref{fig:AandB}.

In fact, in $D=2$ it is possible to go one step further and define a single tensor which can be contracted with itself to construct the partition function.  The details of this construction can be found in \cite{Bazavov:2019qih}.

Within this tensor reformulation it is also possible to define the Polyakov loop straightforwardly.  For $SU(2)$ in the fundamental representation the Polyakov loop at site $x^{*}$ is given by
\begin{align}
  \label{eq:ploop}
  P_{x^{*}} = \Tr \left[ \prod_{n = 0}^{N_{\tau}-1} D^{\frac{1}{2}}(U_{x^{*}+n\hat{\tau},\tau}) \right],
\end{align}
where $\tau$ indicates a direction chosen as time. 
{Here we assume periodic boundary conditions for both directions.}
The expectation value of this operator is,
\begin{align}
  \langle P \rangle = \frac{1}{Z} \int \mathcal{D}U \; P \; e^{-S}.
\end{align}
One can recast this average in terms of local tensors by performing the same steps as before.  The only difference in this case is that for a particular spatial site, $x^{*}$, all the temporal links have an additional $D$-matrix associated with them, altering the integral found in, say, Eq.~\eqref{eq:mu2-const} by the inclusion of a fourth $D$-matrix whose representation is $1/2$.  However, one proceeds as before using Eq.~\eqref{eq:D-cgseries} to make the integral manageable.  The integral on the temporal links of the Polyakov loop have the form,
\begin{align}
  \nonumber
  & \sum_{k} \int dU_{x, 2} \, D^{r_{x, 2}}_{k k} {D^{r_{x,12}}_{m_1 n_1}}^{\dagger} D^{r_{x-\hat{1}, 12}}_{m_2 n_2} D^{\frac{1}{2}}_{ij} = \\ \nonumber
  & \sum_{k,R, M, N} C^{R M}_{r_{x, 2} k \, r_{x-\hat{1}, 12} m_2} C^{R N}_{r_{x, 2} k \, r_{x-\hat{1}, 12} n_2} \times \\ \nonumber
  & \int dU_{x, 2} \, D^{R}_{M N} {D^{r_{x,12}}_{m_1 n_1}}^{\dagger} D^{\frac{1}{2}}_{ij} = \\ \nonumber
  & \sum_{k,R, M, N} d_{r_{x,12}}^{-1} C^{R M}_{r_{x, 2} k \, r_{x-\hat{1}, 12} m_2} C^{R N}_{r_{x, 2} k \, r_{x-\hat{1}, 12} n_2} \times \\
  & C^{r_{x, 12} n_1}_{R M \frac{1}{2} i}
    C^{r_{x, 12} m_1}_{R N \frac{1}{2} j}.
\end{align}
If we define this constraint as $\tilde{\mathcal{C}}_{R_{x, 12}, R_{x-\hat{1}, 12} i j}$ then we can write down the tensor on the Polyakov loop links,
\begin{align}
  \label{eq:ploop_aten}
  \tilde{A}_{R_{x^{*}, 12}, R_{x^{*}-\hat{1}, 12} i j} = \sum_{r_{x^{*}, 2}} F_{r_{x^{*}, 2}}(\kappa) \, \tilde{\mathcal{C}}_{R_{x^{*}, 12}, R_{x^{*}-\hat{1}, 12} i j}.
\end{align}
This tensor has two more indices than the typical $A$ tensor.  This is because of the additional $D$-matrix from the Polyakov loop insertion.  These additional matrix indices are contracted with each other and traced over as in the definition in Eq.~\eqref{eq:ploop}.

\begin{figure}[t]
  \centering
  \includegraphics[width=0.49\textwidth]{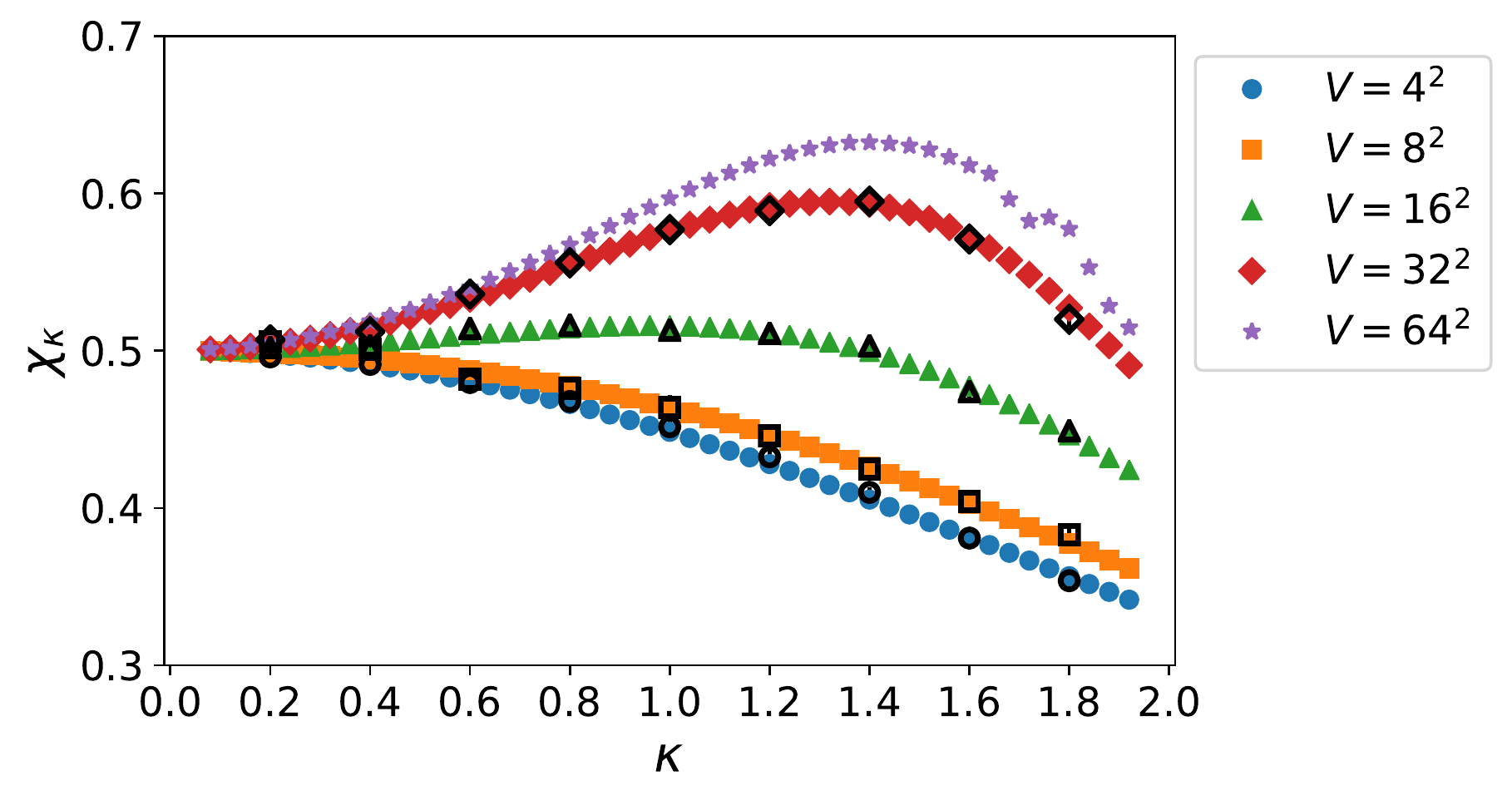}
  \caption{Adapted from~\cite{Bazavov:2019qih}. The susceptibility in the gauge-matter interaction.  There is a peak around $\kappa \approx 1.4$ which seems to indicate a cross-over between a confining regime characteristic of a pure Yang-Mills theory, and a Higgs regime where string-breaking occurs.  Here $\beta/V = 0.01$ was held fixed. The colored symbols are computed using the HOTRG, while the hollow black markers are from Monte Carlo data as a check.  The maximum representation used in the HOTRG calculation was $r = 1$, and the final number of states kept was 50.}
  \label{fig:lphi-chi}
\end{figure}

With the local tensor from \cite{Bazavov:2019qih} and Eq.~\eqref{eq:ploop_aten} it is possible to use coarse-graining schemes to approximate the free energy, and compute expectation values.  Using the higher-order tensor renormalization group for the case of $D = 2$, in  \cite{Bazavov:2019qih} derivatives of the free energy were computed along with the Polyakov loop, and Polyakov loop correlator.  Of the derivatives of the free energy, one of primary interest is the average of the gauge-matter interaction, and its fluctuations,
\begin{align}
  \langle L_{\phi} \rangle  = \frac{1}{V} \frac{\partial \ln Z}{\partial \kappa}, \quad \chi_{L_{\phi}} = \frac{1}{V} \frac{\partial^{2} \ln Z}{\partial \kappa^{2}} .
\end{align}
These were computed while taking the continuum limit.  The continuum limit in this model is controlled by the Yang-Mills coupling and the system volume, since the Yang-Mills coupling is dimensionful in $D = 2$.  By fixing the ratio,
$\beta / V = c$,
with $c$ a constant, and increasing the system volume one approaches the fixed-physical volume continuum limit.
An interesting result from this study was evidence for a cross-over transition between a confining---pure Yang-Mills---regime, and a Higgs-regime.  This can be seen from the expectation value of the squared fluctuations of the gauge-matter interaction in Fig.~\ref{fig:lphi-chi}.
In this figure one can see a gradual convergence as the continuum limit is approached, and the presence of a peak around $\kappa \approx 1.4$ separating the two regimes.

\begin{figure}[t]
  \centering
  \includegraphics[width=0.49\textwidth]{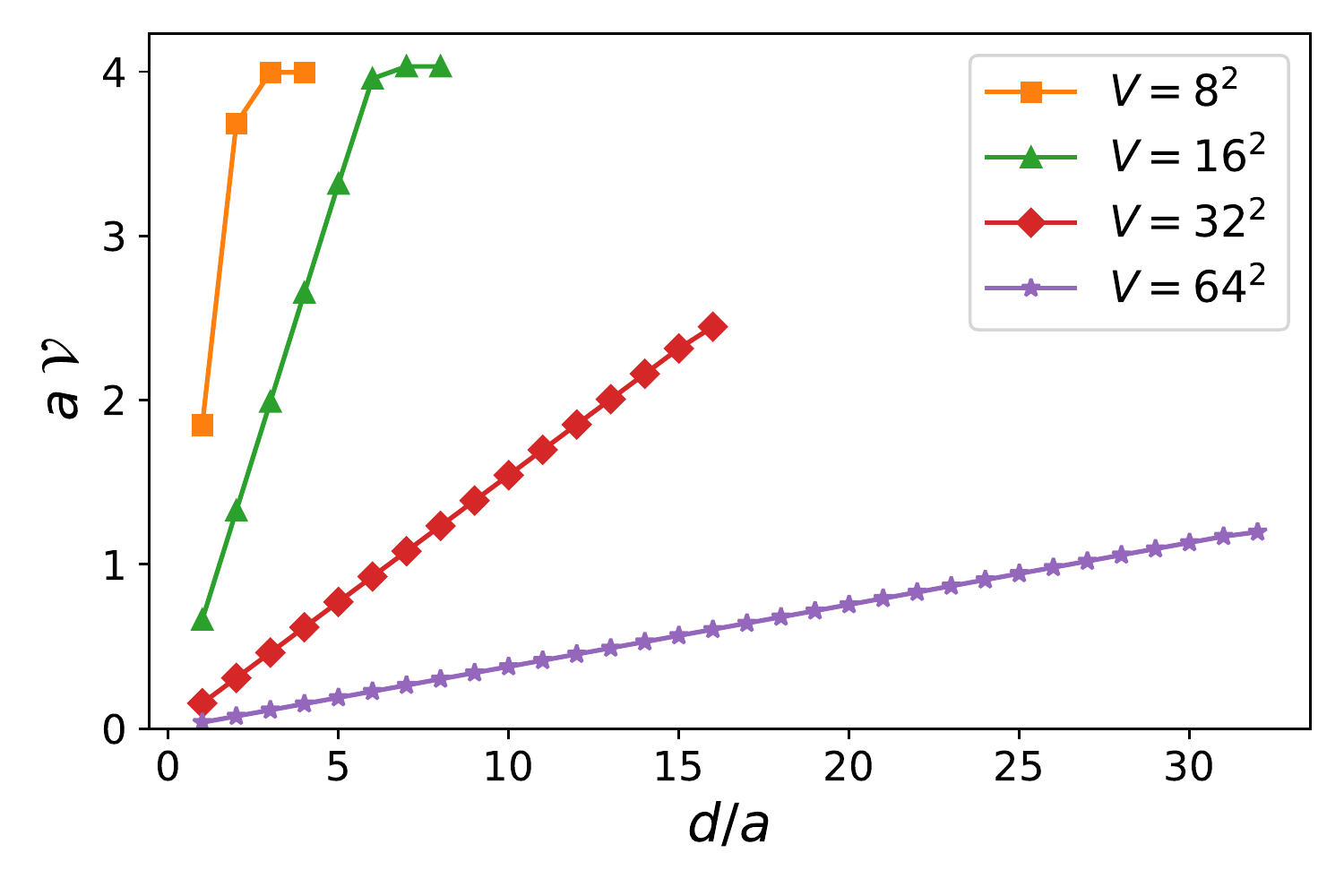}
  \caption{Adapted from~\cite{Bazavov:2019qih}.  The potential, $\mathcal{V}$, between two static charges, for $\kappa = 0.5$, as a function of distance, taking the continuum limit.  For this value of $\kappa$ we see string breaking at small systems but as the large lattice volume limit is taken we find a linear potential across long distances.  Here $\beta/V = 0.01$ was held fixed.  The maximum representation used in the calculation was $r = 1$, and the final number of states kept in the calculation was 50.}
  \label{fig:pp-small-k}
\end{figure}

\begin{figure}[t]
  \centering
  \includegraphics[width=0.49\textwidth]{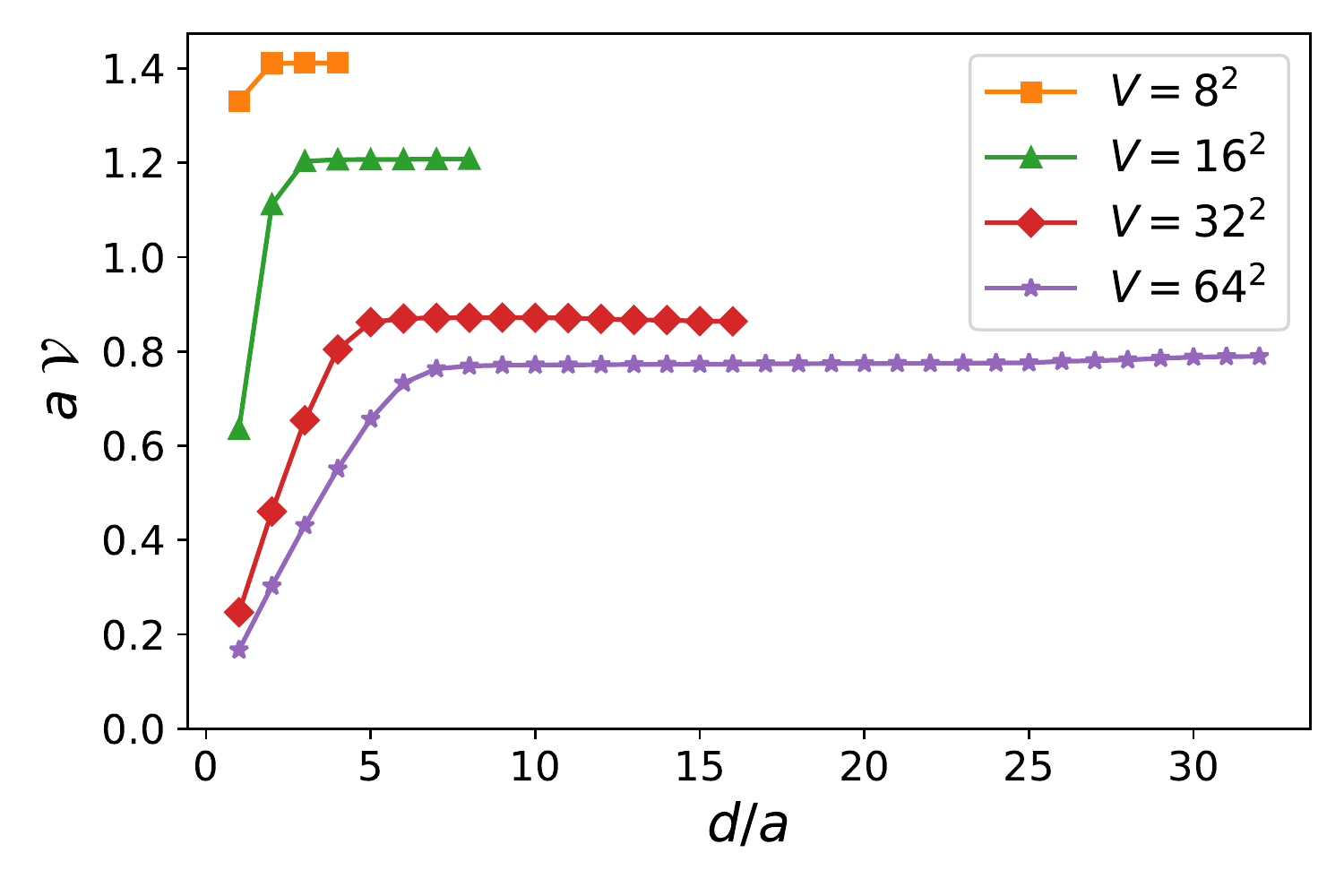}
  \caption{Adapted from~\cite{Bazavov:2019qih}.  The potential, $\mathcal{V}$, between two static charges, for $\kappa = 2$, as a function of distance, taking the continuum limit.  Here we find the phenomenon of string breaking which persists even as we take the continuum limit after some distance.  Here $\beta/V = 0.01$ was held fixed.  The maximum representation used in the calculation was $r = 1$, and the final number of states kept in the calculation was 50.}
  \label{fig:pp-large-k}
\end{figure}

This is further supported by the behavior of the Polyakov loop correlation function on either side of the peak value.  Figs.~\ref{fig:pp-small-k} and~\ref{fig:pp-large-k} show examples of the potential between static charges---$\mathcal{V}$---in the $\kappa < 1.4$ and $\kappa > 1.4$ regimes, respectively.  The potential is found from the logarithm of the correlator.
In Fig.~\ref{fig:pp-small-k} there is a linear confining potential which persists for long distances as the continuum limit is approached.  In the Higgs-like regime seen in Fig.~\ref{fig:pp-large-k} there is a linear potential for short distances, however after a certain distance the potential flattens and the forces between charges is zero (string breaking). The two regimes separated by a peak around $\kappa \approx 1.4$ both appear to be confining, with only a cross-over separating the two regimes.

\section{Tensors for models with (non-compact) scalars}
\label{sec:scalar}

For tensor networks with discrete indices, scalar fields have to be discretized in a proper manner.
There are several ways to apply simple discretization rules to scalar fields to make transfer matrices~\cite{1999PhRvB..6011761C,2001JPhA...3411215N,2001PhRvB..64f4510N,2002PhRvL..88e7203L,Iblisdir:2006tn}.
In this section we discuss the cases of Lagrangian path integrals.
Specifically, we consider a TRG study of the real $\phi^{4}$ theory in two dimensions.

This section is organized as follows.
We first present the definition of the real $\phi^{4}$ model in Sec.~\ref{sec:rphi4}.
In Sec.~\ref{sec:rphi4tnrep}, a tensor network representation of the real $\phi^{4}$ model is made via the Gaussian quadrature rule.

\subsection{Real $\phi^{4}$ theory}
\label{sec:rphi4}

The Euclidean action of the real $\phi^{4}$ theory in two dimensions is
\begin{align}
  \label{eq:rphi4act}
  S_{\mathrm{cont.}} =
  \int d^{2}x
  \left\{
  \frac{1}{2} \left( \partial_{\nu} \phi \right)^{2}
  + \frac{{\mu}_{0}^{2}}{2} \phi^{2}
  + \frac{{\lambda}}{4} \phi^{4}
  \right\},
\end{align}
where $\mu_{0}$ and $\lambda$ are the bare mass and the bare coupling, respectively.
$\phi$ is an one-component real scalar field.
This model possesses the spontaneous breaking of the $Z_{2}$ symmetry,
where the expectation value of the field $\left< \phi \right>$ is an order parameter.

From here on, we treat the model on a square lattice with periodic boundary conditions.
The lattice action is given by
\begin{align}
  \label{eq:rphi4lattact}
  S_{\text{scalar}}
  = \sum_{x} \left\{
  \frac{1}{2} \sum_{\nu=1}^{2} \left( \phi_{x+\hat{\nu}} - \phi_{x} \right)^{2}
  + \frac{\mu_{0}^{2}}{2} \phi_{x}^{2}
  + \frac{\lambda}{4} \phi_{x}^{4} \right\},
\end{align}
where $x$ is the lattice coordinate and $\hat{\nu}$ denotes the unit vector along the $\hat{\nu}$-direction.

In two-dimensional scalar theories, one has to take care of the divergence of the one-loop self energy.
In this section the following renormalization condition for the squared-mass,
\begin{align}
  \label{eq:renormmass2}
  \mu^{2}
  = \mu_{0}^{2}
  + 3 \lambda A\left( \mu^{2} \right),
\end{align}
is used to define the renormalized squared-mass $\mu$.
$A\left( \mu^{2} \right)$ denotes the one-loop self energy on the lattice
\begin{align}
  \label{eq:amulatt}
  A\left( \mu^{2} \right)
  = \frac{1}{V} \sum_{k_{1}, k_{2} = 1}^{N}
  \frac{1}{\mu^{2} + 4 \sin^{2}\left( \pi k_{1} / N \right) + 4 \sin^{2} \left( \pi k_{2} / N \right)}
\end{align}
with the lattice volume $V = N \times N$.
To provide numerical results the nonlinear equation in Eq.~\eqref{eq:renormmass2} is solved to translate the bare squared-mass into the renormalized one.
Note that the coupling constant is free of the renormalization;
this is a common property in two dimensional scalar theories.
Renormalization in scalar field theories and especially in $\phi^4$ theory is discussed in~\cite{Coleman:1974bu,Chang:1976ek}.

\subsection{
  Tensors from Gaussian quadrature}
\label{sec:rphi4tnrep}

In this subsection a tensor network representation of the real $\phi^{4}$ theory is derived by using the Gaussian quadrature rule.
This method is given and used in tensor network studies for Lagrangian path integrals in~\cite{Kadoh:2018hqq,Kadoh:2018tis,Kadoh:2019ube} and as discussed in these references improves the accuracy of an earlier tensor network study of the real $\phi^{4}$ model ~\cite{Shimizu:2012wfa}.

The partition function on the lattice is defined by
\begin{align}
  \label{eq:rphi4pfunc}
  Z = \left(\prod_{x} \int_{-\infty}^{\infty} d \phi_{x} \right) e^{-S_{\text{scalar}} - S_{h}},
\end{align}
where
\begin{align}
  \label{eq:rphi4lattacth}
  S_{h}
  = \sum_{x}
  - h \phi_{x}.
\end{align}
For later use we introduce the external field $h$ here.
An important step to build a tensor network representation is generating discrete degrees of freedom that are the candidates for tensor indices\footnote{
See~\cite{PhysRevB.100.195106,Vanhecke:2019pez}, and a more complete list of references in ~\cite{Kadoh:2018tis} for related approaches.}.
Then, in the following, we mainly discuss how to extract the discrete degrees of freedom from the continuous and non-compact scalar fields.

Since the action has only the nearest neighbor interactions,
the Boltzmann weight can be rewritten as a product of local factors
\begin{align}
  \label{eq:rphi4bweight}
  e^{-S_{\text{scalar}} - S_{h}} = \prod_{x} \prod_{\nu=1}^{2} f\left( \phi_{x}, \phi_{x+\hat{\nu}} \right),
\end{align}
where the local factor is given by
\begin{align}
  \label{eq:rphi4lfactor}
  f\left( \phi_{1}, \phi_{2} \right)
  = \exp \Biggl\{
  & - \frac{1}{2}\left( \phi_{1} - \phi_{2} \right)^{2}
    - \frac{\mu_{0}^{2}}{8}\left( \phi_{1}^{2} + \phi_{2}^{2} \right) \nonumber \\
  & - \frac{\lambda}{16} \left( \phi_{1}^{4} + \phi_{2}^{4} \right)
    + \frac{h}{4}\left( \phi_{1} + \phi_{2} \right)
    \Biggr\}.
\end{align}
To derive the discrete formula,
we briefly summarize the Gaussian quadrature rule for a weighted integral of a single variable function.
We consider to discretize the (well-defined) target integral of a function $g$
\begin{align}
  \label{eq:1dimint}
  I = \int_{-\infty}^{\infty} dx \; W(x) \; g(x),
\end{align}
where $W$ is a weight function.
A successful way to discretize this type of integral is the Gaussian quadrature method.
The quadrature rule gives a simple replacement of the integral with a discrete summation
\begin{align}
  \label{eq:disc1dimint}
  I
  \approx \sum_{i=1}^{K} w_{i} \; g(y_{i}),
\end{align}
where $y_{i}$ and $w_{i}$ are the $i$-th root of the order $K$ orthonormal polynomial, and corresponding weight, respectively.
Comprehensive definitions for the Gaussian quadrature rule (including the definition of weights) are given in~\cite{abramowitz1965handbook}.
The species of the orthonormal polynomial is one's choice and corresponds to the form of the weight function $W$.
Typical choices are the Legendre polynomials, and the Hermite polynomials that correspond to $W\left( y \right) = 1$ and $W\left( y \right) = e^{-y^{2}}$, respectively.
If we consider that the mass term in the action plays the role of the weight function,
it seems to be natural to use the Hermite polynomials.
Indeed, this choice has been used in Refs.~\cite{Kadoh:2018hqq,Kadoh:2018tis,Kadoh:2019ube},
and we exclusively use the Hermite polynomials for the Gaussian quadrature rule in this section~\footnote{
  As mentioned in the main text, the choice does not matter to the numerical accuracy as long as the degree of the orthonormal polynomial is sufficiently large.
}.
When $g$ is a polynomial function of degree $2K - 1$ or less,
the Gaussian quadrature reproduces the exact value.
Even if not, if $g$ is well approximated by a polynomial function of degree $2K - 1$ or less,
the Gaussian quadrature would be accurate.
We apply this quadrature rule to each integral of the scalar field in the path integral.

By applying the Gauss--Hermite quadrature to the partition function, a discrete form is introduced as
\begin{align}
  \label{eq:rphi4discpfunc}
  Z\left( K \right)
  = \sum_{\left\{ \alpha \right\}} \prod_{x} w_{\alpha_{x}} e^{y_{\alpha_{x}}^{2}} \prod_{\nu=1}^{2} f\left( y_{\alpha_{x}}, y_{\alpha_{x+\hat{\nu}}} \right),
\end{align}
where $\sum_{\left\{ \alpha \right\}}$ denotes $\prod_{x} \sum_{\alpha_{x}=1}^{K}$.
The discrete form depends on the order of the Hermite polynomial $K$,
and this parameter would be set to be large for accurate results.
In practice $K \geq 64$ could be regarded as sufficiently large~\cite{Kadoh:2018hqq,Kadoh:2018tis,Kadoh:2019ube}, though,
in numerical results shown later in this section, $K=256$ is taken.

Note that the method is applied numerically and that so far there is
no analog of character expansions and orthogonality relations used in the previous sections. In Sec. \ref{subsec:algebra} we show that for the Gauss-Hermite quadrature, it is possible to interpret the construction in terms of a truncated version of the harmonic oscillator algebra of creation and annihilation operators.

In Eq.~\eqref{eq:rphi4discpfunc} the local Boltzmann factors can be regarded as $K \times K$ matrices,
and one can perform the SVD for them:
\begin{align}
  \label{eq:rphi4svdf}
  f\left( y_{\alpha_{x}}, y_{\beta_{x+\hat{\nu}}} \right)
  = \sum_{i_{x,\nu}=1}^{K} U_{\alpha_{x} i_{x,\nu}} \lambda_{i_{x,\nu}} V^{\dagger}_{i_{x,\nu} \beta_{x+\hat{\nu}}},
\end{align}
where $\left\{ \lambda \right\}$ is the singular values that are assumed to be in descending order ($\lambda_{1} \ge \lambda_{2} \ge \cdots \ge \lambda_{K} \ge 0$),
and $U$ and $V$ are unitary matrices.
Now a tensor network representation of $Z\left( K \right)$ is defined by
\begin{align}
  \label{eq:rphi4tnrep}
  Z\left( K \right)
  =
  \sum_{\left\{ X,T \right\}}
  \prod_{x}T\left( K \right)_{X_{x-\hat{1}} X_{x}  T_{x} T_{x-\hat{2}}},
\end{align}
where $\sum_{\left\{ X,T \right\}}$ denotes $\prod_{x} \sum_{X_{x}=1}^{K} \sum_{T_{x}=1}^{K}$, and we have made the replacement $i_{x,1} \rightarrow X_{x}$ and $i_{x,2} \rightarrow T_{x}$.  The tensor (at any site) is defined by
\begin{align}
  \label{eq:rphi4tensor}
  T\left( K \right)_{i j k l}
  = \sqrt{\lambda_{i} \lambda_{j} \lambda_{k} \lambda_{l}} \sum_{\alpha =1}^{K} w_{\alpha}
  e^{y_{\alpha}^{2}}
   V^{\dagger}_{i \alpha} U_{\alpha j} U_{\alpha k} V^{\dagger}_{\alpha l}.
\end{align}
At this stage the bond dimension of the tensors is $K$.
To take a balance of computational cost and numerical accuracy, one may initially truncate the bond dimension to a certain value $D_{\mathrm{cut}}$ ($\le K$).
In~\cite{Kadoh:2018tis}, $D_{\mathrm{cut}} \le 64$ is taken for actual computations and the sufficiency of this choice is numerically shown~\footnote{
  Note that a fast decay of the singular values guaranties the accuracy of such an approximation.
  Although the decay rate would be weak near the criticality, a notable accuracy of the critical coupling constant is achieved in~\cite{Kadoh:2018tis}.
  This is reviewed later in this section.
}.

Physical quantities can also be expressed as tensor networks.
A key point is to respectively treat the denominator and the numerator in the right-hand side of
\begin{align}
  \label{eq:rphi4exp}
  \left< \phi \right>
  = \frac{Z_{1}}{Z},
\end{align}
where
\begin{align}
  \label{eq:rphi4z1}
  Z_{1}
  = \left( \prod_{x} \int_{-\infty}^{\infty} d \phi_{x} \right) \phi_{\tilde{x}} e^{-S_{h} - S_{\text{scalar}}}
\end{align}
with $\tilde{x} \neq x$~\footnote{
  Because of the translation invariance, a subscript that denotes the coordinate in Eq.~\eqref{eq:rphi4exp} is omitted.
}.
The presence of $\phi_{\tilde{x}}$ does not affect the tensor construction procedure, it merely alters what integral is being approximated,
so Gaussian quadrature rule and the SVD of the local factors works as before.
However, the resultant tensor network representation contains an ``impurity tensor'' owing to $\phi_{\tilde{x}}$.
To perform a coarse-graining of a tensor network that contains impurities, one needs a little ingenuity;
the details are discussed in Sec.~\ref{subsec:tensor-obs}.
Finally, one can calculate $Z$ and $Z_{1}$ separately,
and using them the value of $\left< \phi \right>$ is obtained using Eq.~\eqref{eq:rphi4exp}.

Here the numerical results for the real $\phi^{4}$ theory are shown.
The target quantity in this section is the critical coupling constant and its continuum limit value;
in~\cite{Kadoh:2018tis}, the continuum limit extrapolation proceeds as:
1) Take the thermodynamic and zero-external field limits to get a susceptibility $\chi$ for given (bare) mass and (bare) coupling constant,
2) Find the critical mass where $\chi \to \infty$,
3) Extract renormalized critical squared-mass from Eq.~\eqref{eq:renormmass2},
4) Compute the dimensionless critical coupling constant $\lambda / \mu_{\mathrm{c}}^{2}$,
5) Vary $\lambda$ from $0.1$ to $0.005$ and repeat above procedure to take the continuum limit~\footnote{
  Note that, if we do not omit showing the lattice spacing $a$, $a\lambda \to 0$ means the continuum limit.
},
6) Take a linear extrapolation to find the critical coupling constant at $\lambda = 0$.
As mentioned in the previous sections, the parameters for the tensor network analysis are set to $K = 256$ and $D \le 64$.
The legitimacy of this choice is confirmed numerically in~\cite{Kadoh:2018tis}.

\begin{figure}[htbp]
  \centering
  \includegraphics[width=\hsize]{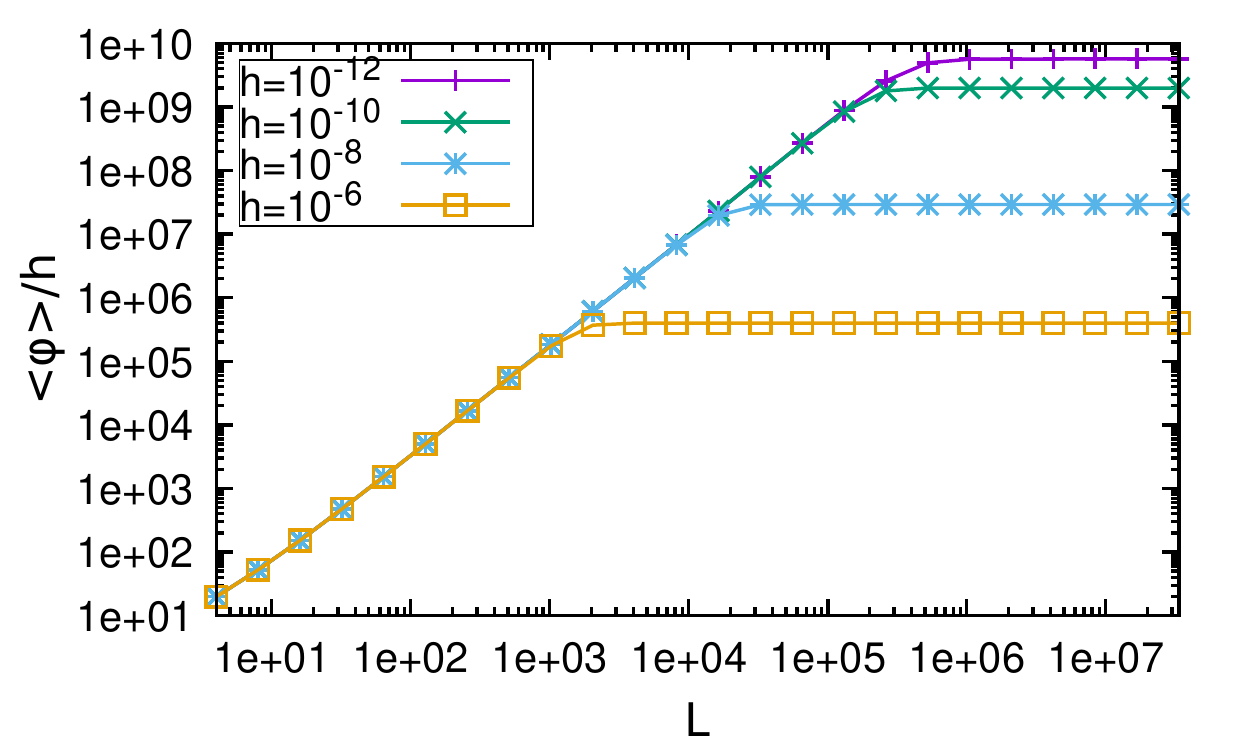}
  \caption{
    Thermodynamic limit of $\left< \phi \right>/h$
    at $\mu_{0}^{2} = -0.1006174$, $\lambda=0.05$, $\ds=32$ for $h \in \left[ 10^{-12}, 10^{-6} \right]$.
  }
  \label{fig:rphi4thermodynamiclimit}
\end{figure}

\begin{figure}[htbp]
  \centering
  \includegraphics[width=\hsize]{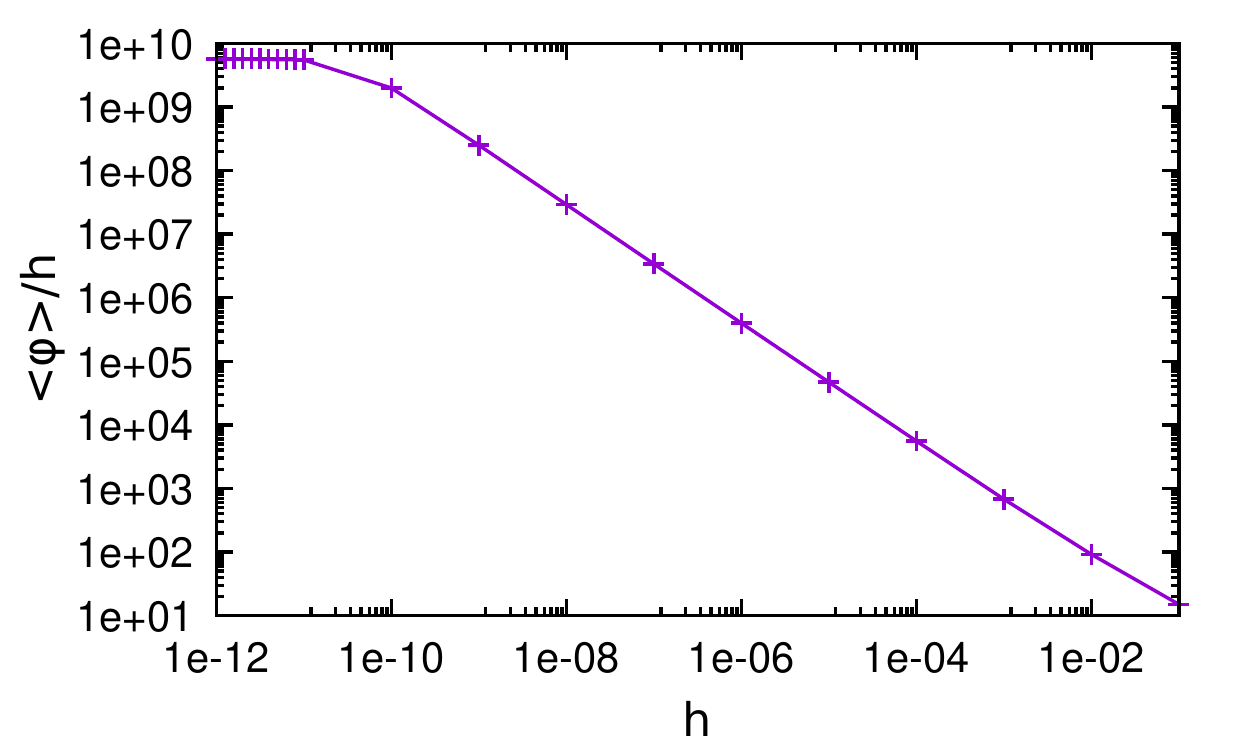}
  \caption{
    Zero-external field limit of $\left< \phi \right>/h$
    at $\mu_{0}^{2} = -0.1006174$, $\lambda=0.05$, $\ds=32$ in the thermodynamic limit.
  }
  \label{fig:rphi4zerohlimit}
\end{figure}

Figures~\ref{fig:rphi4thermodynamiclimit} and~\ref{fig:rphi4zerohlimit} show the results of the thermodynamic limit and zero-external field limit, respectively.
In both cases, (bare) parameters and the bond dimension of the tensor are $\mu_{0}^{2} = -0.1006174$, $\lambda = 0.05$, and $\ds=32$ as an example.
In Fig.~\ref{fig:rphi4thermodynamiclimit} the ratio $\left< \phi \right>/h$ behaves as a constant in the extremely large space-time volume where $L \ge 10^{6}$,
so that one can consider that the system reaches the thermodynamic limit for $L \ge 10^{6}$.
In Fig.~\ref{fig:rphi4zerohlimit} the ratio also behaves as a constant for $h \le 10^{-11}$, so that $\left< \phi \right> \approx \chi h$ holds.
Then the susceptibility $\chi$ can be obtained from the relation~\footnote{
  Actually the ratio shows a quadratic behavior for $h \le 10^{-11}$,
  so it is proper to take the susceptibility using more suitable fitting function.
  In~\cite{Kadoh:2018tis} the susceptibility is defined in such a way.
}.

\begin{figure}[htbp]
  \centering
  \includegraphics[width=\hsize]{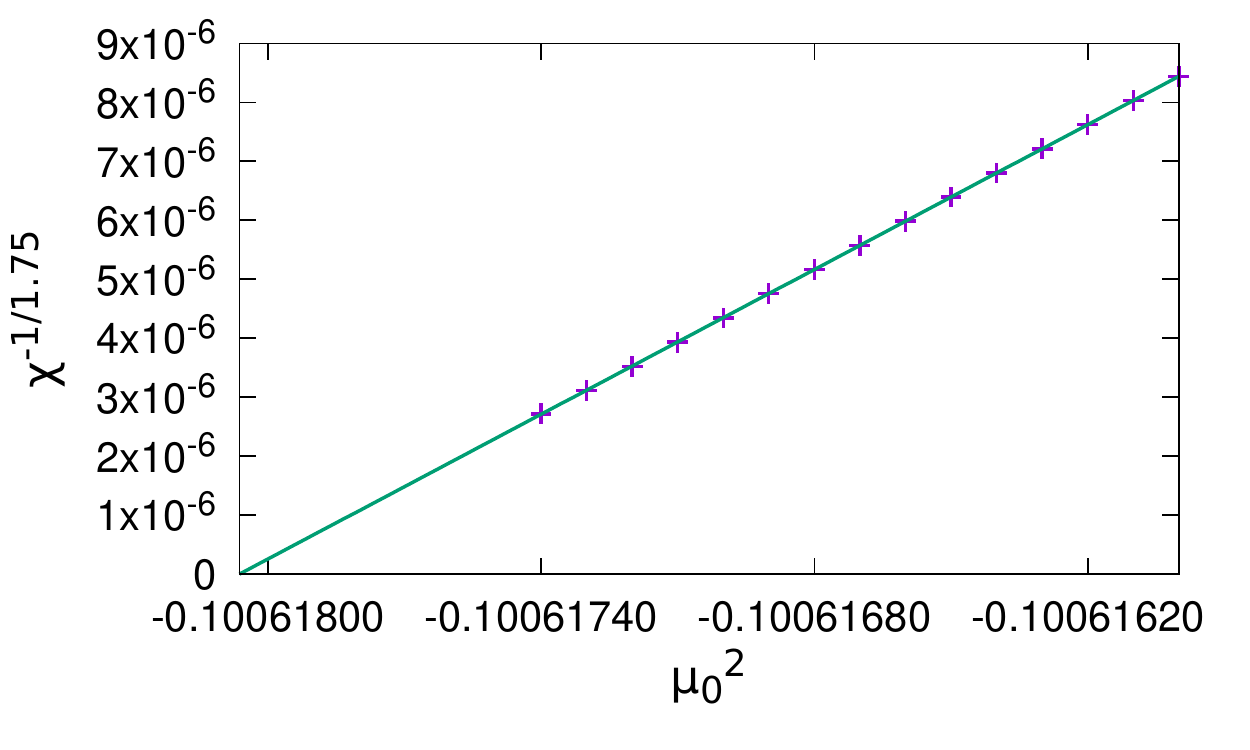}
  \caption{The susceptibility as a function of $\mu_{0}^{2}$ at $\lambda=0.05$ and $\ds=32$.}
  \label{fig:rphi4linearfit}
\end{figure}

From the susceptibilities for several masses, the critical mass where $\chi \to \infty$ is determined.
In~\cite{Kadoh:2018tis} the following fitting formula is used:
\begin{align}
  \label{eq:rphi4fittingform}
  \chi^{-1/1.75}
  = A \left| \mu_{0, \mathrm{c}}^{2} - \mu_{0}^{2} \right|^{\gamma/1.75}.
\end{align}
Figure~\ref{fig:rphi4linearfit} shows the susceptibility with the fit result with fixed $\gamma = \gamma_{\mathrm{Ising}}$~\footnote{
  In~\cite{Kadoh:2018tis} the legitimacy of fixing $\gamma$ to the exact value is supported by reasonable reduced chi-squared values for fittings.
}.
The critical bare squared-mass $\mu_{0,\mathrm{c}}^{2}$ is obtained as $\mu_{0}^{2}$-intercept of the line.
The parameters are set to $\lambda=0.05$ and $\ds=32$.

Taking above procedures for several values of $\lambda$ one can obtain the dimensionless critical coupling constant $\lambda/\mu_{\mathrm{c}}^{2}$ as a function of $\lambda$,
and the remaining procedure is to take the continuum extrapolation.
In~\cite{Kadoh:2018tis}, the $\lambda = 0$ value of the dimensionless critical coupling is calculated by a linear extrapolation with a reasonable chi-squared value $\approx 0.026$.
The result is
\begin{align}
  \label{eq:rphi4result}
  \lim_{\lambda \rightarrow 0} \frac{\lambda}{\mu_{\mathrm{c}}^{2}\left( \lambda \right)} = 10.913(56).
\end{align}
The error is mainly due to a fluctuation in the large $\ds$-region.
In the paper, it is shown that the $\ds$-dependence is the main source of the error.

\begin{figure}[htbp]
  \centering
  \includegraphics[width=\hsize]{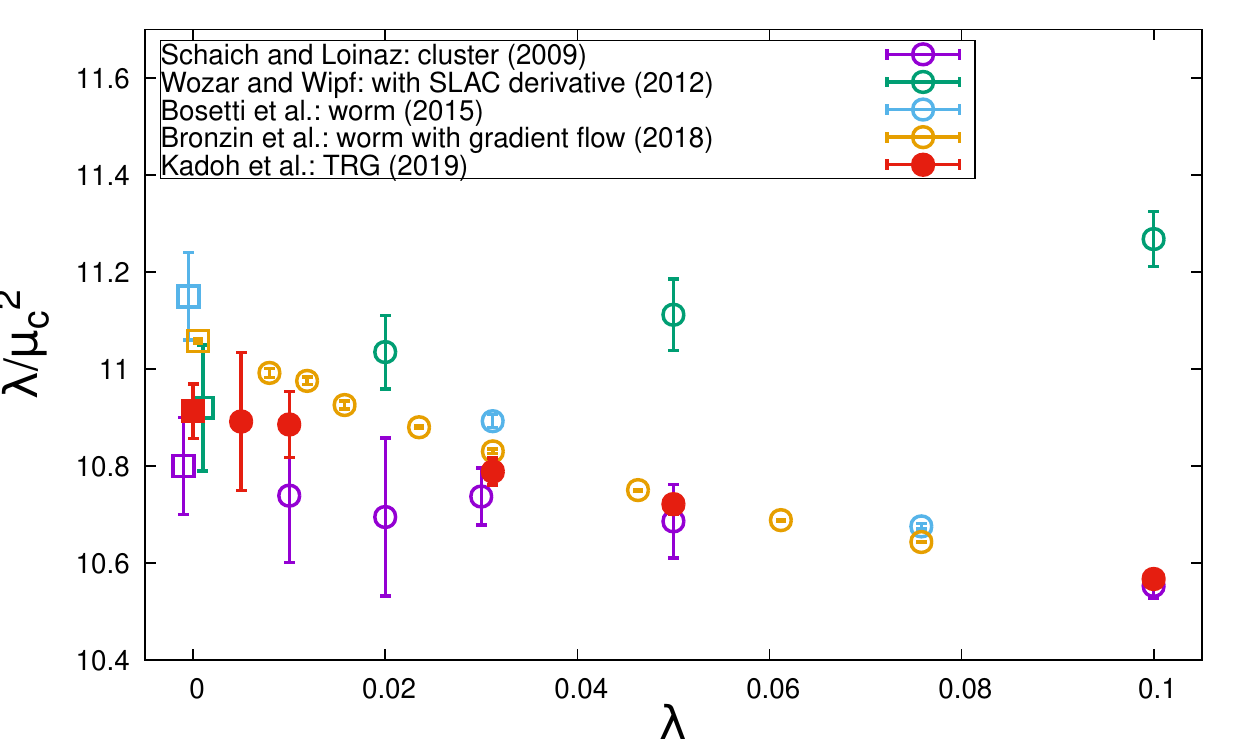}
  \caption{
    Comparison of the continuum extrapolations of the critical coupling $\lambda/\mu_{\mathrm{c}}^{2}$
    given by recent Monte Carlo works~\cite{Schaich:2009jk,Wozar:2011gu,Bosetti:2015lsa,Bronzin:2018tqz} and by the TRG work in~\cite{Kadoh:2018tis}.
    At $\lambda = 0$, data points are horizontally shifted to ensure the visibility.
    Note that the work by Wozar and Wipf was conducted with the SLAC derivative for scalar fields while the others are with the naive discretization.
  }
  \label{fig:rphi4continuumlimit}
\end{figure}

Figure~\ref{fig:rphi4continuumlimit} shows a comparison among recent Monte Carlo works in~\cite{Schaich:2009jk,Wozar:2011gu,Bosetti:2015lsa,Bronzin:2018tqz} and the TRG work in~\cite{Kadoh:2018tis}.
The TRG result shows notable accuracy and has achieved the smallest $\lambda$-value that is essentially important for the continuum extrapolation.
For a comprehensive list of references for the continuum limit value of the coupling constant, see~\cite{SakaiPhDthesis2019}.

The two dimensional complex $\phi^{4}$ theory at finite density, a typical model that suffers from the sign problem, is studied in~\cite{Kadoh:2019ube}.
A complex scalar field is discretized by using the Gaussian quadrature rule for both the real and the imaginary part of the fields.
As in Eq.~\eqref{eq:rphi4discpfunc}, a partition function of a multi-component scalar field theory is discretized as
\begin{align}
\begin{aligned}[t]
    Z\left(K\right)
    = \sum_{\left\{\alpha, \beta, \ldots\right\}} \prod_{x} 
    & w_{\alpha_{x}} w_{\beta_{x}} \cdots e^{y_{\alpha_{x}}^2 + y_{\beta_{x}}^2 + \cdots} \\
    & \prod_{\nu} h\left(y_{\alpha_{x}}, y_{\beta_{x}},\ldots, y_{\alpha_{x+\hat{\nu}}}, y_{\beta_{x+\hat{\nu}}},\ldots\right),
    \end{aligned}
\end{align}
where $h$ is the local Boltzmann factor of the model, and where the definitions of $K$, $w$, $y$ are the same as those that are used earlier in this section\footnote{
Here $K$ is taken the same for each field component just for simplicity;
off course it can be taken differently for each field.}.

Using the TRG, the silver blaze phenomenon\footnote{This refers to the lack of changes in the particle density when the chemical potential is varied below some critical value \cite{silverblaze}.} for thermodynamic quantities is clearly observed.
A remarkable point to note from the work is the ability for the TRG---in a model with a severe sign problem
and multi-component scalar field---to produce robust and sustainable results.
Note that, in the reference, the severeness of the sign problem is measured by the average phase factor $\langle e^{i\theta} \rangle = Z/Z_{\mathrm{pq}}$, where $Z_{\mathrm{pq}}$ is the phase quenched partition function.

\subsection{Additional topics and references}

For more precise works near criticality, it is natural to use improved coarse-graining algorithms such as the (loop-)TNR~\cite{2015PhRvL.115r0405E,yang2017loop}, the graph-independent local truncations (GILT)~\cite{Hauru:2017tne}, and the full environment truncation~\cite{2018PhRvB..98h5155E}.
The common concept is to properly deal with the corner double line (CDL) structure on tensor networks (see~\cite{Gu:2009dr} and Sec.~\ref{sec:cdl}).
Recently the GILT has been applied to the 2D real $\phi^{4}$ model and a precise value of the critical coupling constant has been reported~\cite{Delcamp:2020hzo}.
The result is comparable to that of~\cite{Kadoh:2018tis}, recent Monte Carlo works, and other computational schemes.
When using such a deterministic approach, the systematic error should be properly understood,
and the definition of the error is important for more concrete discussions.

There could be a better scheme for generating tensor networks for scalar bosons although the tensor network representation via the Gaussian quadrature rule seems work quite well.
One big concern is that the Gaussian quadrature rule effectively puts a cut-off for the scalar fields, so that one cannot deal with models whose local Boltzmann factor $f\left( \phi_{1}, \phi_{2} \right)$ has a long (or maybe infinite) tail in the $\phi_{1}$-$\phi_{2}$ space;
\textit{e.g.} massless free scalar bosons could not be suitably treated by the Gaussian quadrature rule, which requires a fast damping of the local Boltzmann factor.
The authors of~\cite{Delcamp:2020hzo} generate discrete degrees of freedom using the Taylor series expansion instead of the Gaussian quadrature rules,
and it also has the same issue.

\section{Models with fermions}
\label{sec:fermions}

In this section we discuss tensor network representations, and coarse-graining algorithms for fermion systems.
An important point is that in fermion tensor networks additional Grassmann variables are generated stemming from the original field variables.
Since there are not Grassmann valued data types on computers, one needs some special treatments for the Grassmann variables on tensor networks.
The coarse-graining procedure for fermion tensor networks was given in~\cite{Gu:2010yh,Gu:2013gba},
and applications for relativistic fermion systems are given in~\cite{Shimizu:2014uva,Takeda:2014vwa}.

\subsection{Tensor representation for free Wilson fermions}
\label{sec:fermiontnrep}

In this section we construct a tensor network representation of the two-dimensional Wilson--Dirac fermion system.
Interactions are not discussed here, but local interaction terms could be easily introduced as in~\cite{Shimizu:2014uva,Yoshimura:2017jpk,Kadoh:2018hqq}, where U(1), four fermion, and Yukawa-type interactions are discussed, respectively.
The Lagrangian density of the target system is given by
\begin{align}
  \label{eq:fermionlag}
  \mathcal{L}_{x}
  = \bar{\psi}_{x} \left( D\psi \right)_{x}
\end{align}
where the Wilson--Dirac operator is defined by
\begin{align}
  \label{eq:fermiondiracop}
  & D_{x x^{\prime}} = \left( m+2 \right) \delta_{x, x^{\prime}} \nonumber \\
  & - \frac{1}{2} \sum_{\mu = 1}^{2} \left\{ \left( 1 + \gamma_{\mu} \right) \delta_{x, x^{\prime}+\hat{\mu}} + \left( 1 - \gamma_{\mu} \right) \delta_{x, x^{\prime}-\hat{\mu}} \right\}.
\end{align}
$\psi$ and $m$ are a two-component spinor field: $\psi_{x} = \left( \psi_{x}^{(1)}, \psi_{x}^{(2)} \right)^{\mathrm{T}}$, and the mass, respectively.
The Grassmann variables satisfy anti-commutation relations.
We assume periodic boundary conditions in all directions in what follows in this section.

The partition function of the system is given by
\begin{align}
  \label{eq:fermionpfunc}
  Z_{\mathrm{F}}
  = \int \mathcal{D}\psi \mathcal{D}\bar{\psi} e^{-\sum_{x} \mathcal{L}_{x}}.
\end{align}
Under this representation of gamma matrices:
\begin{align}
  \label{eq:fermiongamma}
  \gamma_{1} = \sigma_{1} =
  \begin{pmatrix}
    0 & 1 \\
    1 & 0
  \end{pmatrix},
      && \gamma_{2} = \sigma_{3} =
         \begin{pmatrix}
           1 & 0 \\
           0 & -1
         \end{pmatrix},
\end{align}
the hopping factors are expanded as
\begin{align}
  \label{eq:fermionexpansion}
  & e^{-\sum_{x} \bar{\psi}_{x}\left( D\psi \right)_{x}} = \prod_{x} \Biggl\{ e^{-\left( m+2 \right)\bar{\psi}_{x}\psi_{x}} \nonumber \\
  & \begin{aligned}[b]
    & \cdot \sum_{X_{x,1}=0}^{1} \left( \bar{\chi}_{x+\hat{1}}^{(1)}\chi_{x}^{(1)} \right)^{X_{x,1}}
    \sum_{X_{x,2}=0}^{1} \left( \bar{\chi}_{x}^{(2)}\chi_{x+\hat{1}}^{(2)} \right)^{X_{x,2}} \\
    & \cdot \sum_{T_{x,1}=0}^{1} \left( \bar{\psi}_{x+\hat{2}}^{(1)}\psi_{x}^{(1)} \right)^{T_{x,1}}
    \sum_{T_{x,2}=0}^{1} \left( \bar{\psi}_{x}^{(2)}\psi_{x+\hat{2}}^{(2)} \right)^{T_{x,2}} \Biggr\},
  \end{aligned}
\end{align}
where $\chi$ and $\bar{\chi}$ are linear combinations of $\psi$ and $\bar{\psi}$:
$\chi_{x} = \left( 1/\sqrt{2} \right)\left( \psi_{x}^{(1)} + \psi_{x}^{(2)}, \psi_{x}^{(1)} - \psi_{x}^{(2)} \right)$,
$\bar{\chi}_{x} = \left( 1/\sqrt{2} \right)\left( \bar{\psi}_{x}^{(1)} + \bar{\psi}_{x}^{(2)}, \bar{\psi}_{x}^{(1)} - \bar{\psi}_{x}^{(2)} \right)$~\footnote{
Note that $\chi$ and $\bar{\chi}$ are introduced just for notational simplicity, so that the hopping terms are diagonal in spinor space.}.
Each expansion is a binomial because of the nilpotency of Grassmann variables,
and, at this point, discrete indices have arisen at each link.

Next, we integrate out the old degrees of freedom.
An important point here is to break the hopping factors into Grassmann even structures to freely shuffle them between one another.
To do that the following identities are useful:
\begin{align}
  \label{eq:fermionidentity1}
  & \left( \bar{\Psi}_{x+\hat{\mu}}^{(1)}\Psi_{x}^{(1)} \right) \nonumber \\
  & = \int \left( \bar{\Psi}_{x+\hat{\mu}}^{(1)}d\bar{\Phi}_{x+\hat{\mu}}^{(1)} \right) \big( \Psi_{x}^{(1)}d\Phi_{x}^{(1)} \big) \left( \bar{\Phi}_{x+\hat{\mu}}^{(1)}\Phi_{x}^{(1)} \right), \\
  \label{eq:fermionidentity2}
  & \left( \bar{\Psi}_{x}^{(2)}\Psi_{x+\hat{\mu}}^{(2)} \right) \nonumber \\
  & = \int \left( \bar{\Psi}_{x}^{(2)}d\bar{\Phi}_{x}^{(2)} \right) \big( \Psi_{x+\hat{\mu}}^{(2)}d\Phi_{x+\hat{\mu}}^{(2)} \big) \left( \bar{\Phi}_{x}^{(2)}\Phi_{x+\hat{\mu}}^{(2)} \right).
\end{align}
Note that one has to introduce new Grassmann variables here.
Also, during the coarse-graining steps, these varaibles are introduced, and integrated out, iteratively.
This is a key point of the treatment of fermion tensor networks.

Using the above identities, each factor in Eq.~\eqref{eq:fermionexpansion} can be decomposed,
and then the partition function can be deformed to
\begin{align}
  \label{eq:fermionpreptn}
  & Z_{\mathrm{F}} = \sum_{\left\{ X,T \right\}} \int \mathcal{D}\psi \mathcal{D}\bar{\psi} \prod_{x} e^{-\left( m+2 \right)\bar{\psi}_{x} \psi_{x}} \big( \chi_{x}^{(1)}d\eta_{x}^{(1)} \big)^{X_{x,1}} \nonumber \\
  & \cdot \big( \bar{\chi}_{x}^{(2)}d\bar{\eta}_{x}^{(2)} \big)^{X_{x,2}}
    \big( \psi_{x}^{(1)}d\xi_{x}^{(1)} \big)^{T_{x,1}}
    \big( \bar{\psi}_{x}^{(2)}d\bar{\xi}_{x}^{(2)} \big)^{T_{x,2}} \nonumber \\
  & \cdot \big( \bar{\chi}_{x}^{(1)}d\bar{\eta}_{x}^{(1)} \big)^{X_{x-\hat{1},1}}
    \big( \chi_{x}^{(2)}d\eta_{x}^{(2)} \big)^{X_{x-\hat{1},2}}
    \big( \bar{\psi}_{x}^{(1)}d\bar{\xi}_{x}^{(1)} \big)^{T_{x-\hat{2},1}} \nonumber \\
  & \cdot \big( \psi_{x}^{(2)}d\xi_{x}^{(2)} \big)^{T_{x-\hat{2},2}}
    \big( \bar{\eta}_{x+\hat{1}}^{(1)}\eta_{x}^{(1)} \big)^{X_{x,1}}
    \big( \bar{\eta}_{x}^{(2)}\eta_{x+\hat{1}}^{(2)} \big)^{X_{x,2}} \nonumber \\
  & \cdot \big( \bar{\xi}_{x+\hat{2}}^{(1)}\xi_{x}^{(1)} \big)^{T_{x,1}}
    \big( \bar{\xi}_{x}^{(2)}\xi_{x+\hat{2}}^{(2)} \big)^{T_{x,2}},
\end{align}
where $\sum_{\left\{ X,T \right\}}$ means $\prod_{x} \sum_{X_{x,1}, X_{x,2}, T_{x,1} T_{x,2} = 0}^{1}$.
Here the new Grassmann degrees of freedom ($\eta$, $\bar{\eta}$, $\xi$, and $\bar{\xi}$) are introduced in the same manner as Eqs.~\eqref{eq:fermionidentity1}--\eqref{eq:fermionidentity2}.
Note that, thanks to the Grassmann-even decompositions, the old degrees of freedom ($\psi$, $\bar{\psi}$, $\chi$, and $\bar{\chi}$) that belong to the same coordinate are gathered up without involving sign factors.
Then the tensor network representation of the partition function is defined by
\begin{align}
  \label{eq:fermiontn}
  Z_{\mathrm{F}}
  = \sum_{\left\{ X, T \right\}} \int \prod_{x} {T_{\mathrm{F}}}_{X_{x-\hat{1}} X_{x} T_{x} T_{x-\hat{2}}} \mathcal{G}_{x, X_{x-\hat{1}} X_{x} T_{x} T_{x-\hat{2}}},
\end{align}
where
\begin{align}
  \label{eq:fermiontensor}
  {T_{\mathrm{F}}}_{k i j l}
  = & \int d\mathcal{A}^{(1)} d\bar{\mathcal{A}}^{(1)} d\mathcal{A}^{(2)} d\bar{\mathcal{A}}^{(2)}
      e^{-\left( m+2 \right)\bar{\mathcal{A}}\mathcal{A}} \nonumber \\
    & \cdot \mathcal{A}^{(2) l_{2}} \bar{\mathcal{A}}^{(1) l_{1}} \mathcal{B}^{(2) k_{2}} \bar{\mathcal{B}}^{(1) k_{1}} \nonumber\\
    & \cdot \bar{\mathcal{A}}^{(2) j_{2}} \mathcal{A}^{(1) j_{1}} \bar{\mathcal{B}}^{(2) i_{2}} \mathcal{B}^{(1) i_{1}},
\end{align}
with dummy Grassmann variables
\begin{align}
& \mathcal{A} = \left( \mathcal{A}^{(1)}, \mathcal{A}^{(2)} \right)^{\mathrm{T}},
\quad \bar{\mathcal{A}} = \left( \bar{\mathcal{A}}^{(1)}, \bar{\mathcal{A}}^{(2)} \right), \\
& \mathcal{B} = \frac{1}{\sqrt{2}} \left( \mathcal{A}^{(1)} + \mathcal{A}^{(2)}, \mathcal{A}^{(1)} - \mathcal{A}^{(2)} \right)^{\mathrm{T}}, \\
& \bar{\mathcal{B}} = \frac{1}{\sqrt{2}} \left( \bar{\mathcal{A}}^{(1)} + \bar{\mathcal{A}}^{(2)}, \bar{\mathcal{A}}^{(1)} - \bar{\mathcal{A}}^{(2)} \right),
\end{align}
and
\begin{align}
  \label{eq:fermiongpart}
  \mathcal{G}_{x, k i j l}
  = & d\eta_{x}^{(1) i_{1}} d\bar{\eta}_{x}^{(2) i_{2}} d\xi_{x}^{(1) j_{1}} d\bar{\xi}_{x}^{(2) j_{2}} \nonumber \\
    & \cdot d\bar{\eta}_{x}^{(1) k_{1}} d\eta_{x}^{(2) k_{2}} d\bar{\xi}_{x}^{(1) l_{1}} d\xi_{x}^{(2) l_{2}} \nonumber \\
    & \cdot \left( \bar{\eta}_{x+\hat{1}}^{(1)}\eta_{x}^{(1)} \right)^{i_{1}}
      \left( \bar{\eta}_{x}^{(2)}\eta_{x+\hat{1}}^{(2)} \right)^{i_{2}} \nonumber \\
    & \cdot \left( \bar{\xi}_{x+\hat{2}}^{(1)}\xi_{x}^{(1)} \right)^{j_{1}}
      \left( \bar{\xi}_{x}^{(2)}\xi_{x+\hat{2}}^{(2)} \right)^{j_{2}}.
\end{align}

\subsection{Grassmann tensor renormalization group}
\label{sec:gtrg}

In this section, we describe the coarse-graining algorithm for tensor networks including Grassmann variables.
The details are given in~\cite{Takeda:2014vwa};
In this section we put our focus on the treatment of Grassmann variables in the network.
The coarse-graining of the bosonic part of the tensor is assumed to be carried out as in Sec.~\ref{sec:lntrg},
and it would be helpful for readers to see this section with Sec.~\ref{sec:lntrg}.
The coarse-graining of the Grassmann parts yields a phase factor that is to be incorporated into the bosonic part of the tensor.

First, the Grassmann part $\mathcal{G}$ is decomposed into two parts:
\begin{align}
  \label{eq:fermiondecompg13}
  & \mathcal{G}_{x, kijl} \nonumber \\
  = & \int \left( \Theta^{[1]}_{x, ij} d\bar{\eta}_{x^{\star}}^{m_{\mathrm{f}}} \right)
      \left( \Theta^{[3]}_{x, kl} d\eta_{x^{\star}-\hat{1}^{\star}}^{m_{\mathrm{f}}} \right)
      \left( \bar{\eta}_{x^{\star}} \eta_{x^{\star}-\hat{1}^{\star}} \right)^{m_{\mathrm{f}}},
\end{align}
where
\begin{align}
  \label{eq:fermiontheta1}
  &\Theta^{[1]}_{x, ij} =
    \begin{aligned}[t]
      & d\eta_{x}^{(1) i_{1}} d\bar{\eta}_{x}^{(2) i_{2}} d\xi_{x}^{(1) j_{1}} d\bar{\xi}_{x}^{(2) j_{2}} \\
      & \cdot \left( \bar{\eta}_{x+\hat{1}}^{(1)}\eta_{x}^{(1)} \right)^{i_{1}}
      \left( \bar{\eta}_{x}^{(2)}\eta_{x+\hat{1}}^{(2)} \right)^{i_{2}} \\
      & \cdot \left( \bar{\xi}_{x+\hat{2}}^{(1)}\xi_{x}^{(1)} \right)^{j_{1}}
      \left( \bar{\xi}_{x}^{(2)}\xi_{x+\hat{2}}^{(2)} \right)^{j_{2}},
    \end{aligned} \\
  \label{eq:fermiontheta3}
  &\Theta^{[3]}_{x, kl}
    = d\bar{\eta}_{x}^{(1) k_{1}} d\eta_{x}^{(2) k_{2}} d\bar{\xi}_{x}^{(1) l_{1}} d\xi_{x}^{(2) l_{2}}
\end{align}
with the new binary index $m_{\mathrm{f}} = \left( i_{1} + i_{2} + j_{1} + j_{2} \right) \bmod 2 = \left( k_{1} + k_{2} + l_{1} + l_{2} \right) \bmod 2$.  This decomposition is analogous to the decomposition that takes place in the original TRG.
On the right-hand-side of Eq.~(\ref{eq:fermiondecompg13}), each factor is Grassmann-even thanks to the inclusion of the new Grassmann variables and the definition of the new binary index.
This is similar to the construction of the fermion tensor network (see Eqs.~\eqref{eq:fermiondecompg13}--\eqref{eq:fermiondecompg24}).
$x^{\star}$ denotes the new coordinate on the coarse-grained square lattice,
and the unit vectors on the coarse-grained lattice is defined by $\hat{1}^{\star} = \hat{1} + \hat{2}$ and $\hat{2}^{\star} = \hat{1} - \hat{2}$. (See also \ref{sec:lntrg}.)

We have another way of decomposing $\mathcal{G}$
\begin{align}
  \label{eq:fermiondecompg24}
  & \mathcal{G}_{x, kijl} = \left( -1 \right)^{l_{1}+l_{2}} \nonumber \\
  & \cdot \int \left(\Theta^{[2]}_{x, li} d\bar{\xi}_{x^{\star}}^{m_{\mathrm{f}}}\right)
    \left(\Theta^{[4]}_{x, jk} d\eta_{x^{\star} - \hat{2}^{\star}}^{m_{\mathrm{f}}}\right)
    \left(\bar{\eta}_{x^{\star}}\eta_{x^{\star}-\hat{2}^{\star}}\right)^{m_{\mathrm{f}}},
\end{align}
where
\begin{align}
  \label{eq:fermiontheta2}
  \Theta^{[2]}_{x, li} =
  & d\bar{\xi}_{x}^{(1) l_{1}} d\xi_{x}^{(2) l_{2}} d\eta_{x}^{(1) i_{1}} d\bar{\eta}_{x}^{(2) i_{2}} \nonumber \\
  & \cdot \left( \bar{\eta}_{x+\hat{1}}^{(1)}\eta_{x}^{(1)} \right)^{i_{1}}
    \left( \bar{\eta}_{x}^{(2)}\eta_{x+\hat{1}}^{(2)} \right)^{i_{2}}, \\
  \label{eq:fermiontheta4}
  \Theta^{[4]}_{x, jk} =
  & d\xi_{x}^{(1) j_{1}} d\bar{\xi}_{x}^{(2) j_{2}} d\bar{\eta}_{x}^{(1) k_{1}} d\eta_{x}^{(2) k_{2}} \nonumber \\
  & \cdot \left( \bar{\xi}_{x+\hat{2}}^{(1)}\xi_{x}^{(1)} \right)^{j_{1}}
    \left( \bar{\xi}_{x}^{(2)}\xi_{x+\hat{2}}^{(2)} \right)^{j_{2}}
\end{align}
with the new binary index $m_{\mathrm{f}} = \left( l_{1} + l_{2} + i_{1} + i_{2} \right) \bmod 2 = \left( j_{1} + j_{2} + k_{1} + k_{2} \right) \bmod 2$.

Using the above two ways of decomposing $\mathcal{G}$, we can now integrate out the old Grassmann variables, yielding a phase factor,
\begin{align}
  \label{eq:fermionintout}
  & \int \Theta^{[2]}_{x+\hat{2}, T_{x}X_{x+\hat{2}}}
    \Theta^{[1]}_{x, X_{x}T_{x}}
    \Theta^{[4]}_{x+\hat{1}, T_{x+\hat{1}}X_{x}}
    \Theta^{[3]}_{x+\hat{1}+\hat{2}, X_{x+\hat{2}}T_{x+\hat{1}}} \nonumber \\
  & = \left( -1 \right)^{\epsilon_{X_{x} T_{x} X_{x+\hat{2}} T_{x+\hat{1}}}},
\end{align}
where
\begin{align}
  & \epsilon_{X_{x} T_{x} X_{x+\hat{2}} T_{x+\hat{1}}} \nonumber \\
  = & X_{x,2}(X_{x,1}+X_{x,2}) + T_{x,1}(T_{x,1}+T_{x,2}) \nonumber \\
  + & X_{x+\hat{2},2}(X_{x+\hat{2},1}+X_{x+\hat{2},2}) \nonumber \\
  + & T_{x+\hat{1},2}(T_{x+\hat{1},1}+T_{x+\hat{1},2})  \nonumber \\
  + & \begin{aligned}[t]
    & (X_{x,1}+X_{x,2}+X_{x+\hat{2},1}+X_{x+\hat{2},2}) \nonumber \\
    & \cdot (T_{x,1}+T_{x,2}+T_{x+\hat{1},1}+T_{x+\hat{1},2}).
  \end{aligned} \\[-4ex]
\end{align}
Note that the details of the phase factor depend on the ordering of the $\Theta$s in Eq.~\eqref{eq:fermionintout} and is not unique.
Finally, the effect of the coarse-graining of the Grassmann part is interpreted in terms of a (non-Grassmann) phase factor and constraints:
\begin{align}
  & \left( -1 \right)^{T_{x,1}+T_{x,2} + \epsilon_{X_{x} T_{x} X_{x+\hat{2}} T_{x+\hat{1}}}} \nonumber \\
  & \cdot \delta_{(X_{x+\hat{2}, 1} + X_{x+\hat{2}, 2} + T_{x+\hat{1}, 1} + T_{x+\hat{1}, 2}) \bmod 2, X_{x^{\star}, \mathrm{f}}} \nonumber \\
  & \cdot \delta_{(T_{x+\hat{1}, 1} + T_{x+\hat{1}, 2} + X_{x, 1} + X_{x, 2}) \bmod 2, T_{x^{\star}, \mathrm{f}}} \nonumber \\
  & \cdot \delta_{(X_{x, 1} + X_{x, 2} + T_{x, 1} + T_{x, 2}) \bmod 2, X_{x^{\star}-\hat{1}^{\star}, \mathrm{f}}} \nonumber \\
  & \cdot \delta_{(T_{x, 1} + T_{x, 2} + X_{x+\hat{2}, 1} + X_{x+\hat{2}, 2}) \bmod 2, T_{x^{\star}-\hat{2}^{\star}, \mathrm{f}}},
\end{align}
where the indices labeled with ``$\mathrm{f}$'' are the new binary indices introduced above.
The phase and the Kronecker deltas are to be incorporated into the bosonic tensors,
and the coarse-grained Grassmann part $\mathcal{G}^{\star}$ that consists of the new Grassmann variables is defined by
\begin{align}
  \label{eq:fermionreducedgpart}
  & \mathcal{G}^{\star}_{x, k i j l} \nonumber \\
  = & d\eta_{x}^{i_{\mathrm{f}}} d\xi_{x}^{j_{\mathrm{f}}} d\bar{\eta}_{x}^{k_{\mathrm{f}}} d\bar{\xi}_{x}^{l_{\mathrm{f}}}
      \left( \bar{\eta}_{x+\hat{1}^{\star}} \eta_{x} \right)^{i_{\mathrm{f}}}
      \left( \bar{\xi}_{x+\hat{2}^{\star}} \xi_{x} \right)^{j_{\mathrm{f}}}.
\end{align}

The Above procedure is iteratively executed along with the normal coarse-graining steps for the bosonic tensors.

\subsection{2D Schwinger model with Wilson fermions}
\label{sec:fermionresult_schwinger}

In~\cite{Shimizu:2014fsa}, the critical behavior at $\theta=\pi$ of the two dimensional Schwinger model is studied with Wilson fermions.
By studying the Fisher zeros, the authors of~\cite{Shimizu:2014fsa} have confirmed that there is a critical point around $\kappa = 0.2415$ and that the phase transition belongs to the 2D Ising universality class, where $\kappa$ is the inverse of the Wilson fermion mass $m$: $1/\kappa = 2\left( m+2 \right)$.
In addition, they have done the Lee--Yang zero analysis to seriously study the phase structure. In the parameter space of a complex couplings, the Lee-Yang zeros of the partition function are found off of the real-coupling axis at finite volume.  These zeros have their own critical behavior, and in the thermodynamic limit, their approach and condensation along the real-coupling axis causes non-analytic behavior on the real-axis indicating a phase transition.

Assuming that the gauge part of the lattice action is given by the usual Wilson action along with a topological term,
\begin{align}
  S_{\mathrm{G}}
  = & -\frac{1}{g^{2}} \sum_{x} \cos \left( A_{x,1}+A_{x+\hat{1},2}-A_{x+\hat{2},1}-A_{x,2} \right)
      \nonumber \\
    & - i\theta Q,
\end{align}
where $g^2 = 1/\beta_{pl.}$ is the gauge coupling, and $Q$ is the topological charge,
\begin{align}
    Q = \frac{1}{2\pi} \sum_{x} q_{x},
\end{align}
where $q_{x} = \left(
    A_{x,1}+A_{x+\hat{1},2}-A_{x+\hat{2},1}-A_{x,2} \right) \text{mod }2\pi$.
The presence of this term incurs the sign problem in MC calculations.    
The scaling behavior of the partition function zeros in the complex $\theta$-plane is studied with fixed $\mathrm{Re}\theta = \pi$.
At the critical mass $\kappa_{\mathrm{c}}$, the position of a partition function zero $\theta_{0}\left( L \right)$ would obey
\begin{align}
  \mathrm{Im}\theta_{0}\left( L \right) - \mathrm{Im}\theta_{0}\left( \infty \right)
  \propto L^{-\left( 2\nu - \beta \right)/\nu}
\end{align}
with the critical exponents $\nu$ and $\beta$.
If a first order phase transition occurs (conjectured that being at $\kappa < \kappa_{\mathrm{c}}$), it is expected that $\mathrm{Im}\theta_{0}\left( L \right) \propto L^{-2}$ with vanishing $\mathrm{Im}\theta_{0}\left( \infty \right)$.
On the other hand, if there are not phase transitions (conjectured being at $\kappa > \kappa_{\mathrm{c}}$), it is expected that $\mathrm{Im}\theta_{0}\left( \infty \right) \neq 0$.

\begin{figure}[htbp]
  \centering
  \includegraphics[width=\hsize]{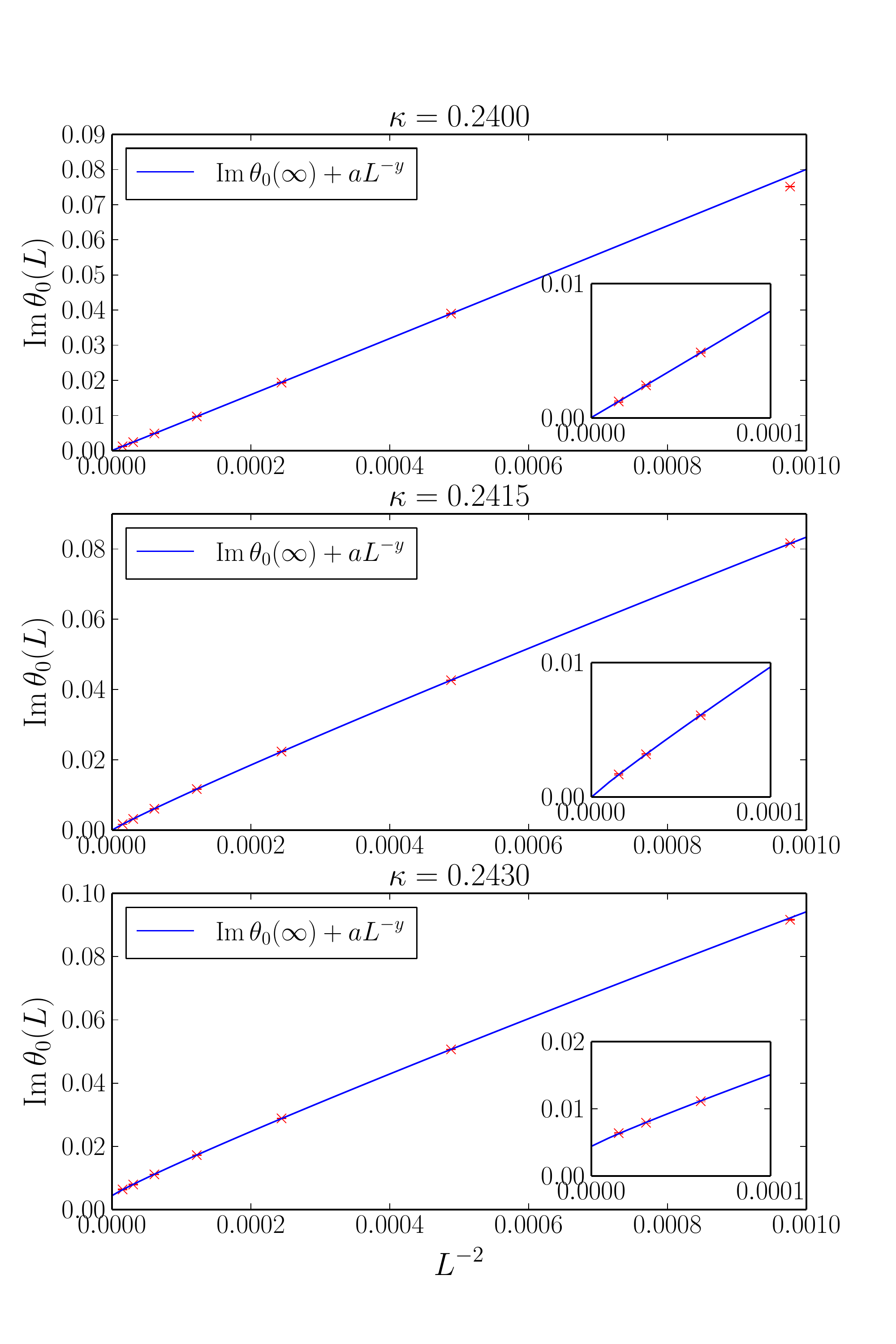}
  \caption{
    Adapted from~\cite{Shimizu:2014fsa}.
    Lee--Yang zeros for $\kappa=0.24$ (top), $0.2415$ (middle), and $0.243$, respectively.
    $1/g^{2} = 10$.
    Solid lines are the fit results with a function $\mathrm{Im}\theta_{0}\left( L \right) = \mathrm{Im}\theta_{0}\left( \infty \right) + aL^{-y}$ via three parameters $\mathrm{Im}\theta_{0}\left( \infty \right)$, $a$, and $y$.
  }
  \label{fig:leeyangzero}
\end{figure}

Figure~\ref{fig:leeyangzero} shows the scaling behaviors of $\mathrm{Im}\theta_{0}\left( L \right)$,
and the fitting results are summarized in table.~\ref{tab:leeyangzero}. The coupling $1/g^{2} = 10$.
The bond dimension of tensors are fixed to $\ds = 160$.
These results clearly show that a) for $\kappa < \kappa_{\mathrm{c}}$, $\mathrm{Im}\theta_{0}\left( \infty \right)$ vanishes, and the exponent $y$ is close to $2$, b) for $\kappa > \kappa_{\mathrm{c}}$, $\mathrm{Im}\left( \infty \right)$ has a non-zero value, c) for $\kappa = 0.2415$ ($\approx \kappa_{\mathrm{c}}$), $y=1.869(10)$ is consistent with $y=1.875$ that is the same as in the 2D Ising universality class.

Summarizing above, on the line $\theta = \pi$, there are no phase transition at $\kappa > \kappa_{\mathrm{c}}$, there is the second order phase transition belonging to the 2D Ising universality class at $\kappa = \kappa_{\mathrm{c}} \approx 0.2415$, and there are first order phase transitions at $\kappa < \kappa_{\mathrm{c}}$.
This is exactly the expected result.

\begin{table}
  \centering
  \caption{
    Adapted from~\cite{Shimizu:2014fsa}.
    Results of the fittings in Fig.~\ref{fig:leeyangzero}.
  }
  \label{tab:leeyangzero}
  \begin{ruledtabular}
    \begin{tabular}{ccccc}
      $\kappa$&$y$&$\text{Im}\,\theta_0(\infty)$&fit range&$\chi^2/{\rm DOF}$\\ \hline
      $0.2400$&$2.009(12)$&$0.000034(59)$&$L\in [32\sqrt{2},256]$&$0.65$ \\
      $0.2415$&$1.869(10)$&$-0.000016(64)$&$L\in [32\sqrt{2},256]$&$0.41$ \\
      $0.2430$&$1.850(15)$&$0.00442(12)$&$L\in [32\sqrt{2},256]$&$0.78$ \\
    \end{tabular}
  \end{ruledtabular}
\end{table}

There is a further study of the Berezinskii–Kosterlitz–Thouless transition in the ($m$, $g$)-plane in the same model by the same authors~\cite{Shimizu:2017onf},
but here we just draw the reader's attention to the paper.

This work is the first application of the Grassmann TRG to a relativistic fermion system.
Following this study, an application to the two dimensional Thirring model~\cite{THIRRING195891}, a pure fermion system, in the presence of a chemical potential was reported~\cite{Takeda:2014vwa}.

\subsection{3D free fermions}
\label{sec:fermionresult_3dfermion}

\begin{figure}[htbp]
  \centering
  \includegraphics[width=\hsize]{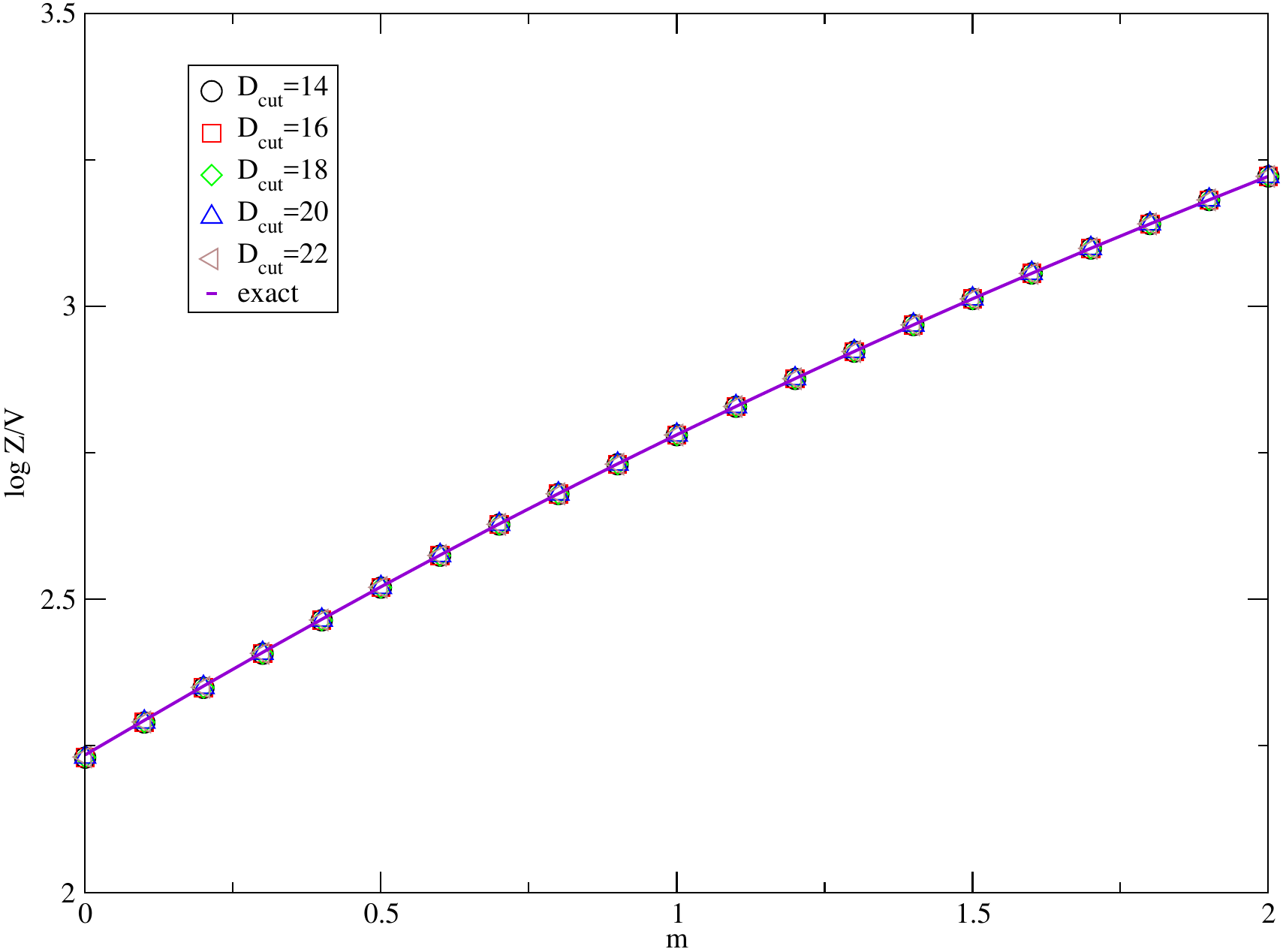}
  \caption{Adapted from~\cite{Yoshimura:2017jpk}. Free energy density of three dimensional free fermions for $V=256^{3}$.}
  \label{fig:3dfreefermionfenergy}
\end{figure}

For three dimensions or higher, the Grassmann parts of tensors can be coarse-grained in a similar way to the higher-order TRG~\cite{Sakai:2017jwp},
and calculations of partition functions and Green's functions are given in~\cite{Yoshimura:2017jpk} with relatively large bond dimensions~\footnote{
HOTRG in the presence of impurity tensors is discussed in detail in~\cite{2019CoPhC.236...65M};
the technique given in the reference would help to increase the accuracy of fermionic Green's functions also.
}.
Figure~\ref{fig:3dfreefermionfenergy} shows the free energy density of the three dimensional free fermion system, where the convergence in the number of states, $\ds$ is extremely rapid, and one cannot see the difference between the Grassmann HOTRG results and the exact values in this resolution.
The treatment of Grassmann variables in the Grassmann HOTRG is straightforwardly applicable to ATRG~\cite{Adachi:2019paf} and triad HOTRG~\cite{Kadoh:2019kqk}; this fact will encourage ones to approach higher dimensional fermion models.

\subsection{2D $\mathcal{N}=1$ Wess--Zumino model}

The interacting two dimensional $\mathcal{N}=1$ Wess--Zumino model, a simple supersymmetric model, displays a vanishing partition function~\cite{Witten:1982df}~\footnote{
  The partition function with periodic boundary conditions is equivalent to the trace of the fermion number operator $\left(-1\right)^{\mathrm{F}}$.
}.
This fact means that the model suffers from a serious sign problem as in the case of other generic supersymmetric models~\footnote{
For a review on lattice supersymmetry see \cite{catterall2009}.
}.

The Euclidean continuum action of the model is given by
\begin{align}
  \label{eq:wzaction}
  S_{\mathrm{cont.}} = \int d^{2}x
  \biggl\{
  & \frac{1}{2}\left(\partial_{\mu}\phi\right)^{2}
    + \frac{1}{2} W^\prime \left(\phi\right)^{2} \nonumber \\
  & + \frac{1}{2}\bar{\psi} \left(\gamma_\mu \partial_\mu
    + W^{\prime\prime}\left(\phi\right)\right)\psi \biggr\},
\end{align}
where $\phi$ and $\psi$ are a one-component real scalar field and a two-component Majorana spinor field, respectively~\footnote{
  The numerical treatment of Majorana fermions on a discrete space-time lattice is discussed in \textit{e.g.}~\cite{Wolff:2007ip}.
}.
$W\left(\phi\right)$ is an arbitrary function of $\phi$ and called the superpotential,
which is the source of the Yukawa- and $\phi^{n}$-interactions.

The spinor field $\psi$ satisfies the Majorana condition
\begin{align}
  \label{eq:majoranacond}
  \bar{\psi} =-\psi^{\mathrm{T}} C^{-1}
\end{align}
with the charge conjugation matrix $C$:
\begin{align}
  \label{eq:cconj}
  C^{\mathrm{T}}=-C, && C^{\dagger}=C^{-1}, && C^{-1}\gamma_{\mu}C=-\gamma_{\mu}^{\mathrm{T}}.
\end{align}
The continuum action~\eqref{eq:wzaction} is invariant under the supersymmetry transformation
\begin{align}
  \label{eq:susytrans}
  &\delta \phi
    = \bar{\epsilon} \psi, \\
  &\delta \psi
    = \left(\gamma_\mu \partial_\mu \phi - W^{\prime}\left(\phi\right)\right) \epsilon,
\end{align}
where $\epsilon$ is a two-component Grassmann number
and $\bar \epsilon$ is defined as in Eq.~\eqref{eq:majoranacond}.

Using the symmetric difference operator $\partial^{\mathrm{S}}_{\mu} = \left( \partial_{\mu} + \partial^{*}_{\mu} \right) / 2$ with the forward difference $\partial$ and the backward difference $\partial^{*}$,
the lattice action is given by
\begin{align}
  \label{eq:wzlattact}
  S
  = \sum_{x}\biggl\{
  & \frac{1}{2}\left(\partial_{\mu}^{\mathrm{S}}\phi_x\right)^{2}
    + \frac{1}{2}\left( W^\prime \left(\phi_{x}\right) -\frac{r}{2} \partial_{\mu}^{\mathrm{}} \partial_{\mu}^{*} \phi_{x} \right)^{2} \nonumber \\
  & + \frac{1}{2}\bar{\psi}_{x} \left( D\psi \right)_{x} \biggr\}.
\end{align}
Note that the lattice action has the Wilson term also in the scalar sector; this is required to retain equal footing for both scalar and fermion sectors.
It is perturbatively proven that the broken supersymmetry on the lattice is restored in the continuum limit for this construction of the action~\cite{Golterman:1988ta}.
The Dirac operator on the lattice is defined by
\begin{align}
  \label{eq:wzdiracop}
  D_{xx^{\prime}}
  = \left( \gamma_\mu \partial_\mu^{\mathrm{S}}
  - \frac{r}{2}\partial_{\mu}\partial_{\mu}^{*}\right)_{xx^{\prime}}
  + W^{\prime\prime} \left(\phi_x\right)\delta_{xx^{\prime}},
\end{align}
where $r$ is a nonzero real parameter that is called the Wilson parameter.

Tensor network representations of both the scalar and the fermion parts are constructed in the same way discussed in Secs.~\ref{sec:rphi4tnrep} and~\ref{sec:fermiontnrep}, respectively.
An important feature of this model is the Wilson term of the scalar part whose square produces next nearest neighbor hopping terms in the lattice action.
This fact prevent ones from simply constructing a tensor network representation.
In~\cite{Kadoh:2018hqq}, auxiliary scalar fields are introduced to make the nearest-neighbor form of the action~\footnote{
This prescription gives multi-component scalar fields.
For the treatment of the multi-components scalars, see also a study of the 2D complex $\phi^{4}$ theory~\cite{Kadoh:2019ube}.}
\begin{align}
  & \tilde S_{\mathrm{B}} = \frac{1}{2}\sum_{x} \big\{ \left( \partial_{\mu} \phi_{x} \right)^{2} + \left( W^{\prime}\left( \phi_{x} \right) \right)^{2} + G_{x}^{2} + H_{x}^{2} \nonumber \\
  & \begin{aligned}[t]
    & -\left( r W' \left(\phi_{x}\right)+\alpha G_x+\beta H_x\right)
    \left(\phi_{x+\hat 1} +\phi_{x-\hat 1}- 2\phi_x\right) \\
    & -\left( r W' \left(\phi_{x}\right)+\alpha G_x-\beta H_x\right)
    \left(\phi_{x+\hat 2} +\phi_{x-\hat 2}- 2\phi_x\right) \big\}
  \end{aligned}
\end{align}
with the auxiliary fields $G$, $H$ and the constants $\alpha=\sqrt{(1-2r^{2})/2}$, $\beta=1/\sqrt{{2}}$.

\begin{figure}[htbp]
  \centering
  \includegraphics[width=\hsize]{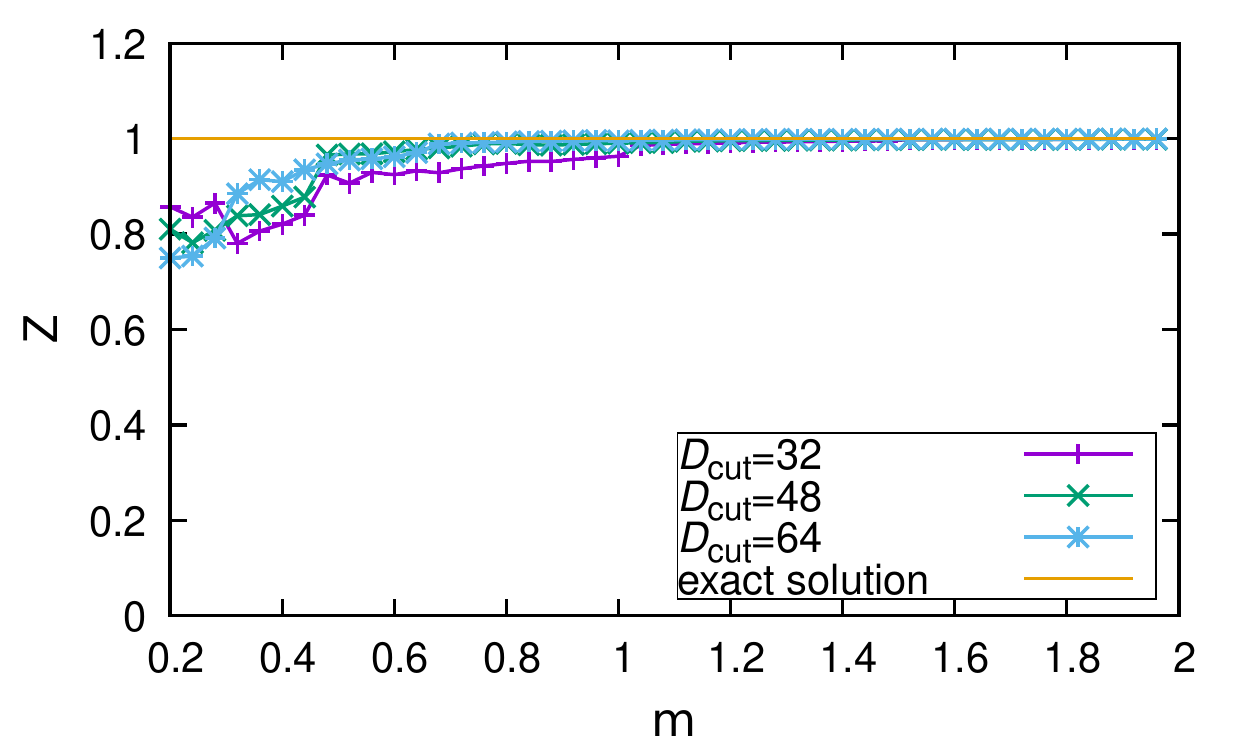}
  \caption{
    The partition function as a function of $m$ on $V=2 \times 2$ lattice.
  }
  \label{fig:wittenindex}
\end{figure}

In Fig.~\ref{fig:wittenindex}, the partition function (called the Witten index in this model) of the free $\mathcal{N}=1$ Wess--Zumino model, whose superpotential is given by $W(\phi) = (1/2)m\phi^{2}$ with the mass parameter $m$, is shown on $V = 2 \times 2$ lattice~\footnote{
In order not to break the supersymmetry, the periodic boundary conditions are assumed in all directions for both fermions and bosons.
The order of the Hermite polynomial used in the Gaussian quadrature is set to $64$ in the paper.
}.
In the free case, the partition function can be analytically obtained, and the exact solution is $Z = \mathrm{sign}\left\{ m\left( m+4r \right) \right\}$.
Thus the exact solution is $1$ for the $m>0$ region shown in the figure.
The TRG results tend to converge to the exact value $1$ with increasing $\ds$, the number of singular values that are kept during the coarse-graining steps.
The less accurate results in the small $m$ region are due to the lack of the fast damping factor in the local Boltzmann weight that is required for the Gaussian quadrature rule to be effective,
but such bad behavior is a special case for the non-interacting model.

When one deals with the interacting Wess--Zumino model that has $\phi^{n}$-interaction terms, they guarantee the presence of fast damping.
Since the tensor network representation given in the paper does not depend on the form of superpotential, the interacting $\mathcal{N}=1$ Wess--Zumino model is within the scope of tensor analyses.
In the interacting case, the restoration of the broken supersymmetry on the lattice should be confirmed numerically first.
However, from a technical point of view, an explicit breaking of the $Z_{2}$ symmetry due to the Wilson term in the lattice action~\cite{Golterman:1988ta} causes singular behavior in the local Boltzmann factor.
This fact makes numerical treatments difficult.
In addition to the Wilson type discretization, other lattice regularizations are worthwhile to consider,
and, in such cases, tensor network analyses would be helpful as long as the lattice model is written in a local way.

Also, because complexities of the field contents do not affect the structure of tensor networks,
more complicated models such as the $\mathcal{N}=\left( 2, 2 \right)$ Wess--Zumino model could be treated in the same way.

\subsection{2D Schwinger model with staggered fermions}
In this subsection a tensor network representation of the Schwinger model with staggered fermions is discussed~\footnote{
While this subsection refers to~\cite{GATTRINGER2015732}, the elimination of fermion fields is also discussed in~\cite{Zohar:2018cwb,Zohar:2019ygc}.
}.
The one-flavor staggered action for the massless Schwinger model on a two-dimensional lattice has the action,
\begin{equation}
  S = S_{F} + S_{g}
\end{equation}
with
\begin{align}
  \nonumber
  S_{F} = \frac{1}{2} \sum_{x=1}^{N} \sum_{\mu = 1}^{2} \eta_{x, \mu} &[ \bar{\psi}_{x} U_{x, \mu} \psi_{x + \hat{\mu}} \\
                                                                      &- \bar{\psi}_{x + \hat{\mu}} U^{\dagger}_{x, \mu} \psi_{x} ]
\end{align}
and $S_{g}$
is the usual Wilson action given by Eq.~\eqref{eq:wilsonaction}.
Here $\eta_{x, \mu}$ is the staggered phase which for $\eta_{x, 1} = 1$ and $\eta_{x, 2} = (-1)^{x_{1}}$ with $x_{1}$ the $1$-component of $x$.
The partition function for this model is then given by
\begin{align}
  \nonumber
  Z  &= \int \mathcal{D}U \mathcal{D}\bar{\psi} \mathcal{D}\psi  \; e^{-S} \\
     &=\int \mathcal{D}U e^{\beta \sum_x {\rm Re} [U_{x, 12}]} Z_F(U)
\end{align}
with $ \int \mathcal{D}U = \prod_{x} \int_{-\pi}^{\pi} dA_{x, \mu} / 2\pi $, $ \int \mathcal{D} \bar{\psi} \mathcal{D} \psi = \prod_{x} \int d\bar{\psi}_{x} d\psi_{x} $, and $Z_{F}$ represents the part of the partition function that depends on the fermion fields.

Following  \cite{GATTRINGER2015732} to formulate the model in terms of discrete degrees of freedom, we first integrate out the
fermions and generate an effective action depending only on the gauge fields.
As a first step we redefine the link variables such that the staggered fermion phases $ \eta_{x, \mu} $ can be absorbed into modified link variables $U_{x, \mu} \to \eta_{x, \mu} U_{x, \mu}$. Under this transformation the
gauge action picks up an overall negative sign but the measure is invariant.
{The Boltzmann factors associated with
each bilinear fermion term can be Taylor expanded yielding
a partition function
\begin{align}
  \nonumber
  Z_F = &\int \mathcal{D} \bar{\psi} \mathcal{D} \psi \times \\ \nonumber
        &\prod_x \prod_{\mu} \sum_{k_{x, \mu} = 0}^{1} \left( -\frac{1}{2}\bar{\psi}_{x} U_{x, \mu} \psi_{x + \hat{\mu}} \right)^{k_{x,\mu}} \times \\
        &\sum_{\bar{k}_{x, \mu} = 0}^{1} \left( \frac{1}{2}\bar{\psi}_{x + \hat{\mu}} U^{\dagger}_{x, \mu} \psi_{x} \right)^{\bar{k}_{x, \mu}}.
\end{align}
Notice that higher order terms in the expansion of the Boltzmann factors vanish because of the Grassmann nature of the fermions.
The partition function is only nonzero when the Grassmann integration is saturated.  This only occurs for closed fermionic loops and dimer configurations.}

For a loop $\ell$ with length $L(\ell)$ one finds a
contribution with absolute value
\begin{equation}
  \left(\frac{1}{2}\right)^{L(\ell)}\prod_{x,\mu\in \ell} \left(U_{x, \mu} \right)^{k_{x, \mu}} \left(U^{\dagger}_{x, \mu} \right)^{\bar{k}_{x, \mu}}
\end{equation}
where on a given link only a single $k$ or $\bar{k}$ is nonzero.
In addition each loop carries a certain $Z_2$ phase which depends on the
length of the loop and its winding along the temporal direction given by
\begin{equation}
  - (-1)^{\frac{1}{2}L(\ell)}(-1)^{W(\ell)}.
\end{equation}
Here, the overall negative sign is the usual one for closed fermion loops while the second
factor keeps track of the number of forward hops which is exactly half the total length of the loop
for a closed loop. Finally the factor $ (-1)^{W(\ell)}$
of the loop will be determined by the number of windings of the loop along the temporal direction assuming
anti-periodic boundary conditions for the fermions.
Using dimers and loops as basic constituents for non-zero contributions to the fermionic partition function we can write
\begin{align}
  \label{eq:z_f}
  \nonumber
  Z_F  = \left( \frac{1}{2} \right)^{V} &\sum_{\{ \ell, d \}} (-1) ^{N_L + \frac{1}{2} \sum_{\ell} L(\ell) + \sum_{\ell} W(\ell)} \times \\
                                        &\prod_{\ell} \left[ \prod_{x,\mu \in \ell} (U_{x, \mu})^{k_{x, \mu}} (U^{\dagger}_{x, \mu})^{\bar{k}_{x, \mu}} \right],
\end{align}
where $\sum_{\{ \ell, d \}}$ indicates a sum over all valid loop and dimer configurations, and $N_{L}$ is the number of loops.
We construct a local tensor which reproduces the nonzero configurations of this partition function.

\begin{figure}[htbp]
  \centering
\includegraphics[width=0.5\hsize]{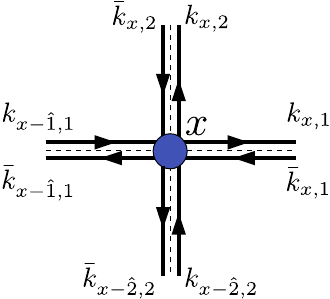}
\vskip10pt
(a) $T$ tensor on a site.
\vskip5pt
\includegraphics[width=0.8\hsize]{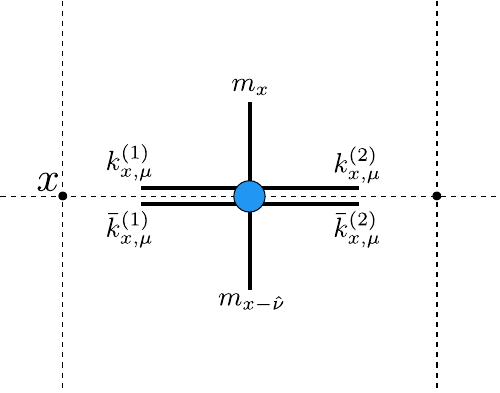}
\vskip5pt
(b) $A$ tensor on a link.
\vskip5pt
\includegraphics[width=0.8\hsize]{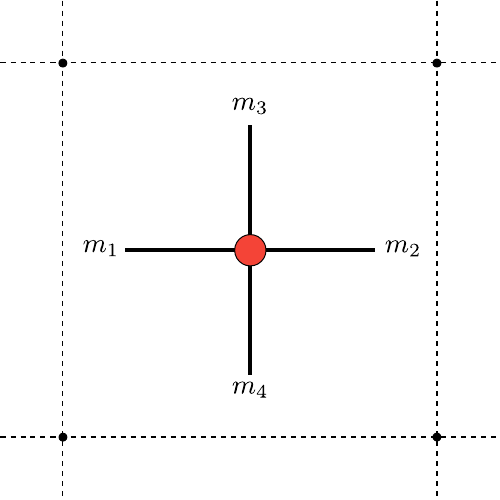}
\vskip5pt
(c) $B$ tensor on a plaquette.
\vskip5pt
  \caption{Graphical representations of tensors.}
  \label{fig:three_tensors}
\end{figure}

Let us ignore the overall sign for now and just deal with the magnitude. We allow two types of indices per link to capture
separately the incoming and outgoing fermion lines making the fermion site tensor a
rank eight object.
To write down a tensor, first, let us fix the coordinates so that right (1-direction) and up (2-direction) are positive (no bar), and left and down are negative (bar).
Since each site is either the endpoint of a dimer, or has fermionic current incoming and outgoing from it,
then we can model this with the tensor structure (we leave off the gauge link factors for now)
\begin{align}
  \nonumber
  & T^{(x)}_{k_{x-\hat{1}, 1} \bar{k}_{x-\hat{1}, 1} k_{x, 1} \bar{k}_{x, 1} k_{x, 2} \bar{k}_{x, 2} k_{x-\hat{2}, 2} \bar{k}_{x-\hat{2}, 2}} =  \\
  &\left\{
    \begin{array}{ll}
      1 & \text{if } k_{x-\hat{1}, 1} + k_{x-\hat{2}, 2} + \bar{k}_{x, 1} + \bar{k}_{x, 2} = 1 \\
        & \text{and } k_{x, 1} + k_{x, 2} + \bar{k}_{x-\hat{1}, 1} + \bar{k}_{x-\hat{2}, 2} = 1     \\
      0  &  \text{otherwise}
    \end{array}
           \right.
\end{align}
where each $ (k_i ,\bar{k}_i ) = 0,1$.  The pairs of indices are ordered (left, right, up, down).  A graphical representation of this tensor is shown in Fig.~\ref{fig:three_tensors}~(a).

By repeatedly contracting this site tensor with copies of itself over the lattice it can be seen that
we generate the full set of closed loops and dimers for the
model at zero gauge coupling {\rm excluding} the overall factor of minus one for each closed fermion loop.

The fermion partition function does not include any contribution or interaction with the gauge fields.  To proceed further we will
employ a character expansion of the Boltzmann factors associated with
the gauge action.
\begin{align}
  \nonumber
  &e^{-\beta \cos{\left[ A_{x, 1} + A_{x + \hat{1}, 2} - A_{x + \hat{2}, 1} - A_{x, 2} \right]} } = \\
  &\sum_{m_{x, 12} = -\infty}^{m_{x, 12} = \infty} I_{m_{x, 12}}(-\beta) e^{i m_{x, 12} \left[ A_{x, 1} + A_{x + \hat{1}, 2} - A_{x + \hat{2}, 1} - A_{x, 2} \right] } .
\end{align}
Each plaquette is now labeled by an integer $m_{x, 12}$ (which we shorten to $m_x$ since there are only temporal plaquettes in two dimensions).  Note that $I_{m_{x}}(-\beta) = (-1)^{m_{x}} I_{m_{x}}(\beta)$. In two dimensions
each link is shared by two plaquettes.  For a link in the $\mu = 1$ direction the two plaquettes give factors of $e^{i m_{x} A_{x, 1}}$ and $e^{-i m_{x-\hat{2}} A_{x, 1}}$.  In the $\mu = 2$ direction, $e^{-i m_{x} A_{x, 2}}$ and $e^{i m_{x-\hat{1}} A_{x, 2}}$.
In addition, the link carries a factor of $e^{i k_{x, \mu} A_{x, \mu}}$
or $e^{-i \bar{k}_{x, \mu} A_{x, \mu}}$ coming from $Z_F$. Thus, in total, links carry two $m$ indices inherited from
their neighboring plaquettes together with a $k$ and a $\bar{k}$ index associated with the fermionic
hopping terms.
The integral over a link variable is given by
\begin{align}
  \label{eq:cnst}
  \nonumber
  \int_{-\pi}^{\pi} &\frac{d A_{x, \mu}}{2 \pi} e^{i (k_{x, \mu} - \bar{k}_{x, \mu}) A_{x, \mu} } \prod_{\nu > \mu} e^{i (m_{x} - m_{x-\hat{\nu}})  A_{x, \mu}} \times \\ \nonumber
                    &\prod_{\nu < \mu} e^{i (m_{x-\hat{\nu}} - m_{x})  A_{x, \mu}}  = \\
                    &\delta^{(x, \mu)}_{\sum_{\nu > \mu} (m_x - m_{x-\hat{\nu}}) + \sum_{\nu < \mu}
                      (m_{x-\hat{\nu}} - m_x) + k_{x, \mu} -\bar{k}_{x, \mu}, 0}.
\end{align}
This allows us to write the partition function as a sum over $m$ and $k, \bar{k}$ variables,
\begin{align}
  \nonumber
  Z &= \sum_{ \{m\} }  \sum_{ \{ k,\bar{k} \}} \left( \prod_{x} I_{m_{x}}(\beta) \right) \times \\ \nonumber
    & \left( \prod_{x} T^{(x)}_{k_{x-\hat{1}, 1} \bar{k}_{x-\hat{1}, 1} k_{x, 1} \bar{k}_{x, 1} k_{x, 2} \bar{k}_{x, 2} k_{x-\hat{2}, 2} \bar{k}_{x-\hat{2}, 2}} \right)\times \\ \nonumber
    &(-1)^{N_{L} + N_{P} + \frac{1}{2}\sum_{\ell} L(\ell) + \sum_{\ell} W(\ell)} \\
    &\prod_{x, \mu} \delta^{(x, \mu)}_{\sum_{\nu > \mu} (m_x - m_{x-\hat{\nu}}) + \sum_{\nu < \mu}
      (m_{x-\hat{\nu}} - m_x) + k_{x, \mu} -\bar{k}_{x, \mu}, 0}
      \label{ZZ}
\end{align}
where $N_{P} = \sum_{x} m_{x}$.  At this
point we have included all the minus signs for completeness.  It was proven in  \cite{GATTRINGER2015732} that every valid contribution to the partition function is positive in the case of periodic boundary conditions, and so from here on we ignore the factor of $(-1)^{N_{L} + N_{P} + \frac{1}{2}\sum_{\ell} L(\ell) + \sum_{\ell} W(\ell)}$.

Now, associated with each link is a constraint between the $k$ and $\bar{k}$ fields on the link and the adjacent $m$ fields on the plaquettes given by Eq.~\eqref{eq:cnst}.  This is a natural object to use to form a tensor.  We define a tensor on each link by,
\begin{align}
  \label{eq:schwinger-a}
  \nonumber
  &A^{(x, \mu)}_{m_x m_{x-\hat{\nu}} k^{1}_{x,\mu} \bar{k}^{1}_{x,\mu} k^{2}_{x,\mu} \bar{k}^{2}_{x, \mu}} \equiv \\ \nonumber
  &\delta_{\sum_{\nu > \mu} (m_x - m_{x-\hat{\nu}}) + \sum_{\nu < \mu}
    (m_{x-\hat{\nu}} - m_x) + k^{1}_{x, \mu} -\bar{k}^{1}_{x, \mu}, 0} \times \\
  &\delta_{k^{1}_{x,\mu}, k^{2}_{x,\mu}} \delta_{\bar{k}^{1}_{x,\mu}, \bar{k}^{2}_{x,\mu}}.
\end{align}
Here the $k^i, \bar{k}^i$ indices are associated with the two ends of a link.  These indices are diagonal as indicated by the Kronecker deltas.
A diagram showing the relative position of the fermion and plaquette indices is shown in Fig.~\ref{fig:three_tensors}~(b).

\begin{figure}[htbp]
  \centering
  \includegraphics[width=0.8\hsize]{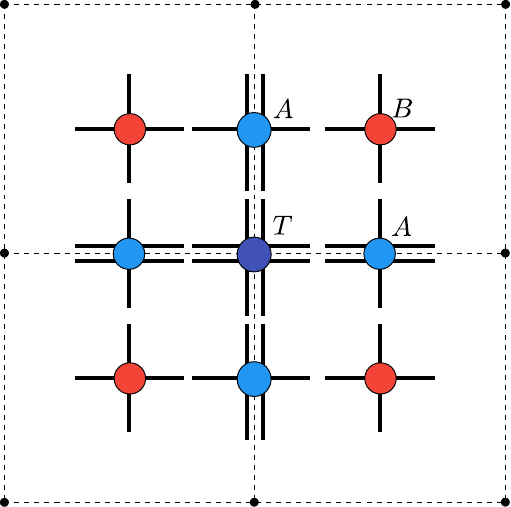}
  \caption{Contraction pattern of basic tensors.}
  \label{fig:tn}
\end{figure}

Finally, we construct a tensor associated with the plaquettes of the lattice.  This is the same tensor used before in previous tensor formulations of Abelian gauge theories, the $B$ tensor (See Eq.~\eqref{eq:ttensorB}). 
A graphical representation for the $ B $ tensor is shown in Fig.~\ref{fig:three_tensors}~(c).
The contraction over these three ($T$, $A$, and $B$) unique tensor types can be represented as the tensor network shown in Fig.~\ref{fig:tn}.

It is possible to include a topological term in the original action with the addition of,
\begin{align}
  S_{\Theta} = \frac{i \Theta}{2 \pi} \sum_{x} \Im[U_{x, 12}].
\end{align}
Taking the staggered phase into account, and expanding the Wilson plaquette term and this term simultaneously,
\begin{align}
  &e^{-\beta \Re[U_{x, 12}] + \frac{i \Theta}{2 \pi} \Im[U_{x, 12}]} = \\
  & \sum_{m_{x} = -\infty}^{\infty} C_{m_{x}}(\beta, \Theta)
    e^{i m_{x}(A_{x, 1} + A_{x+\hat{1}, 2} - A_{x+\hat{2}, 1} - A_{x, 2})},
\end{align}
the previous steps in formulating a tensor network can be followed straightforwardly.  One can solve for the $C$s numerically or analytically \cite{GATTRINGER2015732}.  It amounts to the replacement,
\begin{align}
  I_{m_{x}}(\beta) \rightarrow I_{m_{x}}(2\sqrt{\eta \bar{\eta}})
  \left( \frac{\eta}{\bar{\eta}} \right)^{m_{x} / 2}
\end{align}
in the definition of the $B$ tensor, with $\eta = \beta / 2 - \Theta / 4\pi$ and $\bar{\eta} = \beta / 2 + \Theta / 4\pi$.

\begin{figure}[t]
  \includegraphics[width=0.49\textwidth]{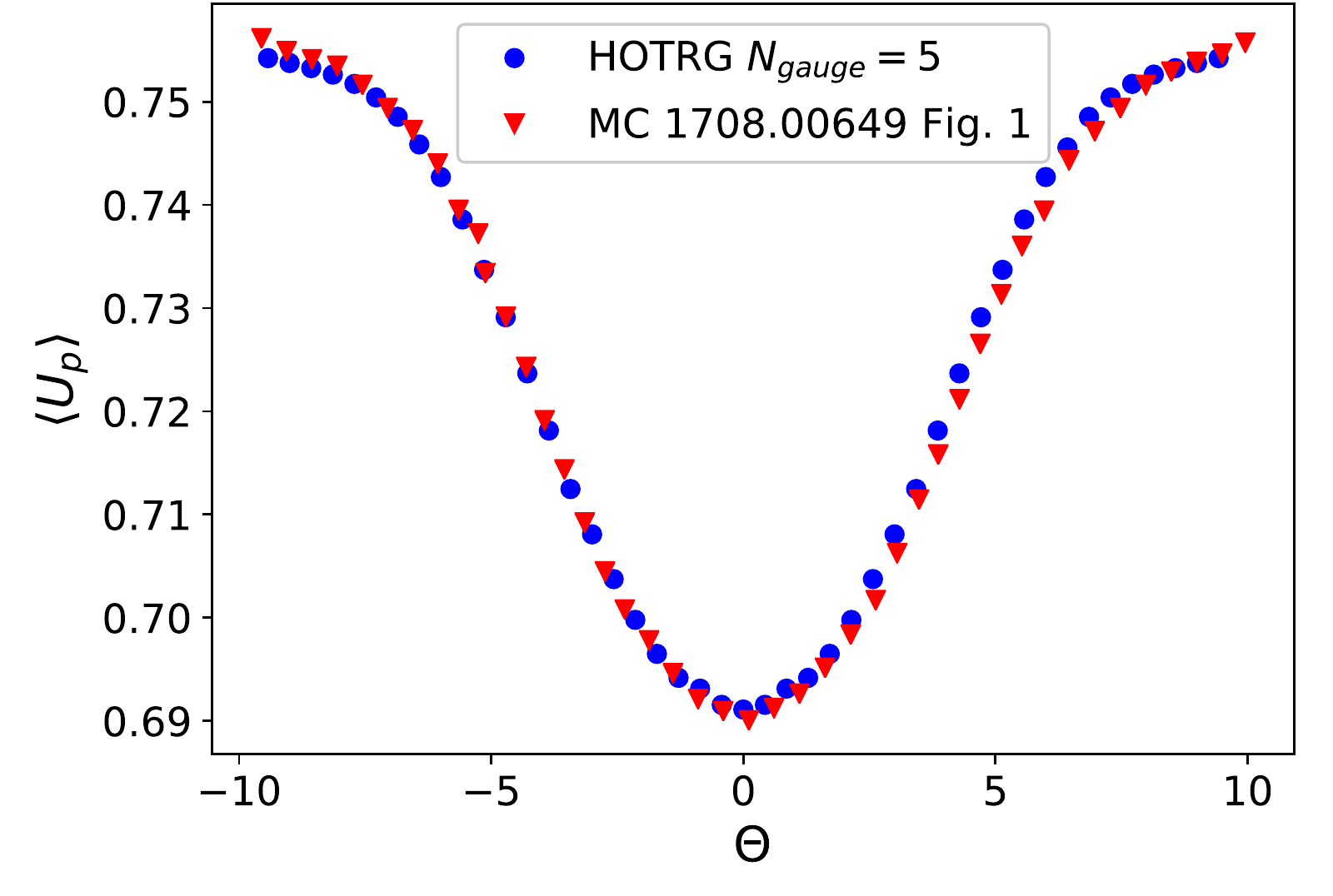}
  \caption{Adapted from~\cite{Butt:2019uul}.  The average plaquette as a function of the $\Theta$ parameter on a $4 \times 4$ lattice.  Here $N_{gauge} = 5$ indicates a truncation on the $m$ numbers such that $m$ runs from $-2$ to 2.  That is to say that $\ds = 5$ on the $B$ tensor initially.}
  \label{fig:schwing-plaq-theta}
\end{figure}

\begin{figure}[t]
  \includegraphics[width=0.49\textwidth]{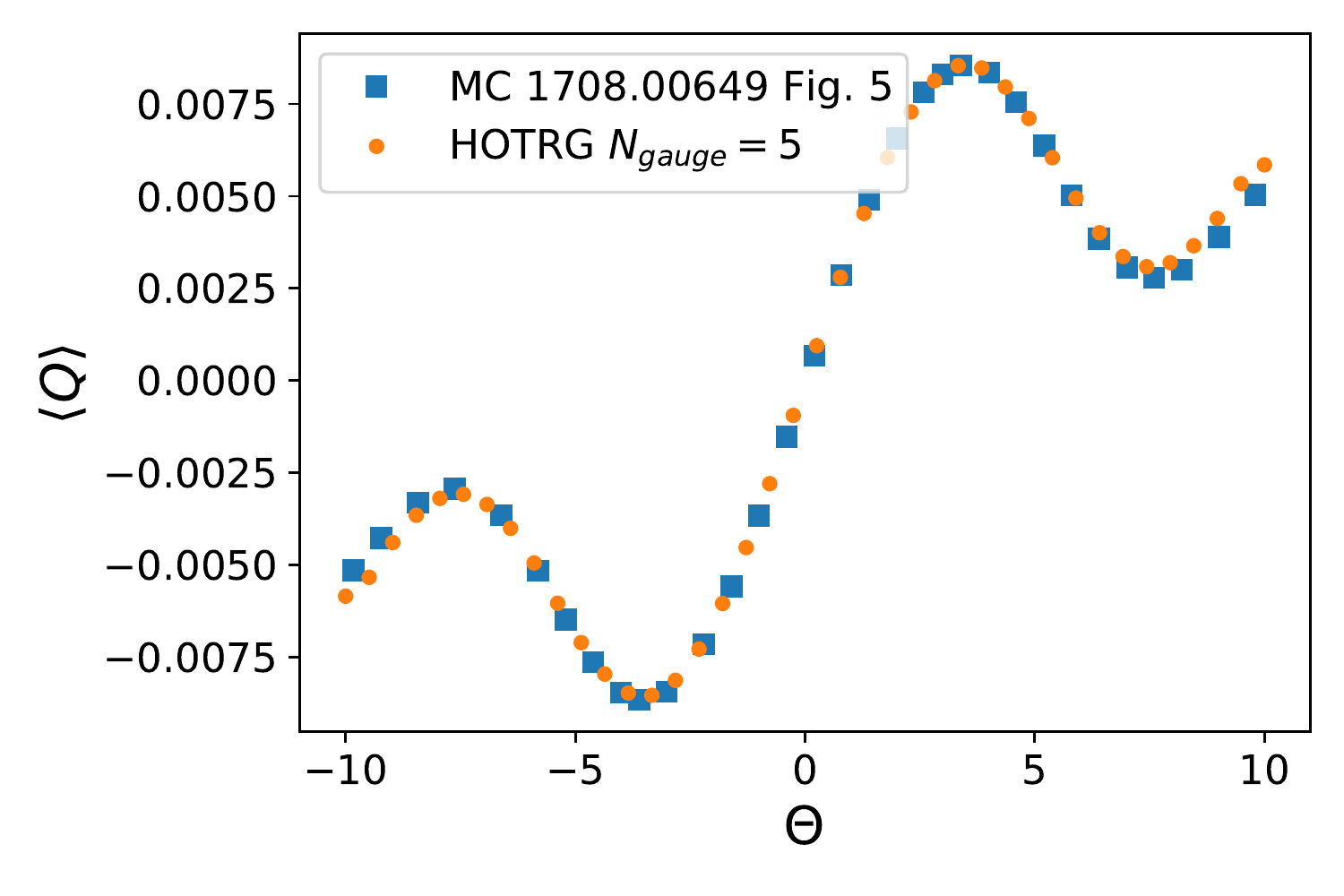}
  \caption{Adapted from~\cite{Butt:2019uul}.  The topological charge as a function of the $\Theta$ parameter on a $4 \times 4$ lattice.  Here $N_{gauge} = 5$ indicates that five states where kept in the $m$ numbers.  The $m$ values were allowed to run from $-2$ to 2 at each plaquette. That is to say that $\ds = 5$ on the $B$ tensor initially.}
  \label{fig:schwing-topo}
\end{figure}

Using these tensors, one can perform numerical calculations using a coarse graining scheme.  In  \cite{Butt:2019uul} the authors used the higher-order tensor renormalization group to calculate the free energy for the massless Schwinger model with and without the presence of a topological term.  From the free energy they calculated the average plaquette, and the topological charge both as a function of the gauge coupling and the $\Theta$ parameter.  The average plaquette and topological charge are given by,
\begin{align}
  \langle U_{p} \rangle = \frac{1}{V} \frac{\partial \ln Z}{\partial \beta}
\end{align}
and
\begin{align}
  \langle Q \rangle = - \frac{1}{V} \frac{\partial \ln Z}{\partial \Theta},
\end{align}
respectively.  These were compared with Monte Carlo calculations from  \cite{GOSCHL201763} when possible.
A figure from  \cite{Butt:2019uul} showing the average plaquette as a function of the the $\Theta$ parameter is shown in Fig.~\ref{fig:schwing-plaq-theta} for a $4\times4$ lattice.
In Fig.~\ref{fig:schwing-topo} we see a comparison between the tensor calculation and Monte Carlo for a fixed volume on a $4 \times 4$ lattice of the topological charge.
The authors reported difficulty at larger volumes, perhaps owing to how the coarse-graining scheme handles which states are kept before knowing the boundary conditions on the lattice.

\subsection{Additional topics and references}
\label{subsec:addfermions}

One of the big goals in lattice gauge theory is the successful simulation of four dimensional QCD at finite density.
However, the computational time of the Grassmann HOTRG in four dimensions is extremely demanding: $\propto \ds^{15}$ with the bond dimension $\ds$.
To achieve the goal, further improvements of the algorithm would be required, such as  Monte Carlo approximation/sampling of the tensors, and effective truncation of bonds.

From an application point of view, non-trivial models in three dimensions would be within the range.
The three dimensional Thirring model that has a non-trivial phase structure
and
$2+1$ dimensional domain-wall fermion systems
would be interesting targets.

As described in Sec.~\ref{sec:fermionresult_3dfermion}, the choice of the unitary matrices used in~\cite{Yoshimura:2017jpk} during coarse-graining is not optimal.
This leaves open the possibility to improve their results by tuning the unitary matrices.

In fermion systems, the spectra of tensors tend to have milder hierarchies than purely bosonic ones.
Indeed, in the works on the Schwinger model~\cite{Shimizu:2014uva,Shimizu:2014fsa,Shimizu:2017onf}, the bond dimension of tensors is taken to be $160$.
This is very large, so that one cannot easily reproduce their results on standard, say desktop or laptop, computers.
Then, even in two dimensions, serious calculations require improvements of the algorithms.
In two dimensions, there are several improved schemes for bosonic tensor networks such as (loop-)TNR~\cite{2015PhRvL.115r0405E,yang2017loop}, graph-independent local truncations~\cite{Hauru:2017tne}, and full environment truncation~\cite{2018PhRvB..98h5155E}.  Studies using these methods in the context of \tft include~\cite{Delcamp:2020hzo,refId0}.
Then Grassmann versions of them would all be possible directions.

Investigations of the Schwinger model with staggered fermions using the MPS formalism
have also been conducted, with and without a topological term. 
In  \cite{Banuls:2013jaa} the mass spectrum of the model is studied and   \cite{PhysRevLett.118.071601} studies the phase diagram of the Schwinger model with two flavors of fermion in the presence of a chemical potential. The authors investigate the isospin as a function of the chemical potential, and map the phase diagram in the chemical potential-mass plane.  
MPS were also used to study the ground state properties \cite{buyens2014} and confinement and string breaking \cite{buyens2016} of the one-flavor model. 
A more recent MPS study of the  Schwinger model with the inclusion of a $\Theta$ term was done in  \cite{PhysRevD.101.054507}.  The authors looked at different thermodynamic quantities as a function of the $\Theta$ parameter, as well as the spectrum of the model.  They also considered the continuum and chiral limits of the model where the $\Theta$ parameter becomes irrelevant.

\section{Transfer matrix and Hamiltonian}
\label{sec:transfer}

We now move on from the topic of reformulating the partition function in terms of tensors, to arranging their contractions so as to deduce a transfer matrix $\mathbb{T}$ which can be used to rewrite the partition function as in Eq.~\eqref{eq:tm}.  Once the partition function has been written entirely in terms of local tensor contractions, it is possible to organize these index contractions
into time layers. The natural choice is to use the
indices attached to time links and/or space-time plaquettes to be the indices of the transfer matrices. Geometrically, the Hilbert space is
located in between two time slices while the transfer matrix is centered on a time slice and connects two copies of the Hilbert space (see Fig. \ref{fig:electric-layer} for an illustration). To the best of our knowledge, interchanging the role of these two types of layers is only possible by going back to configuration space.

In the rest of this section, we target models with continuous Abelian symmetries (the O(2) spin model and $U(1)$ gauge theory) and describe their transfer matrices from a tensor perspective. However, it is not difficult to extend the discussion to other models.

\subsection{Spin models }
\begin{figure}[h]
  \centering
  \includegraphics[width=0.5\hsize]{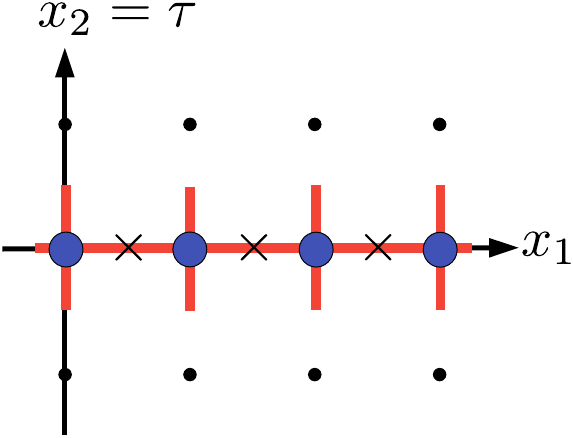}\\[1cm]
  \includegraphics[width=\hsize]{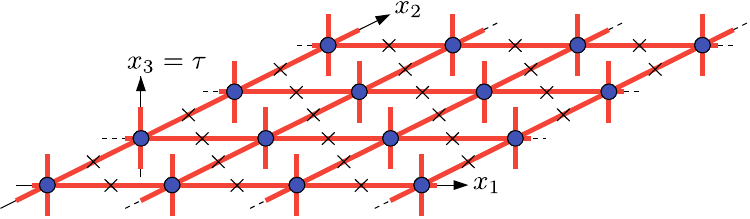}
  \caption{\label{fig:spintransfer}Illustration of the  transfer matrix for spin models in two  and three dimensions. The black crosses means index contraction.}
\end{figure}

For spin models \cite{Zou:2014rha}, the transfer matrix can be constructed by taking all the tensors on a time slice and tracing over the spatial indices. This is illustrated for $D=2$  and $D=3$ in Fig. \ref{fig:spintransfer}.

\begin{figure}[h]
  \includegraphics[width=0.75\columnwidth]{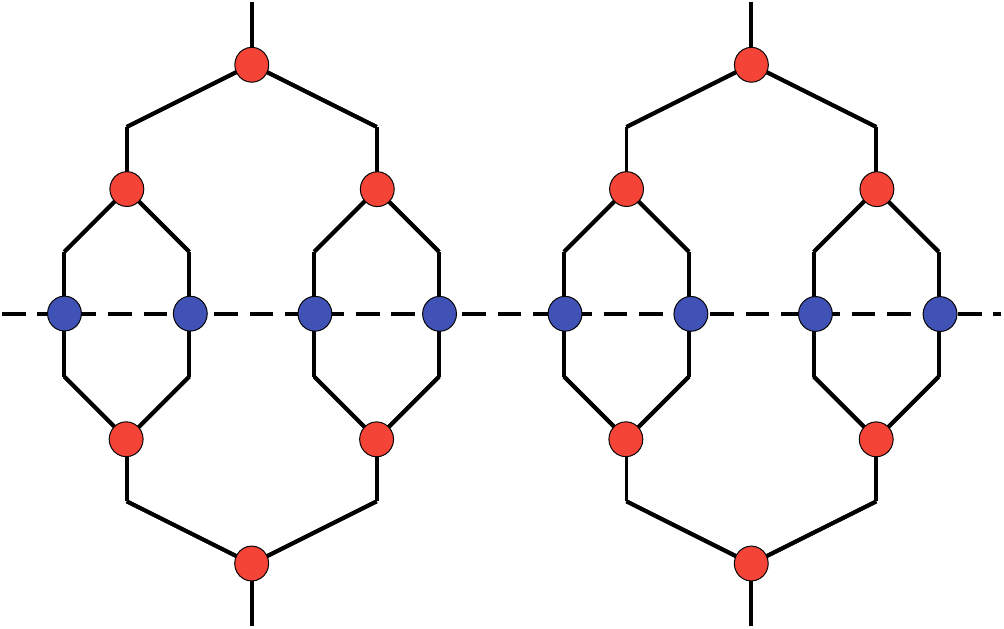}
  \caption{\label{fig:blocked} Graphical representation of the coarse-graining truncation of the transfer matrix described in the text.}
\end{figure}

For the O(2) model, the Hilbert space $\mathfrak{H}$ is the product of integer indices (see Sec.~\ref{subsec:o2}) attached to time links between two time slices.
\beq
\label{eq:hilbertO2}
\mathfrak{H}=\ket{\{ n \} }=\bigotimes_{{\bf x},j}\ket{ n_{{\bf x},j}}.
\enq
For $D=2$ with $N_s$ sites and periodic boundary conditions, the matrix elements of the transfer matrix $\mathbb{T}$  have the explicit form
\begin{align}
  \label{eq:tmspace}
  \nonumber
  \bra{\{n'\}} \mathbb{T} \ket{\{n\}} = \sum_{\bar{n}_{1} \bar{n}_{2} \ldots \bar{n}_{N_{s}}}
  &T^{(1,\tau)}_{\bar{n}_{N_{s}} \bar{n}_{1} n_{1} n'_{1}}
    T^{(2,\tau)}_{\bar{n}_{1} \bar{n}_{2} n_{2} n'_{2}} \\
  &\cdots
    T^{(N_{s},\tau)}_{\bar{n}_{N_{s}-1} \bar{n}_{N_{s}} n_{N_{s}} n'_{N_{s}}}
\end{align}
with the individual tensors provided in Sec.~\ref{sec:abelian}. The transfer matrix can be coarse-grained in the spatial dimension~\cite{Zou:2014rha,pre93} as illustrated in Fig.~\ref{fig:blocked}. This method was used to
perform numerical calculations in Refs.~\cite{Zou:2014rha,pre93,Zhang:2018ufj,prd98}.

The symmetries of the model are completely encoded in Kronecker deltas appearing in the
definition of the tensor~\cite{meurice2019}. This corresponds to a divergenceless condition and
with either periodic or open boundary conditions, the charges carried by the indices cannot flow out in the spatial directions. For the O(2) model,
the sum of the time indices going in the time slice equals the
sum of the indices going out. This conserved quantity can be identified as the charge of the initial and final states and the transfer matrix commutes with the charge
operator which counts the sum of the in or out indices. As we will explain in  Sec.~\ref{subsec:symmetries}, setting some matrix elements to zero if some of the local indices exceed some value, $n_{max}$, in absolute value will not affect this property.

The transfer matrix can be used to define a Hamiltonian by taking an anisotropic limit where $\beta$
becomes large on time links and small on space links \cite{fradkin78,kogut79}. We define
$\tilde{V}=1/(\beta_\tau a_\tau)$, $\tilde{\mu}=\mu/a_\tau$ and $J=\beta_s/a_\tau$. The Hamiltonian is defined
by
\begin{align}
  \mathbb{T} = \mathbb{1} - a_\tau \hat{\mathbb{H}} + \mathcal{O}(a_\tau ^2).
\end{align}
It will inherit the symmetry properties of the transfer matrix.
It explicit form is
\begin{align}
  \hat{\mathbb{H}} = \sum_{\mathbf{x}} \left[ \frac{\tilde{V}}{2} \hat{L}_{\bf x}^2-\tilde{\mu}  \hat{L}_{\bf x} - \frac{J}{2} \sum_{{\bf x},\mu}  (\hat{U}_{{\bf x}+\hat{\mu}}\hat{U}_{{\bf x}}^\dagger + {\rm h.c.}) \right]
  \label{eq:rotor}
\end{align}
with the operator $\hat{L} \ket{n} = n \ket{n}$, and the operator $\cre=\widehat{{\rm e}^{i\varphi}}$ which corresponds to the insertion of ${\rm e}^{i\varphi_x}$  in the path integral and raises the charge
\beq
\cre\ket{n}=\ket{n+1},
\enq
while its Hermitian conjugate lowers it
\beq
(\cre)^\dagger\ket{n}=\ket{n-1}.
\enq
This implies the commutation relations
\beq
[\hat{L},\cre]=\cre, \
[\hat{L},\cre^\dagger ]=-\cre^\dagger,
\label{eq:eu1}
\enq
and
\beq
[\cre,\cre^\dagger ]=0.
\label{eq:zero}
\enq
\def\nmax{n_{max}}

\subsection{Quantum simulations for the O(2) and O(3) model }
For the $O(2)$ nonlinear sigma model,  \cite{Zou:2014rha} uses a mapping between the $O(2)$ model, and the Bose-Hubbard model.  They relate the two phase diagrams in the hopping-chemical potential plane, and give the explicit mapping between variables between the two models.  A similar approach can be seen as the limiting behavior of the Abelian Higgs model in Refs.~\cite{Bazavov:2015kka,Zhang:2018ufj} when the gauge coupling is taken to zero.
In  \cite{pra96} the authors describe a method to measure the second-order R\'{e}nyi entropy for the $O(2)$ model with a chemical potential in the limit where it appears as the Bose-Hubbard model.  They do this in the case of an ultra-cold atomic species trapped in an optical lattice at half-filing.  They also consider the experimental cost to extract the central charge from measurements of the R\'{e}nyi entropy.

There are a couple of results in progress towards quantum simulation of the $O(3)$ nonlinear sigma model, but none on the principal chiral model at the time of writing.  In  \cite{Schutzhold:2005}, a proposal for an analogue quantum simulator for the $O(3)$ nonlinear sigma model in two dimensions is discussed.  The set-up involves an idealized circuit of superconducting and insulating spheres and wires.  The $\sigma$ field is identified with the position of a electron living on the surface of an insulating sphere.  The nearest neighbor potential is discrete in space and is identified with the difference in positions between adjacent electrons.  This is mapped to the spatial gradient of the $\sigma$ field.  These two identifications are used to match couplings between the circuit model and the original nonlinear sigma model.  Possible experimental parameters are discussed as well as an analysis of noise contributions to the simulation.

The reference \cite{PhysRevLett.123.090501} discusses an approach to quantum simulating the $O(3)$ nonlinear sigma model using ``digital'' quantum computers implementing qubits.  The original Hamiltonian is re-expressed in the angular momentum basis.  In this basis (as discussed above) a truncation is made which preserves the $O(3)$ symmetry of the model, but reduces the local state space to four states.  This is a natural truncation to the $l_{\text{max}} = 1$ state which possesses a singlet and triplet state coming from $l = 0$ and $l = 1$ states, respectively. This is precisely what one finds in the addition of angular momentum between the product of two spin-$\frac{1}{2}$ states.  In this truncation and representation the model is cast in terms of two-qubit operators.  Finally the authors use a Suzuki-Trotter decomposition to write the Hamiltonian evolution in short, discrete steps.  Each step is mapped to a quantum circuit over qubits.  The authors simulate the Hamiltonian evolution on a classical computer and discuss results.  They also perform runs on a quantum computer; however, at the time, they find ``mostly noise.''

\subsection{Gauge models}
\label{subsec:gaugehilbert}

\begin{figure}[h]
  \centering
  \includegraphics[width=0.9\hsize]{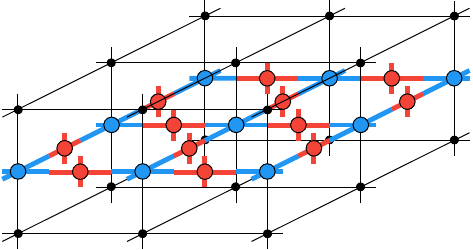} \\[1cm]
  \includegraphics[width=0.6\hsize]{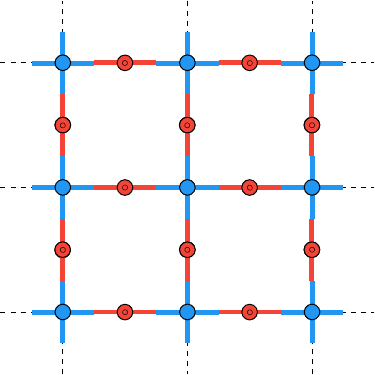}
  \caption{\label{fig:electric-layer}Electric layer of the transfer matrix for $D=3$ between two time slices (top) and ``from above" (bottom).}
\end{figure}

The Hilbert space for the compact Abelian Higgs model, or its pure gauge $U(1)$ limit (see Sec.~\ref{subsec:cahm}), $\mathfrak{H}_G$ can be constructed with the indices associated with space-time plaquettes (see Sec.~\ref{subsec:discrmax})
\beq
\label{eq:hilbertG}
\mathfrak{H}_G=\ket{\{ {\bf e} \} }=\bigotimes_{{\bf x},j}\ket{ e_{{\bf x},j}},
\enq
where the states $\ket{e_{\mathbf{x},j}}$ are eigenstates of $L_{\mathbf{x},j}$, defined below Eq.~\eqref{eq:rotor}.  Here we use $\hat{e}_{\mathbf{x},j}$ for the operator as well for clarity.
The electric layer is a diagonal matrix ${ \mathbb T}_E$ with
matrix elements
\beq
\bra{\{ {\bf e'} \} } { \mathbb T}_E\ket{\{ {\bf e} \} }=\delta_{\{ {\bf e} \} ,\{ {\bf e'} \} } T_E(\{ {\bf e} \}),
\enq
where  $T_E(\{ {\bf e} \})$ are traced products of $A$ tensors on time links with
$B$ tensors on space-time plaquettes
\beq
T_E(\{ {\bf e} \})=\Tr \prod_{\text{time } l.}A^{(l.)}_{m_1, \dots m_{2(D-1)}}\prod_{\text{sp.-time } pl.}B^{(pl.)} ({\bf e}).
\enq
The $A$ tensor of the compact Abelian Higgs model is given in Eq.~\eqref{eq:cahm-Atensor}. It enforces Gauss's law in the pure gauge limit. The electric layer is illustrated in Fig.~\ref{fig:electric-layer}.

\begin{figure}[h]
  \centering
  \includegraphics[width=0.9\hsize]{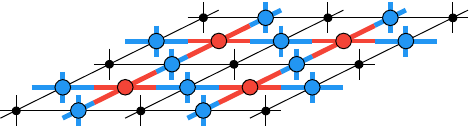} \\[1cm]
  \includegraphics[width=0.6\hsize]{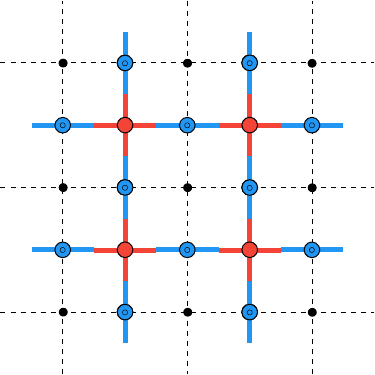}
  \caption{\label{fig:magnetic-layer}magnetic layer of the transfer matrix for $D=3$ in a time slice (top) and ``from above" (bottom).}
\end{figure}

Similarly, we define the magnetic  matrix elements
$\bra{\{ {\bf e} \} } { \mathbb T}_M\ket{\{ {\bf e'} \} }$ with the indices ${\bf e}$ and ${\bf e'}$ carried by the time legs of the $A$-tensors located on time links
\begin{eqnarray}
  &&\bra{\{ {\bf e'} \} } { \mathbb T}_M\ket{\{ {\bf e} \} }=\cr
  &&\Tr \prod_{\text{space } l.}A^{(l.)}_{m_1, \dots m_{2(D-1)}}({\bf e},{\bf e'})\prod_{\text{sp.-sp. } pl.}B^{(pl.)}.
\end{eqnarray}
The traces are taken over the spatial legs of the tensors, while the time legs are left open and carry the the indices ${\bf e}$ and ${\bf e'}$. The magnetic layer is illustrated in Fig.~\ref{fig:magnetic-layer}.

We define the transfer matrix ${ \mathbb T}$ as
\begin{eqnarray}
  \label{eq:transfer}
  { \mathbb T}&\equiv&(e^{-\bpl} I_0(\bpl))^{(V/N_\tau)D(D-1)/2}(e^{-\bl} I_0(\bl))^{(V/N_\tau)D}  \cr
  & &\times { \mathbb T}_E^{1/2}{ \mathbb T}_M{ \mathbb T}_E^{1/2},
\end{eqnarray}
with $N_\tau$ the number of sites in the temporal direction.

Proceeding as for the spin model \cite{fradkin78,kogut79}, we define
\beq
\beta _{\tau pl.} =\frac{1}{a_\tau g^2_{pl.}}, \ {\rm and}\  \beta _{\tau l.}=\frac{1}{a_\tau g^2_{l.}},\enq
for the couplings related to the time direction and
\beq
\beta _{s\  pl.} =a_\tau J_{pl.}, \ {\rm and}\  \beta _{s\  l.}=a_\tau h_{l.},\enq
for the spatial couplings. We then obtain \cite{judah-thesis,meurice2020}
\begin{eqnarray}
\label{eq:ham-cahm}
  \hat{\mathbb H}=&&\frac{1}{2}g^2_{pl.} \sum_{{\bf x},j}(\hat e_{{\bf x},j})^2\cr
  &+& \frac{1}{2}g^2_{l.}\sum_{\bf{x}}(\sum_{j}(\hat e_{{\bf x},j}-\hat e_{{\bf x}-\hat{j},j}))^2\cr
  &-& h_{l.} \sum_{{\bf x},j}(\hat{ U}_{{\bf x},j}+h.c.)\\
                  &-& J_{pl.} \sum_{{\bf x},j<k}(\hat{ U}_{{\bf x},j}\hat{ U}_{{\bf x}+\hat{j},k}\hat{ U}^\dagger _{{\bf x}+\hat{k},j}\hat{ U}^\dagger _{{\bf x},k}+h.c.).\cr\nonumber
\end{eqnarray}

Notice that for the compact Abelian Higgs model the matter fields can absorb non-zero values in Eq.~\eqref{eq:gausslaw} (Gauss's law). However, in the limit where
the link couplings are set to zero, we recover the pure gauge $U(1)$ model where the $A$-tensors for time links enforce Gauss's law.  In Eq.~\eqref{eq:ham-cahm}, if the couplings $h_{l.}$ and $g_{l.}$ are set to zero, we recover the Hamiltonian for $U(1)$ gauge theory in the Kogut-Susskind form.

So far everything we have done has been manifestly gauge-invariant because the tensors resulted from a complete integration over the gauge fields. The partition function remains unchanged if we use a temporal gauge \cite{meurice2020}.
If we gauge away the gauge fields on a time link instead of integrating over them, we lose the Gauss's law enforcement associated with that time link. However, the discrete Maxwell equations of Sec.~\ref{subsec:discrmax} imply that if Gauss's law is satisfied on one electric layer, then it is also satisfied on all the other layers.
With open boundary conditions in time, Gauss's law is trivially satisfied on the first and last layers. With periodic boundary conditions, we cannot gauge away the Polyakov loops and
we need to keep the integration over the temporal links for one layer. This is sufficient to enforce Gauss's law in that layer and consequently everywhere.

If we prepare an initial state which satisfies Gauss's law, the exact time evolution
will preserve this property. However, in the noisy intermediate-scale quantum (NISQ) era, various types of errors can introduce Gauss's law violations. For this reason, it has been argued~\cite{kaplan2018,Unmuth-Yockey:2018xak,meurice2020,juyinprogress,Bender:2020ztu} that it would be desirable to
find a parametrization of the Hilbert space where Gauss's law is automatically satisfied. One possibility
discussed next in Sec.~\ref{subsec:moredual} is to use the the unconstrained variables introduced in Sec.~\ref{subsec:gaugedual}. A simple solution~\cite{meurice2020} for the Hilbert space $\mathfrak{H}_G$
introduced in Eq.~\eqref{eq:hilbertG}, is to write the $e_{{\bf x},i}$ as the discrete divergence of antisymmetric tensors. For $D=3$, we only need one field instead of two and we obtain an optimal representation similar to
what was proposed in Refs.~\cite{kaplan2018,Unmuth-Yockey:2018xak}.
\begin{eqnarray}
  \nonumber
  e_{{\bf x},1}&=&-c_{\bf x}+c_{{\bf x}-\hat{2}} \\
  e_{{\bf x},2}&=&+c_{\bf x}-c_{{\bf x}-\hat{1}}
\end{eqnarray}
For $D=4$, we can write the electric field as the curl of a three component vector \cite{meurice2020}. As this new vector is
defined up to a gradient we can attempt to use this freedom to remove say the first component. This would provide an expression of the form
\begin{eqnarray}
  \nonumber
  e_{{\bf x},1}&=&-c_{{\bf x},3}+c_{{\bf x}-\hat{2},3}+c_{{\bf x},2}-c_{{\bf x}-\hat{3},2}\\
  e_{{\bf x},2}&=&+c_{{\bf x},3}-c_{{\bf x}-\hat{1},3}\\
  \nonumber
  e_{{\bf x},3}&=&-c_{{\bf x},2}+c_{{\bf x}-\hat{1},2}.
\end{eqnarray}
However, the global implementation depends on the boundary conditions \cite{meurice2020}.
A more recent discussion of Gauss's law for PBC and OBC can be found in Ref. \cite{Bender:2020ztu}.
In summary, it is possible to enforce  Gauss's law with no unphysical degrees of freedom that would waste computational resources. This can be done in any dimension and is better understood using the dual formulation discussed in the coming section.

\subsection{Duality revisited and Gauss's law}
\label{subsec:moredual}
The passage to the unconstrained variables discussed in the Lagrangian formalism  Sec.~\ref{subsec:gaugedual} solve Gauss's law in $D = 3$ and remove any gauge freedom from the model.
In the continuous-time limit, when the transfer matrix is close to the identity and one can identify a Hamiltonian, there is no residual gauge freedom, and in fact the model is recast as a spin model.

In $D=4$, the unconstrained variables which solve the divergence-less constraint in Abelian models are left with a redundancy themselves.  That is, there is a local operation which leaves the new Hamiltonian unchanged, and so the question of physical states remains. This arises from,
\begin{align}
  m_{x, \mu \nu} = \epsilon_{\mu \nu \rho \sigma} \Delta_{\rho} C_{x^{*}, \sigma}
\end{align}
which introduces a new ``gauge field'' on the links of the dual lattice.  The field strength tensor for this gauge field possess a similar redundancy to that of the original field, \emph{i.e.} $C_{x, \mu} \rightarrow C'_{x, \mu} = C_{x, \mu} + \Delta_{\mu} \phi_{x}$ leaves the quantum Hamiltonian unchanged.  In the electric basis (the $L^{z}$ basis) this symmetry is manifested in an operator which raises and lowers all angular momentum numbers around a site by one, $G_{x} = \prod_{\mu = 1}^{4} U_{x,\mu}^{+}U_{x-\hat{\mu}, \mu}^{-}$.  This operator commutes with the Hamiltonian.  This identifies physical states as those which do not differ from others by arbitrary applications of $G_{x}$.

\subsection{Quantum simulation of the Abelian Higgs model with cold atoms}
\label{sec:qs-ahm}
There are a few concrete proposals on how to simulate the Abelian Higgs model using cold atoms trapped in an optical lattice.  These methods either make use of the similarity between multi-species Bose-Hubbard models and the Abelian Higgs model, or create an effective model only in terms of gauge degrees of freedom and construct the local Hilbert space of the model directly as a physical dimension and include operators for the new dimension.

In  \cite{Bazavov:2015kka} a two-species Bose-Hubbard model is proposed to simulate the Abelian Higgs model in the limit of infinite Higgs mass in 1+1 dimensions.  The authors use the Fourier expansion for the Abelian fields and rewrite the model in terms of discrete variables.  Then, the matter degrees of freedom are integrated out creating an effective theory only in terms of the discrete gauge-field degrees of freedom.  Finally, the authors make use the mapping between Schwinger Bosons and $SU(2)$ angular momentum operators to write the model in terms of Bosonic creation/annihilation operators and number operators whose form matches that of a two-species Bose-Hubbard model.  The authors compare the energy spectra between the original model and the Bose-Hubbard system and find good agreement.

In  \cite{Gonz_lez_Cuadra_2017} another multi-species Bose-Hubbard model is proposed to simulate the Abelian Higgs model in 2+1 dimensions.  The authors use a six-species Bose-Hubbard Hamiltonaian, and again find a mapping between their electric field and parallel transport operators and Bosonic creation/annihilation and number operators.  They create the plaquette interaction as a higher-order, perturbative, effective corrections to the original Hamiltonian, and report on possible observables which could be seen in the laboratory experiment.

\begin{figure}[t]
  \centering
  \includegraphics[width=8.6cm]{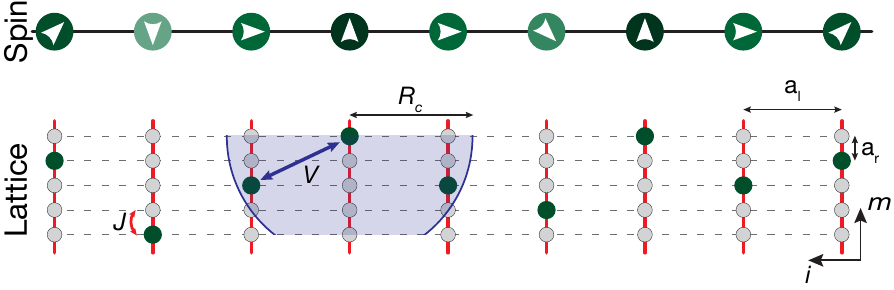}
  \caption{A ladder set-up in an optical lattice for the Abelian-Higgs model in 1+1 dimensions.  Each vertical rung is a single spatial site, and contains only a single atom, whose location along the rung indicates the angular momentum quantum number there.  The nearest neighbor interaction is mapped to the dressed-Rydberg potential $V$ between atoms. From Ref. \cite{Zhang:2018ufj}.
  \label{fig:ahm-ladder}}
\end{figure}

Reference~\cite{Zhang:2018ufj} takes a different approach than the previous two.  Instead of attempting to capture the local Hilbert space with one from a composite multi-species Bose-Hubbard model, the discrete angular momentum quantum numbers associated with the electric field numbers are represented as new physical locations on a higher-dimensional lattice.  For the 1+1 dimensional Abelian Higgs model, a ladder is constructed where one of the lattice directions represents the spatial dimension, and the other direction, the rungs of the ladder, are the different possible angular momentum states.  There is then much freedom of which atomic species to populate the lattice with.  The authors use a dressed-Rydberg potential to describe the two-body interactions.  The authors propose to measure the Polyakov loop and give a prescription of how to do it.  A figure of the lattice set-up in the case of a five-state truncation is shown in Fig.~\ref{fig:ahm-ladder}

\subsection{Algebraic aspects of the Hamiltonian formulation}
\label{subsec:algebra}

As practical implementations require a finite number of states, we need to
discuss the effect of a truncation on the algebra defined by Eqs. (\ref{eq:eu1}) and (\ref{eq:zero}).
By truncation we mean that there exists some $\nmax$ for which
\beq
\label{eq:res}
\cre\ket{\nmax}=0, {\rm and}\
(\cre)^\dagger\ket{-\nmax}=0.
\enq
It is clear that these modifications contradict Eq.~\eqref{eq:zero} because if
$\cre$ and $\cre^\dagger$ commute, we can apply $\cre^\dagger$ on the first equation
(\ref{eq:res}) and obtain that $\cre\ket{\nmax -1}$ is also zero and so on.
If we consider the commutation relations with the restriction (\ref{eq:res}), we see that the only changes are
\begin{eqnarray}
  &\ &\bra{\nmax}[\cre,\cre^\dagger ]\ket{\nmax}=1,\\ \nonumber
  &\ &\bra{-\nmax}[\cre,\cre^\dagger ]\ket{-\nmax}=-1,\\ \nonumber
\end{eqnarray}
instead of 0. The important point is that the truncation does not affect the basic expression of the symmetry in Eq.~\eqref{eq:zero}.
It only affects matrix elements involving the $\cre$ operators but not in a way that contradicts charge conservation. For a related discussion of the
algebra for the O(3) model see Ref. \cite{bruckmann2018}.

Other deformations of the original Hamiltonian algebra defined by Eqs. (\ref{eq:eu1}) and (\ref{eq:zero})appear in the quantum link formulation of
lattice gauge theories \cite{qlink2}. In this approach, one picks a representation of
the $SU(2)$ algebra and replace $\cre$ by the raising operator $S^+$. Eq.~\eqref{eq:res} is then satisfied if the dimension of the representation is $2n_{max} +1$ but Eq.~\eqref{eq:zero} becomes
\beq
[\hat{S}^+, \hat{S}^-]=2 \hat{S}^z.
\enq

Finally, we would like to comment about algebraic aspects of the Gaussian quadrature discussed in Sec. \ref{sec:rphi4tnrep}. This numerical integration method averages over a finite number of sampling points which are the zeros of a Hermite polynomial of sufficiently large order $n_{max}+1$. This can be related to a truncation of the standard harmonic oscillator algebra in the following way. If we use the standard raising and lowering operators on energy eigenstates $\ket{n}$ to calculate $\bra{x}\hat{x}\ket{n}$, we recover the
Hermite polynomial recursion formula. These relations still hold for the
zeros of $H_{n_{max}+1}$ until we reach level $n_{max}$. Iterating one more time provides a relation equivalent to
\beq
\hat{a}^\dagger \ket{n_{max}}=0.\enq
The modified commutation relation become
\beq[\hat{a}, \hat{a}^\dagger]=\mathbb{1}-(n_{max}+1)\ket{n_{max}}\bra{n_{max}}.\enq
A better algebraic understanding of the results of Sec. \ref{sec:rphi4tnrep}
would certainly be of great interest.

\subsection{Additional topics and references}
\label{subsec:additional}

In Sec.~\ref{subsec:tn} tensor network studies in the Hamiltonian limit in 1+1 dimensions were reviewed.  For 2+1, and 3+1 dimensions,
in the Hamiltonian limit, there are relevant studies  using projected entangled pair states, and tree tensor networks.  \cite{Felser:2019xyv} considered 2+1-dimensional electrodynamics at finite density.  Electrodynamics in 3+1 dimensions, again at finite density, was studied using tree-tensor networks in~\cite{Magnifico:2020bqt}. Alternatively, $\mathbb{Z}_{3}$ gauge theory was studied using PEPS in~\cite{robaina2020}, and Abelian $U(1)$ and non-Abelian $SU(2)$ gauge theories were considered in~\cite{Zohar:2015eda,Zohar:2016wcf}.  An entanglement renormalization group approach to 2+1 and 3+1 dimensional gauge theories was explored in~\cite{tag2011}.  Also in the Hamiltonian limit, the idea of a hybrid algorithm between tensor networks and Monte Carlo is discussed in~\cite{Zohar:2017yxl,Emonts:2020drm}. Expansions in representations of continuous groups 
\cite{zohar2015b,celi2014} leading to figures related to Figs. \ref{fig:electric-layer} and \ref{fig:magnetic-layer} can be found in the tensor network literature. For very recent work on $SU(3)$ see \cite{Ciavarella:2021nmj}.

An examination of the the entanglement area-law and how it appears in the PEPS construction, as well as the relationship between PEPS and thermal states of local spin systems was carried out in~\cite{Verstraete:2006mgt}. 
The inherent redundancy in the PEPS construction, and its relationship to symmetry has been discussed in \emph{e.g.}~\cite{2018NJPh...20k3017M}, and furthermore gauging a PEPS with a global symmetry to construct a PEPS with local symmetry was studied in~\cite{haegeman2015,Zohar:2015jnb}.

\section{Additional aspects}
\label{sec:additional}
\subsection{Symmetries and truncations}
\label{subsec:symmetries}
As explained in the Introduction, the implementation of field theory calculations with quantum computers requires discretizations and truncations of the problems considered.
As symmetries play a crucial role in most of these calculations, we need to understand the
effects of discretization on the realization of the original symmetries.
The effects of the discretization of space-time are well understood and
the remaining discrete symmetries---discrete translations and rotations---are used consistently by lattice practitioners.
On the other hand, the fate of internal continuous symmetries in reformulations involving discrete
character expansions and truncations is a more complicated question. We report here recent progress on this question that have a great deal of generality and apply to global, local, continuous and discrete symmetries \cite{meurice2019,meurice2020}.

We consider generic symmetries for a generic lattice model with action $S[\Phi]$
where $\Phi$ denotes a field configuration of fields $\phi_\ell$
attached to locations $\ell$  which can be sites, links or higher dimensional objects.
The partition function reads
\beq
Z=\int {\mathcal D}\Phi {\rm e}^{-S[\Phi]},
\enq
with ${\mathcal D}\Phi$ the measure of integration over the fields.
We define
expectation values of a function of the fields $f$ as
\beq
\langle f(\Phi) \rangle = \frac{1}{Z} \int {\mathcal D}\Phi f(\Phi){\rm e}^{-S[\Phi]} .
\label{eq:sym}
\enq
A symmetry is defined as a field transformation
\beq
\phi_\ell\rightarrow\phi_\ell'= \phi_\ell + \delta \phi_\ell
\enq such that
the action and the integration measure are preserved.
These invariances imply that
\beq
\label{eq:symid}
\langle f(\Phi) \rangle=\langle f(\Phi +\delta\Phi) \rangle.\enq
If the action is not exactly invariant, $\exp(\delta S)$ gets inserted in the expectation value on the right-hand side of the equation.

The O(2) model discussed in Sec.~\ref{subsec:o2}
is invariant under the global shift
\beq
\varphi_x'=\varphi_x+\alpha.
\enq
Assuming that the function $f$ is  $2\pi$-periodic in its $M$ variables, we expand in Fourier modes and after using Eq.~\eqref{eq:symid}, we obtain

\beq
\label{eq:selection}{\rm If}\ \sum_{i=1}^M n_i \neq 0, \  \ {\rm then} \   \   \langle  {\rm e}^{(i(n_1\varphi_{{x}_1}+\dots +n_M\varphi_{{x}_M}))}\rangle=0.\enq

\begin{figure}[h]
  \centering
  \includegraphics[width=5cm]{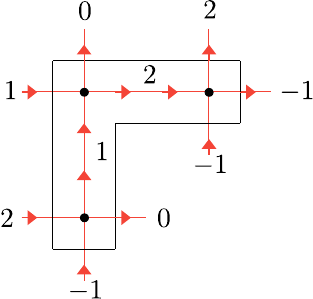}
  \caption{\label{fig:glocons}Example of flux cancellations in $D=2$. The total flux in and out  the upside-down L-shaped domain is +1. }
\end{figure}

This global selection rule can be explained  \cite{meurice2019} in terms of the selection rule of the microscopic tensors at
each site given in Eq.~\eqref{eq:noether}. It is a divergenceless condition and it can be interpreted as a discrete version of Noether's theorem.
If we enclose a site $x$ in a small size (compared to the lattice spacing) $D$-dimensional cube, the sum of indices corresponding to positive directions ($n_{x,\text{out}}$) is the same as the
sum of indices corresponding to negative directions ($n_{x,\text{in}}$).
For instance, in two dimensions the sum of the left and bottom indices equals the sum of the right and top indices.
By assembling such elementary objects (tracing over indices corresponding to their interface) we can construct an arbitrary domain. Each tracing automatically cancels an ``in" index with an ``out" index and consequently, at  the boundary of the domain, the sum of the ``in" indices  remains the same as the sum of the ``out" indices.
This discrete version of Gauss's theorem is illustrated in $D=2$ in Fig.~\ref{fig:glocons}.
We can pursue this process until we reach the boundary. For PBC, the ``in'' and ``out'' cancel and
for OBC, all the indices at the boundary are zero. In both cases, the system is ``isolated'' in the sense that no flux escapes to or comes from the environment. If we now remove one site out of the entire domain, the point-wise conservation inside the rest of the domain implies that the
indices connecting to the missing site satisfy the divergenceless condition independently. This has a very simple interpretation in terms of global symmetry of the model \cite{meurice2020}. We can use the symmetry to fix the value of $\phi$ at the missing site instead of integrating over the possible values because as we just explained, the divergenceless condition resulting from this integration is redundant.

We can now understand the global selection rule of Eq.~\eqref{eq:selection}.
The insertion of various ${\rm e}^{in_Q\varphi_x}$  is equivalent to inserting an ``impure" tensor which differs from the ``pure" tensor by the Kronecker symbol replacement
$\delta_{n_{x,\text{out}},n_{x,\text{in}}} \rightarrow \delta_{n_{x,\text{out}},n_{x,\text{in}}+n_Q}.$
Proceeding as before for PBC or OBC, this implies that the sum of the charges should be zero.

To summarize,  the global selection rule  is a consequence of the selection rule at each site
which is the Kronecker delta in the expression of the tensors.
It is independent of the particular values taken by the tensors (like Bessel functions). So if we set some of the tensor elements to zero as we do in a truncation, this does not affect the
global selection rule and truncation are compatible with symmetries \cite{meurice2019}.

\begin{figure}[h]
  \centering
  \includegraphics[width=6cm]{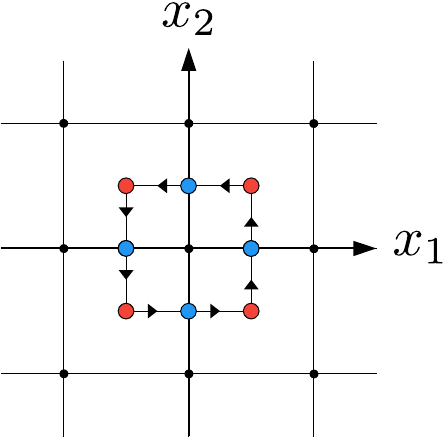}
  \caption{ \label{fig:red}Illustration that one divergenceless condition  is redundant for $D$= 2. }
\end{figure}

The reasoning can be extended to local symmetries \cite{meurice2020}. For the CAHM, the divergenceless condition for the $n_{x,\mu}$ is redundant with the selection rule coming from the integration over the gauge fields as expressed in Eq.~\eqref{eq:discrmax2}. This means that we can eliminate the $\varphi$ field with the unitary gauge.
It  was shown \cite{meurice2019} that in the pure gauge limit the set of equations~\eqref{eq:discrmax} are not independent.
If we pick a site, we can construct an in-out partition for the legs attached to links coming out of this site.  The sum of  ``in'' indices is the same as the sum of the ``out" indices, and if we assemble them on the boundary of a $D$-dimensional cube, one of the divergenceless conditions follows from the other $2D-1$ conditions. This is illustrated for $D=2$  in
Fig.~\ref{fig:red} where three of the delta functions on the $A$-tensors attached to the links imply the fourth one.

The redundancy argument extends to discrete \zq subgroups of $U(1)$ where the divergenceless condition is expressed modulo $q$ and the infinite set of Bessel functions are
replaced by the $q$ discrete ones.
We conclude that Noether's theorem can be expressed in the tensor formulation context as: for each symmetry, there is a corresponding tensor redundancy. This applies to global, local, continuous and discrete Abelian symmetries \cite{meurice2020}.

\subsection{Topological considerations}

In classical field theory, the boundary conditions play an important role in the investigation of topological solutions. As a simple example, if an angle variable $\varphi$ satisfies Laplace equation in
$D=1$, then PBC allow the existence of solutions with any winding number. On the other hand, for arbitrary Dirichlet boundary conditions, the concept of winding number is not applicable because the one-dimensional interval does not have the topology of a circle.

In the $D=1$ $O(2)$ model, we observe features which are reminiscent of this observation.
For PBC, we can assemble tensors with any index $n$,
\begin{align}
  T_{n n'} = \delta_{n n'} \sqrt{I_{n}(\beta) I_{n'}(\beta)}
\end{align}
and
\beq
\label{eq:zpbc}
Z_{\text{PBC}} = \Tr[T^{N_{\tau}}] = \sum_{n=-\infty}^{\infty}I_n(\beta)^{N_\tau}.
\enq
On the other hand for OBC, we have a zero index at the ends and
\beq
Z_{\text{OBC}} = I_0(\beta)^{N_\tau -1}.
\enq
It is often believed that the contributions for various $n$ in $Z_{\text{PBC}}$ correspond to the
topological sectors of the classical equations of motion which become Laplace's equation in the continuum limit (so for large $\beta$). This is not correct because the $I_n(\beta)$ differ
from $I_0$ by corrections of $-n^2/(2\beta)$ as shown in Eq.~\eqref{eq:tn}, while in the semi-classical solution, one expects suppression of the form $\exp(-\beta n^2 C)$ for some calculable constant $C$. However, the two types of behaviors are swapped after the Poisson summation
\beq
\sum_{\ell =-\infty}^{\infty} e^{-\frac{B}{2}\ell^2}=\sqrt{\frac{2\pi}{B}}\sum_{n =-\infty}^{\infty} e^{-\frac{(2\pi)^2}{2B}n^2},
\enq
A detailed analysis of the classical solutions \cite{meurice2020} shows that $B=\beta (2\pi)^2/N_\tau$ and that summation over the winding numbers using Poisson summation and the calculation of the quadratic fluctuations reproduce precisely the leading behavior of Eq.~(\ref{eq:zpbc}) in the large $\beta$ limit. Similar observations were made  in Ref. \cite{akerlund2015} for a version of the O(2) model where the fluctuations are limited.

Similar results were obtained for the $D=2$ pure gauge $U(1)$ model. In these calculations,
the possibility of fixing the values of variables that lead to redundant selection rules to arbitrary values as discussed in  Sec.~\ref{subsec:symmetries} removes the zero modes from the quadratic fluctuation calculations. Note also that it is possible to construct models where the large $\beta$ approximations are exact. The questions of topological
configurations and duality are discussed for Abelian gauge models of this type in various dimensions in Refs. \cite{banks77,savit77,GATTRINGER2018435,sul2019}

\subsection{Quantum gravity}
Tensor networks have also found a use in the study of quantum gravity.  One of the directions where tensor networks have appeared is in the ``spin-foam'' formulation of quantum gravity \cite{perez:2013}.  In  \cite{Dittrich_2016}, a tensor network formulation is developed along with a coarse-graining scheme where the intention is to use it on a spin-foam partition function, although it is applicable in other situations.  The algorithm is sketched out, and some numerical results for the two, and three-dimensional Ising model are presented.  In  \cite{Asaduzzaman:2019mtx} the authors start with the partition function for two-dimensional gravity where the gauge symmetry has been extended to unify the tetrad and spin-connection variables into a single connection.  They present a tensor formulation and study the zeros of the partition function (Fisher's zeros) in the complex-coupling plane.

\section{Conclusions}

In summary, \tft provides new ways to approach models studied by lattice gauge theorists. For models with compact field
variables, character expansions and orthogonality relations provide ways to
do the difficult integrals exactly and replace them by discrete sums. For continuous field variables, the sums are infinite and need to be truncated for practical implementations. These truncations preserve global and local symmetries.

We showed that by combining tensor blocking and truncations, we can obtain  coarse-grained versions of the original model where the new ``effective" tensors are assembled in the same way as the original tensors while taking different values. The TRG flows in the space of tensors replace the RG flows in the space of
effective interactions.
The effective tensors remain local in the coarse-grained system of coordinates. When the Euclidean action is real, TRG calculations can be compared with accurate results obtained using importance sampling in the original Lagrangian formulation. Tensor sampling could also be conducted by generalizations of the worm algorithm.
TRG calculations can be extended to the case of complex actions and evade sign problems.

It is important to realize that if a reasonable control of the truncations can be reached
for values of $D_{\text{cut}}$ that are achievable with current computers, the computation
cost scales logarithmically with the volume of the system which is exceptionally efficient.
The scaling of the cost with the dimension can be seen as an obstacle, however recent
progress in four dimensions provide an optimistic outlook.

The continuum limits of lattice models can be constructed in the vicinity of
RG fixed points. Universal quantities such as the critical exponents can be
extracted by linearizing the RG transformation near the fixed point.
The naive approach for this program has to be amended due to the existence of unphysical fixed points. This requires a detailed understanding of the UV and IR entanglement.

\tft allows for the construction of transfer matrices and smoothly connects the classical Lagrangian approach at Euclidean time to the Hamiltonian approach. The discreteness of the reformulation combined with truncations
provides approximate Hamiltonians which are suitable for quantum simulation experiments and quantum computations.
\tft is a natural tool to design quantum circuits.
In the NISQ era, benchmarking is crucial to assess the progress in this direction.
Hybrid formalisms combining real and imaginary time can be accommodated easily by \tft and may play an important role in the near future.

We think that the \tft program is making good progress towards the long term goal of performing QCD calculations. We expect that it will play an important role in developing practical methods to approach nuclear matter, jet physics and fragmentation. Active collaborations
between the lattice gauge theory and
condensed matter communities seem essential to achieving these goals and will hopefully
provide benefits on both sides.

\begin{acknowledgments}
  Y. M. thanks the late D. Speiser, J. Weyers and C. Itzykson for their teaching on group theory, Pontryagin duality, Peter-Weyl theorem and strong coupling expansions.
  We thank M. C. Ba\~nuls, J. Bloch, S. Catterall, M. Hite, E. Gustafson, Z. Hang, R. Maxton, D. Simons, Z. Y. Xie, J. Zeiher, J. Zhang, H. Zou, and members of the QuLAT collaboration, for useful discussions and comments.
  This work was supported in part by the U.S. Department of Energy (DOE) under Award Numbers DE-SC0010113, and DE-SC0019139.
  This manuscript has been authored by Fermi Research Alliance, LLC under Contract No. DE-AC02-07CH11359 with the U.S. Department of Energy, Office of Science, Office of High Energy Physics.
\end{acknowledgments}

\appendix

\section{Review of mathematical results}
\label{sec:review}
\subsection{Character expansions}
\label{subsec:character-expansion}
One of the most important relations we use in this review is the change of basis called the character expansion.  In the cases considered here, this relates a compact variable (discrete and bounded, or continuous and bounded) to their Pontryagin dual.  The relevant relation can be written,
\begin{align}
  f(x_{\alpha}) = \sum_{k_{\alpha}} \lambda_{k_{\alpha}} \chi^{k_{\alpha}}(x_{\alpha}).
\end{align}
Here $x_{\alpha}$ is the compact variable under consideration (it can be a matrix, or a spin, for example), and $f$ must be a ``class function'', a function that only depends on the trace of the compact variables.  $k_{\alpha}$ is the dual variable, which takes on the values of the irreducible representations of the group that $x_{\alpha}$ belongs.  Finally, $\chi^{k_{\alpha}}$ are the characters of the group, which are complete and orthogonal.  In practice we use the following,
\begin{align}
  \label{eq:all-charac-expansion}
  e^{\beta \sigma} &= \sum_{n = 0}^{1} \lambda_{n}(\beta) \sigma^{n}, \\
  e^{\beta \cos\theta} &= \sum_{n = -\infty}^{\infty} I_{n}(\beta) e^{i n \theta}, \\
  e^{\beta \Tr[U]} &= \sum_{r=0}^{\infty} F_{r}(\beta) \chi^{r}(U),
\end{align}
for the groups $\mathbb{Z}_{2}$, $U(1)$, and $SU(2)$, where $\lambda_{n}$, $I_{n}$, and $F_{r}$ are the expansion coefficients, and $\sigma^{n}$, $e^{i n \theta}$, and $\chi^{r}(U)$ are the characters of the respective groups.  For $Z_{2}$ the expansion coefficients are given by $\lambda_{0} = \cosh\beta$ and $\lambda_{1} = \sinh\beta$.  For $U(1)$ the coefficients are given by the modified Bessel functions.  For $SU(2)$ the coefficients are given by~\cite{itzykson1991statistical}
\begin{align}
    F_{r}(\beta) = I_{2r}(\beta) - I_{2r+2}(\beta)
\end{align}
where $I_{n}$ is again the modified Bessel function of order $n$.

\subsection{Orthogonality \& completeness}
\label{subsec:ortho}
As mentioned above, the expansions in Eq.~\eqref{eq:all-charac-expansion} are examples of the completeness and orthogonality of the characters.  In each of those cases we have,
\begin{align}
  \label{eq:orthogonality}
  \frac{1}{2}\sum_{\sigma} \sigma^{n} \sigma^{m} &= \delta_{n,m}^{(2)} \\
  \int_{-\pi}^{\pi} \frac{d \theta}{2\pi} e^{i n \theta} e^{-i m \theta} &= \delta_{n, m} \\
  \int dU \chi^{r}(U) \chi^{r'}(U) &= (2r+1)^{-1} \delta_{r, r'},
\end{align}
where $\delta^{(2)}$ is a Kronecker delta with the equivalency taken modulo two.  We also use the orthogonality of the matrix representations of group elements under the Haar measure,
\begin{align}
  \int dU D^{r}_{m n}(U) {D^{*}}^{r'}_{m' n'}(U) = (2r+1)^{-1} \delta_{r, r'} \delta_{m, m'} \delta_{n, n'}
\end{align}
with $*$ as complex conjugation \emph{without transposition}.  The $D$-matrices are related to the characters through the trace, $\chi^{r}(U) = \Tr[D^{r}(U)]$.

On the other hand these characters obey a completeness relation.  This is given by the sum over the representations, rather than the original group variables,
\begin{align}
  \frac{1}{2} \sum_{n=0}^{1} \sigma^{n} {\sigma'}^{n} &= \delta_{\sigma, \sigma'} \\
  \sum_{n = -\infty}^{\infty} e^{i n (\theta - \theta')} &= \delta(\theta - \theta') \\
  \sum_{r = 0}^{\infty}(2r+1) \ \Tr( D^{r}(U) D^{r\dagger}(U')) &=  \delta(U,U'). \\
  \sum_{r = 0}^{\infty}(2r+1) \chi^{r}(UU'^{-1})  &=  \delta(U,U'). \\
\end{align}

\subsection{Singular value decomposition}
\label{sec:svd-review}

The singular value decomposition (SVD) plays a central role in many different parts of this review, and when dealing with compact variables it is completely determined by the character expansion.  Here we give bare working knowledge of how it will be used.
One of the basic ideas is to regard Boltzmann factors $ f\left( x_{\alpha}, x_{\beta} \right)$, which appear in the partition function,  as $K \times K$ matrices when $x_{\alpha}$ and $x_{\beta} $ take $K$ values, and with that interpretation perform the SVD:
\begin{align}
  \label{eq:svd-review}
  f\left( x_{\alpha}, x_{\beta} \right)
  = \sum_{j=1}^{K} U_{x_{\alpha} j} \lambda_{j} V^{\dagger}_{j x_{\beta}},
\end{align}
where $\left\{ \lambda \right\}$ are the singular values that are assumed to be in descending order ($\lambda_{1} \ge \lambda_{2} \ge \cdots \ge \lambda_{K} \ge 0$),
and $U$ and $V$ are unitary matrices.  Of course this decomposition can be done for any matrix, and they need not immediately have the interpretation of a Boltzmann factor.

When the $x_\alpha$ can be identified with the elements of an additive group and if the matrix elements depends only on $x_\alpha - x_{\beta}$, the $U$ and $V$ matrices are square matrices which can be expressed in terms of the characters discussed in  Sec.~\ref{subsec:dualities}. For the \zq group, we have $K=q$,
\beq
U_{x_{\alpha} j}=\exp \left( i\frac{2\pi}{q} jx_\alpha \right),
\enq and
\beq V_{x_{\beta} j}=\exp \left( i\frac{2\pi}{q} jx_\beta \right).
\enq

Another case which occurs often is to use the SVD to split a higher-dimensional array in two.  Consider a generic tensor $A_{ijkl}$ with dimensions $(\ds, \ds, \ds, \ds)$.  Suppose one wanted to somehow factorize this tensor into two smaller tensors, one with indices $i$, $j$, and the other with indices $k$, $l$.  Then one can do the following,
\begin{align}
  A_{ijkl} &\rightarrow A_{(i \otimes j)(k \otimes l)} \\
           &= A_{I J} = \sum_{M, M' = 1}^{\ds^2} U_{I M} \lambda_{M} \delta _{M M'} V^{\dagger}_{M' J} \\
           &= \sum_{M = 1}^{\ds^2} (U_{I M}\sqrt{\lambda}_{M}) (\sqrt{\lambda}_{M} V^{\dagger}_{M J}) \\
           &= \sum_{M=1}^{\ds^{2}} B_{i j M} C_{M k l}
\end{align}
which is the split we were looking for.  This allows any tensor to be split exactly into smaller tensors, with the sum over intermediate states as the price.

%

\end{document}